\documentclass[12pt,english,a4paper, 
twoside,BCOR=12mm,abstract=off]{scrreprt} 
\usepackage[english]{babel}
\usepackage{microtype}  		%allegedly makes blocksatz better
\usepackage{graphicx}			%allows pictures
\usepackage{amsmath}			%contains mathmatical environments
\usepackage{mathtools}			%adds some nice math stuff, 
\usepackage[ntheorem]{empheq}	%better endmark placement
\usepackage{bbm}				%needed for ID, maybe load after amsfonts
\usepackage{amsfonts} 			%gives mathmatical symbols
\usepackage{yfonts}				%adds new fonts 
\usepackage{cjhebrew}			%adds Hebrew characters 
\usepackage[thref,amsmath,thmmarks]{ntheorem}			
\usepackage{amsrefs}			%necessary for bibliography
\usepackage[shortlabels]{enumitem}
\usepackage{accents}			%for underlining... 
\usepackage[final]{draftwatermark}
\usepackage{chngcntr}			%needed to let footnote counter 
\newcommand{\abs}[1]{\lvert#1\rvert}	%absolute value, e.g. of complex z
\newcommand{\norm}[1]{\lVert#1\rVert}	%norm, double bar
\newcommand{\set}[2]{\left\lbrace #1 \middle\vert #2 \right\rbrace}
\newcommand{\Lieb}[2]{\left[ #1 , #2 \right]}	
\newcommand{\partd}[2]{\frac{\partial #1}{\partial #2}}	

\renewcommand{\d}{\operatorname{d}\negthinspace}
\renewcommand{\abs}[1]{\left\lvert #1 \right\rvert}
\DeclareMathOperator{\SpanO}{span}
\newcommand{\Span}[1]{\SpanO \left\lbrace #1 \right\rbrace}
\newcommand{\tp}{\otimes}				%tensor product symbol
\newcommand{\cross}{\times}				%cartesian and cross product 
\newcommand{\R}{\mathbb{R}}				%real numbers R
\newcommand{\N}{\mathbb{N}}				%natural numbers
\newcommand{\spti}{\mathcal{Q}}				%spacetime manifold
\DeclareMathOperator{\Frameb}{Fr}
\DeclareMathOperator{\OFrameb}{OFr}
\newcommand{\frameb}[1]{\Frameb \left( #1 \right)}	
\newcommand{\oframeb}[1]{\OFrameb \left( #1 \right)}	
\newcommand{\Id}{\mathbbm 1}			%Identity
\DeclareMathSymbol{\varnothing}{\mathord}{AMSb}{"3F}
\renewcommand{\emptyset}{\varnothing}	%use that for command \emptyset
\DeclareMathSymbol{\upharpoonright} {\mathrel}{AMSa}{"16}

\newcommand{\ubar}[1]{\underaccent{\bar}{#1}} 
\DeclareMathOperator{\dom}{dom}

\DeclareMathOperator{\pr}{pr}

\DeclareMathOperator{\Ric}{R}			%ricci tensor field

\DeclareMathOperator{\CapitalT}{T}		%transpose, tangent bundles
\newcommand{\CapT}{\CapitalT{} \negthinspace} 	
\newcommand{\transp}{^{\CapT}}
\DeclareMathOperator{\Kon}{K}			%connector map
\DeclareMathOperator{\Fl}{Fl}			%canonical flip
\DeclareMathOperator{\Par}{P}

\DeclareMathOperator{\const}{const.}
\DeclareMathOperator{\End}{End}				%Endomorphisms (of vector spaces) 
\DeclareMathOperator{\bigo}{O}				%big o notation
\DeclareMathOperator{\dist}{dist}   		%distance
\DeclareMathOperator{\clos}{clos}    		%closure 
\DeclareMathOperator{\crit}{crit}    		%set of critical points
\DeclareMathOperator{\LieGL}{GL}    		%GL Lie Groups
\DeclareMathOperator{\LieSL}{SL}    		%SL Lie Groups
\newcommand{\lieGL}{\mathfrak{gl}}			%gl Lie algebra
\DeclareMathOperator{\LieO}{O}				%O Lie groups
\DeclareMathOperator{\LieSO}{SO}    			%SO Lie Groups
\DeclareMathOperator{\LieCO}{CO}				%conformal group
\DeclareMathOperator{\Lor}{Lor}				%Identity component of 
\newcommand{\lieLor}{\mathfrak{lor}}		%Lorentz lie algebra
\DeclareMathOperator{\CLor}{CLor}			%Identity component of 
\DeclareMathOperator{\Newtonder}{N} 		
\newcommand{\Nder}{\Newtonder \negthinspace}		
\DeclareMathOperator{\Fermider}{F}			
\newcommand{\Fder}{\Fermider \negthinspace}
\DeclareMathOperator{\baseR}{e}				%standard basis vector 
\newcommand{\cbaseR}{\underline \baseR}		%standard cobasis 
\DeclareMathSymbol{\square}
{\mathord}{AMSa}{"03}
\DeclareMathSymbol{\blacksquare} {\mathord}{AMSa}{"04}
\newcommand{\evat}[1]{\negthickspace\upharpoonright_{#1}}
\newcommand{\Evat}[2]{\left. #1 \right\rvert_{#2}}
\renewcommand{\qedsymbol}{
								$\blacksquare$
							}
\newcommand{\oendmark}{
						$\diamondsuit$
						}
\theoremstyle{break}
\theoremheaderfont{\normalfont\bfseries}
\theorembodyfont{\normalfont %\itshape
				}
\theoremseparator{}
\theoremnumbering{arabic}
\theoremsymbol{\oendmark}
\newtheorem{Proposition}{Proposition}[section]	
\newtheorem{Theorem}[Proposition]{Theorem}
\newtheorem{Lemma}[Proposition]{Lemma}
\newtheorem{Corollary}[Proposition]{Corollary}
\newtheorem{Conjecture}[Proposition]{Conjecture}

\theoremsymbol{\oendmark}
\newtheorem{Definition}[Proposition]{Definition}
\newtheorem{Principle}{Principle}

\newtheorem{Postulate}{Postulate}
\theoremsymbol{\oendmark}
\newtheorem{Remark}[Proposition]{Remark}

\newtheorem{Example}[Proposition]{Example}

\theoremheaderfont{\scshape}
\theorembodyfont{
		\normalfont
				}
\theoremstyle{nonumberplain}
\theoremseparator{}
\theoremsymbol{\qedsymbol} 
\newtheorem{Proof}{Proof}
\numberwithin{equation}{section}
\allowdisplaybreaks					%allows for breaking 
\counterwithout{footnote}{chapter}	
\makeatletter
\title{An Observer's View on Relativity}         
\let\Title\@title             
\subtitle{Space-Time Splitting and Newtonian Limit} 
\let\Subtitle\@subtitle       
\author{Maik Reddiger}     
\let\Author\@author  
\date{22. Mai 2017}              
\let\Date\@date           
\makeatother       

\SetWatermarkText{\textit{DRAFT 17/05/03}}
\SetWatermarkAngle{0}
\SetWatermarkScale{1}    
\SetWatermarkLightness{0}
\SetWatermarkVerCenter{1.2 \textheight}    

\begin{document}   

\pagestyle{empty}
\begin{titlepage} 
\raggedright	%muss durch klammern an und ausgeschaltet werden
{

\begin{minipage}{0.6\textwidth} 
	AG Geometrie und Mathematische Physik \\
	Institut f\"ur Mathematik \\
	Technische Universit\"at Berlin \\
	Str. des 17. Juni 136 \\
	10623 Berlin \\
	
	\vspace{0.5cm}
	AG Relativit\"atstheorie und Quantentheorie \\
	Institut f\"ur Theoretische Physik \\
	Technische Universit\"at Berlin \\
	Hardenbergstr. 36 \\
	10623 Berlin 
	
	\vspace{1cm}
\end{minipage}
%\hspace{0.5cm}
\begin{minipage}[t][][t]{0.3\textwidth} 
\end{minipage}
	
	\begin{center}
	\huge\textbf{
		\Title
	} \\ 
	\vspace{0.5cm}
	\Large\textbf{\Subtitle} \\ 

	\normalsize
	
	\vspace{1cm} 

	Arbeit zur Erlangung des Grades \\
		\vspace{.5cm}
	\LARGE Master of Science \\
		\vspace{.5cm}
	\normalsize an der \\
		\vspace{.5cm}
	\LARGE
	Technischen Universit\"at Berlin \\
	Fakult\"at II Mathematik und Naturwissenschaften \\
	
	\vspace{1cm}
	
	\normalsize
	vorgelegt von \\
	\vspace{.5cm}
	\LARGE\Author \\ \normalsize
	\vspace{.5cm}
	Berlin, \Date
	\end{center} 
	
	\vfill

	\begin{tabbing}  
  		a bit of free space 
   	\= Gutachter: bit \= Prof. Dr. H.-H. von Borzeszkowski 
   	\kill
   	\> Gutachter:	\>		Prof. Dr. Yuri B. Suris 	\\
   	\>					\>		Prof. Dr. Horst-Heino von Borzeszkowski	\\
   	\> Betreuer: 	\>  	Dr. Wolfgang Hasse  
	\end{tabbing}
}	%beendet raggedright
\end{titlepage}

\vspace*{\fill} 
\noindent\rule{\textwidth}{1pt} 
\small{This work is licensed under a Creative 
Commons Attribution-NonCommercial 4.0 International License. 
The license covers all source files of the document (including 
pictures), as well as the output files compiled from 
these source files by arXiv. To the author's 
best knowledge, all external sources have been properly 
attributed and this work does not infringe on the rights 
of any third parties.}

\cleardoublepage

\topskip0pt
\vspace*{\fill}
\begin{center}
	\textit{
	dedicated to \\
	free thought, equality and mutual respect, \\
	the very 
	foundations of a free society 
	} 
\end{center}
\vspace*{\fill}

\cleardoublepage

\setcounter{page}{1} 

~\vspace{3cm}
\renewcommand{\abstractname}{Abstract}

\begin{center}
	\large \textbf{
	\abstractname } 
\end{center}

	%remove German translation -> no need 
	%In particular, we obtain an alternative description for 
	%accelerated observers in Minkowski spacetime 
	%A general formula relating 'actual forces' to 
	%pseudo-forces is derived 

\begin{quote}  
	\noindent
	We motivate and construct a mathematical theory for 
	the separation of 
	space and time in general relativity. The formalism only 
	requires a single observer and an optional choice of 
	reference frame at each instant. As the splitting 
	is done via the observer's past light cone, it is both closer 
	to the experimental situation and mathematically less restrictive
	than the splitting via observer vector fields or spacelike 
	hypersurfaces. Indeed, the theory
	can in principle be applied to all spacetimes and 	
	adapted to other `metric' theories of gravity. Instructive 
	examples are developed along with the general 
	theory. In particular, we obtain an alternative description for 
	accelerated frames of reference in Minkowski spacetime. 
	\par
	Further, we use the splitting formalism to motivate a 
	new mathematical approach to the Newtonian limit of the motion 
	of mass points. This employs a general formula for their  
	observed motion, distinguishing between 
	`actual' forces (i.e. those detectable via an accelerometer) 
	and pseudo-forces. 
	Via this formula we show that for inertial frames of 
	reference in Minkowski spacetime the essential
	laws of non-gravitational Newtonian mechanics can be derived. 
	\par 
	Physically relevant, related, open problems are indicated 
	throughout the text. These include the proof, that the 
	Newtonian limit gives rise to the central 
	pseudo-forces known 
	from Newtonian mechanics (`constant gravity', Euler, Coriolis 
	and centrifugal force) for non-inertial frames of 
	reference in Minkowski spacetime, as well as the derivation 
	of Newton's law of gravitation in the Schwarzschild 
	spacetime under said limit. 
\end{quote} 

\vspace{0.5cm}

\begin{tabbing}
	Keywords:la \= \kill 
	\textit{Keywords:} \> \textit{space-time 
	splitting - Newtonian limit -
	relativistic kinematics -} \\
	\> \textit{frame of reference - gravitational lensing} \\
\end{tabbing}

\newpage

~\vspace{3cm}

\renewcommand{\abstractname}{Zusammenfassung}

\begin{center}
	\large \textbf{
	\abstractname } 
\end{center}

	\begin{quote}
	\noindent 
		Wir motivieren und entwickeln eine mathematische Theorie 
		zur Aufteilung von Raum und Zeit in der Allgemeinen 
		Relativit\"atstheorie. Der Formalismus ben\"otigt lediglich 
		einen einzelnen Beobachter und eine optionale Wahl 
		eines Bezugssystems zu jedem Zeitpunkt. Da die Trennung 
		\"uber den Vergangenheitslichtkegel des Beobachters 
		erfolgt, ist sie sowohl n\"aher an der experimentellen 
		Situation als auch mathematisch weniger restriktiv als die 
		Teilung mittels Beobachtervektorfeldern oder raumartigen 
		Hyperfl\"achen. Tats\"achlich ist die Theorie im Prinzip auf 
		alle Raumzeiten anwendbar und kann an andere `metrische' 
		Gra\-vi\-ta\-tions\-theo\-rien 
		angepasst werden. Instruktive Beispiele 
		werden zusammen mit der allgemeinen Theorie entwickelt. 
		Insbesondere erhalten wir eine alternative Beschreibung 
		beschleunigter Bezugssysteme in der Minkowski-Raumzeit. 
		\par
		Weiter benutzen wir den Trennungsformalismus, um einen neuen 
		mathematischen Zugang zum Newtonschen Grenzfall 
		der Bewegung von Massepunkten zu begr\"unden. Dies wird 
		\"uber den Gebrauch einer allgemeinen Formel zu ihrer 
		beobachteten Bewegung erreicht, welche 
		`echte' Kr\"afte (jene, die sich mit einem 
		Beschleunigungsmesser nachweisen lassen) und Scheinkr\"afte 
		voneinander unterscheidet. Mit Hilfe dieser Formel 
		zeigen wir,  
		dass sich f\"ur inertiale Bezugssysteme in der 
		Minkowski-Raumzeit die wesentlichen 
		Gesetze der gravitationsfreien
		Newtonschen Mechanik herleiten lassen. 
		\par 
		Physikalisch bedeutsame, verwandte, offene Probleme 
		werden im Text angeschnitten. Beispiele daf\"ur sind 
		zum einen 
		der Beweis, dass der Newtonsche Grenzfall tats\"achlich 
		zu den aus der Newtonschen Mechanik bekannten zentralen 
		Scheinkr\"aften (`konstante Gravitation', Euler-, 
		Coriolis- und Zentrifugalkraft) f\"ur nicht-inertiale 
		Bezugssysteme in der Minkowski-Raumzeit f\"uhrt, und zum 
		anderen die Herleitung von Newtons 
		Gra\-vi\-ta\-tions\-ge\-setz 
		in der Schwarzschild-Raumzeit unter diesem Grenzfall.  
	\end{quote}

\vspace{0.5cm}
 
\begin{tabbing}
	Schlusselworter:l \= \kill 
	\textit{Schl\"usselw\"orter:} \> \textit{Raum-Zeit-Trennung
	- Newtonscher Grenzfall - Relativistische} \\
	\> \textit{Kinematik - Bezugssystem - Gravitationslinsen} \\
\end{tabbing}

\newpage 

\pagestyle{headings}	

\tableofcontents

\chapter{Introduction}

With the 1687 publication of his `Philosophiae Naturalis Principia 
Mathematica', Newton not only 
gave birth to modern mathematical physics, but also carved 
in stone a scientific view of the world: 
Space is Euclidean and absolute, time is eternal in both 
directions and everywhere the same. They 
are separate, metaphysical entities, providing 
the stage for all physical occurrences. 
\par
Up to the advent of non-Euclidean geometry in the first half 
of the nineteenth century with the 
works by Lobachevski, Bolyai, Gau\ss{}, Riemann and others, 
the inherent truth of the Newtonian paradigm had been out of 
the question. 
Once, it became apparent, however, that Euclidean geometry 
was not the only one, it could no longer be asserted a priori 
that it represented the true geometry of physical space. 
Among the first to realize this was 
F. K. Schweikart, who, contrary to what one might expect, 
was neither a mathematician, nor a physicist, but 
a professor of law at the university of Marburg 
\cite{Jammer}*{p. 147}. Yet it did not take long for 
some of the founding fathers of non-Euclidean geometry to 
follow suit. In the 1820s Gau\ss{} famously 
decided to settle the issue by measuring the
inner angles of a triangle formed by the mountains 
Brocken, Hoher Hagen and Inselberg in Germany, but, after 
taking their sum, could not detect any clear deviation 
from the anticipated 180 degrees 
\citelist{\cite{Lynch} \cite{Jammer}*{p. 147}}. 
Had his measurements been more precise, he might have 
dealt a serious blow to the view of the world at the time. 
\par 
The Newtonian paradigm remained largely unchallenged for 
almost another 90 years, until the young Einstein published 
his article `On the Electrodynamics of Moving 
Bodies' \cite{Einstein0} in 1905. With his 
foundational work in what is today known as the special theory 
of relativity, he took the courage to scientifically discard the
notion of universal simultaneity, as well as to 
turn length into a relative concept. The mathematical 
axiomatization of his `theory of relativity' was carried 
out by his teacher and mathematician Hermann Minkowski 
\cite{Einstein0}. Minkowski realized that the new 
physics called for a unification of space and time into 
a single concept, named spacetime, and therefore 
completed the new view of the world. 
\par 
It took only 10 years until Einstein again publicly 
defied the prevailing conception of space and time. 
His struggle to include gravitation into the picture forced him 
to impugn the Euclidean nature of space, thereby 
reviving a centuries-old discussion. 
The successful fusion of 
non-Euclidean geometry with Minkowski's spacetime concept 
in his `general theory of relativity' 
ultimately lead to a second revolution in 
our collective understanding of space and time.
\par
Still, the more 
general spacetime concept brought forth new questions. 
The proximity of the special theory of relativity to 
Newtonian mechanics did not lead to any serious issues 
regarding the relation between spacetime and the 
subjectively more familiar notions of space and time. As soon 
as Einstein modeled gravity as a consequence of the curvature 
of spacetime, however, he was unable to clearly identify space 
and time by themselves in the theory. While it was apparent to him
that he had buried the Euclidean conception of space 
\cite{Einstein2}*{p. 69sq.} and that time was only 
a meaningful concept for individual clocks, he needed to 
employ a combination of heuristic reasoning and approximations 
to separate the two again \cite{Einstein2}*{p. 96sq.}. 
Even though his reasoning proved itself to be sufficient 
for the establishment of his new theory of gravity, it 
left a vacancy to be filled by future generations of physicists 
and mathematicians. The main question awaiting to be answered 
was: 
``Does there exist a physically well-motivated, 
mathematically rigorous separation of space and time 
in general relativity?'' 
\par
If one surveys the contemporary literature regarding those 
so called space-time splittings, one finds that 
mathematically well-defined constructions
do indeed exist. A list of references \cite{Bini1} 
was compiled by Bini and Jantzen, but it is not 
exhaustive, of course. Excluding the rather ad hoc coordinate-based 
methods, the two most common splitting formalisms involve a 
choice of global timelike vector field on the spacetime, taken 
to be a Lorentzian manifold for now. In the 
so called `threading approach', originally due to Landau 
and Lifshitz \cite{Landau3}*{\S 10-4}
(see also \cite{Jantzen1} and \cite{EhlersA0}), 
this choice is explicit, 
while in the `slicing approach', being part of the 
Arnowitt-Deser-Misner 
formalism \cite{Misner}*{p. 419sqq. \& \S 21.7} 
(short: ADM formalism), the choice is implicit. 
Philosophically, the approaches are rather similar 
and can be put in mathematical agreement, provided certain 
topological conditions on the spacetime, as well as 
integrability conditions of the vector field, are met.
We refer to the articles \citelist{\cite{Jantzen} 
\cite{Jantzen1}} by Jantzen, Carini and 
Bini for an introduction to the different approaches, 
including a historical review. For a more geometric approach, 
the reader may also find the article \cite{Elst}
by van Elst and Uggla, as well as the book by O'Neill 
\cite{O'Neill}*{p. 358sqq.} beneficial. 
\par 
While from a mathematical perspective 
the aforementioned approaches are in principle 
unproblematic, we 
believe the issue is not yet 
settled. Physically, space-time splittings ought to directly
relate to our individual experience of space and time. Therefore, 
they should be carried out for individual curves representing 
physical motion, not 
via a timelike vector field representing infinitely many such 
curves. So the choice of a timelike vector field is 
ultimately arbitrary and estranged from the experimental situation. 
In addition, focusing on what is observed leads us to 
reject the general philosophy that spacelike 
submanifolds ought to be identified as physical space. We believe 
that this philosophically flawed approach both in the `threading' 
and the `slicing' ansatz is the origin of the rather 
restrictive conditions required for a full splitting to be 
carried out. 
For instance, in the plane wave spacetimes 
one of the major conditions is not met 
(see e.g. the article by Perlick 
\cite{Perlick2}*{\S 5.11} and the original one due to 
Penrose \cite{PenroseA0}). So does it not make sense to 
speak of space and time individually here? 
\par 
Based on an reexamination of the
underlying philosophy and its relation to what is actually 
observed, an alternative approach is proposed here. 
Our original motivation for it came from reading 
the diploma and PhD theses \citelist{\cite{Hasse0} \cite{Hasse1}} 
by Wolfgang Hasse, as well as from the attendance of 
lectures in the philosophy of space and time, given 
by Dennis Dieks at the University of Utrecht. 
Later we discovered that the relativistic 
separation of space and time via observer mappings, as they are 
named here, is not unheard of in the literature. However, we 
are not aware of any source, where the observer mapping 
has explicitly been named as a tool for doing so. 
As far as we know, the first instance, 
where the observer mapping implicitly appears, is a 1938
article \cite{Temple}*{p. 128} due to G. Temple, who 
was one of Eddington's students \cite{Kilmister}*{p. 390}. 
It may also be found in a 1959 article \cite{Mast} by Mast 
and Strathdee, which has been a valuable 
reference to us. 
Further analyses have been carried out, e.g. by 
Kristian and Sachs \cite{Kristian} and Ellis et al. 
\cite{EllisA4}. In addition, the observer mapping 
is closely related to the physical phenomenon of 
gravitational lensing. So in this context, the works by 
Perlick \citelist{\cite{Perlick2}\cite{Perlick1}}, 
Ehlers \cite{Ehlers}, as well as Ellis, Basset and 
Dunsby \cite{Ellis0} should be mentioned. 
\par 
In this work we attempt to answer two main questions: 
\begin{enumerate}[1)]
	\item 	How does our individual perception of the 
			separateness of space and time relate to the 
			spacetime concept 
			on a physical and mathematical level? 
	\item	In what sense is relativity theory a generalization 
			of Newtonian mechanics?
\end{enumerate}
The second question needs to be raised, because the concepts 
of space and time, as defined by the splitting, need to 
reduce to the Newtonian ones in an approximation. In the 
literature, this approximation is referred to as the 
Newtonian limit. Indeed, the mathematical construction 
(and the theory as a whole) is only physically 
tenable, if it can be shown that the 
Newtonian limit exists under assumptions that are compatible 
with the domain of validity of the Newtonian theory. 
We refer to page \pageref{chap:nlimit} for a more detailed 
discussion. 
\par
Accordingly, the structure of this thesis 
follows the main questions: First we give a review of the 
required mathematical machinery and then employ it for
the construction of the splitting formalism in the 
subsequent chapter. The final 
chapter discusses the Newtonian limit in this context and 
shows that it indeed exists for the special theory of 
relativity. We invite the reader to skip 
the technical chapter \ref{chap:prelim} on first 
reading and refer to it when necessary. 
\par
Contrary to what one might expect initially, the existence of 
the Newtonian limit in the special theory of relativity 
already constitutes a non-trivial 
test of the splitting construction. 
There are two other cases, where the Newtonian 
limit needs to be shown to exist, but their treatment here
would blow the size of this thesis out of proportion.  
Nevertheless, they are certainly the most crucial tests 
of the construction and are thus to be considered 
important open problems. They are elaborated upon in section 
\ref{sec:Nlimitgen}. 
\par 
If the theory withstands 
these attempts of falsification, then it may be applied, 
for instance, to elaborate on philosophical issues of 
relativity theory, the subject of 
`gravitoelectromagnetism' and to  
attack the question whether general relativity is 
really unable to account for the internal motion of spiral 
galaxies (being part of the so called `dark matter problem' 
\cite{Trimble}). Of course, this would also necessitate 
the development or application of a variety of 
approximation formalisms, as well as a formulation of the theory, 
which is more suitable for direct application by physicists. 
\par 
We close this introduction with a few important remarks. 
\begin{description}[style=unboxed,leftmargin=0cm]
	\item[Type of work:] This is a thesis in mathematical 
	physics. However, the term mathematical physics is not as 
	well-defined as one might expect - indeed, 
	there appear to be two polar views of the field: 
	One may be named `physical mathematics', where one 
	aims to solve purely mathematical problems, that are 
	either directly or indirectly related to physics. 
	The second approach aims to contribute to 
	the clarification - or even correction - 
	of physical theories 
	and the related solution of physical problems 
	by means of rigorous mathematics. As such, it 
	necessarily requires a certain philosophical understanding of 
	the physical situation at hand and is a supplement as well 
	as a direct competitor \cite{Faddeev} to the 
	field of theoretical physics. In our mind, 
	both approaches to  
	mathematical physics are interdependent, often
	hard to separate and fertilize each 
	other. In fact, mathematics and physics share a common 
	origin in natural philosophy, so it should not come as 
	a surprise that the fields have become so intertwined 
	as to give rise to an own discipline. 
	\par
	The reader might have guessed already that we take the 
	second approach in this work.  
	Therefore, the aim is, at least within the bounds of 
	the theory, to make statements 
	on physical reality, not to elaborate on the 
	underlying mathematical machinery - unless required 
	to achieve this goal. Though we follow the 
	mathematical tradition in giving rigorous definitions 
	and proving theorems, the main emphasis is placed 
	in answering the two questions stated above. 
	So not the most general version of a theorem 
	is stated and proved, but only the one of 
	interest to the model. Moreover, we have tried hard 
	to keep the mathematics separate from the physics, but 
	our empirical data strongly suggests that 
	this is an impossible task 
	for a mathematical physicist of the second type. 
	We have found a middle way in not using 
	undefined terminology in theorems and definitions and, 
	whenever we do so elsewhere 
	in an attempt to reason heuristically, 
	we have usually indicated this with words like 
	`heuristically', `roughly', `intuitively', etc. In cases 
	where we forgot to do this, the context should tell the 
	reader whether the reasoning is mathematical or 
	philosophical in its nature. Consequently, in between 
	theorems and definitions as well as within remarks and 
	examples, the reader may find undefined terminology 
	and heuristic reasoning. While this is almost a crime 
	for mathematicians, it is absolutely necessary when 
	discussing the physical and philosophical aspects 
	of the matter. Again, we discuss phenomena in the physical 
	world and not just purely mathematical structures. 
	These phenomena are of interest precisely because we 
	physicists find them on the interface between the known and 
	the unknown. Therefore, in this research it is a necessary 
	state of affairs that concepts are only rigorously defined 
	once they have passed well into the realms of the known 
	and other new concepts, that have not yet passed this 
	boundary, may show themselves to be ill motivated or 
	even nonsensical after further progress has been made. 
	Nonetheless, founding on top of the known 
	can lead deep into the realms of the unknown. Precisely 
	this is attempted in this work and as such, it is a work 
	in natural philosophy, as much as it is a work in mathematics 
	and physics. 
	\item[Terminology:] The choice of terminology in the 
	field of mathematical and theoretical physics is a problem 
	of its own: As opposed to pure mathematics, where the 
	terminology is often chosen on categorical grounds, 
	there is often a philosophical concept attached to the 
	words, extending beyond the mathematical definition. 
	This inevitably leads to a conflict: On one hand, one would 
	like to capture the physical idea adequately and on the 
	other hand one would like to express this idea as precisely 
	as possible in mathematical terms. 
	\par
	This in turn leads to the problem, that 
	varying the mathematical ansatz for tackling the physical 
	issue leaves many fundamental physical ideas invariant, 
	while requiring a change of their mathematical definitions.  
	This justifies some choice of terminology here, which 
	differs partially from the one used in the physical 
	literature. Whenever such a deviation occurs, we guarantee 
	that it is not without warrant and it is done, because we 
	believe that it captures the underlying physical idea better. 
	A particular example is the word `frame of reference'
	to be discussed later. 
	\par
	In addition, we chose to keep the mathematical terminology, 
	where the introduction of terminology due to physicists 
	is redundant. For example, in the general relativity 
	literature one often reads the word 'tetrad', which is simply 
	a choice of basis in the tangent space of interest. 
	This is a purely mathematical object and mathematicians 
	have already devised the word `frame' for it, hence there 
	is no need for the word 'tetrad'. We believe that the 
	introduction of redundant mathematical terminology in 
	physics contributes to the mutual separation of the 
	fields, which, in our opinion, is harmful to both 
	and should thus be avoided. Of course, 
	if the word carries an additional physical meaning, 
	the matter is different.	
	\item[Choice of examples:] The reader will observe, that 
	the examples we chose in the context of observer mappings 
	are mostly given in flat spacetimes. The reason is 
	not that the construction does not work in the general 
	setting, but due to the fact that examples in curved 
	spacetimes are computationally very challenging 
	and we wished to put the emphasis on the abstract, 
	general theory, rather than on computations. 
	\item[Prerequisites:] As for all works in advanced mathematics, 
	the reader needs a fair amount of background knowledge 
	in order to be able to fully comprehend this thesis. 
	%maybe: For those looking to apply the theory, 
	%we wish to note that the main object of interest is the 
	%so called observer mapping?!? -> no better 
	%put `summary' and refer to here
	Accordingly, she or he should be familiar 
	with the major concepts of differential geometry (e.g. 
	manifolds, tensor fields, pseudo-Riemannian metrics, Lie 
	groups, 
	covariant derivatives, fiber bundles, ...), 
	as well as the basic structure of relativity theory 
	(e.g. its physical motivation, what a spacetime roughly is, 
	equations of free fall motion, the Einstein equation). 
	For the former the books by Lee \cite{Lee}, and Rudolph 
	and Schmidt \cite{Rudolph}*{Chap. 1 to 4} 
	provide a good, rigorous introduction. More advanced differential 
	geometric topics are reviewed in chapter \ref{chap:prelim}, so
	appropriate references are given there. 
	For an introduction to relativity theory, we recommend the 
	books by Carroll \cite{Carroll}*{Chap. 1 to 5} and 
	Wald \cite{Wald}*{Part I}. Moreover, the book by O'Neill 
	\cite{O'Neill} should be mentioned as an excellent work 
	in mathematical relativity. 
	Knowledge of special relativity is explicitly not required. 
	On the contrary, it is conceptually simpler   
	to view special relativity through the lens of
	general relativity as, from our experience, the theory 
	can be very deceiving otherwise.  
	\item[Conventions:] We write definitions in italic in the hope 
	that it will help the reader with a distaste in successive reading 
	to easily spot terminology. 
	Also, we sometimes write words in round
	brackets to prevent misunderstanding or to emphasize that the words 
	can be omitted. For instance, as we are working in the 
	category of smooth manifolds, there is no need to 
	explicitly state that every mapping should be smooth, but 
	sometimes we nonetheless write '(smooth) mapping' instead of 
	just `mapping'. The set of natural numbers $\N$ 
	starts with $1$, not $0$. $c$ is always the speed of light 
	in vacuum, unless it appears as an index.
	Concerning fiber bundles, 
	if it is not clear that it has a global section, 
	then sections are assumed to be local in general. Usually, however, 
	there should be no ambiguity as to what is meant: We 
	explicitly write $s \in 
	\Gamma^\infty \left( \mathcal U, \mathcal E\right)$ 
	to say that $s$ is a (smooth, local) section of the fiber bundle 
	$\mathcal E$ over some $\mathcal M$ with domain 
	$\mathcal U \subseteq \mathcal M$. 
	For trivial 
	bundles we like to drop the distinction between sections 
	and maps into the fiber, though there is technically a difference. 
	That is, if $\mathcal E = \mathcal M \cross \mathcal F$ is 
	trivial, then we sloppily write 
	$s \in C^\infty \left(\mathcal U, \mathcal F \right)$ instead of 
	$s \in 
	\Gamma^\infty \left( \mathcal U, 
	\mathcal M \cross \mathcal F \right)$. 
	Moreover, it should be said that if we use the word `natural', it 
	is meant in the physicist's vague sense of the word, so a priori 
	there is no inherent mathematical meaning to it. 
	\item[Notation:] A joke among mathematicians 
	says `differential geometry is what stays invariant under change 
	of notation' and there is definitely some truth to it. Our 
	notation is a mixture of personal taste and the one used in the 
	book by Rudolph and Schmidt \cite{Rudolph}. In fact, most of the 
	notation is explained in the text or can be inferred from the 
	context. Nevertheless, we shall make some basic remarks. 
	\par
	We use curly letters 
	like $\mathcal M, \mathcal N, \spti, \mathcal U$ for manifolds, 
	except for the classical Lie groups, which get their common 
	declaration $\LieGL, \LieO, \LieSO$ and so on. For a (smooth) 
	map $\varphi$ between manifolds, $\varphi_*$ denotes its 
	differential/pushforward and $\varphi^*$ the respective pullback. 
	If $V$ is a subset of the domain $\dom \varphi$, 
	then $\varphi \evat {V}$ is the restricted map. The upright 
	letter `$\d$' is reserved for the Cartan derivative and ordinary 
	derivatives like $\d/\d t$. $\pr_a$ always denotes projection 
	onto the $a$th factor of some product of sets. The letter 
	$\Id$ is always some kind of identity.  
	$\R_+$ is the open interval $\left( 0, \infty \right)$, 
	$\R^n$ the $n$-fold product of the reals $\R$ with 
	$i$th standard basis vector $\baseR_i$, and 
	$\left( \R^n \right)^*$ denotes the dual (vector) space 
	with $j$th cobasis vector $\cbaseR^j$. 
	Bars under letters always indicate an inverse, e.g. the
	(algebraic) inverse of the matrix $A$ is $\ubar A$. 
	Where appropriate, we use Einstein summation convention 
	with $i, j, k, l, \dots$ going from $0$ to the end and 
	$a,b,c, \dots$ starting at $1$ instead. When we write 
	matrices, they are to be understood as homomorphisms 
	of vector spaces given in a particular basis. So if 
	$A \colon V \to W$ is a linear map with $\dim V = m$ 
	and $\dim W = n$, then 
	\begin{equation*}
		A = \begin{pmatrix}
				A^1{}_1  & \dots & A^1{}_m \\
				\vdots & \ddots & \vdots \\
				A^n{}_1 & \dots & A^n{}_m  
			\end{pmatrix}
	\end{equation*}
	gives the components $A^a{}_b$ of $A$ with respect to named 
	bases in $V$ and $W$. The components of the identity on 
	a vector space with respect to 
	a given basis are $\delta^a_b$.
	\par
	The dot $\cdot$ denotes contraction with the next sensible adjacent 
	entry from the left or right. For instance, when $\theta$ is a 
	$1$-form and $X$ is a vector field over the same manifold, then the 
	contraction is 
	\begin{equation*}
		\theta \left( X \right) = \theta \cdot X = 
		X \cdot \theta \, .
	\end{equation*}
	The left hand views $\theta$ as a covector field, the middle 
	is just $\theta_i \, X^i$ in some coordinates and the 
	right hand side can be read as $ X^i \, \theta_i$. 
	The same formula holds, for example, 
	if we let $\theta$ be an $\mathcal E$-valued $1$-form, 
	where $\mathcal E$ is (real) vector bundle over $\spti$ being 
	neither the cotangent bundle $\CapT^*\spti$ nor any tensor 
	bundle `built' from $\CapT^*\spti$. Otherwise, the first 
	expression would not be defined; the second and third 
	would differ 
	in general. The notation has the advantage that it 
	correctly views contraction as a 
	generalization of ordinary matrix multiplication. 
	Also note that $\varphi_*\cdot X = \varphi_* X$, so we 
	sometimes view $\varphi_*$ point-wise as a matrix. 
	\item[Contact:] Queries regarding this work (e.g. 
	comments, errors, remarks or questions) are  
	received with gratitude and should be 
	submitted to the author's email address:  
	\quad \texttt{maik.reddiger@zoho.com} 
\end{description}

\chapter{Mathematical Preliminaries}  
\label{chap:prelim}

In this chapter we review some mathematical results needed for the 
rigorous formulation of the theory of relativity, including
the construction and analysis of the space-time splitting. 
With the exception of section \ref{ssec:Conf}, the discussion 
here only serves as a reminder and is 
not intended to be an in-depth review. For the latter, 
including detailed proofs, the reader is advised to consult 
the references provided in the respective sections.
Nonetheless, we have attempted to create a coherent 
overview of those mathematical results and do prove some 
propositions, where we have not found a treatment in the 
literature suitable for our purposes or consider it 
pedagogically worthwhile.   
\par
In the first section \ref{sec:pullback} we quickly recall how to 
`pull back' smooth fiber bundles, a concept that 
appears both implicitly and explicitly throughout this thesis. 
Afterwards we give a general treatment of (first order) 
$\mathcal G$-structures. Two examples of particular interest to 
relativity theory are discussed in the subsequent two subsections, 
namely those of Lorentzian 
metrics as well as Lorentzian orientations. Their 
combination will give rise to the mathematical 
definition of spacetime in the next chapter. 
Section \ref{sec:connecgen} introduces the notion of 
connector and tangent bundle connection, first from the
more general point of view of Ehresmann connections and 
then becoming more specific in subsection 
\ref{ssec:connecTQ}. We close the chapter with a treatment of 
Jacobi fields and their relation to the Lorentzian 
exponential map in section \ref{sec:Jacobi}.  

\section{Pullback Bundles}
\label{sec:pullback}

Due to the omnipresence of pullback bundles in this thesis, we 
shall give a brief review. \par 
Let $\pi \colon \mathcal E \to \mathcal N$ be a (smooth) fiber bundle 
and $\xi \colon \mathcal M \to \mathcal N$ be a (smooth) 
mapping between manifolds. Then the set 
\begin{equation}
	\xi^* \mathcal E := \set{\left( x, T \right) \in 
	\mathcal M \cross \mathcal E}{\xi \left( x \right)= 
	\pi \left( T\right)}\, , 
\end{equation} 
has a unique manifold structure (cf. \cite{Lee}*{p. 13} for a 
definition), such that 
it is a (smooth) fiber bundle over $\mathcal N$ with the same typical 
fiber as $\mathcal E$ and $\tilde \pi := 
\pr_1 \evat{\xi^* \mathcal E} \colon \xi^* \mathcal E \to \mathcal M$ 
is its smooth projection (cf. \cite{Baum}*{Satz 2.2; Satz 2.1}). We call $\xi^* \mathcal E $
the \emph{pullback bundle of $\mathcal E$ by $\xi$}.  
Indeed, it is equipped with the subspace topology and is thus 
an embedded submanifold of the product manifold 
$\mathcal M \cross \mathcal E$ (for a proof adapt 
\cite{Rudolph}*{Prop. 2.6.1} to the more general case of fiber 
bundles). We now follow the book by Sachs and Wu 
\cite{Sachs}*{\S 2.0.1}, in defining  
\emph{$T$ (in $\mathcal E$) over $\xi$} to be an element of 
$\mathcal E$ with $\pi\left( T\right) \in \xi \left( \mathcal M\right)$, 
i.e. $T$ is just the projection onto the second factor of an element 
$\left( x, T \right) \in \xi^* \mathcal E$. In this spirit, 
if $\mathcal U \subseteq \mathcal M$ is open and $\left(. , T \right)
\colon \mathcal U \to \xi^*\mathcal E$ is a (smooth, local) section of 
$\xi^* \mathcal E$, then the mapping $T \colon \mathcal U \to \mathcal 
E$ is called a \emph{(smooth, local) section (of $\mathcal E$) 
over $\xi$}. In the 
special case of $\xi$ being a curve, we also call $T$ a 
\emph{(smooth, local) section (of $\mathcal E$) along $\xi$}. 
Note that we always assume domains of curves to be open 
and connected subsets of $\R$, i.e. open intervals. 
\par
The use of these concepts is necessary as the map $\xi$ need not 
be injective, so intuitively we wish to vary $T$ on $\mathcal N$, 
rather than on $\mathcal M$, in order to allow for several $T$s at the 
same point in the image $\xi \left( \mathcal M \right)$. For instance, 
a case of particular interest to us is the one where 
$\mathcal E= \CapT \mathcal N$, $\mathcal M \subseteq \R$ is an 
(open) interval and $\xi \colon \mathcal M 
\to \mathcal N$ is a smooth curve. Intuitively, a mapping 
$T \colon \mathcal M \to \mathcal \CapT \, \mathcal N$ attaches for  
every parameter $s \in \mathcal M$ a vector $T_s \in 
\CapT_{\xi \left( s\right)} \mathcal N$ to the curve $\xi$ 
in $\mathcal N$. Should the curve intersect itself once
at $q \in \mathcal N$, then there will be two usually different vectors 
at $q$. As vector fields are particular instances of sections of 
fiber bundles, we then call $T$ a \emph{vector field over $\xi$} 
or a \emph{vector field along $\xi$}. 
Analogous terminology is used for differential forms, tensor fields, 
frame fields and so on. 
\par
Finally, we remark that fiber bundles over open subsets 
of $\R^n$ with $n \in \N$ are 
always trivial and thus admit global sections. 

\section{(First Order) $\mathcal G$-structures}
\label{sec:gstruct}

The notion of $\mathcal G$-structures allows for an (almost) unified 
view on geometric structures in differential geometry and hence they 
are also of interest to the (mathematical) relativist. 
The fundamental idea is that many geometric structures on a 
manifold are constructed from objects in multi-linear algebra 
(e.g. tensors), which can be brought to a 'standard form' by 
an appropriate choice of basis, unique up to the linear action of 
a 'symmetry group' $\mathcal G$ onto that basis. 
An analogue is 
then constructed locally on a manifold by choosing 'appropriate' 
local frame fields and the structure is 'globalized' by a partition 
of unity argument. However, for non-parallelizable manifolds 
there may be topological obstructions to the existence of 
the geometric 
structure as it is not always possible to choose the frame fields 
appropriately. In subsection \ref{ssec:gstructmath} we will 
give a general definition of (first order) $\mathcal G$-structures, 
as well as dual $\mathcal G$-structures, 
and make some general statements. 
In subsections \ref{ssec:Lorentz} and
\ref{ssec:Conf} we consider two particular examples illustrating the 
relation between the mathematical definition of 
$\mathcal G$-structure and 'actual' geometric structures 
on a manifold. 
\par 
Standard references for $\mathcal G$-structures are for instance 
\cite{Sternberg}*{Chap. VII} and \cite{Kobayashi}. We will not 
recall the mathematical notion and properties of principal 
bundles here as this is not the topic of this thesis and it 
is a standard topic in the differential geometry literature. For 
the German-speaking reader the book by Baum 
\cite{Baum} provides a good reference, for the English-speaking one 
we recommend the books by Poor \cite{Poor} and Rudolph and Schmidt 
\cite{Rudolph1}.
Principal bundles are also necessary for the mathematical formulation 
of so-called gauge theories in particle physics 
\citelist{\cite{Bourguignon} \cite{Rudolph1} \cite{Baum}}. 

\subsection{Mathematical Definitions}
\label{ssec:gstructmath}

Recall that the frame bundle $\frameb{\mathcal E}$ of a 
(smooth, real) vector bundle 
$\mathcal E$ over a (smooth) manifold 
$\spti$ with typical fiber $\mathcal V$
(i.e. a real vector space) is a proper subset of 
$\mathcal E \tp \mathcal V ^*$ and canonically a principal 
$\LieGL\left( \mathcal V\right)$-bundle%
\footnote{For $\dim \mathcal V = n $ one often 
reads that $\frameb{\mathcal E}$ ought to be 
a $\LieGL \left( \R^n\right)$-bundle 
over $\spti$. While this is 
certainly a correct point of view (as all non-trivial 
finite dimensional, real vector spaces are isomorphic to 
$\R^n$ for some $n \in \N$), considering 
$\frameb{\mathcal E}$ as a 
$\LieGL\left( \mathcal V\right)$-bundle is more 
`natural' - at least for most tensor bundles 
$\mathcal E$. As an example, consider the 
endomorphism bundle $\CapT \spti \tp \CapT^* \spti$. }
 over $\spti$ 
(cf. \cite{Poor}*{\S 1.45e}), where 
\begin{equation*}
	\LieGL\left( \mathcal V\right) 
	:= \set{A \in \End\left(\mathcal V \right)
		= \mathcal V \tp {\mathcal V}^*}{
		\exists \ubar A \equiv A^{-1}\colon 
		A \cdot \ubar A = \ubar A \cdot A = \Id_{\mathcal V}
		} \, .
\end{equation*}
In the context of $\mathcal G$-structures, we are interested 
in the (tangent) frame bundle $\frameb{\CapT \spti}$ 
- with $\R^n$ being its typical fiber - 
and the 
co(tangent)frame bundle $\frameb{\CapT^* \spti}$ 
- with typical fiber $\left( \R^n \right)^*$ - 
of an $n$-manifold 
$\spti$. These two principal bundles are dual in the sense that for every 
frame $X \in \frameb{\CapT \spti}$ there exists a unique coframe 
$\ubar X \in \frameb{\CapT^* \spti}$ with the property that 
\begin{equation}
	X \cdot \ubar X = \Id_{\CapT \spti} \quad , \quad 
	 \ubar X \cdot X = \Id_{\R^n} \, . 
	 \label{eq:frduality}
\end{equation}
So the notation $\ubar X = X^{-1}$ is admissible, if 
taken in the algebraic sense. 
Moreover, both $\frameb{\CapT \spti}$ and 
$\frameb{\CapT^* \spti}$ can be viewed as principal 
$\LieGL_n$-bundles, where $\LieGL_n :=
\LieGL \left( \R^n \right)$. 
\begin{Remark} 
\label{Rem:coframeb}
Viewing $\frameb{\CapT^* \spti}$ as a 
principal $\LieGL \left( \R^n \right)$-bundle rather than a 
principal $\LieGL \left( \left( \R^n \right)^* \right)$-bundle 
is more convenient from the point of view of 
$\mathcal G$-structures. The reason for this will 
become apparent later when we introduce dual 
$\mathcal G$-structures. Mathematically, this corresponds 
to replacing the 
right Lie group action 
\begin{equation*}
	\frameb{\CapT^* \spti} \cross 
	\LieGL \left( \left( \R^n \right)^* \right) 
	\colon \left( \ubar X, A \right) 
	\to \ubar X \cdot A 
\end{equation*}
with 
\begin{equation*}
	\frameb{\CapT^* \spti} \cross
	\LieGL \left( \R^n \right) 
	\colon \left( \ubar X, A \right) 
	\to \underline{\left( X \cdot A \right)} = 
	{\ubar{A}} \cdot 
	\ubar X = \ubar X \cdot {\ubar{A}}\transp
	\, ,  
\end{equation*}
which is also a right Lie group action.
\end{Remark}
\par
Assume now we are given a Lie subgroup $\left(\mathcal G, 
\rho\right)$ of $\LieGL_n$. Hence
\begin{equation*}
	\rho \colon \mathcal G \to \LieGL_n 
\end{equation*}
is a (smooth) faithful representation. In general, 
$\mathcal G$ does not need to be a subset of $\LieGL_n$, 
nor do we require $\rho$ to be open onto its image, i.e. 
Lie subgroups do not need to be embedded. 
Then mathematically, a \emph{(first order) $\mathcal G$-structure 
$\mathcal P$ on $\mathcal Q$} is a 
$\mathcal G$-reduction of the frame bundle of $\spti$, where
$\mathcal P \subseteq \frameb{\CapT \spti}$ and
the (right) action of $\mathcal G$ on $\mathcal P$ is induced 
by the canonical action of $\mathcal G$ on $\frameb{\CapT \spti}$ via the representation $\rho$. 
More explicitly, we have a (smooth) principal $\mathcal G$-bundle 
$\left(\mathcal P, \pi, \spti, \mathcal G \right)$ with right action 
\begin{equation*}
	\alpha\colon \mathcal P \cross \mathcal G \to \mathcal P 
	\colon \left( X, A \right) \to \alpha \left(X, A \right)
	= X \cdot \rho \left( A \right) 
\end{equation*}
and the reduction mapping is the inclusion $\iota \colon \mathcal P 
\xhookrightarrow{} \frameb{\CapT \spti}$. 
\begin{Remark}
	\label{Rem:Gstr}
	\begin{enumerate}[i)]
		\item 	The action of $\mathcal G$ on 
				the fibers of $\mathcal P$ 
				is \emph{simply transitive}, i.e. for all 
				$q \in \mathcal Q$
				and $X,Y \in \pi^{-1} \left( 
				\lbrace q \rbrace \right)$ 
				there exists a unique $A \in \mathcal G$
				such that $Y = X \cdot \rho \left( A \right)$.
				\label{sRem:Gstr1}
		\item	Point \ref{sRem:Gstr1} implies that 
				if $X, X'$ are (smooth, local) sections of 
				$\mathcal P$ over $\mathcal U, \mathcal U'$, 
				respectively, with 
				$\mathcal U \cap \mathcal U' \neq \emptyset$, 
				then there exists a unique (smooth) function 
				\begin{equation*}
					A \colon \mathcal U \cap \mathcal U'
					\to \mathcal G 
				\end{equation*}
				such that $X'= X \cdot \rho \left( A\right)$ 
				over $\mathcal U \cap \mathcal U'$. 
				\label{sRem:Gstr3}
		\item 	\label{sRem:Gstr2} 
				Since $\mathcal G$ is only a Lie subgroup of 
				$\LieGL_n$, the pair 
				$\left( \mathcal P, \iota \right)$ is 
				a submanifold of $\frameb{\CapT \spti}$, but it 
				is not necessarily embedded. However, if 
				$\left( \mathcal G, \rho \right)$ is an embedded
				Lie subgroup of $\LieGL_n$, then 
				$\left( \mathcal P, \iota \right)$
				is also an embedded submanifold. The proof of 
				these statements (cf. \cite{Baum}*{p. 66}) 
				employs the fact that locally 
				over some open $\mathcal U \subseteq \mathcal Q$ 
				the mapping $\iota\evat{\pi^{-1} \left( 
				\mathcal U \right)} $ can be viewed as  
				the mapping 
				\begin{equation*}
					\mathcal U \cross \mathcal G 
					\to 
					\mathcal U \cross \LieGL_n \colon 
					\left( q, A \right) \to 
					\left( q, \rho \left( A \right) \right)\, .
				\end{equation*}
	\end{enumerate} 
\end{Remark}
\par
We continue by recalling an important theorem for the theory of 
$\mathcal G$-structures. 
%	In a future version, it might be sensible to use the 
%	theorem for transition functions here and get rid of 
%	corresponding remark -> check for theorems that cite this 
\begin{Theorem}
	\label{Thm:Baum2.14}
	Let $\left( \mathcal H, \rho \right)$ be a Lie subgroup
	of $\mathcal G$ and let 
	$\left(\mathcal P, \pi, \spti, \mathcal G \right)$ 
	be a (smooth) principal $\mathcal G$-bundle over $\spti$ with 
	(smooth) right action $\alpha$. \\
	If $\mathcal P' \subseteq \mathcal P$ satisfies 
	\begin{enumerate}[i)]
		\item	$\alpha \left( \mathcal P', \rho \left( 
					A \right) \right) = \mathcal P'$ 
					for all $A \in \mathcal H$, 
				\label{itm:1Baum2.14}
		\item	for every $q \in \mathcal Q$ and 
				$X,Y \in \mathcal P' \cap \pi^{-1} 
				\left( \lbrace q \rbrace \right)$ with 
				$Y= \alpha \left( X, A \right)$ we have 
				$A \in \rho \left( \mathcal H \right)$, and 
				\label{itm:2Baum2.14}
		\item	\label{itm:3Baum2.14}
				for every $q \in \mathcal Q$ there exists 
				an open neighborhood $\mathcal U$ of $q$ 
				and a smooth, local section $X\colon 
				\mathcal U \to \mathcal P$ such that the image 
				$X_{\mathcal U}$ lies in $\mathcal P'$, 
	\end{enumerate}
	then there exists a unique (smooth) manifold structure 
	on $\mathcal P'$ such that 
	$\left( \mathcal P',  \pi \evat{\mathcal P'}, 
	\spti, \mathcal H \right)$ with the action 
	\begin{equation*}
		\mathcal P' \cross \mathcal H \to 
		\mathcal P' \colon \left( X, A \right) 
		\to \alpha \left( X, \rho \left( A \right) \right)
	\end{equation*}
	is a principal $\mathcal H$-bundle over $\spti$. 
	With respect to this manifold structure $\mathcal P' 
	\xhookrightarrow{} \mathcal P$ is an $\mathcal H$-reduction 
	and smooth submanifold of $\mathcal P$. 
\end{Theorem}
\begin{Proof}
\begin{subequations}
	A proof can be found in the book by Baum 
	\cite{Baum}*{Satz 2.14}.  
	Uniqueness of the manifold structure follows from the fact 
	that \cite{Baum}*{Satz 2.1} was used in the proof. 
\end{subequations}
\end{Proof}
Note that if $\mathcal H$ is embedded, so is $\mathcal P'$
(cf. \thref{Rem:Gstr}/\ref{sRem:Gstr2}). 
Moreover, the manifold structure 
on $\mathcal P'$ is independent of the choice of 
local sections $X$ as for every 
$A \in C^\infty \left( \mathcal U, \mathcal H \right)$ the section 
$\alpha \left( X , A \right)$ gives rise to the same 
manifold structure.
\par 
\thref{Thm:Baum2.14} is used within the 
theory of $\mathcal G$-structures to turn subsets $\mathcal P$ 
of the frame bundle $\frameb{\CapT \spti}$ into 
$\mathcal G$-structures, once one has found an (open, countable) 
cover $\set{\mathcal{U}_\beta \subseteq \spti}{\beta \in I}$ 
with (smooth, local)
frame fields $\overset{\beta}{X} \in \Gamma^\infty 
	\left( \mathcal{ U}_ \beta, \frameb{\CapT \spti} \right)$ taking 
values in $\mathcal P$ for every $\beta \in I$ and an 
appropriate 'symmetry group' $\mathcal G$ such that conditions
\ref{itm:1Baum2.14} and \ref{itm:2Baum2.14} are satisfied. 
To avoid confusion, we emphasize that $\mathcal G$ is the 
`smaller group' here. 
\par
The theorem can also be used in showing that a 
$\mathcal G$-structure $\mathcal P$ on $\mathcal Q$ gives 
rise to a reduction $\mathcal P^*$ of the coframe bundle 
$\frameb{\CapT^* \spti}$ via equation \eqref{eq:frduality}.
\par
This works as follows. For every $X \in \mathcal P \subseteq 
\frameb{\CapT \spti}$ we consider the coframe $\ubar X$ 
and observe that for every $A \in \mathcal G$
\begin{equation}
	X \cdot \ubar X = X \cdot \rho \left( A \right) 
	\cdot \rho \left( \ubar{A} \right) \cdot \ubar X = 
	\Id_{\CapT \spti} \, .
	\label{eq:cofrb1}
\end{equation}
Thus we define the set
\begin{equation}
	\mathcal P^* := \set{\theta \in \frameb{\CapT^* \spti}}{
	\exists X \in \mathcal P 
	\colon \theta = \ubar X}
	\label{eq:cofrb2}
	\end{equation}
and then equation 
\eqref{eq:cofrb1} states that the action of 
$\mathcal G$ on ${\mathcal P}$ induces an action of $\mathcal G$ 
on $\mathcal P^*$. This is in fact the one induced by the (smooth) 
right action of $\LieGL_n$ on $\frameb{\CapT^* \spti}$
	\begin{equation}
	 \tilde \alpha \colon \frameb{\CapT^* \spti} \cross \LieGL_n 
	 \to \frameb{\CapT^* \spti} 
	 \colon \left( \ubar X , A \right) \to  
	 \tilde \alpha 
	 \left( \ubar X , A \right) := \ubar A \cdot \ubar X \, 
	 \label{eq:TstarQaction}
	\end{equation}	
and the dual representation of $\rho \colon \mathcal G \to \LieGL_n$
as given by 
\begin{equation}
	\mathcal G \to \LieGL \left( \left( \R^n\right)^* \right) 
	\colon A \to \left( 
	\rho \left( \ubar A \right)\right) \transp \, .
	\label{eq:rhodual}
\end{equation}
\begin{Corollary}[Dual $\mathcal G$-structures]
	\label{Cor:dualG}
	Let $\left(\mathcal P, \pi, \spti, \mathcal G \right)$ be a 
	$\mathcal G$-structure and $\mathcal P^*$ be a subset of the 
	coframe bundle $\tilde\pi \colon \frameb{\CapT^* \spti} 
	\to \spti$, as defined in
	\eqref{eq:cofrb2}. Then there exists a unique manifold 
	structure on $\mathcal P^*$, such that 
	$\left( {\mathcal P}^*, \tilde \pi \evat{{\mathcal{P}}^ *}, 
	\mathcal Q, \mathcal G \right)$
	together with the action 
	\begin{equation*}
	 {\mathcal P}^* \cross \mathcal G 
	 \to {\mathcal P}^*  
	 \colon \left( \ubar X , A \right) \to 
	 \rho \left( 
	 \ubar A \right) \cdot \ubar X = \ubar X \cdot \left( 
	 \rho \left( \ubar{A} \right)\right)\transp
	 \, , 
	\end{equation*}
	as induced by the action \eqref{eq:TstarQaction} and the dual 
	representation \eqref{eq:rhodual} of $\rho$,  
	is a principal $\mathcal G$-bundle. 
	With respect to this manifold structure $\mathcal P^* 
	\xhookrightarrow{} \frameb{\CapT^* \spti}$ is a 
	$\mathcal G$-reduction and smooth submanifold of 
	$\frameb{\CapT^* \spti}$. 
\end{Corollary}
\begin{Proof}
\begin{subequations}
	As noted above, this is a corollary of \thref{Thm:Baum2.14}, 
	so we need to check the assumptions. 
	Since $\mathcal P$ is a $\mathcal G$-structure, 
	$\left( \mathcal G, \rho \right)$ 
	is a Lie subgroup of $\LieGL_n$. By \thref{Rem:coframeb},
	the latter is the structure group of the principal bundle 
	$\frameb{\CapT^* \spti}$. \\
	``\ref{itm:1Baum2.14}'': 
	We have $\mathcal P \cdot \rho \left( A \right) 
	= \mathcal P$ for all $A \in \mathcal G$ and formally 
	$\ubar {\mathcal P} = \mathcal P^*$ by 
	definition \eqref{eq:cofrb2} of ${\mathcal P}^*$. 
	Upon inversion 
	\begin{equation*}
		X \cdot \rho \left( A \right) \to \underline
		{X \cdot \rho \left( A \right)} = \rho \left( 
		\ubar A \right) \cdot \ubar X
	\end{equation*}
	for $X \in \frameb{\CapT^* \spti}, A \in \mathcal G$, 
	it follows
	\begin{equation*}
	 \rho \left( \ubar A \right) \cdot {\mathcal P}^* = 
	\underline{\mathcal P \cdot \rho \left( A \right)} =
	{\mathcal P}^* \, .
	\end{equation*}
	``\ref{itm:2Baum2.14}'': Again invert and use 
	simple transitivity of the $\mathcal G$-action on 
	$\mathcal P$, see \thref{Rem:Gstr}/\ref{sRem:Gstr1}. \\
	``\ref{itm:3Baum2.14}'': This is true for $\mathcal P$ 
	and upon inversion of the frame field, it is true for 
	${\mathcal P}^*$. 
\end{subequations}
\end{Proof}
We call the principal $\mathcal G$-bundle $\mathcal P^*$ the
\emph{dual $\mathcal G$-structure to $\mathcal P$}. The dual 
point of view is often needed to understand the relation between 
$\mathcal G$-structures and `actual' geometric structures on 
the base manifold. 
\begin{Remark}
	\label{Rem:TransextG}
\begin{enumerate}[i)]
	\item 
	\label{sRem:transitP}
	Recall that if $\set{\left(\mathcal{U}_ \alpha, \phi_\alpha 
	\right)}{\alpha \in I}$ is a system of local principal 
	bundle trivializations of $\mathcal P$ 
	(cf. \cite{Rudolph1}*{p. 5}), then second countability 
	of $\spti$ implies that we can always choose 
	the set $I$ to be countable. We call such a countable cover 
	$\set{\mathcal{U}_ \alpha}{\alpha \in I}$ a 
	\emph{trivializing cover}. 
	Now recall that there is a one-to-one correspondence between 
	the diffeomorphisms $\phi_\alpha$ and smooth 
	local sections 
	\begin{equation*}
		\overset{\alpha}{X}\colon \mathcal{U}_\alpha \to \mathcal P
		\colon q \to \overset{\alpha}{X}_q 
		:=\phi_\alpha^{-1} \left( q, \Id_{\mathcal G} 
		\right) \, .
	\end{equation*}
	Due to \thref{Rem:Gstr}/\ref{sRem:Gstr3} on page 
	\pageref{sRem:Gstr3}, for all $\alpha, \beta \in I$ with 
	$\mathcal{U}_\alpha \cap \mathcal{U}_\beta \neq \emptyset$ 
	there exist (smooth) transition functions 
	\begin{equation*}
		\overset{\alpha \beta}{A} 
		\colon \mathcal{U}_\alpha \cap 
		\mathcal{U}_\beta \to \mathcal G 
		\colon q \to \overset{\alpha \beta}{A}_q \, ,
	\end{equation*}
	uniquely defined by the equation
	\begin{equation*}
		\overset{\beta}{X} =\overset{\alpha}{X} \cdot \rho \bigl(
		\overset{\alpha \beta}{A} \bigr) \, .
	\end{equation*}
	Conversely, if $\set{\mathcal{U}_ \alpha}{\alpha \in I}$ is an 
	(open, countable) cover of $\spti$ with corresponding frame 
	fields $\bigl\lbrace \overset{\alpha}{X} \bigr\rbrace
	_{\alpha \in I}$ satisfying the 
	above equation for some $\overset{\alpha \beta}{A} \in C^\infty 
	\left( \mathcal{U}_\alpha \cap \mathcal{U}_\beta, 
	\mathcal G  \right)$ with $\alpha, \beta \in I$ such that 
	$\mathcal{U}_\alpha \cap \mathcal{U}_\beta \neq \emptyset$, 
	then we can use this to put a unique topology 
	and smooth structure on the union of the images 
	\begin{equation*}
		\mathcal P := \bigcup_{\alpha \in I} \, 
		\left( X_{\mathcal{U}_\alpha} \cdot \rho\left( 
		\mathcal G \right) \right)
	\end{equation*}
	such that $\mathcal P$ is a $\mathcal G$-structure 
	(cf. \cite{Baum}*{Satz 2.31}).
	This yields an alternative proof of \thref{Cor:dualG}. 
	\par
	In many cases the existence of (smooth) frame fields with 
	transition functions $\overset{\alpha \beta}{A}$, 
	taking values in $\rho \left( \mathcal G 
	\right)$ and satisfying the \emph{cocycle condition} 
		\begin{equation}
			\overset{\alpha \beta}{A} \cdot 
			\overset{\beta \gamma}{A} = 
			\overset{\alpha \gamma}{A} 
			\label{eq:cocycle}
		\end{equation}
	for all $\alpha, \beta, \gamma \in I$ 
	with non-empty $\mathcal{U}_\alpha 
	\cap \mathcal{U}_\beta \cap \mathcal{U}_\gamma$, 
	gives a simpler 
	proof that $\mathcal P$
	is a $\mathcal G$-structure, 
	than the use of \thref{Thm:Baum2.14}. See, for instance, 
	\cite{Rudolph1}*{Prop. 1.1.10} for a formal proof. 
	\item 
	\label{sRem:extG} 
	In general, if $\mathcal P$	is a $\mathcal G$-structure on 
	an $n$-manifold $\spti$ and 
	$\left( \mathcal G, \chi\right)$ is also a Lie subgroup of 
	the Lie subgroup $\left(\mathcal G', \rho'\right)$ of 
	$\LieGL_n$ with $\rho = \rho' \circ \chi$, then the argument in
	\ref{sRem:transitP} shows that
	this gives rise to a $\mathcal{G}'$-structure 
	${\mathcal P}'$, where the action of $\mathcal{G}'$ on 
	${\mathcal P}'$ is induced by $\rho'$. $\mathcal P$ is 
	then called a \emph{$\mathcal{G}'$-extension of 
	$\mathcal P$}. For more on extensions in German see 
	\cite{Baum}*{\S 2.5}. 
\end{enumerate}
\end{Remark}
\par
As indicated in the beginning, the mathematical 
definition of $\mathcal G$-structures is not sufficient to 
capture the philosophical 
concept. If one speaks of $\mathcal G$-structures,  
the choice of $\mathcal G$ and 
$\rho$ is not arbitrary, but is thought of as
a 'symmetry group' of an object in linear algebra on 
$\R^n$. This object is
called the `linear model'. 
Among many possible choices, this can be a vector subspace of $\R^n$, 
an orientation on $\R^n$, a particular tensor or a scalar product. 
Geometric structures on $\mathcal Q$ are then constructed 
from frame and coframe fields taking values in $\mathcal P$ 
and its dual $\mathcal P ^*$, respectively, hence the need for 
\thref{Cor:dualG}. As it is difficult to give a general 
definition of how that 
construction of geometric structures on $\mathcal Q$ works 
precisely and as this would be an unnecessary 
abstraction for us, we will consider two particular cases of interest 
in the following two sections. As we will 
observe in the next chapter, both of these cases 
are constitutive for the mathematical theory of relativity.

\subsection{Lorentzian Structures}
\label{ssec:Lorentz}

Though we assume that the reader is 
familiar with the topic of Lorentzian metrics, their discussion 
in this section is of use both for fixing 
conventions and for understanding the philosophical idea 
of $\mathcal G$-structures in terms of a familiar example. 
Specifically, we show how a Lorentzian metric is constructed
from a particular $\mathcal G$-structure on a manifold $\spti$. 
\par
In accordance with the statements made in the beginning of 
section \ref{sec:gstruct}, we first consider a particular object 
in linear algebra and its 'symmetry group' $\mathcal G$ 
that leaves the object invariant. 
\par
Recall that a \emph{(real) Lorentz vector space} is a pair $\left( 
\mathcal V , g \right)$, where $\mathcal V$ is a 
finite dimensional 
(real) vector space equipped with a \emph{Lorentz product} $g$. 
The latter is a bilinear form on $\mathcal V$, with the 
property that there exists a basis such that it takes the form 
\begin{equation}
	g = 
	\begin{pmatrix}
		1 & & & \\
		& -1 & & \\
		& & \ddots & \\
		& & & -1
	\end{pmatrix}
	\label{eq:eta}
\end{equation}
in said basis. Obviously, this is only possible if the dimension of 
$\mathcal V$ is at least $2$. Since \eqref{eq:eta} constitutes 
a diagonalization, $g$ must be non-degenerate, but 
not positive definite. A non-zero 
vector $v \in \mathcal V$ is called \emph{timelike} if 
$g \left( v,v\right)>0$, \emph{lightlike} if $g \left( v,v\right)=0$ 
and \emph{spacelike}%
\footnote{In the book by O'Neill \cite{O'Neill}, which is in a 
sense constitutive for the mathematical theory of relativity, 
the zero vector is taken to be spacelike, but we believe 
this to be an unnatural convention and hence do not follow it.}
 if $g \left( v,v\right)<0$. It is also convenient 
to call a vector $v\neq 0$ \emph{causal}, if it is either 
time- or lightlike. The property of being space-, time- 
or lightlike is sometimes referred to as the 'causal character'
of the vector. \par
The standard example of a Lorentz vector space is 
\emph{Minkowski space} 
$\left( \R^{n+1}, \eta\right)$, where $n \in \N$ and $\eta$ 
has components as in \eqref{eq:eta} with respect to the 
canonical basis on $\R^{n+1}$. 
Trivially, all Lorentz vector spaces of the same dimension 
are linearly isomorphic to 
it via a suitable choice of basis, so we can restrict our 
attention to this instance. 
The 'symmetry group' $\mathcal G$ of Minkowski 
space is the \emph{Lorentz group} $\LieO_{1,n} \equiv \LieO 
\left( \R^{n+1}, \eta \right)
$, defined by 
\begin{equation}
	\LieO_{1,n} :=
	\set{\Lambda \in \LieGL_{n+1}}{\Lambda\transp \cdot \eta 
	\cdot \Lambda = \eta} \, . 
\end{equation}
This group $\mathcal G = \LieO_{1,n}$ acts on $\R^{n+1}$ via the 
standard representation 
\begin{equation*}
	\rho \colon \LieO_{1,n} \xhookrightarrow{} \LieGL_{n+1}
	\colon \Lambda \xhookrightarrow{} \Lambda \, , 
\end{equation*}
which is just an inclusion map. 
In fact, $\LieO_{1,n}$ admits a unique manifold structure
such that $\left(\LieO_{1,n}, \rho\right)$ 
becomes an embedded Lie subgroup of $\LieGL_{n+1}$. 
The proof that it admits such a structure is a standard 
application of Cartan's theorem \cite{Lee}*{Thm. 20.12}
and works in full analogy to the one of \thref{Lem:LieCO} in
the next subsection. 
Uniqueness of the manifold structure follows from the fact 
that the manifold structure of embedded submanifolds is always unique 
(cf. \cite{Lee}*{Prop. 5.18}). 
With respect to this topology, $\LieO_{1, n}$ has $4$ components
\cite{O'Neill}*{Cor. 9.7}. We denote the identity 
component, which is also a Lie group, by $\Lor_{n+1}$. This 
is characterized by 
\begin{equation}
	\Lor_{n+1} = \set{\Lambda = 
	\begin{pmatrix}
		\Lambda^0{}_0 & v\transp \\
		u & \Lambda_S
	\end{pmatrix} \in \LieO_{1,n}}{ \Lambda^0{}_0 > 0\, , 
	\det \Lambda_S > 0 \, \, \text{and} \, \, u, v 
	\in \R^n} 
	\label{eq:Lordef}
\end{equation}
(cf. \cite{O'Neill}*{p. 237 sq.}). The other $3$ 
components are obtained by multiplication 
with the matrices
\begin{equation}
	\label{eq:inversionm}
	\begin{pmatrix}
		-1 & & & \\
		& 1 & & \\
		& & \ddots & \\
		& & & 1
	\end{pmatrix} \quad \text{and} \quad
	\begin{pmatrix}
		1 & & & \\
		& \pm 1 & & \\
		& & \ddots & \\
		& & & \pm 1
	\end{pmatrix} 
\end{equation}	
with an odd number of minus signs. These matrices are 
called the \emph{time inversion matrix} and 
\emph{space inversion matrices}, 
respectively. Obviously, $\LieO_{1,n}$ and $\Lor_{n+1}$ 
have the same Lie algebra $\lieLor_{n+1} = \CapT_{\Id} 
\Lor_{n+1}$. To express it, we first note that, as commonly 
done in mathematics, we canonically identify the Lie 
algebra of the general linear group 
\begin{equation*}
	\lieGL_k \equiv \lieGL \left( \R^k \right) 
	= \CapT_{\Id} \left(\LieGL{\left( \R^k
	\right)}\right)	
\end{equation*} 
in $k$ dimensions with $\End\left( \R^k\right)= \R^k \tp 
	\left( \R^k\right)^*$ 
and hence we may write 
$\lieLor_{n+1} \subset \lieGL_{n+1} = \End\left( \R^{n+1}
\right)$. 
Under this identification, which is basis independent, 
the exponential map of the Lie group 
$\LieO_{1,n}$ is just the matrix exponential and hence 
\begin{equation}
	\lieLor_{n+1} = \set{ 	\lambda \in  \lieGL_{n+1}
						}{ \lambda\transp \cdot \eta + 
						\eta \cdot \lambda = 0}	\, . 
\end{equation}
\par
Now we use this 'linear model' $\left( \R^{n+1}, \eta \right)$
with 'symmetry group' $\LieO_{1,n}$ to construct a geometric 
structure on an $({n+1})$-manifold $\spti$ admitting an 
$\LieO_{1,n}$-structure. An $\LieO_{1,n}$-structure 
$\left( \mathcal P, \pi, \spti, \LieO_{1,n} \right)$ is 
also known as a \emph{Lorentzian structure on $\spti$}. 
\par
The idea is that in the linear case, that is for 
an $({n+1})$-dimensional Lorentz vector space 
$\left( \mathcal V, g \right)$, 
a Lorentz product $g \in {\mathcal V}^* \tp {\mathcal V}^*$
can always be written as 
\begin{equation}
	g = {\ubar X}\transp \cdot \eta \cdot \ubar X 
	\label{eq:linearg}
\end{equation}
where $X \in \mathcal V \tp \left(\R^{n+1}\right)^*$ is a 
basis. This basis is unique up to the action of the group 
$\LieO_{1,n}$ on $X$ from the right. To construct 
a Lorentzian metric we think 
of $\mathcal V$ as the tangent space $\CapT_q \spti$ at a point
$q \in \spti$ and then let $X$ 'vary smoothly' with $q$. 
If the manifold $\spti$ is parallelizable, the tangent 
frame bundle $\frameb{\CapT \spti}$ and hence every 
$\mathcal G$-structure is trivial, so in this case we simply 
choose a global frame field $X \colon \spti \to \mathcal P$
or equivalently a global coframe field 
$\ubar X \colon \spti \to \mathcal P^*$ dual to $X$ 
(see section \ref{ssec:gstructmath}) and define 
a Lorentzian metric $g$ on $\spti$ via \eqref{eq:linearg}.
\par
If $\frameb{\CapT \spti}$ is not trivial, then neither is the 
$\LieO_{1,n}$-structure $\mathcal P$ nor its dual 
${\mathcal P}^*$, but we can still find a (countable) 
trivializing (open) cover $\set{\mathcal{U}_\alpha}
{\alpha \in I}$ with (smooth, local) frame fields 
\begin{equation*}
	\overset{\alpha}{X}\colon \mathcal{U}_\alpha \to \mathcal P
	\subset \frameb{\CapT \spti}
\end{equation*}
for every $\alpha \in I$. Moreover, for $\alpha, \beta 
\in I$ with $\mathcal{U}_\alpha \cap \mathcal{U}_\beta 
\neq \emptyset$ we obtain smooth transition functions 
\begin{equation*}
	\overset{\alpha \beta}{\Lambda} 
	\colon \mathcal{U}_\alpha \cap 
	\mathcal{U}_\beta \to \LieO_{1,n-1} 
	\colon q \to \overset{\alpha \beta}{\Lambda}_q
\end{equation*}
via $\overset{\beta}{X} =\overset{\alpha}{X} \cdot 
\overset{\alpha \beta}{\Lambda}$ on $\mathcal{U}_\alpha 
\cap \mathcal{U}_\beta$. 
If we again use the ansatz of defining $g$ via 
\eqref{eq:linearg}, we can use the freedom in the choice of 
basis, the local frame fields $\overset{\alpha}{X}$ 
and a partition of unity to construct the Lorentzian metric 
$g$. More explicitly, we first define a (smooth) partition 
of unity subordinate to the cover
$\lbrace \mathcal{U}_\alpha\rbrace_{\alpha \in I}$: 
\begin{equation*}
	\rho_\alpha \colon {\mathcal U}_\alpha \to [0,1]
	\quad \forall \alpha \in I \, , \quad
	1 = \sum_{\alpha \in I} \rho_\alpha \, ,
\end{equation*}
and then construct the global tensor field 
\begin{equation}
	\label{eq:gframef}
	g := \sum_{\alpha \in I} \rho_\alpha \, 
	\overset{\alpha}{\ubar {X}}{}\transp \cdot \eta \cdot 
	\overset{\alpha}{\ubar {X}}
	\, .
\end{equation}
For any $\alpha \in I$ and $q \in {\mathcal U}_\alpha$, we calculate: 
\begin{align}
	g_q &= 
	\sum_{\beta \in I} \rho_\beta \left( q \right)
	 \, 
	\overset{\beta}{\ubar {X}}{}\transp_q \cdot \eta \cdot 
	\overset{\beta}{\ubar {X}}_q \nonumber \\
	&= \sum_{\beta \in I} \rho_\beta \left( q \right)
	 \, \bigl( \underline{
	\overset{\alpha}{X} \cdot \overset{\alpha \beta}{\Lambda} }
	\bigr)\transp_q 
	\cdot \eta \cdot 
	\bigl( \underline{
	\overset{\alpha}{X} \cdot \overset{\alpha \beta}{\Lambda} }
	\bigr)_q \nonumber \\
	&= \sum_{\beta \in I} \rho_\beta \left( q \right)
	 \, \overset{\alpha}{\ubar{X}}{}\transp_q 
	 \cdot 
	 \bigl( 
	 	\overset{\alpha \beta}{\ubar{\Lambda}}{}\transp_q 
		\cdot \eta \cdot 
		\overset{\alpha \beta}{\ubar{\Lambda}}_q 
	 \bigr) 
	 \cdot 
	 \overset{\alpha}{\ubar{X}}_q \nonumber \\
	&= \sum_{\beta \in I} \rho_\beta \left( q \right)
	 \, \overset{\alpha}{\ubar{X}}{}\transp_q 
	 \cdot \eta \cdot 
	 \overset{\alpha}{\ubar{X}}_q  \nonumber \\ 
	&= \overset{\alpha}{\ubar{X}}{}\transp_q 
	 \cdot \eta \cdot 
	 \overset{\alpha}{\ubar{X}}_q
	 \label{eq:constrg}
	\,  .
\end{align}
This calculation shows that definition \eqref{eq:gframef}
is independent of the choice of $\left\lbrace 
\rho_\alpha\right\rbrace_{\alpha \in I}$. By a similar argument, 
it is independent of the choice of frame fields taking values in 
$\mathcal P$. 
Now we simply define a \emph{Lorentzian metric} $g$ to be a 
smooth covariant 2-tensor field such that for every $q \in 
\spti$ there exists an open neighborhood $\mathcal U$ of 
$q$ and a local frame field $X \in \Gamma^\infty
\left( \mathcal U, \frameb{\CapT 
\spti} \right)$ with coframe field $\ubar 
X \in \Gamma^\infty \left( \mathcal U, \frameb{\CapT 
\spti} \right)$ dual to $X$ such that 
\begin{equation*}
	g\evat{\mathcal U} = \ubar X \transp \cdot \eta \cdot \ubar X
	\, .
\end{equation*}
As we have shown with \eqref{eq:constrg}, the tensor field defined 
by \eqref{eq:gframef} is a Lorentzian metric and 
conversely every Lorentzian metric can be written in this way. 
We conclude that the existence and choice of a Lorentzian metric 
on a manifold $\mathcal Q$ is equivalent 
to the existence and choice of a 
$\LieO_{1,n}$-structure. As stated in 
the end of section \ref{ssec:gstructmath}, this is not by 
accident, but illustrates the general scheme of 
(first order) $\mathcal G$-structures. Similarly one can 
construct Riemannian metrics, volume forms, symplectic 
forms, etc. from appropriate reductions of the frame bundle 
and vice versa. 
\begin{Remark}[Implicit geometric structures]
	\label{Rem:newgeomstr}
	Extensions of $\mathcal G$-structures as described in 
	\thref{Rem:TransextG}/\ref{sRem:extG} 
	yield new geometric structures on $\spti$, 
	once one has identified the corresponding 'linear model' 
	for which the 'larger group' $\mathcal G'$ acts as a 
	'symmetry group'. If one can argue that the extension 
	is 'natural', these new geometric structures can in turn 
	be used to understand and formulate physical laws without 
	needing to introduce additional postulates or 
	mathematical assumptions. 
	\par
	For instance, assume we are given a $\Lor_{n+1}$-structure 
	on an $({n+1})$-manifold $\spti$. Obviously $\Lor_{n+1}$
	is a Lie subgroup of $\LieO_{1,n}$, hence we can use the 
	$\Lor_{n+1}$-structure to construct a Lorentzian metric on 
	$\spti$ via the procedure described in this section. 
	Moreover, $\Lor_{n+1}$ is also a Lie subgroup of
	the special linear group 
	\begin{equation*}
		\LieSL_{n+1} \equiv 
		\LieSL\left( \R^{n+1}\right) := 
		\set{A \in \LieGL_{n+1}}{\det A = 1} \, , 
	\end{equation*}
	which is the symmetry group of the determinant 
	map. So if $X= X_i \tp \cbaseR^i$ 
	is a basis in $\R^{n+1}$, then for every $A \in 
	\LieSL_{n+1}$ 
	\begin{equation*}
		({\underline{X \cdot A}})^0 \wedge \dots 
		\wedge ({\underline{X \cdot A}})^{n}
		= \det\left( \ubar A \right) \,
		{\ubar X}^0 \wedge \dots \wedge {\ubar X}^{n} 
		= {\ubar X}^0 \wedge \dots \wedge {\ubar X}^{n} \, 
		.
	\end{equation*}
	In full analogy to the procedure of constructing 
	a Lorentz metric $g$, we can use the above 
	calculation to construct an 
	$({n+1})$-form on $\spti$ from sections of the 
	$\Lor_{n+1}$-structure. 
	The resulting $({n+1})$-form
	is in fact the canonical volume
	form with respect to $g$. \par
	In chapter \ref{chap:construction} we will see that 
	our definition of spacetime indeed yields a 
	$\Lor_{n+1}$-structure, namely the so called 
	`frame of reference bundle'. Hence the mathematical theory of 
	extensions of $\mathcal G$-structures gives rise 
	to a natural, physical notion of spacetime volume. 
\end{Remark}
\par
To conclude this section, we add that an 
$\LieO_{1,n}$-structure $\mathcal P$ on an $n$-manifold 
$\spti$ has a natural interpretation in terms of the metric $g$, as 
constructed above. If $X \in \Gamma^\infty\left(\mathcal U, 
\mathcal P  \right)$ is a local frame field over 
$\mathcal U \subseteq \spti$, then we can calculate the 
component functions of $g$ with respect to $X$: 
\begin{align*}
	g_{ij} &:= g \left(X_i, X_j \right) 
	= X_i \cdot \left( {\ubar X}\transp \cdot \eta \cdot \ubar X 
	\right) \cdot X_j \\
	&\phantom{:}= X_i \cdot \left( \eta_{kl} \, {\ubar X}^k \tp 
	{\ubar X}^l \right) \cdot X_j = 
	\eta_{kl}\, \delta^k_i \, \delta^l_j \\ 
	&\phantom{:}= \eta_{ij} \, . 
\end{align*}
In other words, for each $q \in \mathcal U$ the frame 
$X_q$ is an orthonormal basis in $T_q \spti$ with respect to 
$g_q$. Hence we call $X$ an \emph{orthonormal frame field 
(with respect to $g$)} and $\mathcal P = \oframeb{\spti, g}$ the 
\emph{orthonormal frame bundle (on $\spti$ with respect to $g$)}. 

\subsection{Lorentzian Orientations}
\label{ssec:Conf}

In this section we discuss another example of a 
$\mathcal G$-structure on a manifold $\spti$, namely that 
of so called `Lorentzian orientations'. The example is in fact a 
class of examples encompassing space, time, as well as
spacetime orientations on manifolds. Their construction  
again follows the general recipe: First 
find the linear model, second find its 
symmetry group and third use a trivializing cover with 
corresponding frame fields and a partition of unity to construct the 
geometric structure from the linear model. However, 
Lorentzian orientations differ from the example in 
the preceding subsection in the sense that the linear 
model is not a tensor on $\R^{n}$ and hence we will 
not obtain a tensor field on $\spti$ as a result. 
\par 
As noted before, the discussion here is more explicit than 
in the other sections, since we have not been able to find a
suitable reference. Many textbooks, including the one by 
O'Neill \cite{O'Neill}*{Chap. 9} treat the issue of 
Lorentzian orientations implicitly by assuming that a 
Lorentzian metric is given. The treatment here in terms of 
open Lie subgroups of the indefinite conformal group is 
more conceptual and more general.
Nonetheless, we do
recommend O'Neill's treatment as a reference. For more 
information on conformal geometry in the context of 
general relativity, we refer to 
\citelist{\cite{Kuehnel} \cite{Wald}*{Appendix D}}. 
\par
The linear model we wish to consider is an adaption of the concept 
of orientation to a Lorentz vector space 
$\left( \mathcal V, g\right)$, 
which we will later call a '(linear) Lorentzian orientation'. 
First recall that, if $\mathcal V$ is a vector space 
of dimension $n \in \N$, an 
\emph{orientation $O$ on $\mathcal V$} is a choice of basis 
$X = X_i \tp \cbaseR^i$ modulo the canonical action of the 
(Lie) group 
\begin{equation*}
	\LieGL_n^+ := \LieGL^+ \left( \R^n\right)
	:= \set{A \in \LieGL\left(\R^n \right)}{ \det A > 0} \, .
\end{equation*} 
from the right. As a group orbit 
\begin{equation*}
	O := X \cdot \LieGL^+_n = 
	\set{Y \in \mathcal V \tp \left( \R^n \right)^*}{
	\exists A \in \LieGL^+_{n}\colon \, Y= X \cdot A} \, ,  
	\, 
\end{equation*}
of a basis $X$, an orientation defines a subset of the set of 
bases of $\mathcal V$. A basis is called 
\emph{right-handed}, if it is an element of that set
and \emph{left-handed}, if it is not. The terminology stems from 
the 'right-hand rule'. 
\par
However, if we have a Lorentz vector space 
$\left( \mathcal V, g\right)$ and want to find an `adapted' notion of 
orientation on it, it is natural to 
require this notion of orientation to preserve 
the causal character of at least some elements 
$Z \in \mathcal V$, i.e. we ask for time-, space- or 
lightlike vectors (or some combination thereof) to stay 
time-, space- or lightlike under 
the action of the subgroup $\mathcal G \subseteq \LieGL 
\left( \mathcal V \right)$ from the right. The next theorem 
characterizes this group $\mathcal G$. 
\begin{Theorem} 
	\label{Thm:conf}
	Let $\mathcal 
	V$ be a vector space with Lorentz products $g, g'$. \\
	Then the following are equivalent:
	\begin{enumerate}[i)]
	\item 	A vector is timelike with respect to $g$ 
			if and only if it is timelike with respect to $g'$. 
			\label{itm:conf1}
	\item 	A vector is spacelike with respect to $g$ 
			if and only if it is spacelike with respect to $g'$. 
			\label{itm:conf2} 
	\item	A vector is lightlike with respect to $g$ 
			if and only if it is lightlike with respect to $g'$.
			\label{itm:conf3} 
	\item	\label{itm:conf4}
			There exists a constant 
			$\lambda \in \R_+$ such that $g'= \lambda 
			g$. 
	\end{enumerate} 
\end{Theorem}	
\begin{Proof}
	Trivially, \ref{itm:conf4} implies the other 
	statements. Thus it is sufficient to prove
	``\ref{itm:conf3} $\implies$ \ref{itm:conf4}'', 
	``\ref{itm:conf1} $\implies$ \ref{itm:conf3}'' and 
	``\ref{itm:conf2} $\implies$ \ref{itm:conf3}''. \\
	``\ref{itm:conf3} $\implies$ \ref{itm:conf4}'': 
	The main idea is as follows: 
	If we are given a Lorentz product $g$, 
	then for any 
	timelike $X$ and any spacelike $Y$ in
	$\mathcal V$ we can use the 
	parameter $s \in \R$ to 'move' $X+sY$ into and out of the 
	\emph{light cone}  
	\begin{equation*}
		\mathfrak c \left( g \right) 
		:= \set{Z \in \mathcal V}{Z \neq 0 \quad 
	\text{and} \quad g \left( Z,Z\right) = 0 } \, .
	\end{equation*}
	This idea is due to Hawking and Ellis \cite{Hawking}*{p. 61} and 
	was later taken up by Dajczer and Nomizu \cite{Dajczer} to 
	prove a related result.\\ 
	Let $g,g'$ be Lorentz products on 
	$\mathcal V$, $X \in \mathcal V$ 
	be timelike, $Y \in \mathcal V$ be spacelike with respect to $g$ 
	and $s \in \R$. Define $f \colon \R \to \R$ via
	\begin{equation*}
		f \left( s\right) := g \left( X+sY , X+sY \right)
		= g \left( X, X \right) + 2s \, g \left( X, Y \right)
		+ s^2 \,g \left(Y , Y \right), 
	\end{equation*}
	which has zeros $s_+,s_-$ satisfying 
	\begin{equation*}
		s_\pm = - \frac{g \left( X, Y \right)}
		{g \left(Y , Y \right)} \pm \sqrt{\left(\frac{
		g \left( X, Y \right)}
		{g \left(Y , Y \right)}\right)^2 - \frac{
		g \left( X, X \right)}
		{g \left(Y , Y \right)}}
	\end{equation*}
	and $s_- < 0 < s_+$, since $- g \left( X, X \right) /
	g \left(Y , Y \right) > 0$. Analogously, we define $f'$ 
	for the Lorentz product $g'$. \par
	As $X + s_{\pm} Y$ are lightlike with respect to $g$, they 
	are lightlike with respect to $g'$. 
	Hence $f' \left(s_{\pm} \right)=0$ 
	and therefore the zeros of $f'$ are $s'_{\pm}=s_{\pm}$. Thus 
	\begin{equation*}
		s_+ s_- = \frac{
		g \left( X, X \right)}
		{g \left(Y , Y \right)} = s'_+ s'_- 
		= \frac{g' \left( X, X \right)}
		{g' \left(Y , Y \right)}
		\, ,
	\end{equation*}
	which is equivalent to 
	\begin{equation*}
			\frac{g' \left( X, X \right)}
		{g \left(X , X \right)} = 
		\frac{g' \left( Y, Y \right)}
		{g \left(Y , Y \right)} =: \lambda \in \R_+
	\end{equation*}
	for all $g$-timelike $X$ and $g$-spacelike $Y$. Since 
	a vector $Z \in \mathcal V$ is either space-, time-, lightlike 
	or trivial, we obtain:
	\begin{equation*}
		g' \left( Z, Z \right) = \lambda g \left( Z, Z \right) \, .
	\end{equation*}
	The \emph{polarization identity}, given by 
	\begin{equation}
		g \left( X, Y \right) = \frac{1}{2} \left( 
		g \left( X+Y, X+Y \right) - g \left( X, X \right)
		- g \left( Y, Y \right) \right)
	\end{equation}
	for all $X, Y \in \mathcal V$, finishes the proof. \\
	``\ref{itm:conf1} $\implies$ \ref{itm:conf3}'': 
	The idea of proof is to reconstruct the light cone 
	by taking sequences of timelike vectors converging to lightlike 
	ones. So we show that, under some `natural' choice of topology 
	on $\mathcal V$, the light cone $\mathfrak c \left( g \right)$
	with respect to $g$ is the boundary $\partial \mathfrak t$
	of the set of timelike vectors $\mathfrak t$ minus the 
	zero vector. Since the set of timelike vectors coincide 
	for both $g$ and $g'$, 
	the boundary must coincide and so the assertion follows. 
	\par 
	``$\mathfrak c \left( g \right) \subseteq 
	\left( \partial \mathfrak t \setminus 
	\lbrace 0 \rbrace \right)$'':
	Equip $\mathcal V$ with a norm $\norm{.}$. 
	Let $X$ be $g$-lightlike and choose a $g$-timelike $Z$
	such that $g \left(X, Z \right) > 0$. For $k \in \N$ define 
	a sequence via 
	\begin{equation*}
		X_k := X + \frac{1}{k} Z
	\end{equation*}
	and observe that each element is timelike. Since 
	\begin{equation*}
		\norm{X - X_k} = \frac{1}{k} \norm {Z} 
	\end{equation*}
	the sequence converges. 
	\par 
	``$\mathfrak c \left( g \right) \supseteq 
	\left( \partial \mathfrak t \setminus 
	\lbrace 0 \rbrace \right)$'': With respect to the 
	topology induced by the above norm, the quadratic form 
	\begin{equation*}
		p \colon \mathcal V \to \R \colon
		Z \to p \left( Z\right) := g \left( Z, Z\right)
	\end{equation*}
 	is continuous (i.e. bounded with respect to the operator 
 	norm). Thus the set of 
	$g$-spacelike vectors 
	$p^{-1} \left( \left(-\infty, 0 \right)\right)$ is open.
	Since the boundary 
	$\partial \mathfrak t$ is the closure $\bar{\mathfrak{t}}$ 
	without the interior $\mathfrak{t}$, any point $X$ in 
	$\partial \mathfrak t\setminus 
	\lbrace 0 \rbrace $ is the limit of some sequence 
	$\lbrace X_k \rbrace_{k \in \N}$ in $\mathfrak t$. 
	Now $X \notin \mathfrak t$, so by continuity 
	\begin{equation*}
		\lim_{k \to \infty} p \left( X_k \right) = p \left( X \right) 
		\not > 0 \, .
	\end{equation*}
	If $p \left( X \right) < 0$, then it is spacelike, but the set of 
	spacelike vectors is open, so $X$ cannot be a limit point of 
	$\lbrace X_k \rbrace_{k \in \N} \subset \mathfrak t$. 
	Thus $p \left( X \right) = 0$. 
	\par
	Repeating the argument for the set of $g'$-timelike vectors 
	$\mathfrak t'$, we conclude 
	\begin{equation*}
	\mathfrak c \left( g \right) = 
	\left( \partial \mathfrak t \setminus 
	\lbrace 0 \rbrace \right) = 
	\left( \partial \mathfrak t' \setminus 
	\lbrace 0 \rbrace \right) = 
	\mathfrak c \left( g' \right)
	\, . 
	\end{equation*}
	``\ref{itm:conf2} $\implies$ \ref{itm:conf3}'': The proof is 
	entirely analogous to the previous one. 
\end{Proof}
\thref{Thm:conf} shows that asking either for the set of 
time-, space- or lightlike vectors to be preserved 
under the linear action of a group $\mathcal G$ implies that the 
group has to preserve the causal character of every vector
in the Lorentz vector space $\left(\mathcal V, g \right)$. By 
\thref{Thm:conf}/\ref{itm:conf4}, this group $\mathcal G$ is 
given by 
\begin{equation}
		\LieCO{\left(\mathcal V, g \right)} := 
		\set{A \in \LieGL\left( \mathcal V \right)}{\exists 
		\lambda > 0 \colon A\transp \cdot g \cdot A = 
		\lambda \, g} \, ,
		\label{eq:defCO}
\end{equation}
canonically equipped with the restricted multiplication and 
inversion mappings of $\LieGL\left( \mathcal V \right)$. The 
proof that $\LieCO{\left(\mathcal V, g \right)}$ is indeed a 
subgroup of $\LieGL\left( \mathcal V \right)$ is elementary. 
We call $\LieCO{\left(\mathcal V, g \right)}$ the 
\emph{(linear) conformal group of $\left(\mathcal V, g \right)$}.
Note that the notation $\LieCO$ for this group follows the 
one used in the book by Kobayashi \cite{Kobayashi}*{Ex. 2.6} 
and appears to be standard. 
For $n \in \N$ we write $\LieCO_{1,n} := 
\LieCO{\left(\R^{n+1}, \eta \right)}$. As every Lorentz vector 
space $\left(\mathcal V, g \right)$ is linearly isomorphic 
to Minkowski space of the same 
dimension ${n+1}$, the algebraic group 
$\LieCO{\left(\mathcal V, g \right)}$ is isomorphic to the 
algebraic group $\LieCO_{1,n}$. The corresponding group 
isomorphism is obtained by a choice of basis $X= X_i \tp 
\cbaseR^i$ in $\mathcal V$, which needs to be orthonormal with 
respect to $g$ up to a real factor. More precisely, if $X$ 
is a basis of 
$\mathcal V$ with $\lambda g = {\ubar X}\transp \cdot \eta \cdot
\ubar X$ for some $\lambda \in \R_+$, then the map 
\begin{equation*}
	\LieCO_{1,n} \to \LieCO{\left(\mathcal V, g \right)} 
	\colon A \to X \cdot A \cdot \ubar X 
\end{equation*}
is an isomorphism of groups. Conversely, if we are just given a 
vector space $\mathcal V$ together with a basis $X$, we may define 
\begin{equation*}
	O :=  X \cdot \LieCO_{1,n} =
	\set{Y \in \mathcal V \tp \left( \R^{n+1} \right)^*}{
	\exists A \in \LieCO_{1,n}\colon \, Y= X \cdot A} \, ,  
\end{equation*}
i.e. $O$ is the $\LieCO_{1,n}$-orbit of $X$. 
Then every element $Y \in O$ and hence the set $O$ itself uniquely 
defines a Lorentz product $g$ on $\mathcal V$ up to a positive 
factor. We call 
an $({n+1})$-dimensional vector space $\mathcal V$ 
equipped with an orbit $O$ of $\LieCO_{1,n}$ a
\emph{causal vector space (of signature $(1,n)$)} and 
$O$ a \emph{(linear) causal structure}. 
In this setting, it is then natural to define a non-zero vector 
$Z \in \mathcal V$ to be \emph{space-}, 
\emph{time-} or \emph{lightlike}, 
if it is space-, time- or lightlike with respect to some and hence 
every $g := {\ubar X}\transp \cdot \eta \cdot \ubar X$ 
\emph{induced by $X \in O$}. Thus a causal structure is enough 
to define the causal character of any element of 
$\mathcal V$, i.e. we do not need a particular Lorentz product. 
By \thref{Thm:conf}, it is clear 
that this is the most general setting in which it makes sense to
speak of the causal character of vectors. Yet even in this case, we 
have a notion of (hyperbolic) angle 
(cf. \cite{O'Neill}*{Chap. 5 Lem. 30}) between any two non-lightlike 
vectors $Z,Z' \in \mathcal V \setminus \lbrace 0 \rbrace$, 
as for all $\lambda \in \R_+$ it holds that
\begin{equation}
	\frac{g \left( Z, Z' \right)}{\sqrt{\abs{g \left( Z, Z\right)}} 
	\sqrt{\abs{g \left( Z',Z' \right)}}}
	= \frac{\lambda g \left( Z, Z' \right)}
	{\sqrt{\abs{\lambda g \left( Z, Z\right)}} 
	\sqrt{\abs{ \lambda g \left( Z',Z' \right)}}} 
	\, .
	\label{eq:confang}
\end{equation}
Moreover, orthogonality is 
well-defined for any $Z,Z' \in \mathcal V$. This 
angle-preserving property is characteristic of the conformal 
group in the sense that it can also be used as its definition. 
\par 
In a causal vector space $\left( \mathcal V, O\right)$ 
one can classify linear 
subspaces $\mathcal W \subseteq \mathcal V$ in accordance with 
the causal character of the vectors they contain. Adapting the 
definition from O'Neill for Lorentz vector spaces 
\cite{O'Neill}*{pp. 141 sqq.} to the causal case (and our sign 
convention), we call $\mathcal W$ \emph{spacelike}, if 
the restricted product $g \evat{\mathcal W}$ induced by some
$X \in O$ is negative definite. An equivalent condition is that 
$\left( \mathcal W, -g \evat{\mathcal W}\right)$ is an inner 
product space. A subspace $\mathcal W$ is called \emph{timelike}, 
if $\left( \mathcal W, g \evat{\mathcal W}\right)$ is a Lorentz 
vector space, and it is called \emph{lightlike}, if 
$g \evat{\mathcal W}$ is degenerate.%
\footnote{It would be more natural to call a subspace $\mathcal W$
timelike, if $g \evat{\mathcal W}$ is positive definite and 
Lorentzian, if $g \evat{\mathcal W}$ is a Lorentz product on 
$\mathcal W$. We shall submit to O'Neill's convention here, but 
the point deserves to be made.}
Again, those definitions 
do not depend on the choice of $X \in O$. The terminology 
is motivated by the fact that, if $Z \in \mathcal V$ is time-,  
space- or lightlike, then the subspace $\Span{Z} = \R Z$ is 
time-, space- or lightlike, respectively. Using these definitions, 
one can show (cf. \cite{O'Neill}*{p. 141}) that 
a subspace $\mathcal W$ is spacelike if and only if its 
orthogonal subspace 
\begin{equation*}
	{\mathcal W}^\perp = \set{Z \in \mathcal V}
	{\forall Y \in \mathcal W \colon \, 
	g \left(Y,Z \right)= 0}
\end{equation*}
is timelike and a subspace is 
lightlike if and only if its orthogonal subspace is lightlike. 
Note that any lightlike vector is orthogonal to itself and 
hence care must be taken with the terminology 
`orthogonal complement'. 
\par
Before we turn to the issue of how $\LieCO_{1,n}$ 
gives a natural notion of `Lorentzian orientation' on $\R^{n+1}$
and hence more general vector spaces $\mathcal V$, we shall 
prove that $\LieCO_{1,n}$ is in fact a Lie group and have 
a closer look at its properties. 
\begin{Lemma}
	\label{Lem:LieCO} 
	For every $n \in \N$ there exists a 
	unique manifold structure on $\LieCO_{1,n}$ 
	such that, together with this manifold structure, it is an 
	embedded Lie subgroup of $\LieGL_{n+1}$. Moreover, the map 
	\begin{equation}
		\LieCO_{1,n} \to \R_+ \cross \LieO_{1,n} 
		\colon A \to \left(\sqrt[{n+1}]{\abs{\det A}}, A 
		/ \sqrt[{n+1}]{\abs{\det A}} \right)
		\label{eq:COiso}
	\end{equation}
	is a Lie group isomorphism. 
\end{Lemma}
\begin{Proof}
	One of the standard methods to obtain a manifold structure 
	on an algebraic subgroup is to show that \eqref{eq:defCO} gives 
	a 'closed condition' and to use Cartan's theorem 
	\cite{Lee}*{Thm. 20.12}. Uniqueness of the manifold 
	structure then follows from topological embeddedness 
	\cite{Lee}*{Prop. 5.31}. 
	\par
	For any $A \in \LieCO_{1,n}$ we have 
	\begin{equation*}
		\det \left( \ubar \eta \cdot 
		A\transp \cdot \eta \cdot A \right) = 
		\left(\det\left( A\right) \right)^2 
		= \det \left( \lambda \Id \right) = \lambda^{n+1} 
	\end{equation*}
	with $\lambda$ as in \eqref{eq:defCO}. As $\det$ never 
	vanishes and is continuous on $\LieGL_{n+1}$ the function 
	\begin{equation*}
		\xi \colon \LieGL_{n+1} \to \left(\R^{n+1}\right)^* 
		\tp \left(\R^{n+1} \right) \, 
		\colon A \to \frac{A^T \cdot \eta \cdot A}{ 
		\sqrt[{n+1}]{\left(\det A\right)^2} }
	\end{equation*}
	is continuous and hence $\LieCO_{1,n} = \xi^{-1} \left( 
	\left\lbrace \eta \right\rbrace \right)$ is closed. 
	Again, \cite{Lee}*{Thm. 20.12} together with 
	\cite{Lee}*{Prop. 5.31} yields the first assertion. \\ 
	For the second assertion, we note that 
	$\R_+ \cross \LieO_{1,n}$ is the product of 
	the Lie groups $\LieO_{1,n}$ and $\R_+$ together with 
	ordinary multiplication, hence $\R_+ \cross \LieO_{1,n}$
	is canonically a Lie group. Indeed, for every 
	$A \in \LieCO_{1,n}$ we can write 
	\begin{equation*}
		A= \sqrt \lambda \, \frac{A}{\sqrt \lambda}
		= \sqrt[{n+1}]{\abs{\det A}} \, 
		\frac{A}{\left(\sqrt[{n+1}]{\abs{\det A}}\right)} \, .
		\label{eq:COproduct}
	\end{equation*}
	The first factor is a positive number, the second one is 
	an element of $\LieO_{1,n}$ by definition of $\lambda$ in 
	\eqref{eq:defCO}. As the factorization is unique, 
	\eqref{eq:COiso} defines a bijection. It is a group 
	homomorphism, since $\det$ and taking roots of 
	positive numbers are group homomorphisms. As taking 
	absolute values of non-zero reals and roots of positive 
	numbers is smooth,  the map \eqref{eq:COiso} is smooth. 
	The result now follows 
	from the fact that bijective, smooth group homomorphisms 
	between Lie groups are Lie group isomorphisms, see e.g. 
	\cite{Lee}*{Cor. 7.6}. 
\end{Proof}
As a corollary of \thref{Lem:LieCO}, we find that $\LieO_{1,n}$ 
is canonically 
an embedded Lie subgroup of $\LieCO_{1,n}$. Moreover, as 
$\LieCO_{1,n}$ is diffeomorphic to $\R_+ \cross \LieO_{1,n}$
via \eqref{eq:COiso}, $\R_+$ is connected and $\LieO_{1,n}$ has
$4$ components, so does $\LieCO_{1,n}$. In particular,  
its identity component, denoted by $\CLor_{n+1}$, is 
diffeomorphic to $\R_+ \cross \Lor_{n+1}$. 
\par
Now, to define Lorentzian orientations, 
we consider the analogy to ordinary orientations on a 
vector space $\mathcal V$. The general linear group $\LieGL_n$ 
has two connected components, but only 
the component $\LieGL^+_n$ is a (Lie) subgroup of 
$\LieGL_n$. This is the identity 
component and the one used to define an orientation. The 
other component is obtained by multiplication with the 
reflection matrix $-\Id \in \LieGL_n$. 
Carrying this line of thought over to the conformal group 
$\LieCO_{1,n}$, we are lead to the conclusion that there 
should be four kinds of Lorentzian orientations, since there are 
four open 
submanifolds of $\LieCO_{1,n}$, besides $\LieCO_{1,n}$ itself, 
that are also Lie subgroups: 
The identity component $\CLor_{n+1}$, the Lie group generated by 
$\CLor_{n+1}$ together with time inversion, the Lie group 
generated by $\CLor_{n+1}$ together with space inversion 
as well as the Lie group generated by $\CLor_{n+1}$ together 
with 
\begin{equation*}
	\begin{pmatrix}
		- 1 & & & & \\
		& -1 & & & \\
		& & 1 & & \\
		& & & \ddots & \\
		& & & & 1 
	\end{pmatrix}
	\, , 
\end{equation*}
where the $+1$s are dropped for $n+1 =2$. We thus 
rigorously define a \emph{(linear) 
Lorentzian orientation} on a vector space 
$\mathcal V$ to be an orbit of one of these four groups in the set of 
bases of $\mathcal V$ under the canonical action from the right. 
In the first case we call the Lorentzian orientation 
a \emph{(linear) spacetime orientation}, in the second case a 
\emph{(linear) space orientation}, in the third case a \emph{ 
(linear) time orientation} and in 
the last case a \emph{(linear) causal orientation}. The latter 
name is derived from the fact that the group is 
\begin{equation*}
	\LieCO_{1,n} \cap \LieGL^{+}_{n+1}
	= \set{A \in \LieCO_{1,n}}{ \det A > 0}
	 \, ,
\end{equation*}
so its orbit in $\mathcal V$ is just an ordinary orientation 
respecting the causal structure. Clearly, every spacetime 
orientation induces a 
unique space- and a unique time-orientation, but 
causal orientations induce neither. A vector space 
$\mathcal V$ is said to be \emph{spacetime-, space-, time-} or 
\emph{causally oriented}, if it is equipped with the respective 
orientation $O$. In each case it is 
\emph{Lorentz-oriented}. Note that in a Lorentz-oriented 
vector space $\left(\mathcal V, O \right)$, 
the `causal character' of any non-zero element and the notion of
orthogonality is well-defined, since $O$ is contained in an orbit of 
$\LieCO_{1,n}$ and hence $\left( \mathcal V, O \cdot \LieCO_{1,n}
\right)$ is a causal vector space.
\par 
Of course, we would like to use Lorentzian orientations on an 
$({n+1})$-dimensional vector space 
$\mathcal V$ to classify bases or vectors, but this 
classification differs from 
ordinary orientations on $\mathcal V$ and is physically motivated. 
For instance, one does not just define a basis $X$ of 
$\mathcal V$ to be time-oriented, if it is an element of 
$O$. This would be a 
natural definition from a mathematical perspective, but 
physically, we want time-orientations to define whether a 
single vector 
points into the `past' or the `future'. So instead, for a 
time-oriented vector space $\left(\mathcal V, O \right)$, 
we define a timelike vector $Z \in \mathcal V$ 
to be \emph{future-directed}, if there exists a basis 
$X \in O$ such that $Z = X_0$. Else we call $Z$ \emph{past-directed}. 
If a timelike vector is known to be future-directed, there 
exists a convenient way to check whether another timelike vector is 
also future-directed. 
\begin{Proposition}
	\label{Prop:timeorient}
	Let $\left( \mathcal V, O\right)$ be a time-oriented 
	vector space and let $Z \in \mathcal V$ be future directed
	timelike. \\ 
	Then a timelike $Z' \in \mathcal V$ is future directed 
	if and only if 
	\begin{equation*}
		g \left( Z, Z'\right) > 0
	\end{equation*} 
	with respect to some Lorentz product $g$ induced by 
	a basis in $O$. 
\end{Proposition} 
\begin{Proof}
\begin{subequations}
	Denote by $n+1$ the dimension of $\mathcal V$. 
	Obviously the above condition is independent of the choice of 
	$g$.
	\\ 
	`` $\implies$'': If $Z, Z'$ are future directed timelike, 
	then by definition there exist respective bases 
	$X, X' \in O$ such 
	that $Z=X_0$ and $Z'=X'_0$. Since $X,X' \in O$, there 
	exists a time orientation preserving $A 
	\in \LieCO_{1,n}$ such that $X' = X \cdot A$. Now 
	let $g$ be the Lorentz product induced by $X$, then 
	\begin{equation*}
		g \left( Z, Z' \right) = 
		g \left( X_0, X_i \, A^i{}_0 \right) 
		= \eta_{0i}\,  A^i{}_0 = A^0{}_0 \, .
	\end{equation*}
	By \thref{Lem:LieCO}, 
	$A$ is the product of a positive number $\lambda$, 
	an element 
	$\Lambda \in \Lor_{n+1}$ and possibly a space-inversion 
	matrix (cf. \eqref{eq:inversionm} on page 
	\pageref{eq:inversionm} for a definition). 
	Since $\Lambda^0{}_{0}>0$ by the characterization 
	\eqref{eq:Lordef} of $\Lor_{n+1}$ and 
	neither $\lambda$ nor the space inversion 
	matrix change the sign of $A^0{}_{0}$, the assertion 
	is true. \\
	`` $\impliedby$'': Let $Z'$ be a timelike vector with 
	strictly positive $g \left( Z', Z\right)$. Again, 
	choose a 
	basis $X \in O$ such that $Z = X_0$ and assume 
	without loss of generality that $g$ 
	is induced by $X$. 
	Since $Z$ and $Z'$ are both timelike, there exists
	a basis $X'$ in the 
	$\LieCO_{1,n}$-orbit of $X$ such that $Z' = X'_0$. 
	In other words, we have an $A \in \LieCO_{1,n}$ such 
	that $X' = X \cdot A$. Repeating the calculation
	above we find $A^0{}_0 > 0$ and therefore 
	$X' \in O$. 
\end{subequations}
\end{Proof}
\par
Regarding space orientations $O$, we would like to obtain 
an orientation on a spacelike hyperplane $\mathcal W$, i.e. 
a linear $n$-dimensional subspace of $\mathcal V$ with 
positive definite $-g \evat{\mathcal W}$, where 
$g$ is induced by some $X \in O$. We emphasize that a 
spacelike hyperplane is always spanned 
by $n$ linearly independent spacelike vectors, but not every 
hyperplane spanned by $n$ linearly independent spacelike vectors 
is spacelike. Yet for a timelike vector the 
orthogonal subspace is always a 
spacelike hyperplane (cf. 
\cite{O'Neill}*{Chap. 5 Lem. 26}), so, given a spacelike 
subspace $\mathcal W$, we may write 
\begin{equation*}
	{\mathcal{W}}^\perp = \Span{Z}= \R Z \,  
\end{equation*}
for some timelike $Z \in \mathcal V$. 
We may therefore call an ordered set 
of $n$ linearly independent, spacelike vectors 
$\lbrace Y_1, \dots , Y_n\rbrace$ in a 
space-oriented vector space $\left(\mathcal V, O \right)$
\emph{(space-)right handed}, if there exists an $X \in O$ such that 
\begin{equation}  
	\mathcal W := \Span{Y_1, \dots, Y_n} = \Span{X_1, \dots, X_n}
	\label{eq:spaceplane}
\end{equation}
and $Y_a \tp \cbaseR^a$ is right-handed with respect to 
the ordinary orientation induced by $X_a \tp \cbaseR^a$
on $\mathcal W$. An ordered set of $n$ spacelike vectors 
$\lbrace Y_1, \dots , Y_n\rbrace$ spanning a spacelike hyperplane 
is \emph{(space-)left handed}, if it is not (space-)right handed. 
\par
For a spacetime-oriented vector space 
$\left(\mathcal V, O \right)$, we call a 
basis $Y$ of $\mathcal V$ \emph{spacetime-oriented}, 
if $Y_0$ is timelike and 
future-directed, and 
$Y_a \tp \cbaseR^a$ are (space-)right handed. In this 
sense, it can be said that a spacetime-orientation on a vector space
$\mathcal V$ is a space- together with a time-orientation giving 
rise to the same causal structure. In contrast, 
causal orientations are simply ordinary orientations on 
a causal vector space. 
\par 
There also exists an extension of time- and space orientations
to lightlike vectors and lightlike hyperplanes in vector spaces 
$\mathcal V$ carrying the respective Lorentzian orientation 
$O$. As these will be 
employed in chapter \ref{chap:construction}, we shall give a brief 
discussion. First observe that for time- or 
spacelike $Z \in \mathcal V$, we 
can define the 
\emph{parallel projection (endomorphism)} with respect to $Z$: 
\begin{equation}
	\pi^\parallel := \frac{Z \tp Z \cdot g}{g \left(Z,Z \right)}
	\label{eq:piparallel} \, .
\end{equation}
Indeed, \eqref{eq:piparallel} is 
independent of the choice of $g$ as induced by $X \in O$. The 
\emph{orthogonal projection (endomorphism) $\pi^\perp$} is then 
defined via $\Id = \pi^\parallel + \pi^\perp$. As projections, those 
endomorphisms satisfy 
\begin{equation*}
	\pi^\parallel \cdot \pi^\parallel = \pi^\parallel \quad, 
	\quad \text{and} \quad \pi^\perp\cdot \pi^\perp = \pi^\perp \, .
\end{equation*}
The image of $\pi^\parallel$ is $\R Z$ and the image of $\pi^\perp$ 
is $(\R Z)^\perp$, for the kernels the situation is reversed. As 
noted before, if $Z$ is timelike, then $\R Z$ is timelike and 
$(\R Z)^\perp$ is spacelike. Since all non-zero vectors in a 
spacelike subspace are spacelike, it follows that lightlike 
$K \in \mathcal V$ satisfy $K \notin \pi^\perp \left( 
\mathcal V\right) = \ker \pi^\parallel$ and therefore 
$\pi^\parallel \cdot K \neq 0$. Since all non-zero vectors in 
$\pi^\parallel \left( \mathcal V\right) = \R Z$ are timelike, 
$\pi^\parallel \cdot K$ is timelike. So if $O$ is a 
time-orientation on $\mathcal V$, we may define 
lightlike $K \in \mathcal V$ to be 
\emph{future-directed}, if for some timelike 
vector $Z \in \mathcal V$, the vector 
$\pi^\parallel \cdot K$ is future-directed. 
Else $K$ is \emph{past-directed}. 
For the case of a space-oriented vector space 
$\left( \mathcal V, O\right)$, assume $K_1, \dots, K_{n} \in 
\mathcal V$ span a lightlike hyperplane $\mathcal W$ in 
$\mathcal V$. Now take some $X \in O$, consider the 
corresponding spacelike hyperplane 
\begin{equation*}
	\mathcal{W} ' := \Span{X_1, \dots, X_n}
\end{equation*}
and the parallel projection $\pi^\parallel$ with respect to $X_0$. 
Since $\mathcal W$ is lightlike, it does not contain any 
timelike vector \cite{O'Neill}*{Chap. 5 Lem. 28}, so each 
$K_a$ with $a \in \lbrace 1, \dots, n\rbrace$ is either 
light- or spacelike. In either case, $K_a \notin \pi^\parallel 
\left( \mathcal V\right) = \ker \pi^\perp$ and thus 
$\pi^\perp \cdot K_a \neq 0$ for each $a$. Thus the restriction 
$\pi^\perp \evat{\mathcal W} \colon \mathcal W \to \mathcal{W}'$ 
is a vector space isomorphism and hence 
$\pi^\perp \cdot K_1, \dots, \pi^\perp \cdot K_{n}$ form a basis 
of $\mathcal {W}'$. It is thus natural to define an ordered tuple 
of vectors $\left\lbrace K_1, \dots, K_{n} \right\rbrace$ 
spanning a lightlike hyperplane to be \emph{(space-)right handed}, 
if for some $X \in O$ the ordered set 
$\left\lbrace \pi^\perp \cdot K_1, \dots, \pi^\perp \cdot K_{n}
\right\rbrace$ yields a space-right-handed basis on $(\R X_0)^\perp$. 
Else it is \emph{(space-)left handed}. Formally, we still need to 
show that the definitions are independent of the choice of $Z$ 
and $X$. 
\begin{Theorem}[Lightlike 
				time \& space orientations are well-defined]
	\label{Thm:welldeforient}
	\NoEndMark
	\begin{enumerate}[i)]
	\item \label{itm:orient1}
	Let $\left( \mathcal V, O\right)$ be a time-oriented vector 
	space and let $K \in \mathcal V$ be lightlike. Then $K$ is 
	future-directed with respect to a timelike
	vector $Z \in \mathcal V$ if and only if it is future-directed 
	with respect to any other timelike 
	$Z' \in \mathcal V$. 
	\item \label{itm:orient2} 
	Let $\left( \mathcal V, O\right)$ be a space-oriented vector 
	space and let $K_1, \dots, K_{n} \in \mathcal V$ span 
	a lightlike hyperplane. Then $K_1, \dots, K_{n}$ are 
	right-handed with respect to $X \in O$ if and only if 
	they are right-handed with respect to any other 
	$X' \in O$.	
	\end{enumerate}
	\hfill \LemmaSymbol
\end{Theorem}
\begin{Proof}
	It is enough to show one 
	direction in each case. We set $n+1 = \dim 
	\mathcal V$. \\
	``\ref{itm:orient1}'': The idea is to decompose $K$ 
	and $Z'$ into `spatial' and `temporal' parts 
	with respect to $Z$
	and then use the Cauchy-Schwarz inequality. We 
	borrowed it from a related proof by 
	O'Neill \cite{O'Neill}*{Lem. 5.29}. \\  
	Choose some $X \in O$ such that $Z=X_0$ and let 
	$g$ be the induced Lorentz product.
	\\ 
	We introduce some notation: Any $Y \in 
	\mathcal V$ may be decomposed into a timelike and 
	spacelike part via 
	\begin{equation*}
		Y = Y^0 \, X_0 + \vec Y
	\end{equation*}
	where $\vec Y$ is a linear combination 
	of the $X_a$s with $a \in \lbrace 1, \dots , n \rbrace$. 
	Applying the 
	same decomposition on $Y' \in \mathcal V$, 
	we write 
	\begin{equation*}
		g \left( Y, Y' \right) = 
		Y^0 \, {Y'}^0 - \vec {Y} \cdot {\vec Y}'
	\end{equation*}
	with $\cdot$ denoting the standard inner product in the 
	spacelike hyperplane spanned by the $X_a$s. We write
	the length of $\vec Y$ with respect to this inner product
	as $\vec {\abs{Y}} := \sqrt{\vec Y \cdot \vec Y}$. 
	Now employ this notation for $K$ and $Z'$. 
	\\
	Since $K$ is future directed with respect 
	to $X_0$, \thref{Prop:timeorient} implies that 
	$K^0 > 0$. Lightlikeness of $K$ yields
	\begin{equation*}
		K^0 = \vec {\abs{K}} \, .
	\end{equation*}
	Similarly \thref{Prop:timeorient} and timelikeness of 
	$Z'$ implies that 
	\begin{equation*}
		Z'^0 > \vec {\abs{Z'}}\, .
	\end{equation*}
	Now by definition of the projection $K^\parallel$ of 
	$K$ onto the subspace 
	$\R Z'$	we have 
	\begin{equation*}
		g \left(K^\parallel, Z' \right) 
		= g \left( K, Z' \right) 
		= K^0 \, Z'^0 - \vec K \cdot \vec Z' \, .
	\end{equation*}
	Again by \thref{Prop:timeorient}, 
	$K$ is future directed with respect to 
	$Z'$ if and only if the expression is strictly positive. 
	So the Cauchy-Schwartz inequality applied to 
	\begin{equation*}
		\vec K \cdot \vec Z \leq 
		\abs{\vec K \cdot \vec Z}
		\leq \vec {\abs{K}} \, \vec {\abs{Z'}} 
		< K^0 \, Z'^0 
	\end{equation*}
	yields the assertion. \\
	``\ref{itm:orient2}'': The idea of proof is to 
	continuously transform the projections of the
	$K_a$s onto the orthogonal subspaces into each other and 
	observe that no reflection can occur in this process.\\ 
	The basis $X$ induces a 
	spacetime orientation $O'$ on $\mathcal V$. Since
	time inversion leaves the space orientation invariant, 
	we may assume that $X' \in O'$. Then there exists a 
	matrix $A \in \CLor_{n+1}$ such that $X'= X \cdot A$. 
	Now, since $\CLor_{n+1}$ is (path-)connected, there 
	exists a continuous function 
	\begin{equation*}
		B \colon [0,1] \to \CLor_{n+1} 
		\colon s \to B \left( s \right)
	\end{equation*}
	such that $B \left( 0\right) = \Id$ and 
	$B \left( 1\right) = A$. Define now the `intermediate' 
	bases $Y \colon [0,1] \to O'$ via 
	$Y \left( s\right) := X \cdot B \left( s\right)$
	for $s \in [0,1]$. This gives rise to corresponding 
	orthogonal projections 
	\begin{equation*}
		\pi^{\perp} \colon [0,1] \to \End{\mathcal V} 
		\colon s \to \pi^{\perp} \left( s\right)
	\end{equation*}
	with respect to $Y_0$. These can be written as 
	\begin{equation}
		\pi^\perp = Y_{a} \tp \ubar{Y}^a \, ,
		\label{eq:perpproj}
	\end{equation}
	where we sum over $a \in \lbrace 1, \dots, n \rbrace$. 
	Now collect the $K_a$s into 
	\begin{equation*}
		K:= K_{a} \tp \cbaseR^a \, 
	\end{equation*} 
	in order to define 
	\begin{equation*}
		K^\perp \left( s\right) := 
		\pi^\perp \left( s\right) \cdot K 
		= K^\perp _a \left( s\right) \, 
			\tp \cbaseR^a
	\end{equation*}
	for each $s \in [0,1]$. By assumption, 
	$K^\perp \left( 0\right)$ is right handed with 
	respect to $X= Y \left( 0 \right)$ in the sense 
	that there exists a $C \in \LieGL^+_{n}$ 
	such that 
	\begin{equation*}
		K^\perp \left( 0\right) = X \cdot 
		\begin{pmatrix}
			0 & 0 \\
			0 & C
		\end{pmatrix} \, .
	\end{equation*} 
	This motivates us to define the (matrix) 
	function $D$ via 
	\begin{equation*}
		K^\perp = Y \cdot 
		\begin{pmatrix}
				0 & 0 \\
				0 & D
		\end{pmatrix} \, ,
	\end{equation*} 
	which necessarily exists as a transformation in the 
	$Y_1, \dots, Y_n$ hyperplane. Use \eqref{eq:perpproj}
	to show this algebraically. Due to 
	\begin{equation*}
		\begin{pmatrix}
				0 & 0 \\
				0 & D
		\end{pmatrix}
		= \ubar Y \cdot \pi^\perp \cdot K \, ,
	\end{equation*}
	$D$ is continuous and invertible, i.e. we may write
	$D \colon [0,1] \to \LieGL_n$. Since 
	$D	\left( 0\right) = C \in \LieGL^+_n$, 
	$\LieGL^+_n$ is connected in $\LieGL_n$, 
	and 
	$D$ is continuous,  we have $D \left( 1\right) 
	\in \LieGL^+_n$. This proves the assertion. 
\end{Proof}
\par
If $\left(\mathcal V, g \right)$ is a Lorentzian vector 
space that is also Lorentz oriented via $O$, then we call $O$ 
\emph{compatible with} $g$, if the causal character of vectors 
with respect to $g$ and $O$ coincide. That is, 
if $g'$ is one of the Lorentz products induced by $O$, then 
$O$ is compatible with $g$, if and only if 
one of the equivalent conditions of
\thref{Thm:conf} holds. So a 
\emph{Lorentz-oriented Lorentz vector space} 
$\left( \mathcal V, g, O \right)$ is a Lorentz vector space 
$\left( \mathcal V, g \right)$ together with a Lorentzian 
orientation $O$ compatible with $g$. \emph{Space oriented, 
time oriented, spacetime oriented} and 
\emph{oriented Lorentz vector spaces} 
are defined accordingly, hence ignoring the possibility that 
Lorentzian orientations and Lorentz products need not 
be compatible. 
If $\left( \mathcal V, g, O \right)$ is 
a time-oriented Lorentz vector space and $c \in \R_+$ is a particular
distinguished number, e.g. $1$ or the speed of light 
(in vacuum), then a vector
$Z \in \mathcal V$ is called an \emph{observer vector}, if $Z$ is 
future directed timelike and 
$g \left( Z,Z\right)= c^2$. The motivation for 
this definition will become apparent in chapter 
\ref{chap:construction}. At this point we are finished 
with the discussion of the 
linear model.
\par
Knowing this, the 
notions of Lorentzian orientations $\mathcal O$ on manifolds 
$\spti$ follow immediately from the definition of a
$\mathcal G$-structure for the respective 
open Lie subgroups $\mathcal G$ of $\LieCO_{1,n}$. In this 
manner we obtain \emph{space-, time-, spacetime-} and 
\emph{causal orientations} on 
manifolds. Since the fiber $\mathcal O_q$ 
of $\mathcal O$ at each $q \in \spti$ is a linear 
Lorentzian orientation on the tangent space $\CapT_q \spti$, 
a Lorentzian orientation on a manifold yields a classification 
of tangent vectors into timelike, spacelike, future 
directed, etc. depending on the respective group $\mathcal G$. 
Common to all these particular cases is that they induce
a \emph{causal structure}, that is a $\LieCO_{1,n}$-structure 
on $\spti$. 
The following lemma gives a condition for the
existence thereof and hence a necessary condition for the 
existence of Lorentzian orientations. 
\begin{Lemma}[Existence of causal structures on manifolds]
	\label{Lem:COstruct}
	Let $\spti$ be a smooth $(n+1)$-manifold. Then $\spti$ 
	admits a $\LieCO_{1,n}$-structure if and only if it 
	admits an $\LieO_{1,n}$-structure. 
\end{Lemma}
\begin{Proof}
	``$\impliedby$'': This is trivial, since 
	$\LieO_{1,n} \subset \LieCO_{1,n}$ is a Lie subgroup. \\
	``$\implies$'': We need to show the existence of an 
	$\LieO_{1,n}$-reduction $\mathcal P '$
	of some $\LieCO_{1,n}$-structure $\mathcal P
	\subset \frameb{\CapT \spti}$. 
	By \thref{Rem:TransextG}\ref{sRem:transitP} on 
	page \pageref{sRem:transitP}, we can construct 
	$\mathcal P$ by 
	taking appropriate local sections and making 
	sure that the transition functions take values in 
	$\LieO_{1,n}$. So we choose a trivializing 
	cover $\set{\mathcal{U}_\alpha}{\alpha \in I}$ with 
	frame fields $\overset{\alpha}{X} \in \Gamma^\infty 
	\left(\mathcal{U}_\alpha, \mathcal P \right)$
	and transition functions $\overset{\alpha \beta}{A} \in 
	C^\infty \left(\mathcal{U}_\alpha 
	\cap \mathcal{U}_\beta, \LieCO_{1,n}  \right)$ 
	for all $\alpha, \beta \in I$ with non-empty
	$\mathcal{U}_\alpha \cap \mathcal{U}_\beta$. 
	From the Lie group isomorphism between 
	$\LieCO_{1,n}$ and $\R_+ \cross \LieO_{1,n}$ 
	(cf. \eqref{eq:COiso} on page \pageref{eq:COiso}), 
	we are motivated to define 
	\begin{equation*}
		\overset{\alpha \beta}{\Lambda} := 
		{\abs{\det \overset{\alpha \beta}{A} 
		}}^{-1/(n+1)} \, \, \overset{\alpha \beta}{A} 
	\end{equation*}
	taking values in $\LieO_{1,n}$. Employing the 
	$\overset{\alpha}{X}$ as sections of $\mathcal P'$ 
	and checking the cocycle condition \eqref{eq:cocycle}
	for the $\overset{\alpha \beta}{\Lambda}$ yields the result. 
\end{Proof}
Since the choice of transition functions for the 
$\LieCO_{1,n}$-structure in the above proof 
was arbitrary, a causal structure on $\spti$ 
does in general not induce a unique Lorentzian metric $g$. 
Yet as in the linear case, any causal structure uniquely 
determines the 
metric up to a strictly positive factor, i.e. if 
$\mathcal P, \mathcal P'$ are $\LieO_{1,n}$-reductions 
of a causal structure on an $(n+1)$-manifold $\spti$ 
with respective Lorentzian metrics $g$ and $g'$, then 
there exists a strictly positive function 
$f \in C^\infty \left( \spti, \R_+\right)$ such that 
\begin{equation*}
	g'= f \, g \, .
\end{equation*}
The transformation from $g$ to $g'$ is called 
a \emph{conformal transformation} and then 
$g, g'$ are called \emph{conformally equivalent}. 
It is elementary to show that this is indeed an equivalence 
relation on the set of Lorentzian metrics on a manifold 
$\spti$. Conversely, if a Lorentzian metric $g$ on 
$\spti$ is given up to conformal equivalence, this 
uniquely determines a causal structure on $\spti$. 
If we consider Lorentzian orientations instead, 
then a Lorentzian metric alone is not enough to reconstruct 
the orientation. However, we may ask for 
a given Lorentzian orientation $\mathcal O$ to be 
\emph{compatible} with a Lorentzian metric $g$ by 
requiring that some (and hence every) metric induced by 
$\mathcal O$ is conformally equivalent to $g$. So 
we define a \emph{time, space} or \emph{spacetime oriented 
Lorentzian manifold} to be a tuple 
$\left( \spti, g, \mathcal O\right)$ such 
that $\left( \spti, g\right)$ is a Lorentzian manifold and 
$\mathcal O$ is a compatible time, space or spacetime 
orientation on $\spti$, respectively. Similarly, 
an oriented Lorentzian manifold can be viewed 
as a Lorentzian manifold equipped with a compatible 
causal orientation. 
\par
Spacetime oriented Lorentzian manifolds are the main 
objects of interest in chapter \ref{chap:construction}, 
so we complete this section with a statement on their 
existence. 
\begin{Proposition}[Existence of spacetime orientations]
	\label{Prop:existspacetime}
	Let $\left( \spti, g\right)$ be a Lorentzian manifold.\\ 
	Then it admits a compatible spacetime orientation 
	if and only if it is orientable and 
	there exists a global time-like vector field.  
\end{Proposition}
\begin{Proof}
	'' $\implies$ ´´: Since $\CLor_{n+1} \subset \LieGL^+_{n+1}$ 
	for each $n \in \N$, the spacetime orientation $\mathcal O$
	on the $(n+1)$-manifold $\spti$
	can be extended to an ordinary orientation on $\spti$. For 
	the existence of the vector field we note that 
	$\mathcal O$ yields a time-orientation on $\spti$ 
	and then a folklore theorem in relativity states that 
	the existence of a time orientation is equivalent to the 
	existence of a global timelike vector field
	(see e.g. \cite{O'Neill}*{Lem. 5.32}). The proof thereof 
	is essentially a partition of unity argument employing 
	the fact that the sum of two future-directed timelike 
	vector fields is again future-directed timelike. \\
	'' $\impliedby$ ´´: Without loss of generality, we may 
	assume that the global timelike vector field $Z$ is 
	normalized with respect to $g$. Now choose an 
	at most countable trivializing cover 
	$\set{\mathcal U_\alpha}{\alpha \in I}$ such that 
	$\overset{\alpha}{X}$ are smooth orthonormal 
	frame fields over each $\mathcal U_\alpha$. By applying
	appropriate Lorentz transformations, we may 
	assume
	\begin{equation}
		Z \evat{\mathcal U_\alpha}= 
		\overset{\alpha}{X}_0
		\label{eq:Zbase}
	\end{equation} 
	for each 
	$\alpha \in I$. Moreover, if $\overset{\alpha}{X}$ is 
	not right handed, we may multiply by $-\Id 
	\in \LieO_{1,n}$ to make it right handed. Now, for each 
	$\alpha, \beta \in I$ with non-empty 
	$\mathcal U_\alpha \cap \mathcal U_\beta$ 
	there exists an 
	$\overset{\alpha \beta}{\Lambda} 
	\in C^\infty \left( 
	\mathcal U_\alpha \cap \mathcal U_\beta, 
	\LieO_{1, n}\right)$ such that 
	\begin{equation*}
		\overset{\beta}{X} = 
		\overset{\alpha}{X} \cdot \overset{\alpha \beta}{\Lambda}
		\, .
	\end{equation*}
	Since each $\overset{\alpha}{X}$ is right-handed, 
	$\det \Lambda > 0$. By \eqref{eq:Zbase}, 
	$\Lambda^i{}_0 = \delta^i_0$ and as it maps into 
	$\LieO_{1, n}$, we also have $\Lambda^0_i = \delta^0_i$. 
	Thus $\Lambda$ takes values in the product Lie group
	$\lbrace 1 \rbrace \cross \LieSO_n$. As this is a 
	subgroup of $\CLor_{n+1}$, extension yields the assertion. 
\end{Proof}
Indeed, the proof shows that a spacetime oriented 
Lorentzian $(n+1)$-manifold admits a
$(\lbrace 1 \rbrace \cross \LieSO_n)$-reduction of
its frame bundle. It needs to be stressed, however, that
this reduction is highly non-unique and thus, 
at least without further physical motivation of the choice of
global vector field, a mere mathematical tool 
without intrinsic meaning. Moreover, we may combine 
\thref{Prop:existspacetime} 
with \thref{Lem:COstruct} to conclude that a manifold can be
equipped with a Lorentzian metric and a compatible spacetime 
orientation if and only if it is orientable and admits a 
global nowhere-vanishing vector field.%
\footnote{In the reverse implication we employed the 
fact, that the existence of a nowhere-vanishing vector field 
$X$ on $\spti$ 
implies the existence of a Lorentzian metric $g$. To show 
this, choose a Riemannian metric $h$ on $\spti$ and assume 
for convenience that $h \left( X, X\right) = 1$. Then 
$g := 2 \, h \cdot X \tp X \cdot h - h$ is a Lorentzian metric with 
respect to which $X$ is timelike and all vectors  
in $\ker \left( X \cdot h\right)$ are orthogonal to 
$X$.}

\section{Connections on the Tangent Bundle}
\label{sec:connecgen}

In this section we discuss the geometry of tangent 
bundles equipped with a covariant derivative. We approach 
the subject by first recapitulating the theory of 
Ehresmann connections on fiber bundles and then 
relating this general point of view to 
covariant derivatives on the tangent bundle. By taking this 
top-down perspective, we intend to convey an adequate
understanding of the notion of connectors. 
This in turn is a prerequisite for understanding the detailed 
geometry of the space-time splitting in terms of 
Jacobi fields. Appropriate references are given in the
respective subsections. 

\subsection{Ehresmann Connections}
\label{ssec:Ehresmann}

As stated above, we recollect the main theory of 
Ehresmann connections on fiber bundles here. 
A more in-depth treatment 
of connections and the closely related concept of 
parallel transport 
can be found in the German book by Baum \cite{Baum}
as well as the English ones by Poor \cite{Poor} and Rudolph et al 
\cite{Rudolph1}. 
\par 
So let $\mathcal E$ be a fiber bundle over a manifold 
$\spti$ with bundle 
projection $\pi \colon \mathcal E \to \spti$. 
Then for every $q \in \mathcal Q$ the fiber 
$\pi^{-1} \left(  \lbrace q \rbrace \right)$ is an 
embedded submanifold of $\mathcal E$ 
(diffeomorphic to the typical fiber) 
due to the regular value theorem. Moreover, the fact that 
$\pi$ is a submersion also implies that the kernel of $\pi_*$,
denoted by $\ker \pi_*$, defines a (smooth, 
regular geometric) distribution on the manifold $\mathcal E$. 
$\mathcal V := \ker \pi_*$ is called the 
\emph{vertical distribution}. 
In accordance, we call a vector $X \in \CapT \mathcal E$ 
\emph{vertical}, if it is tangent to the distribution $\mathcal V$. 
Again by the regular value theorem, vertical vectors 
in $\CapT \mathcal E$ are precisely those that are tangent to 
the fiber over their base point in $\spti$. Comparing dimensions, 
this shows that the fibers are the integral manifolds of the vertical 
distribution, i.e. $\ker \pi_*$ is integrable. One can now ask 
for a complementary distribution on $\mathcal E$, i.e. 
a smooth distribution $\mathcal H$ on $\mathcal E$ such that 
\begin{equation}
	\CapT \mathcal E = \mathcal V \oplus \mathcal H \, ,
	\label{eq:splitE}
\end{equation}
where $\oplus$ is the Whitney sum. The Whitney sum amounts to 
taking the fiber-wise direct sum of vector spaces and equipping 
the resulting set with a `natural' manifold structure. 
A distribution $\mathcal H$ satisfying \eqref{eq:splitE}
is called an \emph{Ehresmann connection} or, equivalently, 
a \emph{horizontal distribution}. They are highly non-unique
and in the generic case not integrable. Obviously, 
we call a vector \emph{horizontal}, 
if it is tangent to $\mathcal H$. 
\par
If we wish to take a less abstract 
perspective, an Ehresmann connection 
$\mathcal H$ 
can be constructed from a
\emph{vertical projection (endomorphism field) 
$\pi^{\mathcal V}$}. This is a (smooth) 
tensor field on $\mathcal E$ taking values in 
$\End \left( \CapT \mathcal E \right) = \CapT \mathcal E \tp 
\CapT^* \mathcal E$ and satisfying $\pi^{\mathcal V} 
\cdot \pi^{\mathcal V} = \pi^{\mathcal V}$ as well as 
$\pi^{\mathcal V} \cdot X \in \mathcal V$ for every
$X \in \CapT \mathcal E$. The \emph{horizontal projection (endomorphism field) 
$\pi^{\mathcal H}$} is then just 
\begin{equation*}
	\pi^{\mathcal H} := \Id - 
	\pi^{\mathcal V} \, .
\end{equation*}
Since $\pi^{\mathcal V}$ has constant rank, so does 
$\pi^{\mathcal H}$ and therefore the equation 
\begin{equation*}
	\mathcal H := \pi^{\mathcal H} 
\left( \CapT \mathcal E \right) = \ker \pi^{\mathcal V}
\end{equation*}
defines an Ehresmann connection on $\mathcal E$.  
\par
We thus conclude that connections exist on general 
fiber bundles $\mathcal E$. However, if the fiber bundle 
itself has a 'symmetry', one usually requires the induced 
'infinitesimal symmetry' to carry over to $\mathcal H$. 
We give meaning to this statement in the following 
section. 

\subsection{Covariant Derivatives \& Connectors}
\label{ssec:connecTQ}

As the tangent bundle $\CapT \mathcal Q$ of a manifold 
$\spti$ is a fiber bundle, we may equip it with an Ehresmann 
connection. Our choice is, however, not arbitrary, but we
wish to relate it to the concept of covariant derivative in 
this section. Note that the following treatment can 
be generalized to arbitrary vector 
bundles in a straightforward manner, but we are primarily 
interested in the specific case of the tangent bundle. 
We recommend the books by Poor \cite{Poor}*{2.49 ff.}, 
Burns and Gidea \cite{Burns}*{\S 5.8} as well as the one 
by the group of French mathematicians under the 
pseudonym 'Arthur Besse' \cite{Besse0}*{1.59 ff.} as 
references for connectors and vector bundle connections. 
The first book \cite{Burns} provides a coherent motivation, 
the second one \cite{Poor} gives 
a good abstract definition and embeds it into the general theory, 
while the third one \cite{Besse0} shows the relation to 
sprays and the symplectic point of view.
\par 
To start off, consider a curve 
$X \colon \mathcal I \to \CapT \spti$ in 
the tangent bundle of an $n$-manifold $\spti$, 
equipped with a covariant derivative $\nabla$, and let 
$\pi \colon \CapT \spti \to \spti$
be the bundle projection. Intuitively, $X$ can be thought 
of as the curve 
\begin{equation*}
	\gamma := \pi \circ X \colon \mathcal I 
	\to \spti \colon \tau \to \gamma \left( \tau \right) 
	= \pi \left( X_{\tau} \right)
\end{equation*} 
with a vector $X_\tau$ attached to it for every 
$\tau$. Recalling the terminology introduced in section 
\ref{sec:pullback}, $X$ is a vector field 
along $\gamma$. Note that $X_\tau$ does not need to be tangent to 
$\gamma$. If $X$ is parallel transported along $\gamma$, then 
by definition it satisfies 
\begin{equation}
	\left(\frac{\nabla X}{\d \tau} \right)_\tau = 0 
	\label{eq:Xparallel}
\end{equation}
for every $\tau \in \mathcal I$. Moreover, if the image of 
$\gamma$ is contained in the domain $\mathcal U$ of a 
coordinate map $\kappa$, then equation 
\eqref{eq:Xparallel} reads in those coordinates 
\begin{equation}
	\dot v^k \left( \tau \right) + 
	\Gamma^k{}_{ij} \left( \gamma \left( \tau \right) \right)
	\, \dot \kappa^i \left( \tau \right) \, 
	v^j \left( \tau \right) = 0
	\, . 
	\label{eq:Xparallelcoord}
\end{equation}
Here the $\Gamma^k{}_{ij}$s are the connection coefficients 
with indices $i,j,k \in \lbrace 1, \dots, n  
\rbrace$, the $\kappa^i$s and $v^k$s are the components of 
$X$ with respect to the coordinate map $\kappa^i$ and the 
respective coordinate vector field $\partial/\partial \kappa^k$, 
and the dot denotes differentiation of the components with 
respect to $\tau$. 
Employing an analogy in Newtonian physics, 
equation \eqref{eq:Xparallelcoord} shows 
that we can tell whether a vector $X_\tau$ is (infinitesimally) 
parallel at time 
$\tau$ by knowing the position 
$\kappa \left( \tau \right)$ and velocity $\dot \kappa 
\left( \tau \right)$ of $\gamma$ as well as the vector  
$v\left( \tau \right) \in \R^{n}$ and its rate of change $ \dot 
v\left( \tau \right) \in \R^{n}$ at time $\tau$. 
\par
Now recall that the chart $\left( \mathcal U, \kappa \right)$ 
on $\spti$ induces the canonical bundle chart 
$( \pi^{-1} ( \mathcal U ), \allowbreak
( \kappa \circ \pi, \allowbreak v ) )$ on 
$\CapT \spti$ as follows: If $Y = Y^i \, \Evat{
\partial_i}{q}$ is a vector 
in $\pi^{-1} \left( \mathcal U \right) \subseteq \CapT \spti$ 
with base point $\pi \left( Y\right) = q \in \mathcal U$, then 
$\left( \kappa \circ \pi, v \right) \left(Y \right) = \left( \kappa 
\left( q\right), Y^i \, \baseR_i \right) \in \R^{2n}$. 
Note that 
we do not notationally distinguish between 
$\kappa$ and $\kappa \circ \pi$ in order to avoid 
cluttery formulas. It is also common to write $\dot \kappa$ 
instead of $v$ for obvious reasons. 
\par 
If we repeat this procedure for the double tangent bundle 
$\pi' \colon \CapT\CapT \spti \to \CapT \spti$, we get 
canonical bundle coordinates $\left(\kappa, v, \dot \kappa, 
\dot v \right) $ on 
${\pi'} ^{-1} \left( \pi^{-1} \left( \mathcal U \right)
\right) \subseteq \CapT \CapT \spti $
induced by the chart $\left( \mathcal U, \kappa \right)$. 
With respect to these coordinates and for every 
$\tau \in \mathcal I$ the vector $\dot X _\tau :=  X_{*} 
 \left(\left( \partial/\partial \tau \right)_\tau\right) 
\in \CapT \CapT \spti$ takes precisely the form 
$\left( \kappa \left( \tau\right), v \left( \tau\right)
, \allowbreak \dot \kappa \left( \tau\right), \dot v 
\left( \tau\right)\right)$ as before. This is true for 
any canonical bundle chart induced 
by a chart on $\mathcal U \subseteq \spti$.
So from an invariant perspective, the vector 
$ \dot X _\tau $ in the 
double tangent bundle $\CapT\CapT \spti$ suffices to check 
whether $X_\tau$ is (infinitesimally) parallel at time 
$\tau$. Namely, we evaluate $\dot X _\tau $
in arbitrary canonical bundle coordinates and check whether it 
satisfies 
\begin{equation}
	\dot v^k + 
	\Gamma^k{}_{ij} \left( q \right)
	\, \dot \kappa^i \, 
	v^j = 0
	\, ,
	\label{eq:Xparallelcoord1}
\end{equation}
where $ q := \pi \circ \pi' \bigl(  \dot X _\tau  
\bigr) $. 
As every $Z \in \CapT \CapT \spti$ can be 
considered the tangent vector of such a curve $X$ in 
$\CapT \spti$ at some $\tau$, we obtain a condition for
elements of $\CapT \CapT \spti$ to be '(infinitesimally) 
parallel'. 
\par 
In fact, we have implicitly constructed a smooth map 
$\Kon \colon \CapT \CapT \spti 
\to \CapT \spti$, which is locally 
given by 
\begin{equation}
	\left( \kappa, v, \dot \kappa, \dot v \right)
	\to \left( \kappa, \left( \dot v^k + \Gamma^k{}_{ij} 
	\left( q \right) \,  
	\dot \kappa^i 
	v^j \right) \baseR_k 
	\right)
	\, 
	\label{eq:Kloc1}
\end{equation} 
with $q:= \pi \circ \pi' \left( Z \right)$. Moreover, for 
arbitrary curves $X \colon \mathcal I \to \CapT \spti$ 
and all $\tau \in \mathcal I$ it satisfies 
\begin{equation}
	 \Kon \bigl( \dot X_ \tau \bigr) = \left( 
	 \frac{\nabla X}{\d \tau}
	 \right)_\tau \, . 
	 \label{eq:DefK}
\end{equation}
Condition \eqref{eq:DefK} gives 
an indirect global definition of $\Kon$ and conversely, 
if such a $\Kon$ is given, this uniquely determines a covariant 
derivative or, equivalently, a \emph{Koszul connection} $\nabla$
on $\CapT \spti$. The converse follows from 
the fact that $X$ is arbitrary. 
So we find that $\Kon$ defined via \eqref{eq:DefK} 
is a uniquely defined (smooth) $\CapT \spti$-valued 
$1$-form on $\CapT \spti$, locally given by
\begin{equation}
	\Kon =  \Gamma^k{}_{ij} \, v^j \,  
	\partd{}{\kappa^k} \tp \d \kappa^i 
	+ \partd{}{\kappa^i} \tp \d v^i
	\label{eq:Kloc2}
\end{equation}
in accordance with \eqref{eq:Kloc1}. We call 
$\Kon \in \Omega^1 \left(\CapT \mathcal Q, 
\CapT \mathcal Q\right)$ as defined by \eqref{eq:DefK}
a \emph{connector (on the tangent bundle 
$\CapT \spti$ induced by $\nabla$)}. If $\nabla$ is the 
Levi-Civita connection, then $\Kon$ is called the 
\emph{Levi-Civita connector}. 
\par
If the connection $\nabla$ is 
torsion-free, then this can also be expressed in terms of $\Kon$
via the \emph{canonical flip} $\Fl \colon \CapT\CapT 
\spti \to \CapT\CapT \spti$. In canonical bundle 
coordinates it is given by the smooth mapping 
\begin{equation*}
	\left(\kappa, v, \dot \kappa, \dot v \right)
	\to \left( \kappa, \dot \kappa, v, \dot v\right) \, .
\end{equation*} 
Since both $\dot \kappa$ 
and $v$ can be viewed as tangent vectors, this definition of 
$\Fl$ is invariant. Moreover, 
$\Fl$ is a diffeomorphism due to the fact
that $\Fl \circ \Fl$ is the identity. 
Now the vanishing torsion of $\nabla$ 
implies that 
$\Gamma^k{}_{ij}\equiv \Gamma^k{}_{ji}$ for the connection
coefficients, so 
the equation $\Kon \circ \Fl = \Kon$ indeed defines a 
\emph{torsion-free connector}. 
\par
From the global \eqref{eq:DefK} or the local expression 
\eqref{eq:Kloc2} of $\Kon$, we directly observe that it 
has constant rank $n$. Hence its kernel determines a smooth 
rank $n$ distribution $\ker \Kon$ on $\CapT \spti$. From the local 
expression of 
\begin{equation*}
	\pi_* = \partd{}{\kappa^i} \tp \d \kappa^i 
\end{equation*} 
we deduce that a vertical vector $Y \in {\pi'} ^{-1} 
\left( \pi^{-1} \left( \mathcal U \right) \right) 
\subseteq \CapT \CapT \spti$ can be written as 
\begin{equation*}
	Y = \overset{v}{Y}{}^i \, \Evat{\partd{}{v^i}}
	{\pi' \left( Y \right)}
\end{equation*}
with $\overset{v}{Y}{}^i := \dot v ^i \left( Y \right)$
and $\overset{\kappa}{Y}{}^i := \dot \kappa ^i \left( Y \right)
=0$
for each $i \in \lbrace 1, \dots, n \rbrace$. 
As canonical bundle charts exist everywhere on 
$\CapT \CapT \spti$, a vector 
$Y \in \mathcal V \subset \CapT \CapT \spti$ 
is vertical and satisfies 
$\Kon \left(Y \right) = 0$ if and only if $Y=0$. 
This proves that $\mathcal H := \ker \Kon$ defines a horizontal 
distribution on $\CapT \spti$. 
In canonical bundle coordinates horizontal vectors 
$Y \in {\pi'} ^{-1} 
\left( \pi^{-1} \left( \mathcal U \right) \right)$
with ${\pi'} \left( Y \right) = y$ and $\pi \left( y \right) = q$
take the form 
\begin{equation}
	Y= 	\overset{\kappa}{Y}{}^i \, \Evat{\partd{}{\kappa^k}}{y}
		- \Gamma^k{}_{ij} \left( q \right) \, 
		\overset{\kappa}{Y}{}^i
		y^j \, \Evat{\partd{}{v^k}}{y} \, ,
	\label{eq:Yhor}
\end{equation}
and thus precisely those vectors 
$Y \in \CapT \CapT \spti$ are horizontal, that are 
'(infinitesimally) parallel' in the aforementioned sense 
of satisfying \eqref{eq:Xparallelcoord1}. This is the reason 
why $\mathcal H$ is called a connection. Moreover, 
since $\ker \Kon = \ker \pi^{\mathcal V}= \mathcal H$
we can explicitly determine the local expression of the 
vertical projection in canonical bundle coordinates 
\begin{equation}
	\pi^{\mathcal V} = 
	\Gamma^k{}_{ij} \, v^j \, \partd{}{v^k}\tp \d \kappa^i 
	+ \partd{}{v^i}\tp \d v^i 
	\label{eq:piVloc}
\end{equation}
and thus the local expression for the horizontal projection reads 
\begin{equation}
	\pi^{\mathcal H} = \Id - \pi^{\mathcal V} 
	= 	\partd{}{\kappa^i}\tp \d \kappa^i 
		- \Gamma^k{}_{ij} \, v^j \, \partd{}{v^k}\tp \d \kappa^i 
	\, . 
	\label{eq:piHloc}
\end{equation}
\par
We would now like to consider the reverse construction. 
So assume we are given a horizontal distribution $\mathcal H$ in the 
sense of section \ref{ssec:Ehresmann} on page 
\pageref{ssec:Ehresmann} 
with vertical projection $\pi^{\mathcal V}$ on the tangent bundle. 
We would like 
$\mathcal H$ to determine a unique $1$-form
$\Kon \in \Omega^1\left( \CapT \spti, \CapT \spti 
\right)$, which should be a 'connector' in the sense 
that we can use it to construct a Koszul connection on 
$\CapT \spti$ via equation \eqref{eq:DefK}. In general 
an element $\Kon \in \Omega^1\left( \CapT \spti, \CapT \spti 
\right)$ can be locally written as 
\begin{equation}
	\Kon = \overset{\kappa}{\Kon}{}^i{}_j \, 
	\partd{}{\kappa^i} \tp \d \kappa^j +
	\overset{v}{\Kon}{}^i{}_j \, 
	\partd{}{\kappa^i} \tp \d v^j
	\label{eq:Kgen}
\end{equation} 
with smooth functions $\overset{\kappa}{\Kon}{}^i{}_j$, 
$\overset{v}{\Kon}{}^i{}_j$ on 
$\pi^{-1} \left( \mathcal U \right)$ 
and indicies $i,j \in \lbrace 1, \dots, n \rbrace$. 
Obviously, we would like to find invariant conditions such 
that $\Kon$ takes the form \eqref{eq:Kloc2} -- which is equivalent
to $\pi^{\mathcal V}$ taking the form \eqref{eq:piVloc} 
(as $\ker \pi^{\mathcal V} = \ker \Kon$) and 
then we just set $\mathcal H = \ker \pi^{\mathcal V} = \ker 
\Kon$. 
Those conditions on $\Kon$ might then in turn lead to 
restrictions on $\mathcal H$, i.e. we only accept those 
Ehresmann connections $\mathcal H$ on $\CapT \spti$ that 
can be used to construct a sensible $\Kon$. 
\par 
To find such invariant conditions on $\Kon$, consider the 
multiplication map 
\begin{equation}
	M \colon \R \cross \CapT \spti \to \CapT \spti
	\colon \left( \lambda, X \right) \to 
	M_\lambda \left( X \right):= \lambda X
	\label{eq:multop}
\end{equation}
and the \emph{vertical lift $\tilde {X}_Y$ of $X \in \CapT \spti$ 
at $Y \in \pi^{-1} \left(  \lbrace \pi
\left( X \right) \rbrace \right)$},  
as defined by 
\begin{equation}
	\tilde {X}_Y \left( f \right) 
	:= \Evat{\partd{}{s}}{0} f \left( Y + s \, X \right)
	\label{eq:vertlift}
\end{equation}
for every $f \in C^\infty \left( \CapT \spti, \R \right)$. 
It is straightforward to show that \eqref{eq:vertlift} 
defines a vector field $\tilde {X}$ over the fiber 
$\pi^{-1} \left(  \lbrace \pi \left( X \right) \rbrace \right)$. 
In particular $\tilde X$ is vertical at each point, 
hence the name \emph{vertical vector field}. Fixing 
$\lambda \in \R$ and looking 
at $M_\lambda$ in canonical bundle coordinates induced by the 
chart $\left( \mathcal U, \kappa \right)$ on $\spti$, 
it is locally given by 
\begin{equation*}
	\left( \kappa, v \right) \to \left( \kappa, \lambda v 
	\right) 
\end{equation*}
and thus over $\pi^{-1} \left( \mathcal U \right)$
\begin{equation*}
	\left( M_\lambda \right)_*	= 
	\partd{}{\kappa^i} \tp \kappa^i +
	\lambda \, \partd{}{v^i} \tp v^i 
	\, .
\end{equation*}
Combining this with our desired expression 
\eqref{eq:Kloc2} for $\Kon$, we find $M_\lambda^*
\Kon = \lambda 
\Kon$. Similarly, we find for $q \in \mathcal U$,  
$X = X^i \, \Evat{\left(\partial/ \partial \kappa^i \right)}{q}
 \in \CapT_q \spti$ 
\begin{equation*}
	\tilde X = X^i \, \partd{}{v^i} \, ,
\end{equation*}
and combining this again with \eqref{eq:Kloc2}, we get 
$\Kon ( \tilde X ) =X$. Thus to get our desired 
expression \eqref{eq:Kloc2} from the general one 
\eqref{eq:Kgen}, $\Kon$ necessarily has to satisfy
\begin{equation}
	M_\lambda^*\Kon = \lambda \Kon \quad \quad \forall \, 
	\lambda \in \R \text{, and} \quad  
	\Kon( \tilde X ) = X \quad \quad \forall \, X \in 
	\CapT \spti \, . 
	\label{eq:Kcond}
\end{equation}
In fact, starting from \eqref{eq:Kgen}
the second condition yields 
$\overset{v}{\Kon}{}^i{}_j = \delta^i_j$ and the 
first condition implies that each 
$\overset{\kappa}{\Kon}{}^i{}_j$ is a first degree 
homogeneous polynomial in the components of $v$. Hence these two 
conditions specify $\Kon$ modulo the (consistent) choice of 
the functions $\Gamma^k{}_{ij}$ in each chart. As the 
second condition is only relevant for vertical vectors, it 
does not put any restriction on $\mathcal H = \ker \Kon$. 
Regarding the first one, we observe that for every horizontal 
$Y \in \CapT \CapT \spti$ and $\lambda \in \R$ 
\begin{equation*}
	M_\lambda^* \Kon \left( Y \right) 
	= \Kon \left( \left( M_\lambda \right)_* Y \right)
	= \lambda \, \Kon \left( Y \right) = 0 \, ,
\end{equation*}
hence $\left(M_\lambda\right)_* Y \in \mathcal H$ and thus 
$\left(M_\lambda \right)_* \mathcal H = \mathcal H$. This 
is the sought-after condition. In particular, for 
$X \in \CapT \spti$ we can define 
${\mathcal H} _X := \mathcal H \cap \CapT_X \CapT \spti$ and 
then 
\begin{equation}
	\left( M_\lambda \right)_* {\mathcal H} _X = 
	{\mathcal H}_{\lambda X}	
	\label{eq:Hhom}
\end{equation}
for every $\lambda \in \R$.
\par 
We have thus motivated a natural definition of the word 
'tangent bundle connection' and of 'connectors' induced by them. 
\begin{Definition}[Tangent bundle connection]
	\label{Def:tbconnect}
\begin{subequations}
	Let $\mathcal H$ be an Ehresmann connection on 
	the tangent bundle
	$\pi \colon \mathcal \CapT \spti \to 
	\spti$ of a manifold $\spti$. 
	Then $\mathcal H$ is called a \emph{tangent bundle 
	connection}, if \eqref{eq:Hhom} holds 
	for every $X \in \CapT \spti$ and
	$\lambda \in \R$. An element  
	$\Kon \in \Omega^1\left( \CapT \spti, 
	\CapT \spti \right)$ is called a \emph{connector (on the 
	tangent bundle)}, if it satisfies \eqref{eq:Kcond}. A 
	connector is said to be \emph{induced by a tangent bundle 
	connection $\mathcal H$}, if $\mathcal H = \ker \Kon$. 
\end{subequations}
\end{Definition}
If we replace $\CapT \spti$ by a general (real) vector bundle 
$\mathcal E$ in \thref{Def:tbconnect}, we obtain a definition 
of vector bundle connections and connectors on vector bundles. 
As noted before, one commonly requires Ehresmann 
connections on particular fiber bundles to satisfy an 
additional 'infinitesimal symmetry condition'. For the tangent 
bundle and analogously for general vector bundles this 
condition is given by \eqref{eq:Hhom}. Colloquially speaking, 
the condition guarantees that the Ehresmann connection respects 
the vector bundle structure of $\CapT \spti$.
\par 
Summing up, a Koszul connection $\nabla$ on $\CapT \spti$ 
gives rise to a connector $\Kon$, whose kernel defines a 
tangent bundle connection $\mathcal H$. Conversely, a tangent 
bundle connection $\mathcal H$ can be considered 
as a choice of vertical 
projection $\pi^{\mathcal V}$, which directly induces a 
corresponding connector $\Kon$ on $\CapT \spti$. Locally this 
works via \eqref{eq:piVloc} and \eqref{eq:Kloc2}. 
The connector then gives rise to a Koszul 
connection $\nabla$ on $\CapT \spti$ via \eqref{eq:DefK} and 
a corresponding notion of parallel transport. 

\section{Jacobi Fields and the Lorentzian Exponential}
\label{sec:Jacobi}

In the following we review the notion of geodesic 
variation, its relation to Jacobi fields and the 
exponential map within Lorentzian geometry. This will be 
essential in our construction later. 
For a discussion of Jacobi fields in Lorentzian geometry 
going beyond the material presented here, we refer to the 
books by Beem, Ehrlich and Easley \cite{Beem}*{Chap. 10}, 
O'Neill \cite{O'Neill}*{Chap. 8 \& 10}, and 
Hawking and Ellis \cite{Hawking}*{Chap. 4 \& 8}. 
A coherent introduction to the topic in the context 
of Riemannian geometry can also be found 
in the book by Burns and Gidea 
\cite{Burns}*{\S 4.5, \S 5.5, \S 5.7, \S 5.9}, as well as 
the one by Sakai \cite{Sakai}*{Chap. III \& IV}.
\par 
Throughout this section $\left( \mathcal Q, g\right)$ is a 
(smooth) Lorentzian manifold. Although the 
discussion applies more generally to 
pseudo-Riemannian manifolds, we are 
only interested in the Lorentzian case. 
\par 
Recall that for any Koszul connection $\nabla$ on the tangent 
bundle, we call a (smooth) curve $\gamma \colon \mathcal I 
\to \spti \colon 
r \to \gamma \left( r\right)$ an \emph{autoparallel}, if its 
tangent vector is parallel to itself, i.e. 
\begin{equation}
	\left( \frac{\nabla \dot \gamma}{\d r} \right)_r = 0 
	\quad \quad \forall r \in \mathcal I \, .
\end{equation}
In the Lorentzian case, we commonly take $\nabla$
to be the Levi-Civita connection with respect to $g$ 
and, neglecting conceptual subtleties, we call autoparallels 
\emph{geodesics}. Intuitively, they are straight lines 
in a curved geometry. Metricity of $\nabla$ implies that for
geodesics $\gamma$ the function $g \left( \dot \gamma, 
\dot \gamma\right)$ is a constant. By definition, a
\emph{curve} $\gamma$ in $\spti$ is 
\emph{time-, light-} or \emph{spacelike}, if each tangent vector 
$\dot \gamma _r$ is time-, light- or spacelike, 
respectively. So non-trivial geodesics are either 
time-, light- or spacelike.
\par
Often we are not just interested in one single geodesic, 
but also in its behavior with respect to a parameter: 
If $\gamma\colon \mathcal I \to \spti$ is a curve 
and for some $\epsilon \in \R_+$ the (smooth) map
\begin{equation}
	\theta \colon \left(- \epsilon, \epsilon 
	\right) \cross \mathcal I \to \spti 
	\colon \left( s,r \right) \to \theta_s \left(r \right)
	\label{eq:gvariation}
\end{equation}
satisfies $\theta_0 = \gamma$, then we call 
$\theta$ a \emph{variation of $\gamma$}. If in addition 
$\gamma$ itself and $\theta_s$ are geodesics for all 
$s \in \left(- \epsilon, \epsilon \right)$, we call 
$\theta$ a \emph{geodesic variation of $\gamma$}. 
More generally, one might want to let the domain of $\theta$ 
be an open, connected subset of $\R^2$, but, as the 
definition \eqref{eq:gvariation} is fairly standard in 
the literature and this is only a brief review, we shall not 
consider this case.
\par 
Given a geodesic variation $\theta$ of (a geodesic) $\gamma$, 
we are naturally led to consider two particular tangent vectors 
at the point $\gamma \left( r\right) = \theta_0 \left( r\right)$, 
namely  
$\theta_* \left(
\partial/\partial r\right)_{\left( 0, r\right)}$ and 
$\theta_* \left(
\partial/\partial s\right)_{\left( s, 0\right)}$. The former 
gives the tangent vector $\dot \gamma_r$ at parameter value $r$
and the second one gives its `infinitesimal displacement' with 
respect to the variation. Recalling the 
terminology from section \ref{sec:pullback}, the smooth mapping 
\begin{equation*}
	J \colon \mathcal I \to \CapT \spti
	\colon r \to J_r := \left(\theta_* \partd{}{s} \right)
	_{\left( 0,r\right)}
\end{equation*}
is vector field along $\gamma$. If we define the Riemann tensor 
field $\mathcal R$ via 
\begin{equation*}
	\mathcal{R} \left(X,Y \right)Z 
	:= \nabla_X\nabla_Y Z -\nabla_Y\nabla_X Z -\nabla_{\Lieb{X}{Y}}Z
\end{equation*}
for $X,Y,Z \in \mathfrak{X}\left( \spti\right)$, then one can 
show (cf. \citelist{\cite{O'Neill}*{Chap. 8, Lem. 3}
 \cite{Burns}*{Thm. 5.5.3} }) that $J$ satisfies 
\begin{equation}
	\frac{\nabla^2 J}{\d r ^2} 
	+ \mathcal R \left(J , 
		\dot \gamma \right) \dot \gamma = 0 \, .
		\label{eq:Jacobi} 
\end{equation}
\par
Equation \eqref{eq:Jacobi} is known as the \emph{Jacobi equation} 
and vector fields $J$ over a geodesic $\gamma$ satisfying it are 
known as \emph{Jacobi fields}. As we have chosen the domain 
$\dom \theta$ to be in accordance 
with \eqref{eq:gvariation}, not every Jacobi field along $\gamma$ 
gives rise to a geodesic variation of $\gamma$, but the statement 
is true for finite intervals $\mathcal I$. We refer to 
\cite{Burns}*{p. 207} for a counterexample and remark that 
equation \eqref{eq:solJac} below gives an explicit formula 
for this variation. 
Since \eqref{eq:Jacobi} is a second order, linear 
ordinary differential equation, knowing $J_{r_0}$ and its 
derivative $(\nabla J / \d r)_{r_0}$ at some $r_0 \in 
\mathcal I$ entirely determines the Jacobi field $J$ 
(cf. \cite{O'Neill}*{Chap. 8, Lem. 5}). Moreover, knowing these 
tangent vectors we can easily compute 
$g \left( J, \dot \gamma \right)$ without 
solving the Jacobi equation. 
\begin{Lemma}
	\label{Lem:Jac}
	Let $\left( \spti, g \right)$ be a Lorentzian manifold and let 
	$J$ be a Jacobi field over the geodesic 
	$\gamma\colon \mathcal I \to \spti$ with $r_{0} \in \mathcal 
	I$ and $\gamma \left( r_0\right) = q$. Then 
	\begin{equation}
		g_{\gamma \left( r \right)} \left( J_r, \dot 
		\gamma _r \right) = g_q \left(  
		\left( \frac{\nabla J}{\d r} \right)_{r_0} , \dot \gamma 
		_{r_0} 
		\right) \left( r- r_0 \right) +  
		g_q \left( J_{r_0} , \dot \gamma _ {r_0} \right)
		\label{eq:prodJ}
	\end{equation}
	for all $r \in \mathcal I$. 
\end{Lemma}
\begin{Proof}
	This is an adaption of Lemma 10.9 found in the book by 
	Beem et al. \cite{Beem}. We consider the left hand side of 
	\eqref{eq:prodJ} and derive twice with respect to $r$,  
	keeping in mind the metricity of $g$ and the fact that
	$\gamma$ is a geodesic. Then using the Jacobi equation 
	and the symmetry properties of $\mathcal R$ 
	(cf. \cite{O'Neill}*{Chap. 3, Prop. 36(2)}), we get 
	\begin{equation*}
		\partd{^2}{r^2} \left( g_{\gamma \left( r \right)} 
		\left( J_r, \dot 
		\gamma _r \right) \right) = 
		g_{\gamma \left( r \right)} \left( \left( 
		\frac{\nabla^2 J}{\d r ^2}
		\right)_r, \dot 
		\gamma _r \right) = - g \left( 
		\mathcal R _{\gamma \left( r \right)} \left(J_r, 
		\dot \gamma _r\right) \dot \gamma_r , \dot \gamma_r
		\right) \equiv  0 \, .
	\end{equation*}
	Integrating twice yields \eqref{eq:prodJ}. 
\end{Proof}
\par
As the Jacobi equation \eqref{eq:Jacobi} depends on the Riemann 
tensor field, Jacobi fields yield implicit information 
on the curvature of the Lorentzian manifold $\spti$. 
Indeed, there is an elaborate theory on the precise nature of 
this relation. We again refer to the books by Beem et al. 
\cite{Beem}*{Chap. 10} and by O'Neill
\cite{O'Neill}*{Chap. 8} for further reading. The 
booklet by Penrose \cite{PenroseB0} should also 
be mentioned here. 
\par
Of particular interest in this theory are 
so-called conjugate points. Two points $q,q'$ on the image 
of a non-trivial geodesic $\gamma\colon \mathcal I \to \spti$ are 
called \emph{conjugate along $\gamma$}, if there exists a 
non-trivial Jacobi 
field $J$ and $r_1, r_2 \in \mathcal I$ with $r_1 \neq r_2$ such 
that $J$ vanishes at $r_1$ and $r_2$. In that case, $r_1$, $r_2$ 
are called \emph{conjugate values along $\gamma$}. More generally, 
$q,q' \in \spti$ are called \emph{conjugate points}, if there
exists a non-trivial geodesic $\gamma$ such that $q,q'$ are conjugate 
along $\gamma$. It is possible for a point to be conjugate to itself. 
Intuitively, conjugate points are intersection points of 
nearby geodesics starting at the same initial point. 
Mathematically, a second geodesic intersecting the original one 
twice need not exist.
\par
We may use \thref{Lem:Jac} to determine some properties of the 
Jacobi fields giving rise to conjugate points. 
\begin{Corollary}
	\label{Cor:Jac}
	Let $\left( \spti, g \right)$ be a Lorentzian manifold 
	and $\gamma$ be a geodesic. If $q$ is conjugate to $q'$ 
	along $\gamma$, then the corresponding Jacobi field $J$ 
	satisfies 
	\begin{equation*}
		J \perp \dot \gamma \quad , \quad \frac{\nabla J}{\d r}
		\perp \dot \gamma \, .
	\end{equation*}
\end{Corollary}
\begin{Proof}
	Without loss of generality, assume $\gamma \left( 0\right) 
	=q$ and $\gamma \left( r' \right)= q'$ with $r'>0$. 
	Now use \eqref{eq:prodJ} for $J_0 =0$ and $J_{r'}=0$ to 
	determine the slope and summand, then derive once. 
\end{Proof}
It is noteworthy that for lightlike geodesics a Jacobi field 
$J$ can be both parallel and orthogonal to $\dot \gamma$. 
However, if $J_r$ is timelike at any point $r \in \mathcal I$, 
then it is not orthogonal to $\dot \gamma_r$ and hence 
\thref{Cor:Jac} states that it cannot vanish at two separate 
parameter values $r_1, r_2 \in \mathcal I$. 
\par
Jacobi fields frequently occur in the context of the 
\emph{exponential map}
\begin{equation}
	\exp \colon \dom \exp \to \spti \colon Z \to 
	\exp \left( Z \right) := \gamma _Z \left( 1\right)
\end{equation}
with $(\dom \exp)$ being the maximal set in $\CapT \spti$ such that 
the autoparallel $\tilde \gamma_Z$ with $\dot{\gamma}_0 = Z$
is defined. By re-parametrization, one shows that 
$\exp \left( s Z\right) = \gamma_Z \left( s\right)$ 
for all $s \in \R$ such that $s Z \in \dom \exp$. It is 
often convenient to restrict $\exp$ to the fiber $\CapT_q 
\spti$ at $q \in \spti$. Correspondingly, we call 
$\exp_q := \exp \evat{\CapT _q \spti}$ the 
\emph{exponential map at $q$}. Recall now the definition of the 
vertical lift $\tilde X _Y$ of a vector $X \in \CapT_q \spti$
at $Y \in \CapT_q \spti$ 
(cf. \eqref{eq:vertlift} on page 
\pageref{eq:vertlift}) and define the (smooth) addition map 
$P_Y$ by $Y$ on $\CapT _q \spti$ via 
\begin{equation*}
	P_Y \left( X\right) := X + Y \, ,
\end{equation*}
which has inverse $P_{-Y}$. Then for any $f \in C^\infty 
\left( \CapT_q \spti , \R\right)$ we compute 
\begin{align*}
	\bigl( \left( \exp \circ P_{-Y} \right)_* 
	\tilde X_Y \bigr) \left( f\right)
	&= \tilde X _Y \left( f \circ 
	\exp \circ P_{-Y}\right) \\
	&= \Evat{\partd{}{s} \left( f \circ 
	\exp \circ P_{-Y} \right) \left( Y + s X \right)}{s=0} \\ 
	&= \Evat{\partd{}{s} f \circ 
	\tilde \gamma_X \left( s\right) }{s=0} = X \left( f \right) \, .
\end{align*}
This proves that the vertical lift at $Y$ is a (smooth) 
linear isomorphism and yields a direct inverse in terms of 
the exponential map, which is independent of the particular 
choice of $g$. Moreover, it shows that 
$\left(\exp_q\right)_*$ has full rank at $0$ and hence 
(cf. \cite{Lee}*{Prop. 4.1 \& Thm. 4.5}) it is
a diffeomorphism from an open neighborhood of $0$ onto its 
image in $\spti$. Thus any coordinates around $0$ 
in the tangent space $\CapT_q \spti$ can be viewed as coordinates 
on $\spti$. An important instance are \emph{normal 
coordinates at $q$}, which are given by linear coordinates on 
$\CapT_q \spti$ with respect to an orthonormal basis. 
Normal coordinates are useful for `approximating the 
manifold around $q$'. For a more precise statement, 
including an explicit definition of normal coordinates, 
we refer to the book by O'Neill 
\cite{O'Neill}*{Chap. 3, Prop. 33}. One may also 
adapt the treatment in reference 
\cite{Sakai}*{Chap. II, Prop. 3.1} to the Lorentzian 
case. Later we will use the existence of normal coordinates 
in the heuristic construction of the space-time splitting. 
\par
We continue with an analysis of the exponential map. 
\begin{Proposition}[Domain of exponential map]
	\label{Prop:domexp}
	Let $\left( \spti, g \right)$ be a Lorentzian manifold 
	with exponential $\exp$ as defined by the Levi-Civita 
	connection. Then the domain $\dom \exp$ is open in $\CapT \spti$ 
	and for every $q \in \spti$ the domain 
	$\dom \exp_q$ is open in $\CapT_q \spti$. 
	Moreover, $\dom \exp_q$ is star-shaped around $0$. 
\end{Proposition}
\begin{Proof}
	The full proof can be found in the book by O'Neill 
	\cite{O'Neill}*{Chap. 5, Cor. 4}. The idea is that 
	$\exp$ can be defined in terms of a flow on the tangent 
	bundle, called the geodesic flow, and flow domains are 
	open. \\ 
	Star-shapedness follows from the fact that 
	$\exp_q \left( s Z\right)$ is defined $\forall s 
	\in [0,1]$ whenever $\exp_q \left(Z\right)$ is defined. 
\end{Proof}
The relation to Jacobi fields arises when one asks for 
the differential of the exponential map and is given by 
the following theorem. We refer to section 
\ref{ssec:connecTQ} for a discussion of the concept 
of connectors. 
\begin{Theorem}[Differential of exponential map]
	\label{Thm:diffexp}
	Let $\left( \spti, g \right)$ be a Lorentzian manifold 
	with exponential $\exp$ and connector $\Kon$, 
	as defined by the Levi-Civita connection. Denote by 
	$\pi \colon \CapT \spti \to \spti$,  
	$\tilde \pi \colon \CapT \CapT \spti 
	\to \CapT \spti$ the respective bundle projections. \\
	Then for all $Z \in \CapT \CapT \spti$ 
	such that $\tilde \pi \left( Z\right) \in \dom \exp$, 
	we have 
	\begin{equation}
		\exp_* Z = J_1 \, ,
		\label{eq:diffexp}
	\end{equation}
	where $J\colon r \to J_r$ is the unique Jacobi 
	field along the geodesic 
	$r \to \exp \left( r \, \tilde \pi \left( Z\right) 
	\right)$ with $J_0 = \pi_* Z$ and $(\nabla J / \d r)_0 = \Kon 
	\left( Z \right)$.
\end{Theorem}
\begin{Proof}
	A special case of this theorem can be found in 
	the book by O'Neill \cite{O'Neill}*{Chap. 8, Prop. 6}. 
	The full theorem can be found in the book by Sakai 
	\cite{Sakai}*{Lem. 2.2 \& Lem. 4.3} and Burns et al 
	\cite{Burns}*{5.9.2}. %REF !  
	We shall give an intrinsic proof here. \\ 
	Since $Z \in \CapT \CapT \spti$, there exists an 
	$\epsilon >0$ and a smooth curve 
	\begin{equation*}
		Y\colon \left( - \epsilon, \epsilon \right) 
		\to \CapT \spti \colon s \to Y_s
	\end{equation*}
	with $\dot Y _ 0= Z $
	projecting to the curve $\gamma := \pi \circ Y$ on 
	$\spti$. Therefore the map 
	\begin{equation}
		\theta \colon  
		\left( - \epsilon, \epsilon \right) \cross 
		\mathcal I \to \spti 
		\colon \left( s, r\right) \to 
		\exp \left( r Y_s\right)
		\label{eq:solJac}
	\end{equation}
	with 
	\begin{equation*}
		\mathcal I := \set{r \in \R}{\forall s \in 
	\left( - \epsilon, \epsilon \right)\colon \, 
	r Y_s \in \dom \exp} \neq \emptyset
	\end{equation*}
	is a geodesic 
	variation of $r \to  \exp \left( r \, 
	\tilde \pi \left( Z\right)\right)$. 
	Hence $J_r := \theta_* 
	\left( \partial/\partial s\right)_{(0,r)}$
	defines a Jacobi field $J$. Recalling 
	the multiplication map $M$ (cf. 
	\eqref{eq:multop} on page \pageref{eq:multop}), we 
	see that $M_1 = \Id_{\CapT \spti}$ and thus 
	\begin{equation*}
		J_1 = \left( \exp \circ M_1 \circ Y \right)_*
			\Evat{\partd{}{s}}{0} 
		= \exp_* Y_* \Evat{\partd{}{s}}{0} 
		= \exp_* \dot Y_0
	\end{equation*}
	yields 
	\eqref{eq:diffexp}. \\ 
	It remains to express $J$ in terms of $Z$. For any 
	$f \in C^\infty \left( \spti, \R \right)$ we calculate 
	\begin{equation*}
		J_{0} \left( f \right) = 
		\Evat{\partd{}{s}}{0} f \circ \theta_{s} 
		\left( 0\right) 
		= \Evat{\partd{}{s}}{0} 
		f \circ \exp_{\gamma \left( s\right)} 
		\left( 0\right) 
		= \dot \gamma _0 \left( f \right) \, ,
	\end{equation*}
	so $J_0 = \dot \gamma_0 = \dot{(\pi \circ Y)}_0 
	= \pi_* \dot Y_0 = \pi_* Z$. On the other hand 
	\begin{equation*}
		\left( \pi_* \dot J _0 \right)\left( f 
		\right) = \Evat{\partd{}{r}}{0} f \circ 
		\pi \circ J _r 
		= \Evat{\partd{}{r}}{0} f \circ \exp 
		\left(r Y_ 0 \right) = Y_0 \left( f\right) \, ,
	\end{equation*}
	so $\pi_* \dot J_0 = Y_0 = \tilde \pi 
	\left( Z \right)$. To get the vertical 
	parts, we calculate 
	\begin{align*}
		\dot J _0 \left( \d f \right) &= 
		\Evat{\partd{}{r}}{0} \d f \left( J _ r\right)
		= \Evat{\partd{}{r}}{0}  J _ r \left( f \right) \\
		&= \Evat{\partd{}{r}}{0} \Evat{\partd{}{s}}{0} 
		f \circ \theta_s \left( r \right) = 
		\Evat{\partd{}{s}}{0} \Evat{\partd{}{r}}{0} 
		f \circ \theta_s \left( r \right) \\
		&= \Evat{\partd{}{s}}{0} Y_s \left( f\right) 
		= \Evat{\partd{}{s}}{0} \d f \left( Y_s \right) 
		= \dot Y_0 \left( \d f\right) \, , 
	\end{align*} 
	hence $\Fl \left( \dot J _0  \right) = \dot Y_0 $\, .
	Finally, since the Levi-Civita connector is 
	torsion-free, 
	\begin{equation*}
		\Kon \left( Z \right) 
		= \Kon \left( \dot Y_0\right) 
		= \Kon \left( \Fl \left( \dot J _0  \right) 
		\right) 
		= \Kon \left( \dot J _0 \right) = 
		\left( \frac{\nabla J}{\d r}\right)_0 
		\, .
	\end{equation*}
\end{Proof}
By \thref{Thm:diffexp}, two points $q$ and $q'$ in $Q$ 
are conjugate to each other if and only if there exists a 
vertical $Z \in \CapT \CapT \spti$ with base point 
$Y \in \CapT_q \spti$ such that 
\begin{equation*}
	q' = \exp Y \quad 
	\text{and} \quad Z \in \ker \exp_* \, .
\end{equation*} 
In other words, the
set of critical points of the exponential at $q$
\begin{equation*}
	\crit \exp_q := 
	\set{Y \in \dom \exp_q \subseteq \CapT_q \spti
	}{\, \ker \left( 
	\Evat{(\exp_{q})_*}{Y} \right) \neq 
	\lbrace 0 \rbrace}
\end{equation*}
is given by 
\begin{equation}
	\crit \exp_q = 
	\set{Y \in \dom \exp_q}{\, \exp_q \left( Y\right) \, 
	\text{is conjugate to} \, q} \, .
	\label{eq:critexp}
\end{equation}
For this reason, we call 
$\crit \exp_q$ the \emph{conjugate locus at $q$}. In 
the Riemannian case an analysis of this set has been 
carried out by Warner \cite{Warner0}. In the 
Lorentzian case one needs to distinguish between the 
\emph{time-, light-} and \emph{spacelike conjugate locus}, 
which is defined as the intersection of the conjugate 
locus at $q$ with the respective subsets of 
$\CapT_q \spti$. The following Lemma indicates their 
structure.   
\begin{Lemma}[Causal conjugate values are isolated]
	\label{Lem:isoconj}
\begin{subequations}
	Let $\left( \spti, g\right)$ be a Lorentzian manifold 
	equipped with the Levi-Civita connection. \\
	Then conjugate values along any time- 
	or lightlike geodesic 
	$\gamma \colon \mathcal I \to \spti$ 
	are \emph{isolated}, i.e. for any $r$ in the set
	of conjugate values $S_0 \subset \mathcal I$ 
	to some $r_0 \in \mathcal I$ there 
	exists an open neighborhood $\mathcal J$ of $r$ 
	with $\mathcal J \cap S_0 = \lbrace r \rbrace$. 
\end{subequations}
\end{Lemma}
\begin{Proof}
	The statement is a corollary of the fact that 
	along any finite causal geodesic, the number 
	of conjugate values with respect to a given 
	value is finite. 
	The proof thereof is part of 
	so called Morse index theory 
	and rather elaborate. 
	It can be found in the
	book by Beem et al: See
	\cite{Beem}*{Slem. 10.26 \& Thm. 10.27} 
	for the timelike case and 
	\cite{Beem}*{Prop. 10.76 \& Thm. 10.77}
	for the lightlike case. 
	\par 
	If $S_0$ is empty, we are done. So take any 
	$r_1 \in S_0$ conjugate to $r_0$. By the above 
	statement, on any finite open neighborhood 
	$\mathcal J'$ of $r_1$ there exist at most 
	finitely many points in $S_0 \cap \mathcal J'$. 
	If there is none, set $\mathcal J = \mathcal J'$. 
	If there is at least one, Hausdorffness of $\R$ 
	(and thus of $\mathcal J'$) says that $r_1$
	and any $r_2 \in S_0 \cap \mathcal J'$ 
	admit mutually disjoint open neighborhoods. 
	Since there are only finitely many such $r_2$, 
	the assertion follows. 
\end{Proof}
Regarding the adaption of the above statement to spacelike 
geodesics, Helfer has constructed a 
counterexample in reference \cite{Helfer}.  

\chapter{The Splitting Construction}
\label{chap:construction}

The main objective of this chapter is to philosophically 
motivate and mathematically define the construction of splitting 
relativistic spacetimes 
into their spatial and temporal components. 
We give consistency proofs 
and examples along with the general theory.
\par 
In the first section, we give a definition of the word 
`spacetime', introduce some elementary concepts 
required for the mathematical theory of relativity and provide 
some physically relevant examples. 
Section \ref{sec:motivation} is devoted to the heuristic 
motivation of the splitting construction. The reader only 
interested in the mathematical machinery is invited to 
skip this section, but the underlying philosophy is intended 
to convince the reader that the construction is `natural' 
rather than ad hoc. Afterwards, we erect the mathematical 
theory in two steps: First, the 
`static splitting' is considered in section 
\ref{sec:space}. It derives its name from the fact 
that there is no time evolution in this setting. The second 
step is done in section \ref{sec:kinematic} with the 
`kinematic splitting', which allows for time evolution and 
thus constitutes an actual `space-time splitting'. 
Mathematically, the static splitting lays the foundation 
for the kinematic one, so we recommend to read them in this order. 

\section{General Considerations}     
\label{sec:generalspti}  

As a brief introduction to the mathematical theory of 
relativity, this section provides a mathematical definition and 
motivation of the relativistic concept of spacetime along 
with the two physically most important examples. We also 
introduce observers, light cones and frames of reference. 
Apart from their general relevance within the theory of 
relativity, those are needed for the space-time splitting 
developed in the following sections. 
\par 
The mathematical concept of 
spacetime admits a condensed definition, if we employ 
our findings from section \ref{ssec:Lorentz} on 
Lorentzian structures, as well as the ones from 
section \ref{ssec:Conf} on Lorentzian orientations. 
\begin{Definition}[Spacetime]
	\label{Def:spti}
	A \emph{spacetime} is a spacetime oriented Lorentzian 
	manifold $\left(\spti, g, \mathcal O \right)$
	equipped with the Levi-Civita connection. 
\end{Definition}
It should be stressed that the spacetime orientation 
$\mathcal O$ is implicitly assumed to be compatible with 
the metric $g$, i.e. they give rise to the same 
causal structure on $\spti$. For simplicity, we often 
call $\spti$ the spacetime rather than 
$\left(\spti, g , \mathcal O \right)$. 
Furthermore, \thref{Def:spti} should be read from a categorical 
perspective in the sense that we do not distinguish between 
\emph{isomorphic spacetimes}. Clearly, two (smooth) spacetimes 
$\left(\spti, g, \mathcal O \right)$ and 
$\left(\spti', g', \mathcal O' \right)$ are called isomorphic, 
if there exists a (smooth, global) diffeomorphism
$\varphi \colon \spti \to 
\spti'$ such that 
\begin{equation*}
	\varphi^*g' = g \quad \text{and} \quad 
	\varphi_* \mathcal O = \mathcal O' \, .
\end{equation*}
The first condition means that $\varphi$ is an 
\emph{isometry} and the second one says that it is 
\emph{spacetime-orientation preserving}. We remark that 
for a frame $X$ on $\spti$ and a diffeomorphism 
$\varphi \colon \mathcal Q \to \mathcal Q'$, 
$\varphi_* X$ is always a frame on $\spti'$, so the second 
condition is well-defined. 
\par
The choice of \thref{Def:spti} is the middle path between 
physical sensibility and mathematical generality. In the 
following we shall give some justification to this claim along 
with a brief physical motivation. 
A complete physical 
justification of the mathematical concept of spacetime, if 
even possible, would derail the content of this work, so 
we limit ourselves to a few remarks regarding the main points. 
Nonetheless we wish to indicate the underlying principles 
and ideas that lead to this mathematical formulation. 
We refer to the books by Carroll 
\cite{Carroll}, Kriele \cite{Kriele} and Wald \cite{Wald}
for similar motivations. 
\begin{description}[style=unboxed,leftmargin=0cm]
	\item[Dimension:]
			It is an experiment we leave to the reader 
			that, at least within humanly accessible 
			realms, space is $3$-dimensional and 
			time is $1$-dimensional. So if we wish to 
			mathematically represent space and time as 
			one `object' called spacetime, the definition 
			should respect this empirical fact. 
			Obviously, in general relativity this is done 
			by assuming the manifold $\spti$ to be 
			$4$-dimensional. 
			\par 
			Despite this, we have chosen not to fix the dimension of 
			$\spti$ in the definition, since it is 
			mathematically inconvenient, the dimension is not 
			important in any of the general definitions or 
			proofs here and it is sometimes useful to 
			consider lower dimensional `toy models'. 
			Funnily enough, dimension $4$ also has  
			some particular mathematical significance, 
			as it is the only dimension $n \in \N$ for which 
			$\R^n$, equipped with the standard topology, 
			does not have a unique smooth 
			structure (up to diffeomorphism, cf. 
			\cite{Lee}*{p. 39sq.}). 
			Though we do not use this here, it is 
			certainly noteworthy. We refer to 
			the books by Scorpan \cite{Scorpan}
			and Asselmeyer-Maluga \& Brans 
			\cite{Asselmeyer} regarding this fact.   
	\item[Manifold:]
			The choice to model space and time on a 
			manifold is a mathematical expression of the 
			old principle ``Natura non facit saltus'', 
			roughly translating into ``nature does not make 
			jumps''. This refers to both the 
			need to model space and time on a 
			continuum of points as well as the assumption that 
			the dimension of space and time is fixed. 
			While it appears to have become fashionable in these 
			days to doubt this principle (see e.g. 
			\cites{Callender,Woit}), there is, at least to 
			our knowledge, no evidence to the contrary 
			(see e.g. \cite{Forrest}). 
			For instance, the discrete spectral lines of light
			emitted by atoms 
			do provide conclusive evidence in favor of the 
			theory of quantum mechanics, but it is a fallacy to 
			read this as evidence in favor of the discreteness 
			of space and time itself. Indeed, quantum mechanics 
			itself is formulated on a so called 
			Newtonian spacetime (see e.g. 
			\cite{Reddiger0}*{\S 2}), so it can hardly be 
			taken as a justification for discarding the 
			principle (along with the manifold model). 
	\item[Lorentzian metric:] 
			A manifold alone cannot be taken as 
			a physical model of space and time, as
			it lacks an appropriate notion of `distance'. 
			The choice to use a Lorentzian metric for this 
			purpose primarily evolved out of a mathematical 
			generalization of the special theory of 
			relativity. In special 
			relativity, one equips the spacetime 
			manifold $\spti = \R^4$ with a flat 
			Lorentzian metric $g$, but Einstein realized that the 
			assumption of flatness is ad hoc and the 
			phenomenon of gravity could be explained by 
			dropping it. So instead he had to 
			formulate a law for the 
			curvature of $g$, which is today known 
			as the Einstein (field) equation. The flat case 
			could then be recovered locally 
			by considering it as the `tangent space approximation' 
			of the `true geometry' 
			via Lorentzian normal coordinates. 
			Of course, if 
			compared with Newtonian physics, this 
			constitutes a serious weakening of the 
			former principles that space is Euclidean 
			and the rate of time is the same everywhere, 
			regardless of how the two are implemented 
			into the theory in detail. 
	\item[Spacetime orientation:] Once the manifold 
			$\spti$ and the metric $g$ is chosen, it is 
			still not possible to distinguish between 
			past and future in time or between 
			right-handed and left-handed
			in space. As indicated in section 
			\ref{ssec:Conf}, a spacetime orientation 
			$\mathcal O$ is to be used to make mathematical 
			sense of both of these concepts in 
			each tangent space of $\spti$. Due to the 
			characterization of the conformal group in
			\thref{Thm:conf}, spacetime orientations 
			are the most general $\mathcal G$-structures
			able to give rise to time and 
			space orientations in a philosophical 
			sense and which also respect the 
			causal structure induced by 
			$g$. While relativistic 
			spacetimes are commonly assumed to 
			be time-oriented in the physics literature, 
			space orientations are often not included in 
			the definition. 
			Yet a space orientation is 
			indispensable to differentiate between a 
			relativistic model and its 
			`spatial mirror image'. It 
			therefore needs to be included in 
			the general definition of `spacetime'. 
			Later in this section we will see that 
			space orientations 
			are also required in order to give a mathematical
			definition of the physical concept 
			`frame of reference'.  
	\item[Levi-Civita connection:] As in Riemannian geometry, 
			the metric $g$ naturally gives rise to 
			the class of metric connections, that is the set 
			of those covariant derivatives 
			$\nabla$ for which $g$ is covariantly constant: 
			\begin{equation*}
					\nabla g = 0 \, .
			\end{equation*} 
			For given initial conditions, (maximal)
			autoparallels with respect to any of these 
			$\nabla$ coincide, so 
			we may philosophically 
			understand these autoparallels to be intrinsic 
			to the Lorentzian manifold 
			$\left( \spti, g\right)$. Again in full analogy 
			to the Riemannian case, $\nabla$ is uniquely defined 
			by the choice of a torsion tensor field 
			$T$, given by  
			\begin{equation*}
				T \left(Y, Z \right) = 
				\nabla_Y Z - \nabla_Z Y 
				- \Lieb{Y}{Z}
			\end{equation*}
			for vector fields $Y,Z$ on $\spti$. While 
			the choice of $T$
			has no influence on the shape of autoparallels, 
			it is important for parallel 
			transport of tangent vectors along curves. 
			Thus the choice of
			the Levi-Civita connection for $\nabla$, i.e. 
			$T=0$, constitutes an additional 
			assumption on the geometry of the spacetime and cannot 
			be taken for granted. Physically, non-vanishing 
			torsion would lead to a spinning of (infinitesimal) 
			rigid bodies in free fall with 
			initially vanishing angular momentum 
			(as measured from the center of mass) -- 
			to our knowledge this has not been 
			observed. Nevertheless the 
			case of $T \neq 0$ is considered, for instance, 
			within so called Einstein-Cartan 
			theory. See e.g. the article by 
			Hehl, von der Heyde and Kerlick \cite{Hehl} 
			for a discussion thereof. 
	\item[Connectedness:] One may define spacetimes to be 
	connected, but we decide not to do so. This is not 
	due to a need to consider `multiple universes', which 
	would be physically inaccessible even if they existed, 
	but rather for practical, modeling reasons. 
	\par
	For instance, if $f$ is a smooth, real-valued function 
	on a connected spacetime $\spti$ and we wish 
	to consider the function $1/f$, then 
	\begin{equation*}
		\spti' := f^{-1} \left( \R \setminus \lbrace 
		0 \rbrace \right)
	\end{equation*}
	is an open submanifold of $\spti$ and $1/f$ is smooth 
	on $\spti'$. If we do not require spacetimes to be 
	connected, then $\spti'$ is also canonically a 
	spacetime. 
	\item[Inextendibility:] 
	It is sensible to ask for a certain 
	'maximality condition' on the connected 
	components of the spacetime. 
	This is a condition that can be defined for any 
	Lorentzian manifold: 
	An 
	\emph{extension} of a connected Lorentzian 
	manifold $\left( \spti, g\right)$ 
	is a connected Lorentzian manifold 
	$\left( \spti', g'\right)$ together with a (smooth) 
	mapping $\varphi \colon \spti \to \spti'$ 
	such that $\left( \spti, \varphi \right)$ is an 
	open submanifold 
	of $\spti'$ with 
	$\varphi^* g' =g$. 
	A connected Lorentzian manifold 
	$\left( \spti, g\right)$ is said to be 
	\emph{inextendible}, if there 
	does not exist an extension with 
	$\varphi \left( \spti\right) \subset \spti'$ 
	(cf. 
	\citelist{\cite{Hawking}*{p. 85 sq.;} 
	\cite{O'Neill}*{Def. 5.44}
	}). A general Lorentzian manifold is 
	\emph{inextendible}, if each of its connected 
	components is inextendible (as a connected 
	Lorentzian manifold equipped with the restricted metric). 
	\par 
	Clearly, 
	the terminology carries over to spacetimes. 
	As in the case of connectedness, modeling arguments 
	speak against defining spacetimes to be inextendible. 
	\item[Causality:] 
	The last requirement one might want to add to the 
	definition of a spacetime is a 
	so called `causality condition'. They are usually 
	topological (hence global) restrictions on the 
	manifold $\spti$ and derive their name from the 
	fact that, among others, they have implications for the 
	possible trajectories of point masses, i.e. 
	for (future directed) timelike curves on $\spti$. As 
	there are quite a few 
	possible choices, we mention only one example 
	and refer to the books by O'Neill 
	\cite{O'Neill}*{Chap. 14}, 
	Beem at al \cite{Beem}*{\S 3.2 \& \S 3.3}, 
	Penrose \cite{PenroseB0}, 
	as well as the article by Minguzzi and Sanchez 
	\cite{Minguzzi} for further reading.  
	\par 
	One of the weakest causality conditions is 
	non-viciousness. By definition, a spacetime is called 
	\emph{vicious}, if there exists a periodic timelike 
	curve, i.e. one for which the image is compact in $\spti$. 
	If such a spacetime were considered a serious 
	physical model, it would imply the possibility 
	of time travel into the past as well as time 
	repeating itself over and over. The name derives itself 
	from the phrase `vicious circle' and it is arguably 
	quite judgmental terminology. In the author's opinion,  
	the physicist's judgment on 
	the acceptability of these models depends more on 
	cultural background than scientific evidence. 
	Independent of where one stands on this issue, 
	we are not aware of any good reason for 
	not calling these models spacetimes.
\end{description}
\par 
We continue by giving two physically relevant 
examples of spacetimes. 
\begin{Example}[Minkowski spacetime]
	\label{Ex:Mspti}
	Consider the manifold $\R^4$ with standard topology and smooth 
	stucture. In canonical global coordinates $y = \left( y^0, \vec y
	\right)$, we may define 
	the Lorentzian metric 
	\begin{equation*}
		g := \eta_{ij} \, \d y^i \tp \d y^j \, 
	\end{equation*}
	with respect to which $\partial$ is an orthonormal frame 
	field. As a global frame field, this gives rise to a 
	$\CLor_4$-reduction $\mathcal O$ of the frame bundle 
	$\frameb{\CapT \R^4}$
	(cf. \thref{Rem:TransextG}\ref{sRem:transitP} on page 
	\pageref{sRem:transitP}) 
	and thus a spacetime orientation. Hence 
	$\left(\R^4, g, \mathcal O \right)$ is a spacetime, known as 
	\emph{Minkowski spacetime}. It is obviously flat and 
	connected. In addition, open submanifolds of Minkowski 
	spacetime can be canonically turned into flat spacetimes, 
	but these are not necessarily connected. 
	\par
	Minkowski spacetime is the mathematical setting 
	of the special theory of relativity. We refer to 
	\cite{Einstein0} for the original articles on 
	the theory due to Einstein and Minkowski. 
\end{Example}
The following lemma is of use in defining spacetime 
orientations, if one has 
an `almost global' orthonormal frame field on a 
parallelizable Lorentzian manifold. 
\begin{Lemma}
	\label{Lem:sptiO}
	Let $\left( \spti, g\right)$ be a parallelizable  
	Lorentzian manifold, $\mathcal U \subseteq \spti$ 
	be open and let 
	$X \colon \mathcal U \to \oframeb{\spti, g}$ 
	be a local, orthonormal frame field, such that the 
	closure $(\clos \mathcal U)$ of $\mathcal U$ is $\spti$. 
	\\
	Then there exists a unique spacetime orientation 
	$\mathcal O$ on $\spti$ such that $X$ is a local 
	section of $\mathcal O$ and 
	$\left( \spti, g , \mathcal O\right)$ is a spacetime. 
\end{Lemma}
\begin{Proof}
	Since the $(n+1)$-manifold $\spti$ is parallelizable 
	and $g$ is Lorentzian, there 
	exists a global, orthonormal frame field $Y$ and a 
	$\Lambda \in C^\infty \left(\mathcal U, 
	\Lor_{n+1} \right)$ such that $Y = X \cdot \Lambda$ on 
	$\mathcal U$. Extension of this $\Lor_{n+1}$-structure 
	to $\CLor_{n+1}$ proves existence 
	(see \thref{Rem:TransextG}/\ref{sRem:extG} 
	on page \pageref{sRem:extG} and 
	\thref{Lem:COstruct} on page \pageref{Lem:COstruct}).\\
	To get uniqueness, consider a second such
	frame field $Y'$ with $Y' = X \cdot \Lambda'$ on 
	$\mathcal U$. Then 
	$Y' = Y \cdot \ubar \Lambda \cdot \Lambda'$ on $\mathcal U$. 
	On the other hand, there must exist an 
	$A \in C^\infty \left( \spti, \LieO_{1,n} \right)$ with 
	$Y'=Y \cdot A$ and $A = \ubar \Lambda \cdot \Lambda'$ on 
	$\mathcal U$. Choosing a sequence 
	$\left( q_i\right)_{i \in \N}$ in $\mathcal U$ converging to 
	$q \in \spti \setminus \mathcal U$, we obtain
	\begin{equation*}
		\lim_{i \to \infty} \left(\ubar \Lambda \cdot \Lambda' 
		\right)_{q_i} = \lim_{i \to \infty} A_{q_i} = A_q \, .
	\end{equation*}
	As a connected component, $\Lor_{n+1}$ is closed 
	in $\LieO_{1,n}$ and hence
	$A_q \in \Lor_{n+1} \subset \CLor_{n+1}$. As the sequence was 
	arbitrary, $Y$ and $Y'$ induce the same spacetime orientation 
	$\mathcal O$. 
\end{Proof}
We apply \thref{Lem:sptiO} in the next example of a spacetime. 
\begin{Example}[Exterior Schwarzschild spacetime]
	\label{Ex:exschwspti}
	Let $R \in \R_+$, consider 
	$\R \cross \left( R, \infty \right)$ equipped with the 
	Lorentzian metric $g'$, given by 
	\begin{equation*}
		g'_{\left(ct,r \right)} :=
		\left( 1 - \frac{R}{r} \right) \, \d \left( c t\right)\tp 
		\d \left( c t\right) 
		- \left( 1 - \frac{R}{r} \right)^{-1} \, 
		\d r \tp \d r
	\end{equation*}     
	in canonical global coordinates $\left( ct, r \right)$, as well as 
	the $2$-sphere $\mathbb{S}^2 \subset \R^3$ equipped with the 
	standard Riemannian metric $g''$. We define a new Lorentzian 
	manifold $\left( \spti, g \right)$ by taking the 
	pseudo-Riemannian product of $\left( \R \cross 
	\left( R, \infty \right), g' \right)$ with 
	$\left( \mathbb{S}^2 , - g''\right)$ 
	(cf. \citelist{\cite{O'Neill}*{Chap. 3, Lem. 5}
	\cite{Sachs}*{Ex. 1.4.2}}). $g$ is called the 
	\emph{Schwarzschild metric} and $R$ is called the 
	\emph{Schwarzschild radius}. Note that the word `radius'
	is potentially misleading. Choosing spherical coordinates 
	$\left( \theta, \phi \right)$ on $\mathbb{S}^2 \subset \R^3$, 
	we obtain \emph{Schwarzschild coordinates} 
	$\kappa := \left(ct, r, \theta, \phi \right)$: 
	\begin{align*}
		\R \cross \left( R, \infty \right)
		\cross \left(\mathbb{S}^2 \setminus  
		\set{\vec y \in \R^3}{ y^1 = 0, y^2 \geq 0}
		\right) &\to \R \cross \left( R, \infty\right)
		\cross \left( 0, \pi\right) \cross \left(0, 2 \pi \right) \\ 
		\left( c t, r, \vec y \right) &\to 
		\left( ct, r, \theta \left( \vec y\right), 
		\phi \left( \vec y \right) \right)
	\end{align*}
	on $\spti$, and in these coordinates the metric reads 
	\begin{equation}
		g_{\left( ct, r, \theta, \phi\right)} 
		= \begin{pmatrix}
			\left( 1 - \frac{R}{r} \right) & & & \\
			& - \left( 1 - \frac{R}{r} \right)^{-1} & & \\
			& & - r^2 & \\
			& & & - r^2 \sin ^2 \theta
		\end{pmatrix} \, .
	\end{equation}
	To obtain a compatible spacetime orientation, we use $\kappa$ to 
	define the orthonormal frame field $X$ on 
	$\mathcal U := \dom \kappa \subset \spti$ via 
	\begin{equation*}
		\left(X \right)_\kappa := 
		\begin{pmatrix}
			\left( 1 - \frac{R}{r} \right)^{-1/2} & & & \\
			& \left( 1 - \frac{R}{r} \right)^{1/2} & & \\
			& & \frac{1}{r} & \\
			& & & \frac{1}{r \sin \theta} 
		\end{pmatrix} \, .
	\end{equation*}
	Contrary to \thref{Ex:Mspti}, this is not a global frame 
	field and, since $\mathbb{S}^2$ is not parallelizable, it 
	cannot be smoothly extended to one. However, 
	$\left( R, \infty \right) \cross \mathbb{S}^2$ is 
	diffeomorphic to $\set{\vec y \in \R^3}{\abs{\vec y} > R}$
	via $\left( r, \vec y\right) \to r \vec y$ and hence 
	parallelizable. Thus $\spti$ is parallelizable 
	and since $\clos \mathcal U = \spti$, we may apply 
	\thref{Lem:sptiO} to obtain a compatible 
	spacetime orientation induced by $X$.\\ 
	We call $\left( \spti, g, \mathcal O\right)$ the
	\emph{exterior Schwarzschild spacetime}. It was 
	discovered independently by Schwarzschild and 
	Droste as a solution of Einstein's vacuum (field) 
	equation in the year 1916. Being a `vacuum solution' 
	means that its Ricci tensor 
	field $\Ric$ vanishes, i.e. it is \emph{Ricci-flat}. 
	See \cite{Perlick2}*{p. 55} 
	for a historical overview including references to 
	the original works. 
	The Schwarzschild spacetime is used to 
	model black holes and stars (see e.g. \cite{Wald}*{I \S 6}). 
	Mathematically, it is interesting since 
	the Lie group $\R \cross \LieSO_3$ acts on it canonically 
	by spacetime-orientation preserving isometries, i.e. by 
	spacetime automorphisms. 
\end{Example}
\par
In the remaining part of this section, we 
consider additional mathematical structures on 
spacetimes. These structures are relevant for the theory of relativity 
and our splitting construction. 
\par 
First we recall the definition of observers and discuss 
their physical relevance. Within this thesis, the constant 
$c \in \R_+$ is always the speed of light (in vacuum). 
\begin{Definition}[Observer]
	\label{Def:obs}
	Let $\left( \spti, g, \mathcal O\right)$ be a spacetime. \\
	A (smooth) \emph{observer} 
	is a (smooth) curve 
	\begin{equation*}
		\gamma \colon \mathcal I \to \spti 
		\colon \tau \to \gamma \left( \tau \right)
	\end{equation*}
	for which each tangent vector 
	$\dot \gamma _\tau := 
	\gamma_* \left( \left( \partial/ \partial 
	\tau \right)_\tau \right)$ is an observer vector. That is,  
	$\gamma$ is a timelike, future-directed 
	curve satisfying 
	\begin{equation}
		g \left( \dot \gamma_ \tau, \dot \gamma_\tau \right) 
		= c^2 \, .
		\label{eq:normalc}
	\end{equation}
	for all $\tau \in \mathcal I$. 
\end{Definition}
In the theory of relativity, future directed, timelike curves 
are of particular importance, as they describe physical 
motion of point masses. The normalization 
condition \eqref{eq:normalc} fixes the parametrization: 
We say that $\gamma$ is 
\emph{proper time parametrized}. This choice 
of parametrization is taken in accordance with the so called 
`clock hypothesis'. As general relativity is currently 
a well-established theory of nature, it is a principle, 
rather than a postulate or hypothesis. 
\begin{Principle}[Clock Principle]
	\label{Prin:clock}
\begin{subequations}
	The time difference measured by an (ideal) clock 
	moving along a future-directed 
	timelike curve in spacetime is given by its
	Lorentzian arc-length divided by the speed of light. 
\end{subequations}
\end{Principle}
In formulas \thref{Prin:clock} states the following: 
If $\gamma'\colon \mathcal I' \to \spti$ is a 
future directed timelike curve physically representing the 
motion of a point-like, ideal clock, then the 
\emph{(proper) time} passed between the 
parameter values $s_1, s_2 \in \mathcal I'$ with $s_1 < s_2$ 
is 
\begin{equation}
	\label{eq:propertime}
	\Delta \tau := \frac{1}{c}
	\int_{s_1}^{s_2} 
	\sqrt{g 
	\left( \dot \gamma'_ s , \dot \gamma' _ s
	\right) } \, \, \d s \,  
\end{equation}
according to the clock. Indeed, it is often convenient and 
always physically correct to picture observers $\gamma$
(in the sense of \thref{Def:obs}) as 
moving, point-like, ideal clocks. The 
force applied to such a clock is given by the 
general-relativistic generalization of Newton's second law: 
\begin{equation}
	m \frac{\nabla \dot \gamma}{\d \tau} = F \, ,
	\label{eq:Newton2}
\end{equation}
where $m \in \R_+$ is the inertial mass and the force $F$ 
is a (necessarily spacelike) vector field over $\gamma$. We 
strongly emphasize that gravity is not a force. 
Hence the relativistic generalization 
of Newton's first law states that point masses move geodesically
in the absence of forces, i.e.
the \emph{(proper/absolute) acceleration 
$\nabla \dot \gamma / \d \tau$} 
vanishes entirely or equivalently, $\gamma$ is \emph{unaccelerated}. 
In the presence of forces $F \neq 0$, however, $\gamma$ is 
\emph{accelerated}. In both cases, the function 
\begin{equation}
	a := \sqrt{ - g \left( \frac{\nabla \dot \gamma}{\d \tau}, 
	\frac{\nabla \dot \gamma}{\d \tau} \right) }
\end{equation}
is called the \emph{absolute value of the (proper/absolute) 
acceleration}, 
accordingly. We shall consider two examples of observers. 
\begin{Example}[Observers in Minkowski spacetime]
	\label{Ex:obsMink}
\begin{subequations}
	In the following, let $\left( \R^4, g, \spti\right)$ be 
	$4$-dimensional Minkowski spacetime from \thref{Ex:Mspti}. 
	\begin{enumerate}[i)]
	\item 	The observer $\gamma \colon 
			\R \to \R^4 \colon \tau \to \gamma \left( \tau\right)$, 
			defined by 
			\begin{equation}
				\gamma \left( \tau \right) 
				= \left( c \tau, 0,0,0\right) \, ,
				\label{eq:stndrtgam}
			\end{equation}
			is prototypical and often implicitly assumed to 
			be given in 
			discussions on special relativity. As all Christoffel 
			symbols vanish in standard coordinates, it is 
			unaccelerated. 
	\item 	If we look for \emph{constantly accelerated observers}, 
			then $\gamma \colon \tau \to \gamma \left( \tau\right)$ 
			must satisfy: 
			\begin{equation*}
				0 \neq - a^2 = \eta_{ij} \, \ddot{\gamma}^i 
				\ddot{\gamma}^j = 
				\left ( \ddot{\gamma}^0 \right)^2 - 
				\bigl( \ddot{\vec{\gamma}}\bigr)^2 = \const \, \, ,
			\end{equation*}
			where the dot denotes the derivative with respect to the
			parameter $\tau$. 
			Setting $\gamma^2 
			\left( \tau \right) = \gamma^3 
			\left( \tau \right) \equiv 0$, we obtain 
			\begin{equation*}
				- a^2 = (\ddot{\gamma}^0)^2 - (\ddot{\gamma}^1)^2 
				\quad \text{and} \quad 
				c^2 = (\dot{\gamma}^0)^2 - (\dot{\gamma}^1)^2 \, \, .
			\end{equation*}
			We solve the latter equation for $\dot \gamma^0$, observe
			that $\dot \gamma^0 > 0$, since 
			$\gamma$ is future-directed, 
			and then plug the derivative into the former equation. 
			After some rearrangement we get 
			\begin{equation*}
				c^2 + (\dot \gamma^1)^2 = 
				\left(\frac{c \, \ddot \gamma^1}{a}\right)^2 \, , 
			\end{equation*}
			which may be transformed to a first-order equation 
			by defining $ u := \dot \gamma^1 / c $. After 
			re-parametrization via $s := a \tau / c$, we obtain
			\begin{equation*}
					1 + u^2 = \left( \frac{\d u}{\d s} \right)^2 \, .
			\end{equation*}
			By comparing this with the identity 
			\begin{equation*}
					1 + \sinh^2 s = \cosh^2 s \, ,
			\end{equation*}
			we get $u \left( s\right) = 
			\sinh \left( s + \const \right)$. Thus for the initial 
			condition $\dot \gamma^1 \left( 0\right) = 0$, the 
			tangent vector of $\gamma$ at $\tau$ reads:
			\begin{equation}
					\dot \gamma_ \tau 
					= c \cosh \left(\frac{a \tau}{c} \right) 
						\Evat{\partd{}{y^0}}{
						\gamma \left( \tau\right)}
					 	+
						c\sinh \left(\frac{a \tau}{c} \right)
					\Evat{\partd{}{y^1}}{\gamma \left( \tau\right)} 
					\, .
					\label{eq:accelgamdot}
			\end{equation}
			So for $\gamma \left( 0\right)= 0$, we conclude 
			\begin{equation}
				\gamma \left( \tau \right) 
				= 
				\left( \frac{c^2}{a} \sinh 
				\left(\frac{a \tau}{c} \right), 
				\frac{c^2}{a} \left( \cosh 
				\left(\frac{a \tau}{c}  \right) -1 \right), 0, 0
				\right) \, ,
				\label{eq:accelgam}
			\end{equation}
			which is defined for all $\tau \in \R$.
			\par
			The need to solve non-linear 
			differential equations even for physically 
			simple situations is the norm in general-relativity, 
			not the exception. 
	\end{enumerate}
\end{subequations}
\end{Example}
\par
Light cones are the next mathematical structures 
we consider here.  
\begin{Definition}[Light cones]
	\label{Def:cones}
	Let $\left( \spti,g, \mathcal O\right)$ be a spacetime and 
	let $q$ be a point on $\spti$. \\
	The \emph{tangent light cone 
	$\mathfrak{c}_q$ at $q$} is the set of lightlike vectors 
	in $\CapT_q \spti$. The \emph{future tangent light cone 
	$\mathfrak{c}_q^+$ at $q$} is the set of future-directed 
	lightlike tangent vectors in $\CapT_q \spti$. Analogously, 
	we define the \emph{past tangent light cone 
	$\mathfrak{c}_q^-$ at $q$}. If $\exp \colon \dom \exp 
	\to \spti$ is the exponential map induced by the Levi-Civita 
	connection, we define the \emph{light cone 
	$\mathcal C_q$ at $q$} to be the image of $\dom \exp 
	\cap \mathfrak{c}_q$ under $\exp$. Similarly, 
	\begin{equation*}
		\mathcal C_q^+ := 
		\exp \left( \dom \exp \cap \mathfrak{c}_q^+ \right)
	\end{equation*}
	is the \emph{future light cone at $q$} and 
	\begin{equation*}
	\mathcal C_q^- := \exp \left( \dom \exp \cap \mathfrak{c}_q^-
	\right)
	\end{equation*}
	is the \emph{past light cone at $q$}. 
\end{Definition}
The light cones in the tangent spaces are manifolds in a 
natural way. 
\begin{Proposition}[Tangent light cone as manifold]
	\label{Prop:lightcone}
	\begin{subequations}
	Let $\left( \spti,g, \mathcal O\right)$ be a spacetime of 
	dimension $n+1$ and let $q \in \spti$. Then there is a unique 
	manifold structure 
	on the tangent light cone $\mathfrak{c}_q$, such that 
	it is an embedded submanifold of $\CapT_q \spti$. 
	With respect to this manifold structure, $\mathfrak{c}_q$ 
	splits into two connected components $\mathfrak{c}_q^+$ and 
	$\mathfrak{c}_q^-$. Moreover, for an orthonormal frame 
	$X \in \oframeb{\spti, g}$ at $q$, the maps 
	\begin{equation}
		\vec x_{\pm}  \colon \quad 
		\mathfrak{c}_q^\pm \to 
		\R^{n} \setminus \lbrace 0 \rbrace \quad
		\colon \quad 
		K \to \left( {\ubar X}^1 \cdot K, \dots , {\ubar X}^n 
		\cdot K \right)
		\label{eq:pmcoordinates}
	\end{equation}
	define (compatible) coordinates on $\mathfrak{c}_q^+$ and 
	$\mathfrak{c}_q^-$, respectively, having inverses 
	\begin{equation}
		\left( \vec x_{\pm}\right)^{-1}  \colon \quad 
		\R^{n} \setminus \lbrace 0 \rbrace \to 
		\mathfrak{c}_q^\pm \quad \colon \quad  
		\vec y \to \pm \abs{\vec y} X_0 
		+ y^a \, X_a \, 
		\label{eq:pmcoordinatesinv}
	\end{equation}
	with $a \in \lbrace 1, \dots, n \rbrace$. 
\end{subequations}
\end{Proposition}
\begin{Proof}
\begin{subequations}
	As an open submanifold, $\CapT_q \spti \setminus \lbrace 0 
	\rbrace$ is embedded in $\CapT_q \spti$. Now consider 
	\begin{equation}
		p \colon \CapT_q \spti \setminus \lbrace 0 \rbrace 
		\to \R \colon
		Y \to g_q \left( Y, Y\right) \, ,
		\label{eq:lightp}
	\end{equation}
	which is smooth and 
	$p^{-1}\left( \lbrace 0 \rbrace \right) = \mathfrak{c}_q$. 
	Since the vertical lift is a linear isomorphism and $p$ takes 
	values in $\R$, we may compute the differential 
	$p_*$ for $Y,Z \in \CapT_q \spti \setminus  \lbrace 0 \rbrace$
	via: 
	\begin{equation}
		p_* \tilde Z _Y = 
		\Evat{\partd{}{s}}{0} p \left( Y + s Z \right)
		= 2 g_q \left( Z,Y \right) \, . 
		\label{eq:difflightp}  
	\end{equation}
	Since $g_q$ is non-degenerate, ${\left(p_*\right)}_Y$ 
	is non-degenerate for each $Y
	\in \CapT_q \spti \setminus  \lbrace 0 \rbrace$ and hence 
	a submersion on $\mathfrak{c}_q$. Thus by the regular 
	value theorem, $\mathfrak{c}_q$ (together with the inclusion) 
	is an embedded submanifold of $\CapT_q \spti$. \\
	To get the components, choose any future-directed 
	timelike $Z \in \CapT_q \spti$ and define the continuous 
	function 
	\begin{equation}
		p' \colon \mathfrak{c}_q \to \R \colon K \to 
		g_q \left(Z,K \right) \, .
		\label{eq:lightpprim}
	\end{equation}
	As $p' \left( \mathfrak{c}_q^+ \right) = 
	(0, \infty)$, $p' \left( \mathfrak{c}_q^-  \right) = 
	(-\infty, 0)$, $\mathfrak{c}_q^\pm$ are mutually 
	disjoint. \\
	Connectedness of $\mathfrak{c}_q^\pm$ follows immediately, 
	if we can prove that
	\eqref{eq:pmcoordinates} are indeed coordinates. 
	If we consider $\vec x_\pm$ as maps from $\CapT_q \spti$ 
	to $\R^n$, 
	then smoothness is trivial. Since $\mathfrak{c}_q^\pm$ are 
	open submanifolds of the submanifold 
	$\mathfrak{c}_q$, the restriction of this map is also 
	smooth. The image under $\vec x_\pm$ is indeed contained in 
	$\R^n \setminus  \lbrace 0 \rbrace$, since 
	$K^0 \, X_0$ is not lightlike for any $K^0 \in \R$. 
	We now consider the map 
	\eqref{eq:pmcoordinatesinv}. Recalling that $X$ is 
	orthonormal, one verifies with $g_q$ and 
	$p'$ that \eqref{eq:pmcoordinatesinv}
	indeed maps into $\mathfrak{c}_q^\pm$. 
	\eqref{eq:pmcoordinatesinv}
	is also smooth, since taking absolute values for 
	$\vec y \in \R^n \setminus \lbrace 0 \rbrace$ is 
	smooth, and multiplication is smooth on 
	$\CapT_q \spti$ and hence on $\mathfrak{c}_q^\pm$. 
	Composing the maps in each direction, we get the identity. 
	Hence $\vec x_\pm$ are diffeomorphisms with inverses 
	\eqref{eq:pmcoordinatesinv}. 
\end{subequations}
\end{Proof}
Clearly, the coordinates $\vec x _ \pm$ depend highly on 
the choice of $X \in \oframeb{\spti, g}$. 
From a mathematical perspective, one would prefer to take 
those frames $X$, that are not just orthonormal but also 
spacetime-oriented, i.e. $X \in \mathcal O \cap 
\oframeb{\spti, g}$. 
\par
This leads us to the third set of important mathematical 
structures on spacetimes. 
\begin{Definition}[Frame of Reference]
	\label{Def:FR}
\begin{subequations}
	Let $\left( \spti , g, \mathcal O \right)$ be a 
	spacetime of dimension $n+1$.\\
	The \emph{frame of reference bundle%
	\footnote{German: ``Bezugssystemb\"undel''} 
	(over the spacetime $\left( \spti , g, \mathcal O \right)$)}
	or \emph{reference frame bundle (over 
	$\left( \spti , g, \mathcal O \right)$)} is the tuple 
	$\left( \mathcal P, \tilde \pi, \spti, \Lor_{n+1} 
	 \right)$, where $\mathcal P$ is the set
	\begin{equation}
		\mathcal P := \mathcal O \cap 
		\left( \oframeb{\spti, g} \right) \, ,
	\end{equation}
	$\tilde\pi := \pi \evat{\mathcal P} \colon \mathcal P 
	\to \mathcal Q$ is the restriction of $\pi \colon 
	\frameb{\CapT \spti} \to \spti$ to $\mathcal P$, and 
	$\Lor_{n+1}$ acts 
	canonically on $\mathcal{P}$ from the 
	right via 
	\begin{equation}
			\mathcal P \cross \Lor_{n+1} \to 
			\mathcal P \quad \colon \quad
			\left( X, \Lambda \right) \to X \cdot \Lambda \, .
			\label{eq:Loraction}
	\end{equation}
	A \emph{frame of reference} is an element of 
	$\mathcal P$ and a \emph{frame 
	of reference at $q \in \spti$} is an element of 
	$\mathcal{P}_q := \tilde \pi ^{-1} \left( q \right)$.  
\end{subequations}
\end{Definition}
The vector $X_0$ of a frame of reference 
$X \in \mathcal P$ may be identified as the tangent vector 
(divided by $c$) of some observer $\gamma$ at time 
$\tau$ on $\spti$. The vectors 
$X_1, X_2, X_3$ represent the orientation in `space' 
of a physical observer at time $\tau$ moving along $\gamma$. 
\begin{Remark}
	\label{Rem:framecredit}
	To our knowledge, the identification of 
	particular orthonormal frames $X$ with physical
	frames of reference is due to Walker 
	\cite{Walker1}. While he assumed $X_0$ to be future 
	directed, he did not explicitly assume 
	$X_1, X_1, X_3$ to be right-handed. As far as we know, 
	the use of these frames of reference to define 
	coordinates on the tangent past light cone 
	(as in \eqref{eq:pmcoordinates} on page 
	\pageref{eq:pmcoordinates}) is due to 
	Mast and Strathdee \cite{Mast}. 
\end{Remark}
We continue by showing that the frame of reference bundle 
is an embedded $\Lor_{n+1}$-structure on $\spti$. 
\begin{Theorem}
	\label{Thm:Psmoothness}
		Let $\left( \mathcal P, \tilde \pi, \spti, 
		\Lor_{n+1} \right)$ be the frame of reference 
		bundle over a spacetime 
		$\left( \spti, g, \mathcal O \right)$. \\
		Then there is a unique manifold structure on 
		$\mathcal P$ 
		such that $\left( \mathcal P, \tilde \pi, \spti, 
		\Lor_{n+1} \right)$ with the group 
		action \eqref{eq:Loraction}
		is a principal $\Lor_{n+1}$-bundle. 
		With respect to this manifold structure, 
		$\mathcal P$ is an embedded submanifold both of
		the spacetime orientation $\mathcal O$ and 
		the orthonormal frame bundle 
		$\oframeb{\spti, g}$. 
\end{Theorem} 	
\begin{Proof}
	As in the proof of 
	\thref{Prop:existspacetime} on page 
	\pageref{Prop:existspacetime}, let $\set{\mathcal U_ 
	\alpha}{\alpha \in I}$ be a trivializing 
	cover of $\spti$ with respective (smooth, local) 
	orthonormal frame fields $\overset{\alpha}{X}$. 
	If necessary, apply the time inversion matrix and 
	a space inversion matrix to make them 
	spacetime-oriented. 
	Then the $\overset{\alpha}{X}$s map into 
	$\mathcal P$ and the resulting transition matricies 
	are elements of $\Lor_{n+1}$. So by 
	\thref{Rem:TransextG}\ref{sRem:transitP} on 
	page \pageref{sRem:transitP}, $\mathcal P$ indeed 
	carries a unique manifold structure that turns it 
	into a $\Lor_{n+1}$-structure.
	Since $\Lor_{n+1}$ is embedded both in
	$\CLor_{n+1}$ and $\LieO_{1,n}$, an analogous 
	reasoning to the one in 
	\thref{Rem:Gstr}\ref{sRem:Gstr2} implies that 
	$\mathcal P$ is embedded both in 
	$\mathcal O$ and $\oframeb{\spti, g}$. We also 
	refer to the book by Baum \cite{Baum}*{p. 66} 
	regarding the latter argument. 
\end{Proof}
	We close this section with the remark that the frame
	of referenc bundle is a very `natural' mathematical 
	object. Indeed, if we only consider the bundle 
	$\mathcal P$, then we can recover both the metric 
	$g$ and the spacetime orientation $\mathcal O$ 
	on $\spti$ by extension. 
	This in turn leads to the point of view that, at least 
	mathematically, a spacetime is a 
	$\Lor_{n+1}$-structure 
	$\mathcal P$ on a manifold $\spti$ 
	equipped with the Levi-Civita
	connection (considered either as a tangent bundle 
	connection on 
	$\CapT \spti$ or as a particular Ehresmann connection on 
	$\mathcal P$). This approach is geometrically more 
	coherent, but physically less accessible. 
	
\section{Heuristic Motivation of the Space-Time Splitting}
\label{sec:motivation}

%Hasse said: velocity of light independent of velocity of source
%			can be used to define radial dist...

In the following we motivate the mathematical formalism 
of separating space and time in general relativity. 
We restrict ourselves 
to presenting the underlying philosophy, 
the detailed mathematical implementation is 
postponed to the two consequent sections. 
Contrary to other such splitting formalisms, we will discover 
that no further restrictions on the spacetime are 
required apart from the ones already imposed by 
the mathematical definition. 
\par
It needs to be said in 
advance, that the mathematical machinery, as outlined 
in sections \ref{sec:space} and \ref{sec:kinematic}, is 
self-contained and can be applied without taking notice of 
the underlying philosophy as presented here - provided 
the mathematical quantities are interpreted appropriately. 
Yet we would not have been able to construct the splitting 
formalism without these philosophical considerations and 
thus believe them to be as much part of the theory as 
the mathematical formalism. In our mind, a physical theory is 
ideally build upon an ontology consisting of principles 
and postulates, not just mathematics. The relation between 
the mathematical formalism and the measured quantities is 
then derived from this ontology. We refer to the book by 
Frisch \cite{Frisch} for an in-depth discussion of this approach 
in the context of the theory of electrodynamics. Of course, 
regardless of one's view 
towards the role of metaphysics in physics,  
any theory of nature needs to be assessed by the quantity and 
quality of its empirical predictions. 
\par
The main objective of this section is to provide an 
answer to the following questions: 
How does the mathematical formalism of relativistic spacetimes
relate to our individual experience 
of the separateness of time and space? In other words, 
what is time and what is space in general relativity? 
\par
Indeed, we have already touched upon the role of time in 
the theory: On page \pageref{Prin:clock} 
in the previous section, we stated the clock principle: 
The motion of physical objects in spacetime 
is represented by future directed timelike curves 
and the time measured by a clock moving along such 
a curve is the proper time \eqref{eq:propertime}. In order 
to give a definite 
answer to the question of what constitutes time in relativity, 
it is sufficient to realize that proper time is the only 
physically measurable time and all other concepts of 
time are derived thereof. We invite the reader to ponder 
this claim for some time. Once we accept it, we must 
conclude that \thref{Prin:clock} reduces the 
problem of separating space and time to giving meaning 
to the physical concept of space in general relativity. 
\par
In the literature on general relativity, 
there exist several mutually conflicting answers 
to the question what `space' is in the theory. 
In the following, we list what we believe to be the three most 
common misconceptions of space in general relativity. 
\begin{description}[style=unboxed,leftmargin=0cm]
	\item[Local rest spaces:] Given a point 
	$q$ on the spacetime $\spti$ and an observer 
	vector $Z$ in the tangent space $\CapT_q \spti$, 
	the orthogonal complement $\left( \R Z\right)^\perp$
	with respect to $g_q$ is sometimes referred to 
	as the `local rest space' with respect to $Z$ 
	(see e.g. \cite{Sachs}*{\S 2.1.4}). Philosophically, 
	the vector $Z$ should be viewed as a tangent vector 
	of an observer in the spacetime. 
	\par
	While this conception of space has the advantage 
	of being well-defined and relating 
	directly to observers, it is not clear 
	how it is connected to the dynamics 
	on $\spti$ and observed quantities like relative 
	position, velocity, etc. One may apply the 
	exponential map to $\left( \R Z\right)^\perp$ in 
	order to relate it to $\spti$, but, as there can be no 
	causal interaction between these points and the 
	observer, this image is unrelated to what we 
	experience as `space'. 
	\item[Spacelike hypersurfaces:] By definition, 
	a \emph{spacelike hypersurface 
	$\mathcal S$}, if given as a subset of a spacetime 
	$\spti$, is a submanifold of $\spti$ such that for every 
	$q \in \mathcal S$ the tangent space $\CapT_q \mathcal S$
	is a spacelike hyperplane in the Lorentz vector
	space $\left(\CapT_q \spti, g_q \right)$. As noted 
	in the introduction, this is the view of `space' taken
	in the so called ADM formalism (cf. 
	\cite{Misner}*{p. 419sqq. \& \S 21.7}). 
	\par
	Trivially, spacelike hyperplanes are highly 
	non-unique and thus one needs to motivate 
	why and how one chooses one or more particular ones 
	as `space'. There do exist approaches to this issue, 
	but the fundamental problem remains 
	that this conception of space is substantially 
	unrelated to our individual experience. 
	We believe that, as in the previous case, 
	the researchers have been mislead by the word 
	`spacelike'. In addition, identifying physical
	`space' as spacelike hypersurfaces seems only 
	natural to us if we take 
	an outside perspective towards the theory
	- an approach that is subtly, yet warningly 
	reminiscent of Newtonian thinking. 
	\item[Coordinate hyperplanes:] In the practical 
	application of the theory, spacetimes are usually 
	given implicitly by writing down the component functions
	of the metric in a single chart. In addition, the 
	coordinates often carry suggestive names such as 
	$t$ and $x$, which can be misleading 
	to those physicists lacking a formal training in 
	differential geometry and being accustomed to 
	Newtonian thinking. Those are then tempted to 
	identify $(t = \const)$-hyperplanes as space. 
	\par
	In contrast, it is evident 
	to the physically versed geometer 
	that coordinates may only carry direct physical 
	meaning, if they are `adapted' to the underlying 
	geometric structures. This means that the 
	structures take a special form in those coordinates, 
	as is, for instance, the case for normal 
	coordinates with respect to a metric. 
	In most cases, however, coordinates are 
	meaningless ways of labeling points. The fact 
	that geometric 
	structures have usually been implicitly given 
	in Newtonian physics has certainly assisted 
	the fallacy of identifying $(t = \const)$-hyperplanes 
	with physical space. To put it bluntly, 
	calling a coordinate $t$ does not make it a physical 
	measure of time just like calling a coordinate 
	$x$ does not make it a physical 
	measure of spatial distance. 
\end{description}
\par 
Therefore, to understand what space is in general 
relativity, we need to let go of Newtonian 
concepts and grasp the role played by the different 
geometric structures. 
\par
The abandonment of Newtonian thinking is eased by the 
appreciation of the fact that the Newtonian way 
of seeing the world has already proven itself to be 
an inadequate, approximative at best, 
path of gaining insight into the inner workings of nature. 
So the function of Einstein's theory 
in this inquiry needs to be openly 
exposed: It is nothing short of 
a revolutionary, scientific 
alternative to outdated metaphysical ways of thinking about 
space and time. 
\par
Hence, to make a step in understanding the role played by the 
geometric structures, we need to view the physical world from 
within the theory. If we ask for our individual experience 
of physical space, it has to be defined mathematically with 
respect to a single observer 
$\gamma \colon \mathcal I \to \spti$ at fixed time 
$\tau \in \mathcal I$. That is, we ask what a 
single observer would identify as space at a fixed time. 
\par
A natural answer to this question is that space is what an 
observer sees at an instant. However, this `definition' is 
more subtle than it may appear at first sight. In general 
relativity, light is commonly modeled with a closed $2$-form 
on the spacetime, known as the Faraday-form or electromagnetic 
field. If the Einstein equation is correct, then 
the presence of this electromagnetic field has a direct effect 
on the spacetime geometry and vice versa. But even if one 
ignores this issue, it requires some more assumptions and 
a lengthy heuristic argument (cf. \citelist{
\cite{Misner}*{\S 22.5} \cite{Straumann}*{\S 1.8}}) 
to arrive at the so-called geometric optics approximation 
\cite{Perlick2}*{p. 7sq.}, where light rays may be 
identified with lightlike geodesics. In addition, it 
is conceptually problematic to implicitly presuppose the 
existence of light in an identification of `space' within 
the mathematical framework, since `space' should also be 
present in the absence of light. Hence the concept of 
space ought not to be based on the concept of light, 
despite the observation that our subjective 
experience of space always involves 
light. For these reasons, we avoid the 
subject of light here altogether with the well-meant advice that the 
specific physical situation has to be analyzed when light 
is part of the model. 
\par
Fortunately, we may circumvent the issue 
by identifying space with the set of all points 
in the spacetime that can causally interact with the 
observer at an instant, i.e. those points that can be linked 
to the observer's position via a future directed lightlike 
geodesic. This is physically vague, yet functional, 
as in the geometric optics 
approximation those are precisely the points $q \in 
\spti$ that can send light to the observer $\gamma$ at that 
instant $\tau$ of his or her proper time. Thus we identify the
`space' for an observer at time $\tau$ with the past light 
cone $\mathcal C ^- _{\gamma \left( \tau \right)}$ of the 
point $\gamma \left( \tau \right)$. So for 
convenience, from now on we say that an observer $\gamma$
\emph{sees} a point $q\in \spti$ at time $\tau 
\in \dom \gamma$, if $q$ is in the past light cone 
$\mathcal C ^- _{\gamma \left( \tau \right)}$ of 
$\gamma \left( \tau \right)$ - well-knowing that this 
physically refers to potential causal interactions, 
rather than the emission and reception of light. This subtlety 
makes it possible in principle to apply the theory of 
space-time splitting as presented here also in conjunction with 
other models of light and beyond the geometric optics approximation.
\par
However, the identification of space at an instant $\tau$ 
with the past light cone $\mathcal C ^- _{\gamma \left( 
\tau \right)}$ leads to another problem: How do we 
define a natural distance function on $\mathcal C ^- _{\gamma \left( 
\tau \right)}$? Clearly, we would like this distance 
function to be closely related to the empirically 
measured distance we observe in
the presence of light and where the geometric optics approximation
is admissible. Yet, mathematically, 
lightlike geodesics have vanishing length, so 
it is inadmissible to directly use the Lorentzian distance 
to measure lengths on the past light cone. 
\par
The bad news is, that the careful motivation of this distance 
function is quite elaborate and will occupy the rest of this section. 
The good news is, that once this is done, figuratively 
speaking, everything else falls into place and we will then have 
obtained a well motivated, and, as we believe, physically correct 
splitting of spacetime into space and time. 
\par
Let us specify what precisely we mean with the word
`empirically measured distance'. 
If we idealize a physical observer at an instant to be 
point-like, as it is commonly done in the theory of relativity, 
then we can abstractly 
think of its spatial alignment to be given by 
three vectors: The first one pointing in the forward direction, 
the second one pointing to the left and the third one 
pointing upwards in accordance with the right-hand rule. 
This is the original concept behind 
the words `frame of reference', i.e. a point in `space' to 
which three mutually orthogonal (as perceived by the 
physical observer), right-handed vectors are attached. 
Even if the `surrounding geometry' is very complicated, the 
physical observer may assign to everything it sees 
(in the aforementioned sense) a polar angle $\phi$ and an 
azimuthal angle 
$\theta$ with respect to its frame of reference. 
The measurement of these angles is done with 
an (idealized, infinitely small) goniometer. Now, if we were to find 
an appropriate notion of radial distance $r$, then we could 
use the definition of spherical coordinates in $\R^3$ to assign 
to each observed point a value $\vec x \in \R^3$ with 
respect to the frame of reference. Then the distances between two 
points $\vec x, \vec x' \in \R^3$ would be given by the usual 
Euclidean distance formula: 
\begin{equation}
	\dist \left( \vec x, \vec x' \right) 
	:= \abs {\vec x - \vec x' } \, .
	\label{eq:distR3}
\end{equation}
Please note that this expression is invariant under rotation, i.e. 
it does not change under the action
\begin{equation*}
	 \LieSO_3 \cross \, \R^3 \to \R^3
	\colon \left( A, \vec x\right) \to A \cdot \vec x \, . 
\end{equation*}
\par
One may object to our discussion that we have implicitly assumed the 
observed geometry as Euclidean at the onset and hence this is what 
we ultimately arrive at. This criticism is justified, but 
again the issue 
is more subtle: How a physical observer measures 
distance is in fact convention, 
i.e. up to its subjective choice, but it is objective, 
what is ultimately measured empirically in accordance with 
this convention. 
Hence the task of finding the physically correct distance function 
on the past light cone $\mathcal C ^- _{\gamma \left( 
\tau \right)}$ at time $\tau$ is linking this choice of viewing 
the world from the (perhaps due to 
our cultural imprint) subjectively chosen perspective of Euclidean 
geometry to the actual spacetime geometry. In fact, the 
conventionality of the metrization of physical space 
was already emphasized by Poincar\'e, Reichenbach and 
Gr\"unbaum (cf. \citelist{\cite{Jammer}*{p. 207 sqq.;} 
\cite{Reichenbach}}).
\par
It is therefore our task to establish this link 
between the chosen Euclidean geometry and the actual 
geometry of spacetime. A priori it may depend on 
the physical observer's state of motion, so we 
require an appropriate postulate. The following one 
is due to Mashoon \cites{Mashoon1,Mashoon2}, which he termed 
``the hypothesis of locality''. While we do not follow 
his wording and do not agree 
with his implementation, the postulate does appear to be 
implicit in the general theory of relativity. 
\begin{Postulate}[Mashoon]
	\label{Post:Mashoon}
	The spatial distances an observer measures at an instant 
	are independent both of its absolute acceleration and 
	rotation. 
\end{Postulate}	
The word `absolute' is used to distinguish the acceleration of 
the observer and the rotation of its frame of reference from 
relative acceleration and rotation, which are unrelated concepts. 
The presence of absolute acceleration and rotation can be 
empirically verified on in principle arbitrarily small 
spatiotemporal scales, as measured by the observer, 
with an accelerometer and gyroscope, respectively. 
That is, ideal accelerometers and 
ideal gyroscopes are pointlike and it makes sense to speak of them 
for a physical observer without knowing how the observer's 
geometry relates to the spacetime geometry. 
\thref{Post:Mashoon} now states that independent 
of whether the accelerometer and 
gyroscope show the presence of acceleration or rotation, the 
distances, as measured by that observer at that instant, remain 
unaffected. In particular, for the definition of spatial 
distances at $\gamma\left( \tau \right)$ and on arbitrarily large 
length scales, we may assume the observer's frame of reference at 
the event is inertial, i.e. non-accelerating and non-rotating.
Moreover, scientific realism dictates that the distances 
are also independent of its spatial orientation. Thus in 
the mathematical theory of relativity, the definition of a 
distance function for an observer $\gamma$ on the past light 
cone at proper time $\tau \in \dom \gamma$ may only depend on the 
point $\gamma \left( \tau \right)$ and the observer's so-called 
$4$-velocity $\dot \gamma _ \tau$ at the point. Hence we 
do not need any additional mathematical structures for a so-called 
spacetime splitting. As observers exist on arbitrary spacetimes, 
no further constraints need to be put on it
and we are not faced with 
the uncomfortable question of why we have chosen this 
particular mathematical structure among the possibly 
infinitely many possibilities. 
\par
Of course, \thref{Post:Mashoon} does not say which angles and 
radial distance an observer should measure in $\spti$. 
To answer this question, we will employ the Einstein equivalence 
principle, which directly gives us the angles $\theta$
and $\phi$, and to obtain the radial distances we employ 
an analogy in Riemannian geometry. 
\par 
The following formulation of the Einstein equivalence 
principle% 
\footnote{Dr. Hasse pointed out to us that there exist other 
formulations of the Einstein equivalence principle 
(see e.g. \cite{LaemmerzahlA0}). For our purposes we 
only require, that the measurement 
of spatial distances at an instant 
by an inertial, physical observer works in 
approximately the same way as in special relativity theory, 
provided the measured distances are `small enough'.}
 has been 
taken from the book by Carroll \cite{Carroll}*{p. 50}. 
For an original 
discussion due to Einstein, see e.g. his lecture notes
\cite{Einstein2}. It is also 
worth a note, that in the original anticipation of the 
theory Einstein 
identified what an observer `sees' directly with coordinates 
in spacetime. However, since what an observer `sees' 
is its past light cone and one can prove that in the presence of 
curvature this may intersect itself, this point of view is only 
justified on small enough spatiotemporal scales. 
\begin{Principle}[Einstein Equivalence Principle]
	\label{Prin:Einst}
\begin{subequations}
	It is impossible to detect the existence of a gravitational field 
	by means of local experiments. Hence for inertial frames of 
	reference within sufficiently small spatiotemporal scales, the 
	laws of special relativity are approximately valid. 
\end{subequations}
\end{Principle}
As observed by Mashoon, \thref{Prin:Einst} in conjunction with 
\thref{Post:Mashoon} imply that at an instant and on 
sufficiently small
spatial scales, a rotating and accelerating frame of reference under 
the influence of gravitation measures approximately the same 
distances as one which is neither rotating, accelerating nor 
under the influence of gravitation. 
So intuitively, general relativistic distances are locally 
approximated by special relativistic ones. Mathematically, we 
approximate the sought-after distance function on
$\mathcal C^- _{\gamma \left( \tau\right)}$ for general observers 
$\gamma$ on $\spti$ at proper time $\tau$ with 
the special-relativistic distance function in the tangent past light cone 
$\mathfrak c _{\gamma \left( \tau\right)}$ at 
$\gamma \left( \tau \right)$, since a neighborhood of the origin in the 
tangent past light cone of $\gamma \left( \tau \right)$ 
approximates the past light cone at $\gamma \left( \tau \right)$ 
via the exponential map. The latter statement 
is precisely the proposition \cite{O'Neill}*{Chap. 3, Prop. 33}
on the existence and properties of normal coordinates at the point 
$\gamma \left( \tau\right)$. Therefore, we may identify 
$\CapT _{\gamma \left( \tau \right)} \spti$, along 
with the natural geometric structures, as dictated by this 
approximation, as Minkowski spacetime and $\dot \gamma_\tau$ 
as the $4$-velocity vector of a special-relativistic observer at rest
at the origin. 
\par
So let us recall how distances are 
measured in special relativity. Accordingly, let
$\left( \R^4, g, \mathcal O \right)$ be 
Minkowski spacetime (cf. \thref{Ex:Mspti} on page 
\pageref{Ex:Mspti}) and let $\gamma'$ be the standard 
observer from \thref{Ex:obsMink}, given by 
\begin{equation*}
	\gamma' \colon \R \to \R^4 \colon \tau \to \gamma \left( 
	\tau \right) := \left( c \tau, 0, 0, 0\right) = c \tau \baseR_0
	\, ,
\end{equation*}
which is heuristically considered to be `at rest' at the origin and 
sees physical occurrences in the spacetime. For two points 
$x= \left( c t, \vec x \right)$, $x'= \left( c t', \vec x' \right)$ 
in $\R^4$ the distance is then given by 
\begin{equation}
	\dist \left(x, x' \right) = \abs {\vec x - \vec x' } \, ,
	\label{eq:distSR}
\end{equation}
in analogy to equation \eqref{eq:distR3}. Since we are only 
interested in the distances measured by the observer 
$\gamma'$ at one particular time, we may restrict 
the distance function \eqref{eq:distSR} to the past light cone 
$\mathcal C^-_{\gamma' \left( 0\right)} = \mathcal C^-_{0}$ 
in $\R^4$ at time $0$ and have thus obtained what we asked for.
\par
\begin{figure}[htp]  
\centering
\includegraphics[scale=0.25]{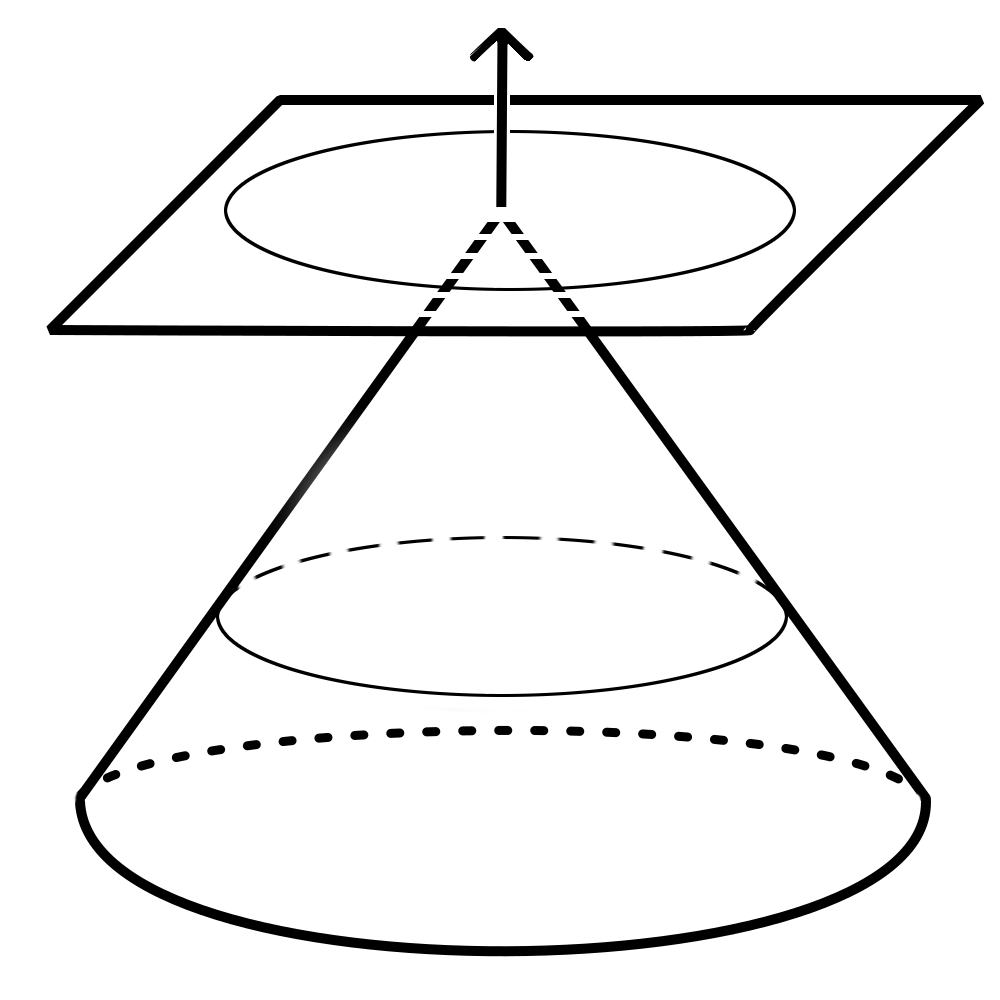}
\caption{	The picture indicates the 
			measurement of distances on the 
			past tangent light cone 
			$\mathfrak{c}_q^-$ at a point $q$ in 
			a three dimensional spacetime by embedding the 
			tangent space $\CapT_q \spti$ into 
			Euclidean $3$-space. 
			At the top we see the 
			tangent vector of the observer 
			at some fixed time, which gives rise to an 
			orthogonal hyperplane. The cone $\mathfrak{c}_q^-$
			is situated below. 
			The surface of constant radial distance $r$ 
			on $\mathfrak{c}_q^-$, as measured by the observer, 
			is orthonally projected to 
			a circle of radius $r$ on the hyperplane. 
			The observer's measurement of angles is analogous. 
			%maybe don't use standard vector in future
			}
\label{fig:distSR}
\end{figure}

 
So how precisely do we carry this 
construction over to the general relativistic 
case? Instead of considering an observer $\gamma$ in 
$\spti$, \thref{Post:Mashoon} allows us to simplify the situation 
by only considering an 
observer vector $c X_0$ at $q \in \spti$ and interpreting it 
as the tangent vector of $\gamma$ at time 
$\tau=0$. Since we would 
like to consider 
Minkowski spacetime as the `tangent cone approximation 
to the geometry on $\spti$' at $q$, we need to construct an 
analogous distance function on $\mathfrak{c_q^-}$.  
Obviously, $\CapT_q \spti$ is equipped 
with the Lorentz product $g_q$ and for any $Y,Z 
\in \CapT_q \spti$ we can use the vertical lift 
$\tilde Z_Y \in \CapT_Y \CapT _q \spti$ of $Z$ at $Y$ 
(see \eqref{eq:vertlift} on page \pageref{eq:vertlift}) 
to define a Lorentzian metric $\tilde g$ on $\CapT _q \spti$ 
via
\begin{equation*}
	\tilde g _ Y \left(\tilde Z_ Y, \tilde{Z}'_Y \right) := 
	g_q \left( Z, Z' \right)
\end{equation*}
for any $Y, Z, Z' \in \CapT _q \spti$. Simliarly, one may
use the vertical lift together with the fiber $\mathcal O _q$ 
to define a spacetime orientation on $\CapT_q \spti$. This 
procedure turns $\CapT_q \spti$ into a spacetime categorically
isomorphic to Minkowski spacetime, i.e. 
they are essentially the same mathematical object. 
Indeed, this procedure is precisely the 
one imposed by viewing $\CapT_q \spti$ as an approximation to
the spacetime $\spti$ at $q$ via the exponential map. 
Hence the curve
\begin{equation*}
	\gamma'' \colon \R \to \CapT_q \spti \colon \tau \to 
	\gamma'' \left( \tau\right) := 
	\tau c X_0 \, ,
\end{equation*} 
can be identified with $\gamma'$ above, 
and so $c X_0 \in \CapT_q \spti$ can be identified with 
$c \baseR_0 \in \R^4$. Observing now that in Minkowski 
spacetime $\R^4$, we can reverse the procedure and 
carry the Lorentzian metric down to the standard Lorentz product 
$\eta$ in the vector space $\R^4$, we may use $\eta$ to 
identify the $t = 0$ hyperplane as 
$\left( \R c \, \baseR_0 \right)^\perp$. The distance from 
equation \eqref{eq:distSR} is then just the Euclidean distance 
of the orthogonal projections of $x$ and $x' \in \R^4$ along 
$c \, \baseR_0$, that is 
\begin{align*}
	\dist \left( x, x'\right) 
	&= \sqrt{\delta \left(\vec x - \vec x',  
	\vec x - \vec x'\right)} \\
	&= \sqrt{ - \eta \left( \left( x -  x'\right)^\perp, 
	\left( x -  x'\right)^\perp \right) } \\ 
	&=: \sqrt{ \left( - \eta^\perp\right) 
	\left( x -  x',  
	x - x'\right) }
	\, \, .
\end{align*}
Therefore, the observer vector $c X_0 \in \CapT_q \spti$ 
defines a spacelike hyperplane $\left( \R X_0\right)^\perp$ 
via the Lorentz product $g_q$ and the distances on the tangent
past light cone $\mathfrak{c}^-_q$ are the ones induced by 
restricting the degenerate product $- g_q^\perp$ to the cone. 
The construction is graphically depicted in figure 
\ref{fig:distSR}. 
Explicitly, the mutual distances of any $K,K' \in \mathfrak{c}^-_q$ 
are given by 
\begin{equation}
	\dist \left( K, K'\right) 
	:= \sqrt{\left( - g_q^\perp \right) 
	\left( K-K', K-K' \right)} \, .
	\label{eq:distcone}
\end{equation}
\par
Since the exponential is invertible on an open 
neighborhood of $0 \in \CapT_q \spti$, we have thus 
obtained a distance for any two points $q',q''$ in the image
of the restricted exponential $\exp_q$:  
\begin{equation*}
	\dist \left( q', q''\right) 
	\approx \dist \left(\exp_q^{-1}\left( q' \right),  
	\exp_q^{-1} \left( q''\right)
	\right) \, \, . 
\end{equation*}
The left hand side refers to the sought-after 
physical distance on $\mathcal C^-_q$ for an observer 
with tangent vector $c X_0$ at $q$ and the right hand side 
refers to the just constructed distance on 
$\mathfrak{c}^-_q$.
\par 
By assumption, the approximation gets 
better the closer $\exp_q^{-1}\left( q' \right)$ and 
$\exp_q^{-1}\left( q'' \right)$ are to the origin $0 \in 
\CapT _q \spti$. Thus, if we are only interested in the mutual 
angles as seen by the the observer $\gamma$ at $q$, we can make 
the approximation arbitrarily precise by decreasing the radial 
distance (in $\CapT_q \spti$ with respect to $- g_q^\perp$) of 
$\exp_q^{-1}\left( q' \right)$ and 
$\exp_q^{-1}\left( q'' \right)$ to the origin. 
In the limit 
we get a precise value. Moreover, in the more general case where 
$q' = \exp_q \left( K\right), q'' = \exp_q \left(K' \right) 
\in \mathcal C^-_q$ with $K,K' \in \mathfrak{c}^-_q$ given, 
we can also
decrease the radial distance of $K$ and $K'$ in $\mathfrak{c}^-_q$ 
from the origin to get the mutual angles of arbitrary observed 
points $q'$ and $q''$. 
\par
It is, however, important to note that the
fact that $\mathcal C^-_q$ can intersect itself implies that 
an observer can see one and the same event $q'$ from more 
than one direction. This corresponds to two different 
radial curves in $\mathfrak{c}^-_q$ intersecting $q'$ under 
the exponential map. Physically, the phenomenon is known as (strong) 
gravitational lensing and its existence has been 
empirically confirmed. We refer to the article by Perlick 
\cite{Perlick2} for an introductory discussion 
of this phenomenon from the perspective of general relativity. 
This notion of angle has also been employed by Hasse 
in an analysis \cite{HasseA1} of the observed size 
of astronomical objects in the geometric optics approximation. 
\par
Having determined how to measure angles 
with respect to a chosen frame of reference, 
we are left with identifying the radial distances. 
As this refers to the situation where the angles 
$\phi, \theta$ are fixed, the radial distance has to be 
defined on past directed lightlike geodesics `starting' 
at $q$ and, as a distance between a point on the geodesic and 
the observer at $q$, it should be monotonically increasing 
along the geodesic. Thus we actually ask for a particular 
(continuous) parametrization of each such geodesic. In 
addition, the equivalence principle requires that, 
at least `close' to the observer, this parametrization 
is approximately given by 
\begin{equation*} 
	\set{s \in \R_+}{s K \in \dom \exp_q}
	\to \spti \quad \colon \quad 
	s \to \exp_q \left( s K\right)  
\end{equation*}
for a unit vector $K$ in $\mathfrak{c}^-_q$ (with 
respect to $- g^\perp_q$). 
The analogue in Riemannian geometry supports the conjecture 
that this \emph{affine parameter distance} is not merely 
an approximate, but an exact radial distance. 
\begin{Lemma}[Exponential preserves radial distances]
	\label{Lem:RiemExp}
	Let $\left( \spti, g\right)$ be a Riemannian manifold 
	with standard exponential $\exp_q$ at $q$. 
	Further denote by $\tilde Z_Y \in \CapT_Y \CapT _q \spti$
	the vertical lift of $Z \in \CapT _q \spti$ at 
	$Y \in \CapT _q \spti$ (cf. \eqref{eq:vertlift} 
	on page \pageref{eq:vertlift}). \\
	Then the equation 
	\begin{equation*}
		\tilde g _ Y \left(\tilde Z _ Y, \tilde{Z}'_Y \right) := 
		g_q \left( Z, Z' \right)
	\end{equation*}
	for any $Y, Z, Z' \in \CapT _q \spti$ 
	defines a Riemannian metric $\tilde g$
	on $\CapT _q \spti$. 
	Moreover, for all radial curves 
	\begin{equation*}
		\theta \colon \left(0, r \right)
		\to \CapT_q \spti \quad
		\colon \quad
		s \to \theta \left( s \right) :=	s Y
	\end{equation*}
	with $Y \in \CapT _q \spti$ and $r \in \R_+$
	such that $\exp_q \circ \, \theta$ 
	is defined, the corresponding Riemannian lengths of 
	$\theta$ and $\exp_q \circ \, \theta$ coincide. In particular, 
	if $Y$ is a unit vector, the length of $\exp_q \circ \, 
	\theta$ is $r$. 
\end{Lemma}
\begin{Proof}
\begin{subequations}
	$\tilde g$ is a smooth Riemannian metric, since 
	$\left( Y,Z\right) \to \tilde Z_Y$ is smooth and 
	$Z \to \tilde Z_Y$ is a linear isomorphism for all
	$Y \in \CapT_q \spti$. It is now sufficient to calculate 
	\begin{align*}
		\tilde g _{\theta \left( s \right)} 
		\left( \dot \theta _s , \dot \theta _s \right)
		&= 
		\tilde g _{s Y} 
		\left( \tilde Y _{s Y } , \tilde Y _{s Y } \right) 
		= g_q \left( Y, Y\right) \\
		&= g_{\exp_q \left( 0 \, Y \right)}
		\left( \dot{\left(\exp \circ \,\theta \right)}_{0}, 
		\dot{\left(\exp \circ \,\theta \right)}_{0} \right) \\ 
		&= g_{\left(\exp \circ \, \theta \right) \left( s\right)
		} \left( \dot{\left(\exp \circ \,\theta \right)}_{s}, 
		\dot{\left(\exp \circ \,\theta \right)}_{s}
		\right) 
		\, , 
	\end{align*}
	where the last equality follows from the fact that 
	tangent vectors of geodesics have constant length. 
\end{subequations}
\end{Proof}
\thref{Lem:RiemExp} is a special case of the 
Gau\ss' lemma (see e.g. \cite{O'Neill}*{Lem. 5.1}). From 
a mathematical perspective, it is therefore natural 
to assume that the Lorentzian exponential preserves 
the radial distances on the past tangent light 
cone with respect to $- g_q ^\perp$. 
\par
According to Perlick 
\cite{Perlick2}*{p. 21}, the affine parameter distance 
was discovered by Kermack, M'Crea 
and Whittacker \cite{Kermack}. 
Unfortunately, they did not give any 
physical interpretation of it. 
\begin{Remark}[Parallax Distance]
	\label{Rem:par}
	There exist several common astronomical 
	distance measures, we refer for instance to the articles 
	by Hogg \cite{Hogg} and Perlick 
	\cite{Perlick2}*{\S 2.4}. 
	\par 
	Only few of them can serve as a parametrization 
	of lightlike geodesics in the general case, but 
	one might be tempted to employ the so called 
	parallax distance for this purpose:  
	If a physical observer sees an extended 
	massive object and neither the  
	observed angular size, shape nor location 
	changes measurably, then the observer may 
	accelerate without rotating 
	in a direction orthogonal to the observed 
	center of the object. After travelling a 
	distance $s$ (as determined by the 
	acceleration), this yields an 
	angular displacement of the center by an angle 
	$\alpha$ and thus by triangulation in 
	the Euclidean plane, we may define 
	the \emph{parallax distance} to be 
	\begin{equation*}
		r = s \tan \alpha \, . 
	\end{equation*}
	In the process of taking the limit where $s$ tends to $0$, 
	$s$ becomes a better approximation to the 
	affine parameter distance by the equivalence principle 
	and $r$ should, at least intuitively, remain constant. 
	Furthermore, by taking the limit, it might be possible 
	to circumvent 
	the problem that quite a few assumptions 
	are necessary to make such an idealized situation 
	mathematically and physically feasible. 
	\par
	However, in a personal correspondence 
	W. Hasse provided a counterexample 
	to the claim that the affine parameter distance 
	and parallax distance coincide. We give a slight 
	adaption of his argument here: 
	Consider a spacetime 
	region between the observer and the 
	seen object and assume the parallax distance 
	has been determined in accordance with the above 
	procedure. Now apply a conformal transformation 
	$g \to f g$ such that the strictly positive 
	function $f$ is precisely one outside the region 
	and greater than $1$ inside the region. 
	It can be shown (see e.g. \cite{Wald}*{p. 446}) 
	that this leaves lightlike geodesics invariant 
	and leads to an increase in the affine parameter 
	distance. Yet the invariance of lightlike geodesics 
	under this transformation together with the fact that 
	it acts neither on the observer nor the object implies
	that the parallax distance also stays invariant. 
	Therefore the two distances are conceptually 
	different.
\end{Remark}
In his article \cite{Perlick2}*{p. 21} Perlick also claims 
that the affine parameter distance is ``not an observable''. 
Indeed, the argument in \thref{Rem:par} supports his claim 
by indicating that the distance is  
not directly measurable. Nonetheless, we believe that the 
affine parameter distance is not devoid of 
physical meaning. In fact, we are not aware of any 
other (non-ad hoc) 
radial distance measure satisfying the aforementioned 
requirements and behaving sensibly under conformal transformations.  
Beyond these arguments, its mathematical 
`naturalness' even suggests it to be a fundamental physical 
distance measure, despite the problem that its empirical 
measurement needs to be indirect in the presence of 
curvature. Its empirical evaluation depends strongly on 
the physical model and shall not concern us here. References 
\citelist{\cite{Perlick2}*{\S 2.4} \cite{HasseA1} 
\cite{EllisA4} \cite{Hogg} \cite{Kristian} \cite{Ellis1} 
\cite{Etherington} \cite{Ellis2}} provide further reading 
on the subject. 
\par 
Returning to our original discussion, 
if the physical observer has determined the angles 
$\theta, \phi$ and radial distances $r$ of each event
relative to its instantaneous frame of reference, 
it may employ the standard formula for spherical 
coordinates to relabel these in Cartesian coordinates 
$\vec x=(x^1, x^2, x^3)$ and the distance between the 
these events is then simply given by the Euclidean 
distance in accordance with equation
\eqref{eq:distR3} on page \pageref{eq:distR3}. 
By doing this for every time 
$\tau$, the observer can assign to each observed event 
a point $(\tau, \vec x )$ in its `observer spacetime' and
even assign distances to events at different times relative
to its frame of reference. 
Therefore, the space-time splitting philosophically 
gives rise to a second `spacetime' with its own geometric 
structures, which is of course
not a spacetime in the mathematical sense, but de facto 
the spacetime of Newtonian mechanics: Its geometry is Euclidean 
and time may be treated as a simple parameter. As noted 
before, this is not by accident, but a result of convention. 
Ultimately, the separation of space and time in general 
relativity is the question of how the Newtonian conception of 
the physical world relates to the general relativistic one. 
Thus a space-time splitting construction is also a prerequisite 
for showing the precise mathematical relation between the theories
including the so called Newtonian limit. For a discussion of the 
latter, we refer to chapter \ref{chap:nlimit}. 
\par
Summing up, we have identified the primary concept of 
time in general relativity to be the one measured by 
individual clocks along future directed timelike curves, 
in accordance with the clock principle. Therefore, we 
found that the concept of space within the theory also 
had to be defined with respect to individual observers and 
determined it to be the past light cone 
at the point where the physical observer is 
located at an instant of its time. 
We discussed that the choice of geometry with which 
the observer views `space' is conventional, 
but the distances on the past light cone in accordance with 
that convention are not conventional. 
In an attempt to link the two, we applied Mashoon's postulate 
and the Einstein equivalence principle to conclude 
that the measured angles with respect to an 
instantaneous frame of reference 
are determined solely by the observer's tangent vector at 
the event. We then argued that the radial distance is obtained in 
a similar manner, even though it is not as empirically accessible 
as the measurement of angles. Finally, we added that this 
indeed yields a Euclidean conception of space, which can be 
recombined with the time dimension to give rise to a so 
called `observer spacetime'.

\section{Static Splitting}
\label{sec:space}

After having laid out the philosophical foundation of 
the splitting formalism, we may now implement it 
mathematically. As for an observer $\gamma \colon 
\mathcal I \to \spti \colon \tau \to \gamma \left( \tau 
\right)$, the spatial distance function at time $\tau$ on 
the past light cone 
$\mathcal C_{\gamma \left( \tau\right)}^-$ only depends 
on the tangent vector $\dot \gamma _\tau$ 
(\thref{Post:Mashoon}), we may 
construct the splitting in two steps. First 
we consider the `static case', where we are only given 
a point in the spacetime with `attached' observer vector. 
This will be the content of 
this section. We then carry this construction 
over in the concluding section to the `dynamic case', 
where we are actually 
given an observer $\gamma$ and hence `add the time 
dimension'. 
\par
As noted before, the mathematical machinery does 
formally not require a philosophical basis,  
provided the identification of the physical 
concepts with the mathematical ones is given. 
Nonetheless, we invite the reader to return to 
section \ref{sec:motivation} for 
a motivation of the specific definitions. 
\par
So in this section, let $\left( \spti, g , \mathcal O\right)$ 
be a spacetime, $q \in \spti$ and let 
$c X_0$ be an observer vector in $\CapT_q \spti$. 
$c X_0$ is to be interpreted as the 
tangent vector of an observer at fixed 
(proper) time. 
\par
\begin{figure}[htp]
\centering
\includegraphics[scale=0.2]{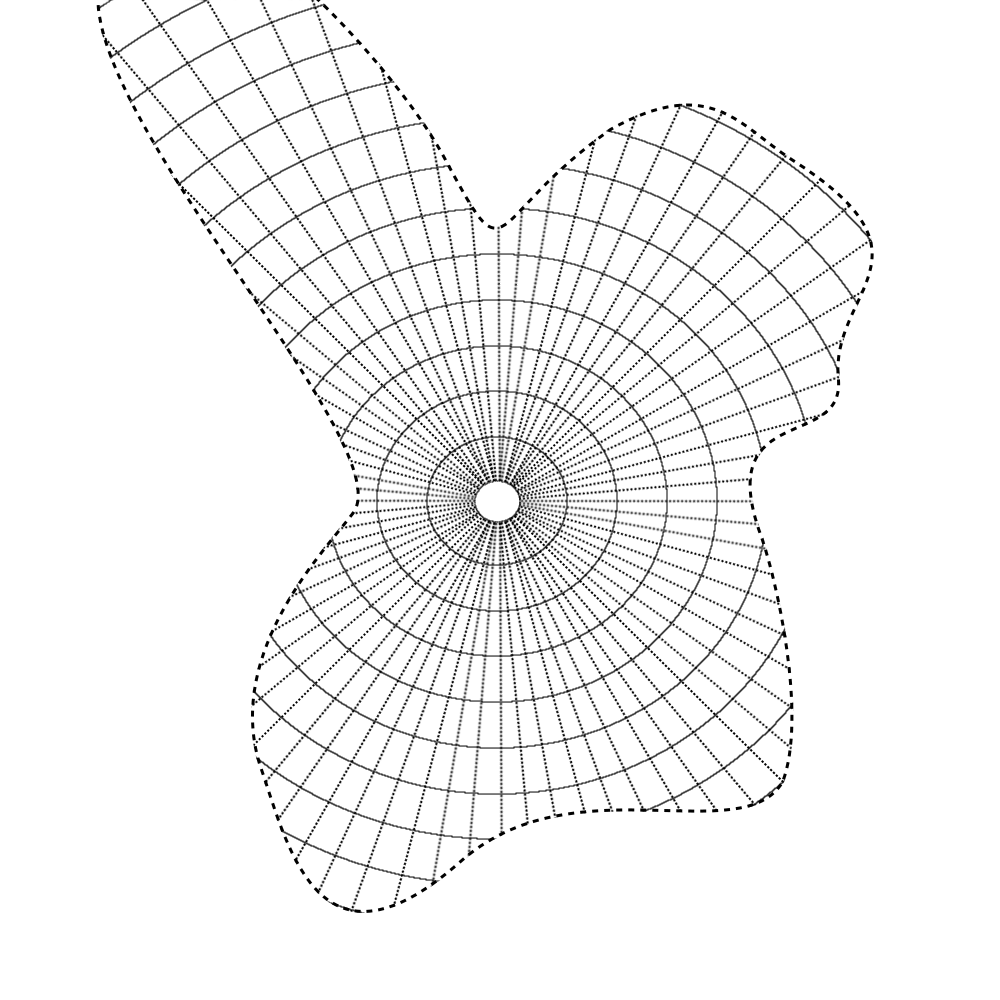}
\caption{	The picture shows a typical domain of 
			the static observer mapping 
			in $3$-spacetimes. We chose to transform 
			the coordinates \eqref{eq:Mqcoord} to
			polar coordinates $(r, \phi)$. The
			origin is accentuated by the ring in the 
			middle, the straight lines diverging away are 
			$(\phi = \const)$-lines, the circles are
			$(r = \const)$-lines, both in equal spacings. 
			The outer dotted line indicates the region 
			where the mapping is undefined. 
		}
\label{fig:domain}
\end{figure}

 
The central object of the static 
splitting is the (static) observer mapping, which 
intuitively maps the world as the observer `sees' it 
`to the world as it is'.\footnote{
\label{ftn:Perlick}
This is 
a phrase we borrowed from Perlick 
\cite{Perlick2}*{p.10}.} 
\begin{Definition}[Static observer mapping]
	\label{Def:somapping}
	Let $\left( \spti , g , \mathcal O\right)$ be a 
	spacetime and $q \in \spti$.\\ 
	Define $\mathcal M_q := \mathfrak{c}^-_q \cap 
	\dom \exp_q$. Then the \emph{(static) 
	observer mapping at $q \in Q$} is 
		\begin{equation}
			 \xi_q  \colon \mathcal M_q 
			  \to \spti \colon 
			 K \to \xi_q \left( K\right) := \exp K	\, .
		\end{equation}
\end{Definition}
As required, the image of $\xi_q$ is the past light cone 
$\mathcal C_q^-$. It remains to show that $\xi_q$ is 
smooth. 
\begin{Proposition}[Domain \& smoothness of static observer mapping]
	\label{Prop:sospaces}
	Let $\left( \spti, g, \mathcal O \right)$ be a 
	spacetime and $\xi_q$ be the static observer mapping 
	at $q \in \spti$. \\
	Then there is a unique manifold structure on the 
	domain $\mathcal M _q$ of $\xi_q$, such that it is 
	a smooth embedded submanifold in $\CapT _q \spti$ 
	of dimension $n = \dim \spti -1$. 
	With respect to this manifold structure $\xi_q$ is 
	smooth. Moreover, $\mathcal M _q \cup 
	\lbrace 0 \rbrace$ is star-like about 
	$0 \in \CapT_q \spti$ and the coordinate map 
	$\vec x _-$ for some orthonormal frame $X$ at $q$, 
	as defined in 
	\eqref{eq:pmcoordinates} on page 
	\pageref{eq:pmcoordinates}, restricts to a 
	global coordinate map 
	\begin{equation}
		\vec x \colon \quad \mathcal M _q 
		\to \vec x_- \left( \mathcal M _q \right) 
		\subseteq{\R^n \setminus \lbrace 0 \rbrace} 
		\quad \colon \quad K \to 
		\vec x \left( K \right) := \vec x_- 
		\left( K\right)
		\label{eq:Mqcoord}
	\end{equation}
	on $\mathcal M _q$. 
\end{Proposition} 
\begin{Proof}
	By \thref{Prop:lightcone}, $\mathfrak{c}_q^-$ is an embedded 
	submanifold of $\CapT_q \spti$ and, by \thref{Prop:domexp},  
	$\dom \exp_q$ is open and star-like about $0$. 
	Openness of $\dom \exp_q$ together with non-emptyness of 
	$\dom \exp_q \cap \mathfrak{c}_q^-$ gives that 
	$\mathcal M_q$ 
	is an open submanifold of $\mathfrak{c}_q$, and thus an 
	embedded submanifold of $\CapT_q \spti$. 
	Smoothness of $\exp_q$ then implies 
	smoothness of $\xi_q$. \\
	$\mathfrak{c}_q^- \cup 
	\lbrace 0 \rbrace$ is star-like about $0$. 
	To show this, recall $p$ and $p'$, as defined in 
	\eqref{eq:lightp} and \eqref{eq:lightpprim} 
	on page \pageref{eq:lightpprim}, and
	compute for all $K \in \mathfrak{c}_q^- \cup 
	\lbrace 0 \rbrace$, $\lambda \in [0,1]$:
	\begin{equation*}
		p \left( \lambda K \right) = \lambda^2 p 
		\left( K \right) = 0 \quad , \quad 
		p' \left( \lambda K \right)= \lambda 
		p' \left(  K \right) \geq 0 \, . 
	\end{equation*}
	Since the intersection of star-like sets about the 
	same point is starlike and 
	\begin{equation*}
		 \mathcal M _q  \cup 
	\lbrace 0 \rbrace = \left( \dom \exp_q \right) \cap
	\left( \mathfrak{c}_q^- \cup 
	\lbrace 0 \rbrace \right) \, ,
	\end{equation*} 
	$\mathcal M _q  \cup 
	\lbrace 0 \rbrace$ is star-like about $0$. \\
	Finally, as $\vec x_-$ are global coordinates on 
	$\mathfrak{c}_q^-$ and $\mathcal M _q$ is open, the 
	restriction $\vec x$ defines global coordinates on 
	$\mathcal M _q$. 
\end{Proof}
It follows from \thref{Prop:sospaces} that 
$\mathcal M_q$ is diffeomorphic to an open subset of 
$\R^n \setminus \lbrace 0 \rbrace$,  
that is star-like about $0$ (if one adds the point) and 
hence connected. An example of such a domain is shown in 
figure \ref{fig:domain}. 
Note again that the image of the 
static observer mapping $\mathcal C^-_q$ is usually not a 
submanifold of $\spti$, since the conjugate locus 
at $q$ may intersect $\mathcal M _q$. 
\par
So far we have not used the observer vector $c X_0$. The 
reason is that every physical 
observer positioned at the event $q$ sees the same (tangent) 
past light cone. The $4$-velocity $c X_0$ at $q$ is 
only needed to determine the distances the observer 
measures. 
We again refer to figure \ref{fig:distSR} on page 
\pageref{fig:distSR}. In accordance with equation 
\eqref{eq:distcone} on page \pageref{eq:distcone}, 
this distance is given by 
\begin{align}
	\dist & \colon &\mathcal M_q 
	\cross \mathcal M_q & \to \quad [0, \infty) &  
	\label{eq:distspace} 
	\\
	 & \colon & \left( K, K' \right) & \to \quad
	\dist \left( K, K' \right) 
	:= \sqrt{- g_q 
	\left( \left( K -K' \right)^\perp, 
	\left( K -K' \right)^\perp \right)} \, , &
	\notag
\end{align}
where $\perp$ denotes the orthogonal projection 
along $c X_0$. Moreover, if we normalize the vector 
to $X_0$, we may `complete' it to an orthonormal frame 
$X$ at $q$. According to \eqref{eq:Mqcoord} above, this 
in turn yields coordinates $\vec x$ on $\mathcal M_q$. 
Then from a straightforward 
computation we indeed find that 
\begin{equation*}
	\dist \left( K, K' \right)
	= \abs{\vec x \left( K\right) - 
	\vec x \left( K' \right)}
	\, . 
\end{equation*}
This also gives a quick proof 
that $\left( \mathcal M_q, \dist \right)$ is 
a metric space, so $\dist$ is 
indeed a distance function in the mathematical sense. 
In addition, it proves that $\dist$ induces 
a flat Riemannian metric $h$ on $\CapT \mathcal M_q$, 
given by 
\begin{equation}
	h = \delta_{ab} \, \d x^a \tp \d x^b \, .
	\label{eq:deltametric}
\end{equation}
This metric only depends on $g_q$ and $c X_0$. 
In fact, we may 
express $h$ invariantly by using the vertical lift: 
By a coordinate calculation one verifies directly 
that for every $\tilde Y_K, \tilde Z_K \in \CapT_K  
\mathcal M_q$: 
\begin{equation}
	h \left( \tilde Y_ K, \tilde Z_K \right) 
	= - g_q^\perp \left(Y, Z \right) \, , 
	\label{eq:invarianth}
\end{equation}
so formally 
$h = \widetilde{ - g _q ^\perp} 
\evat{\CapT \mathcal M_q 
\oplus \CapT \mathcal M_q}$. 
\par
Apart from the ability to measure distances, it 
should also be possible for the observer to distinguish 
the space it sees from its mirror image. So we 
have to find a `natural' way to use the space 
orientation on $\spti$ to define an orientation on 
$\mathcal M _q$. First recall that we have applied the regular 
value theorem on the quadratic form $p$ for 
the proof that $\mathfrak{c}_q$
is a submanifold of $\CapT _q \spti$ 
(cf. \thref{Prop:lightcone} on page \pageref{Prop:lightcone}), 
so the tangent space $\CapT _K \mathcal M_q$ at each 
$K \in \mathcal M_q$ is simply the kernel of 
$\left(p_* \right)_K$. Identifying the tangent spaces 
in $\CapT_q \spti$ with the space itself under 
the vertical lift, we obtain from 
the computed expression \eqref{eq:difflightp} 
for $\left(p_* \right)_K$ that 
\begin{equation}
	\CapT _K \mathcal M_q = \ker \left( K \cdot g_q
	\right) = 
	\left( \R K \right)^\perp \, ,
	\label{eq:tangentobs}
\end{equation}
i.e. the tangent space $\CapT _K \mathcal M_q$ 
is then identified with the lightlike hyperplane 
$\left( \R K \right)^\perp \subset \CapT_q \spti$. 
Now from our discussion on page \pageref{eq:piparallel} 
sqq. on orientations on lightlike hyperplanes, we know 
that the fiber $\mathcal O _q$ defines a vector 
space orientation on 
$\left( \R K \right)^\perp$. Since the vertical 
lift is smooth, this construction 
indeed defines a smooth orientation 
on $\mathcal M_q$ (in the sense of a 
$\LieGL_n^+$-structure), which is what we asked for. 
\par
Summing up, we have a `space' $\mathcal M_q$ equipped 
with a smooth metric $h$ and a smooth orientation 
$O \subset \frameb{\CapT \mathcal M _q}$. 
This construction is formalized in the next definition. 
\begin{Definition}[Observer space]
	\label{Def:obsspace}
\begin{subequations}
	Let $\left( \spti, g, \mathcal O \right)$ be a 
	spacetime, $\mathcal M_q$ be the domain of the 
	static observer mapping at $q \in \spti$ 
	and $c X_0 \in \CapT_q \spti$ be an observer 
	vector. \\ 
	We define the \emph{(static) observer metric $h$ 
	induced by $c X_0$} as 
	$h = \widetilde{ - g _q ^\perp} \evat{\CapT 
	\mathcal M_q \oplus \CapT 
	\mathcal M_q}$, in accordance with 
	\eqref{eq:invarianth}. 
	Furthermore, the 
	\emph{observer space orientation $O$ (induced by 
	$\mathcal O$)} is the orientation on $\mathcal M_q$
	induced by the orientation $\mathcal O_q$
	in each tangent space $\left( \R K \right)^\perp$ 
	at $K \in \mathcal M_q$ (under the identification 
	\eqref{eq:tangentobs}). \\ 
	The tuple $\left( \mathcal M_q, h, O \right)$ is 
	called the \emph{observer space (at 
	$q \in \spti$ induced by $c X_0$}). 
\end{subequations}
\end{Definition}
The coordinates $\vec x$ induced by an orthonormal 
frame $X$ at $q$ allow us to identify $\mathcal M_q$ 
with an open subset of Euclidean space $\R^n$. 
Mathematically speaking, every observer space 
is isomorphic to an open submanifold of Euclidean space 
of the same dimension in the category of oriented 
Riemannian manifolds. 
\par
One may now ask if and for which 
choices of $X \in \oframeb{\spti, g}$, the 
coordinates $\vec x$ on $\mathcal M_q$ are adapted to the 
geometric structures on $\mathcal M_q$, in the
sense that in these coordinates the geometric structures 
take the standard form. From \eqref{eq:invarianth} and 
\eqref{eq:deltametric}, we conclude that $X_0$ needs 
to be the zeroth frame vector of $X$, so  
we only need to identify those $X$ for which 
the vector fields 
$\partial/\partial x^1, \dots, \partial/\partial x^n$ 
on $\mathcal M_q$ are right-handed. Due to the fact 
that any frame of reference $X \in \mathcal P _q$ induces 
the linear spacetime orientation $\mathcal O _q$ 
on $\CapT_q \spti$ (see section \ref{ssec:Conf}), 
$\vec x$ are adapted coordinates precisely when 
$X$ is a frame of reference at $q$ and the observer metric 
is induced by $c X_0$. 
\begin{Definition}[Static observer coordinates]
	\label{Def:obscoord}
\begin{subequations}
	Let $\left( \spti, g, \mathcal O\right)$ be a 
	spacetime and $\left( \mathcal M_q, h, O \right)$ 
	be the observer space at $q \in \spti$ 
	induced by the observer vector $Z \in \CapT_q 
	\spti$. \\ 
	If $X$ is a frame of reference at $q$ with 
	$X_0 = Z/c$, then the coordinates 
	$\vec x$ on $\mathcal M_q$, as defined by 
	\eqref{eq:Mqcoord}, are called \emph{(static) 
	observer coordinates (with respect to $X$)}. 
\end{subequations}
\end{Definition}
The usage of the term `observer coordinates' was inspired by 
G.F.R. Ellis, who considered the spherical 
coordinate analog of $\vec x$ and named 
these `observational coordinates' 
(cf. \cite{Ellis2}*{p. 324}). As indicated in the 
previous section, 
the choice of the frame of reference $X$ at $q$ has 
a physical meaning: It specifies the position, the 
$4$-velocity and directions `forward', `left' and `up' 
(in accordance with the right-hand rule) of the 
physical observer in spacetime and 
can thus be understood as an `infinitesimal rigid body' at 
the event $q$. Observer coordinates with respect to 
$X$ then label each observed point in space in accordance 
with this choice.  
\par
With respect to these coordinates, the observer 
mapping is given by 
\begin{equation}
	\varphi_X \left( \vec x \right) := 
	\exp_q \left(- \abs{\vec x} X_0 + x^a X_a  \right) \, .
	\label{eq:localsom}
\end{equation}
We call this map $\varphi_X := \xi_q \circ \vec x ^{-1}$ the 
\emph{(static) observer mapping with respect to $X$}. In fact, 
to compute the observer mapping at a point $q \in Q$ in practice, 
one takes a chart $\left( \mathcal U, \kappa \right)$ around $q$, 
a frame of reference $X$ at $q$ and computes the past 
lightlike geodesics in the chart with initial conditions 
\begin{equation*}
	\kappa_0 = \kappa \left( q \right) \quad \text{and} \quad 
	\dot{\kappa}{}^i_0 = - \abs{\vec x } X^i{}_{0} 
	+ x^a \, X^i{}_a 
\end{equation*}
for $i \in \lbrace 0, \dots ,n \rbrace$ and all possible 
$\vec x \in \R^n \setminus \lbrace 0 \rbrace$. To illustrate 
our construction, we give a simple example. 
\begin{Example}[Static observer mapping in Minkowski spacetime]
	\label{Ex:statoMink}
Consider Minkowski spacetime $\left( \R, g, \mathcal O\right)$, 
as defined in \thref{Ex:Mspti}, a point 
$(c t_0, \vec y_0) = y_0 \in \R^4$ and a frame 
of reference $Y \in \mathcal P _{y_0}$. In canonical 
coordinates $y = \left( ct, \vec y 
\right)$ on $\R^4$, the geodesic equation is 
simply $\ddot y^k = 0$. Hence for a tangent vector 
$K \in \CapT_{y_0} \R^4$, we get 
\begin{equation}
	\left( \exp_{y_0} \left( K \right)\right)^k 
	= \Evat{(K^k s + y_0^k)}{s=1} = K^k + y_0^k \, .
	\label{eq:expMinkow}
\end{equation}
Thus, according to \eqref{eq:localsom}, the 
observer mapping with respect to $Y$ is 
\begin{equation*}
	\varphi_Y \left( \vec x \right) 
	= 
	\begin{pmatrix}
		c t_0 - \abs{\vec x} Y^0{}_0 + x^a \, 
		Y^0{}_a \\ 
		y_0^1 - \abs{\vec x} Y^1{}_0 + x^a \, 
		Y^1{}_a \\
		y_0^2 - \abs{\vec x} Y^2{}_0 + x^a \, 
		Y^2{}_a \\
		y_0^3 - \abs{\vec x} Y^3{}_0 + x^a \, 
		Y^3{}_a
	\end{pmatrix}
	\, . 
\end{equation*}
Since $\partial$ is a global frame of reference 
field and $Y \in \mathcal P$, the matrix  
$\Lambda \in \End \left( \R^4\right)$ with components 
$\Lambda^i{}_j \equiv Y^i{}_j$ is an element of 
$\Lor_4$. Under a coordinate change via 
the Poincar\'e transformation 
$y' :=  \ubar \Lambda \cdot \left( y  - y_0 \right)$, 
the observer mapping takes the form 
\begin{equation*}
	\varphi'_Y \left( \vec x\right) 
	= 
	\begin{pmatrix}
		- \abs{\vec x} \\ 
		\vec x
	\end{pmatrix}
	\, . 
\end{equation*} 
This shows that distances are indeed given by 
\eqref{eq:distSR}, as 
dictated by special relativity. Moreover, we see 
that $\varphi'_Y$ (and hence $\varphi_Y$) 
is invertible, if restricted to its image. In 
practice, we determine $\vec x = {\vec y}\mkern2mu\vphantom{y}'$ 
and then 
check whether $ct' = - \abs {{\vec y}\mkern2mu\vphantom{y}'}$ 
%needed to add this to make prime not overlap vector arrow
is satisfied 
to guarantee that the observer actually sees the event. 
\end{Example}
To compute the observer mapping in general spacetimes, 
we need to solve the geodesic equation 
for arbitrary initial conditions. Following
Perlick (cf. \cite{Perlick2}), Langrangian and 
Hamiltonian techniques may be applied to do so. See 
\cite{Perlick2}*{\S 5} for specific examples and 
further references. It should be noted that 
solving the geodesic equations analytically is usually 
a very difficult, if not impossible task. 
\par
A further difficulty is imposed by the fact 
that in practical applications it is often 
required to invert the observer mapping. An inverse is 
needed, since we would like to assign relative 
positions and mutual distances to different observed 
events on the past light cone, not on the tangent 
past light cone. The following lemma states that this is 
possible, if one only considers events `close 
enough' to the observer. 
\begin{Lemma}
	\label{Lem:sinverse0}
	Let $\left( \spti, g, \mathcal O\right)$ be a 
	spacetime and $\xi_q \colon \mathcal M _q 
	\to \spti$ be the static observer mapping. \\
	Then there exists an open neighborhood $\mathcal V$ 
	of $0 \in \CapT_q \spti$ such that the restriction 
	of $\xi_q$ to $\mathcal V \cap \mathcal M _q$ 
	is a diffeomorphism onto its image. 
\end{Lemma}
\begin{Proof}
	As already shown, $\exp_q$ has full rank at $0$, hence
	there exists an open neighborhood $\mathcal V$ of $0$ 
	such that $\exp_q \evat{\mathcal V}$ 
	is a diffeomorphism onto its image. 
	Again, since $\mathfrak{c}_q^-$ is an embedded submanifold 
	of $\CapT_q \spti$ and $\mathcal V \cap 
	\mathfrak{c}_q^- \neq \emptyset$, the latter is 
	also an embedded submanifold of $\CapT_q \spti$ 
	having the same dimension as $\mathfrak{c}_q^-$. 
	Restricting $\exp_q$ yields the result. 
\end{Proof}
\par
Of course, one would like to be able to (smoothly) invert 
$\xi_q$ globally to define directions and 
distances on the whole of $\mathcal{C}_q^-$. Physically, the 
phenomenon of (strong) gravitational lensing 
implies that there cannot be a $1$-$1$ map between `the 
world as the observer sees it' and `the world as it is' 
(cf. footnote \ref{ftn:Perlick} on page \pageref{ftn:Perlick}). 
Mathematically, this `non-invertibility' of the observer 
mapping is due to two effects: First
the conjugate locus in $\CapT_q \spti$ may intersect 
the tangent past light cone, preventing the map 
from being an immersion, and second the light 
cone may intersect itself, preventing it 
from being injective. We refer to the article by 
Perlick \cite{Perlick2}*{\S 2.6 \& \S 2.7} for an 
introductory discussion on this issue. References  
\citelist{\cite{Ehlers} \cite{Ellis0} \cite{HasseA3}} 
provide further reading on the relativistic description 
of gravitational lensing. 
\par
Despite this problem, we can still prove a weak 
kind of invertibility for the static observer mapping. First 
we recollect some definitions. 
\begin{Definition}[Almost everywhere locally 
					invertible maps]
	\label{Def:aelim}
	Let $\mathcal M, \mathcal N$ be smooth manifolds.
	\begin{enumerate}[i)]
	\item	A subset $S$ of $\mathcal M$ 
			is said to have \emph{(Lebesque) measure zero}, 
			if it has Lebesque measure zero 
			in each chart. 
	\item	A smooth map 
			$\varphi \colon \mathcal M \to \mathcal N$
			is said to be 
			\emph{almost everywhere locally invertible}, 
			if there exists a set $S$ of (Lebesque) 
			measure zero 
			in $\mathcal M$ such that for every 
			$m \in \mathcal M \setminus S$ there is an 
			open neighborhood $\mathcal V$ of 
			$m$ in $\mathcal M$ for which 
			$\varphi\evat{\mathcal V}$ is a diffeomorphism 
			onto its image. 
	\end{enumerate}
\end{Definition}
One can show (cf. \cite{Lee}*{Lem. 6.6}) that for a subset 
$S$ of a manifold to have Lebesque measure zero, it is 
sufficient to prove this for a collection of charts whose 
domains cover $S$. Similarly, there exists an 
arguably simpler 
condition for a map to be almost everywhere locally 
invertible. 
\begin{Lemma}[Condition for local invertibility almost 
				everywhere]
	\label{Lem:locinv}
	A (smooth) map is almost everywhere 
	locally invertible if and only if its set of 
	critical points has measure zero and 
	the dimension of the target manifold is 
	greater than 
	or equal to the dimension of the domain. 
\end{Lemma}
\begin{Proof}
\begin{subequations}
	First observe that, if the dimension of the 
	target manifold $\mathcal N$ of any smooth 
	map $\varphi \colon \mathcal M \to \mathcal N$
	is less than the dimension of the domain 
	manifold $\mathcal M$, the 
	map cannot be an immersion at any point 
	$m \in \mathcal M$ and thus cannot be a local 
	diffeomorphism in a neighborhood of $m$. 
	Consequently, we obtain this as a necessary 
	condition for local invertibility. 
	\par
	Next we show that the set $S$ from above is 
	the set of critical points. Let 
	$m$ be a regular value of $\varphi$. 
	Then there exists an open neighborhood $\mathcal V$ 
	of $m$ in
	$\mathcal M$ such that $\varphi 
	\evat{\mathcal V}$ is an immersion and by the 
	constant rank theorem (see e.g. 
	\cite{Lee}*{Thm. 4.12}), we can choose 
	$\mathcal V$ such that $\varphi 
	\evat{\mathcal V}$ is a diffeomorphism onto 
	its image. Hence $S \subseteq \crit \varphi$. 
	On the other hand, if $m \in \crit \varphi$, 
	then $(\varphi_*)_m$ does not have full 
	rank and hence there cannot exist an open 
	neighborhood $\mathcal V$ around $m$ on which 
	$\varphi \evat{\mathcal V}$ is a diffeomorphism 
	onto its image. So $\crit \varphi 
	\subseteq S$ and thus $\crit \varphi = S$. 
	\par
	Finally, 
	Lebesque-measurability of 
	$\crit \varphi$ is 
	guaranteed by the fact that $\crit \varphi$ 
	is closed (cf. \cite{Lee}*{Prop. 4.1}) and 
	thus a Borel set. 
	\end{subequations}
\end{Proof}
\begin{Example}[Almost everywhere locally invertible map]
	\label{Ex:parab}
	Consider the smooth function 
	\begin{equation*}
		\varphi \colon 
		\R \to \R^2 \colon s \to \left( s^2, s^2 \right) \, . 
	\end{equation*}
	Except at the origin $\lbrace 0 \rbrace$, it is an 
	immersion everywhere. Thus 
	the map is almost everywhere locally invertible and we may
	even analytically express its two (maximal) local inverses 
	\begin{align*}
		\ubar{\varphi}^\pm & \colon &
		\set{\left( x^1,x^2\right) \in \R^2}{x^1=x^2 > 0}
		&\to \R_\pm & \\
		& \colon & \left( x^1,x^2\right) 
		& \to \ubar{\varphi}^\pm \left( x^1, x^2 
		\right) = \pm \sqrt{x^1}  \quad . & 
	\end{align*}
\end{Example}
Of course, we intend to show that the  
static observer mapping is almost everywhere
locally invertible. 
From our discussion of the Lorentzian exponential 
in section \ref{sec:Jacobi}, we know 
that the set of critical points of the static 
observer mapping $\xi_q$ at $q \in \spti$
is the \emph{past lightlike conjugate locus}, i.e.  
the intersection of the conjugate locus 
$\crit \exp _q$ with the tangent past light cone 
$\mathfrak{c}_q^-$. So by \thref{Lem:locinv}, we 
require the following for the static observer mapping 
to be almost everywhere locally invertible. 
\begin{Proposition}[Past lightlike conjugate locus 
					has measure zero]
	\label{Prop:measzero}
	Let $\left( \spti, g, \mathcal O\right)$ 
	be a spacetime and $\xi_q$ be the static 
	observer mapping at $q \in \spti$. \\
	Then the past lightlike 
	conjugate locus $\crit \xi_q$ has measure zero in the 
	domain $\mathcal M _q$ of $\xi_q$. 
\end{Proposition}
\begin{Proof}
	The idea is to use the isolatedness of conjugate 
	points along lightlike geodesics (cf. 
	\thref{Lem:isoconj} on page \pageref{Lem:isoconj}) 
	to show that 
	the integral of the past lightlike 
	conjugate locus $\crit \xi_q$ vanishes 
	in radial direction. This in turn implies that 
	the entire volume of $\crit \xi_q$ needs to vanish.
	\par
	We shall first give meaning to the 
	word `radial direction': Take $n+1$ to be the 
	dimension of $\spti$. 
	Since $\mathcal M _q$ is an open submanifold of 
	the past light cone ${\mathfrak{c}}^-_q$ and it is more 
	convenient to work on the latter, we choose some 
	timelike vector $Z$ in $\CapT_q \spti$ and
	consider the map 
	\begin{equation*}
		\norm{.} \colon \mathfrak{c}_q^- 
		\to \R \colon K
		\to \norm{K} := \sqrt{- g_q 
		\left( K^\perp , K^\perp \right)} \, . 
	\end{equation*}
	Here $\perp$ denotes the orthogonal projection with 
	respect to $Z$. For all $K \in 
	\mathfrak{c}_q^-$ the vector 
	$K^\perp$ is spacelike and $\norm{K}>0$, so 
	$\norm{K}$ may be interpreted as the `length' 
	of $K$. Using this length, identify the $(n-1)$-sphere 
	$\mathbb{S}^{n-1}$ as a (possibly 
	$0$-dimensional) submanifold of 
	${\mathfrak{c}}^-_q$. Observe that 
	the map
	\begin{equation*}
		\mathfrak{c}_q^- \to \R_+ \cross 
		\mathbb{S}^{n-1} \colon K \to 
		\left( \norm{K}, \frac{K}{\norm{K}} \right)
	\end{equation*}
	is a smooth bijection. Since the inverse is 
	just multiplication by a strictly positive number, 
	it is a
	diffeomorphism. So we have indeed obtained a 
	splitting of the past light cone into `radial' 
	and `angular' parts. 
	\par 
	It remains to do the integration: 
	Consider the induced Borel product 
	measure $B := B_{\R_+} \cross 
	B_{\mathbb{S}^{n-1}}$ on $\R_+ \cross 
	\mathbb{S}^{n-1}$ as a measure on 
	$\mathfrak{c}_q^-$. The Borel measure coincides 
	with the Lebesque measure on Borel sets, so 
	a Borel set $A$ on $\mathfrak{c}_q^-$ has 
	Lebesque measure zero if and only if it 
	has measure zero with respect to the product 
	measure $B$. Again note that the set of 
	critical points $\crit \xi_q$ is closed and thus 
	Borel-measurable. Now, for every 
	$Y \in \mathbb{S}^{n-1} \subset \mathfrak{c}_q^-$, 
	the map 
	\begin{equation*}
		\set{r \in \R_+}{rY \in \mathfrak{c}_q^-} \to \spti
		\colon r \to \exp_q \left( r Y\right)
	\end{equation*}
	is a geodesic, so we use the 
	isolatedness of lightlike conjugate values
	(\thref{Lem:isoconj}) to conclude that the set 
	\begin{equation*}
		S_Y := \set{ r \in \R_+}{r Y \in \crit \xi_q}
	\end{equation*}
	is (at most) countable.  Thus
	$B_{\R_+} \left( S_Y \right) \equiv 0$. 
	So if we denote by $\d Y$ the `volume element' for 
	the measure $B_{\mathbb{S}^{n-1}}$ and 
	apply `Fubini's theorem in measure theory' 
	(cf. \cite{Bogachev}*{Thm. 3.4.1}), we indeed find 
	\begin{equation*}
		B \left( \crit \xi_q \right) 
		= \int_{\mathbb{S}^{n-1}} 
		B_{\R_+} \left( S_Y \right) \, \d Y = 0 \, . 
	\end{equation*}
\end{Proof}
Therefore the set, where the static observer mapping is not 
locally invertible, is `negligible'. For the sake of 
coherence and ease of referencing, we state the 
proven theorem below.
\begin{Corollary}[Static observer mapping inversion theorem]
	\label{Thm:sobsinv}
	The static observer mapping is 
	almost everywhere locally invertible. The critical 
	set is the past lightlike conjugate locus. 
\end{Corollary}
\begin{Proof}
	Again, the second sentence follows from the 
	characterization \eqref{eq:critexp} 
	of the critical points of the 
	exponential on page 
	\pageref{eq:critexp}. Now 
	\thref{Lem:locinv} 
	in conjunction with \thref{Prop:measzero}
	yields the assertion. 
\end{Proof}
\par
We conclude our treatment of the static splitting 
with a computation of the differential of $\xi_q$ in
terms of Jacobi fields. 
\begin{Proposition}[Differential of static observer mapping]
	\label{Prop:diffsobm}
\begin{subequations}
	Let $\left( \spti, g, \mathcal O\right)$ be 
	a spacetime of dimension $n+1$ 
	and $\varphi_X$ be the static observer 
	mapping with respect to the frame of reference $X$ at 
	$q$. \\
	Then for all $\vec x \in \dom \varphi_X \subseteq 
	\R^n \setminus \lbrace 0 \rbrace$ and 
	$a \in \lbrace 1, \dots n \rbrace$: 
	\begin{equation}
		\left(\left(\varphi_X\right)_* \partd{}{x^a} 
		\right)_{\vec x} = J^a_1 \left( \vec x \right)
		\, , 
	\end{equation}
	where $J^a \left( \vec x \right) 
	\colon s \to J^a_s \left( \vec x \right)$ 
	is the unique Jacobi field along the geodesic 
	\begin{equation*}
		s \to \exp_q \left( s \left( - \abs{\vec x} X_0 
		+ x^a \, X_a \right)
		\right) 
	\end{equation*} 
	with $J^a_0 \left( \vec x \right) = 0$ and 
	\begin{equation}
		\left( \frac{\nabla J^a \left( \vec x \right)
		}{\d s} \right)_0 = 
		- \frac{\delta_{ab} x^b}{\abs{\vec x}} X_0 
		+ X_a
		\, .
		\label{eq:sJIC2}
	\end{equation}
\end{subequations}
\end{Proposition}
\begin{Proof}
	We consider the curve 
	$x^a \to - \abs{\vec x} X_0 + x^b \, X_b$ in 
	$\CapT \spti$, its tangent vector field 
	$Z \colon x^a \to Z_{x^a}$ and apply 
	\thref{Thm:diffexp} from page \pageref{Thm:diffexp}.  
	Since the curve stays within the fiber 
	$\CapT_q \spti$, we have $J^a_0 \left( \vec x \right) 
	\equiv 0$. If $\Kon$ is the Levi-Civita connector, 
	then 
	\begin{equation*}
		\left( \frac{\nabla J^a \left( \vec x \right)
		}{\d s} \right)_0 = 
		\Kon \left( Z_{x^a}\right) 
		= \frac{\nabla}{\d x^a} 
		\left( - \abs{\vec x} X_0 + x^b \, X_b \right)
	\end{equation*}
	yielding \eqref{eq:sJIC2}. 
\end{Proof}
\thref{Prop:diffsobm} might be useful for finding 
an approximate expression of $\xi_q^*g$ in observer 
coordinates. We refer to the book by Sakai 
\cite{Sakai}*{\S 3.1} for an analogous expression in 
Riemannian geometry, and to the article by Klein 
and Collas \cite{Klein} for references to similar work 
already done in relativity theory. 
Also note that the differential of $\xi_q$
 in radial direction never vanishes, hence 
the kernel of $(\xi_q)_*$ is at most $(n-1)$-dimensional for 
$(n+1)$-dimensional $\spti$. In particular, for $2$-spacetimes 
the static observer mapping, if restricted to a small enough domain, 
is always a diffeomorphism onto its image. 

\section{Kinematic Splitting}
\label{sec:kinematic}  

Based on the findings of sections \ref{sec:generalspti} and 
\ref{sec:space}, we obtain a natural definition of a 
spacetime-splitting by 'adding the time-dimension'. 
However, going over to the kinematic case requires 
additional considerations, as the kinematics on the 
spacetime needs to be related to the kinematics in the 
`observer spacetime' and appropriately interpreted. To 
do this, we first define and analyze the kinematic 
observer mapping in section \ref{ssec:kingen} and then 
consider so called moving frames of 
reference in section \ref{ssec:movframes} in order to 
introduce so called `observer spacetimes' thereafter. 
Observer spacetimes 
are needed to rigorously relate the dynamics and 
kinematics on the spacetime 
$\spti$ to the observed kinematics, as seen by an actual physical 
observer. 

\subsection{Kinematic Observer Mapping}
\label{ssec:kingen}

We now adapt the static splitting to the kinematic case, 
i.e. where motion comes into play. 
In the spirit of \thref{Def:somapping} and 
\thref{Prop:sospaces}, we first define the kinematic 
observer mapping and then prove its smoothness. 
Afterwards, we treat the measurement of time in the
formalism and the question whether the kinematic 
observer mapping can be inverted. 
\par
As already indicated, the kinematic splitting is 
obtained by taking an observer 
$\gamma\colon \mathcal I \to \spti 
\colon \tau \to \gamma \left( \tau \right)$ and 
considering for each time $\tau \in \mathcal I$ 
the observer space 
$\mathcal M _{\gamma \left( \tau\right)}$ induced 
by $\dot \gamma _ \tau$, as well as the observer 
mapping $\xi_{\gamma \left( \tau \right)}$. 
\begin{Definition}[Kinematic observer mapping]
	\label{Def:domapping}
	Let $\left( \spti , g , \mathcal O\right)$ be 
	a spacetime and $\gamma \colon \mathcal I \to \spti$ 
	be an observer. \\
	The set 
	\begin{equation*}
		{\mathfrak {C}}_\gamma^- := \bigsqcup_{\tau \in \mathcal I} 
		\mathfrak{c}_{\gamma \left( \tau \right)}^- = 
		\bigcup_{\tau \in \mathcal I} \lbrace \tau \rbrace \cross 
		\mathfrak{c}_{\gamma \left( \tau \right)}^- 
		\subset \gamma^* \CapT \spti 
	\end{equation*}
	is called the \emph{past tangent light cone along $\gamma$}. 
	For each $\tau \in \mathcal I$ define 
	\begin{equation*}
		{\mathcal M}^\gamma_ \tau := 
		\mathcal M_{\gamma \left( \tau \right)} = 
		\mathfrak{c}_{\gamma \left( \tau \right)}^- \cap 
		\left( \dom \exp_{\gamma \left( \tau \right)} 
		\right) 
		\, ,  
	\end{equation*}
	and 
	\begin{equation*}
		{\mathcal M}^\gamma := 
		\bigsqcup_{\tau \in \mathcal I} {\mathcal M}^\gamma_ \tau 
		\subseteq \mathfrak {C}_\gamma^- \, .
	\end{equation*}
	Then the \emph{(kinematic) observer mapping (for $\gamma$)} 
	is 
	\begin{equation}
		\xi^\gamma \colon {\mathcal M}^\gamma 
		\to \spti  \colon \left( \tau, K \right) 
		\to \xi^\gamma \left( \tau, K \right):=
		\xi_{\gamma \left( \tau \right)} \left( K \right) = 
		\exp_{\gamma \left( \tau \right)} 
		\left( K\right) \, . 
	\end{equation}
\end{Definition}
The image of the kinematic observer mapping $\xi^\gamma$ 
is the union of the past light cones 
$\mathcal C _{\gamma \left( \tau\right)}^-$ over all 
$\tau \in \mathcal I$. In general $\xi^\gamma$ is not surjective, 
which corresponds to the physical situation that the observer 
does not see the entire spacetime in the temporal interval 
$\mathcal I$. One may now object to this construction, that it 
does not yield a `full splitting' of the spacetime. However, 
this criticism is physically unwarranted, as, 
following the discussion in section \ref{sec:motivation}, 
the separation between space and time is only sensible 
for individual physical observers, and so it would be an 
inadmissible assumption to demand that it `sees' the entire 
spacetime. 
\par
\begin{figure}[htp]
\centering
\includegraphics[scale=0.37]{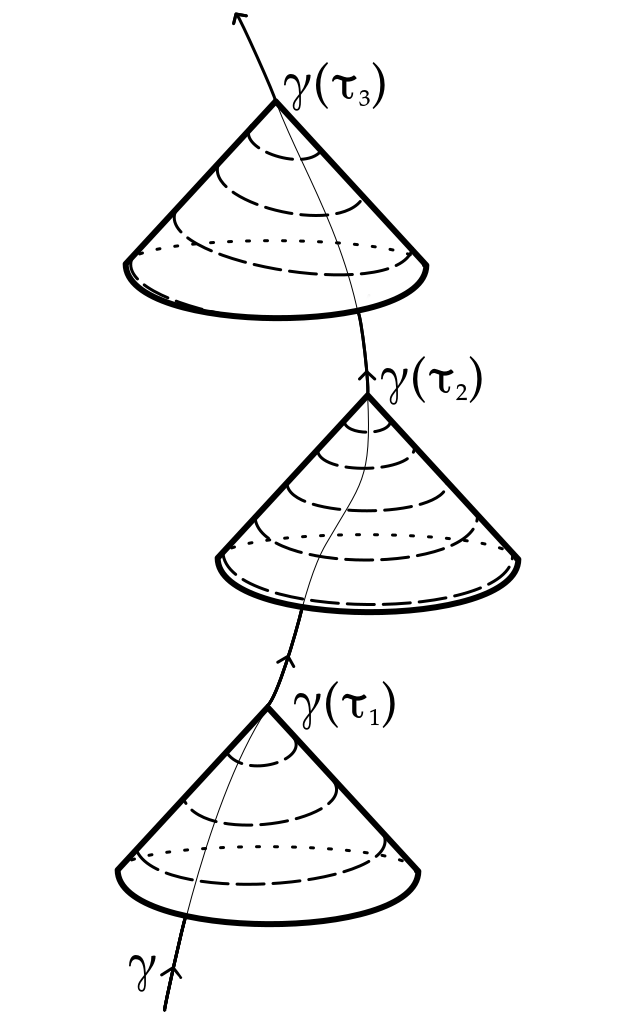}
\caption{	In this image we see a portion of $3$-dimensional 
			Minkowski spacetime embedded in Euclidean $3$-space. 
			The accelerated observer $\gamma$ 
			is indicated by a curved 
			line going upwards. At three different instances 
			$\tau_1, \tau_2, \tau_3$ of $\gamma$'s proper 
			time, a part of the respective 
			past light cone is shown. Since the spacetime is 
			flat, the cones do not intersect themselves. 
			Lines of constant radial distance, 
			as measured by the observer, are schematically 
			drawn in accordance with the direction of 
			$\gamma$'s tangent vector. We remind the reader that 
			orthogonality in Minkowski space differs 
			from orthogonality in Euclidean space. 
			}
\label{fig:kinematic}
\end{figure}

To prove that the map $\xi^\gamma$ is smooth we employ
the following Lemma. 
\begin{Lemma}[Global sections are closed maps]
	\label{Lem:globsec}
\begin{subequations}
	Let $\mathcal M, \mathcal E$ be manifolds and 
	let $\pi \colon \mathcal E \to \mathcal M$ be a 
	surjective submersion. Then every global 
	section $s \colon \mathcal M \to \mathcal E$ 
	is a closed map. 
\end{subequations}
\end{Lemma} 
\begin{Proof}
	Let $V$ be closed in $\mathcal M$ and consider an arbitrary 
	sequence $\set{y_i \in s \left( V\right)}{i \in \N}$ 
	converging to $y \in \mathcal E$. Then 
	$x_i := \pi \left( y_i \right)$ with $i \in \N$ defines 
	a sequence in $V$ with $y_i = s \left( x_i\right)$. 
	As $V$ is closed in $\mathcal M$, its 
	limit $x= \pi \left( y \right)$ lies 
	in $V \subseteq \dom s$. Continuity of $s$ yields: 
	\begin{equation*}
		y= \lim_{i \to \infty} s \left(x_i\right)
		= s \left( x\right) \in s \left( V \right) 
		\subseteq \mathcal E \, .
	\end{equation*}
\end{Proof}
We remark that the statement is false for local sections, 
since their domains are not always closed in $\mathcal M$ 
and thus one may have $\pi \left( y \right) \notin \dom s$.
\begin{Proposition}[Domain \& smoothness of kinem. observer mapping]
	\label{Prop:domkin}
	Let $\left( \spti , g , \mathcal O\right)$ be 
	a spacetime of dimension $n+1$
	and let $\xi^\gamma$ be the kinematic observer mapping
	for the observer $\gamma \colon \mathcal I \to \spti$. \\
	Then there is a unique manifold structure on the domain 
	$\mathcal M ^\gamma$ of $\xi^\gamma$, such that it is an 
	embedded submanifold 
	of $\gamma^* \CapT \spti$. With respect to 
	this manifold structure, $\mathcal M ^\gamma$ is 
	connected and $\xi^\gamma$ is smooth. 
	Moreover, if $X \colon \mathcal I \to \oframeb{\spti, g} 
	\colon \tau \to (X)_\tau$ is an orthonormal frame field along 
	$\gamma$, then the map 
	\begin{align}
		x &\colon \mathcal M ^\gamma & &\to &
		\bigcup_{ \tau \in \mathcal I} 
		\lbrace c \tau \rbrace \cross \vec x _- 
		\left( \mathcal M^\gamma _\tau \right) \quad
		\subseteq \, \left(c \, \mathcal I\right) 
		\cross \left( \R^n 
		\setminus \lbrace 0 \rbrace \right) 
		\notag \\
		& \colon 
		\left( \tau, K\right) & &\to & x 
		\left( \tau, K  \right) 
		= \left( c \tau, \ubar{X}^1_\tau \cdot K, \dots, 
		\ubar{X}^n_\tau \cdot K \right) 
		\label{eq:chartMgam}
	\end{align} 
	defines global coordinates on the manifold 
	$\mathcal M^\gamma$. 
\end{Proposition}
Please note that the factor of 
$c$ in \eqref{eq:chartMgam} is a convention, which 
guarantees that all coordinate values have the physical 
dimension of length. 
\begin{Proof}
	First we show that the \emph{tangent light cone along 
	$\gamma$} 
	\begin{equation*}
		\mathfrak{C}_\gamma = \bigsqcup_{\tau \in 
		\mathcal I} \mathfrak{c}_{\gamma \left( \tau
		\right)}
	\end{equation*}
	is an embedded submanifold of $\gamma^* \CapT \spti$. \\
	In full analogy to the proof of 
	\thref{Prop:lightcone} on page \pageref{Prop:lightcone}, 
	we exclude from $\gamma^* \CapT \spti$ the image 
	of $\mathcal I$ under 
	the zero-section along $\gamma$ to obtain the 
	set $\mathcal N$. 
	By \thref{Lem:globsec}, this image is closed, 
	hence $\mathcal N$ is 
	a (non-empty) open submanifold of $\gamma^* \CapT \spti$. 
	Now consider the map 
	\begin{equation*}
		p \colon \mathcal N \subset \mathcal I 
		\cross \CapT \spti 
		\to \R \colon \left( \tau, Z \right)
		\to p \left( \tau, Z\right) := 
		g_{\gamma \left( \tau \right)} \left(Z, Z \right)
	\end{equation*}
	and observe that $p^{-1} 
	\left( \lbrace 0 \rbrace \right) = 
	\mathfrak{C}_\gamma \subset \gamma^* \CapT \spti$. 
	As in \eqref{eq:difflightp}, we want to compute the 
	differential of $p$ and apply the regular value theorem. 
	So first we define the vertical 
	lift on $\gamma^* \CapT \spti$ via 
	\begin{equation*}
		\tilde Z _{\left( \tau, Y\right)} \left(f  \right)
		:= \Evat{\partd{}{s}}{0} 
		f \left( \tau, Y + s Z\right)
	\end{equation*}
	for each $\left( \tau, Y\right) \in \gamma^* \CapT \spti$,  
	$Z \in \CapT_{\gamma \left( \tau \right)} \spti$ and 
	$f \in C^\infty \left(\gamma^* \CapT \spti, \R\right)$. 
	Then for 
	each $\left(\tau, K \right) \in \mathfrak{C}_\gamma $ 
	and $Z \in \CapT_{\gamma 
	\left( \tau \right)} \spti$ we find 
	\begin{equation*}
		p_* \tilde{Z}_{\left( \tau, K\right)}
		= 2 g_{\gamma \left( \tau \right)} 
		\left(K,Z \right) \, .
	\end{equation*}
	Again, non-degeneracy of 
	$g$ and $K \neq 0$ implies that $p$ is a submersion on 
	$\mathfrak{C}_\gamma$. Thus 
	$\mathfrak{C}_\gamma$ is an embedded submanifold 
	of $\mathcal N$ and of $\gamma^* \CapT \spti$ 
	of dimension 
	$\left( 1+(n+1)\right)-1 =n+1$. As such, its 
	manifold structure is unique. \\ 
	To obtain $\mathfrak{C}_\gamma^-$, we proceed as 
	in \thref{Prop:lightcone} and 
	use the continuous function 
	\begin{equation*}
		p' \colon \mathfrak{C}_\gamma \to 
		\R \colon \left( \tau, K\right)
		\to p' \left( \tau, K \right):=
		g_{\gamma \left( \tau\right)} 
		\left( \dot \gamma_\tau, K\right)
	\end{equation*}
	to show that it splits into 
	$\mathfrak{C}_\gamma^- = 
	p' \left( \left( - \infty, 0\right)\right)$ 
	and $\mathfrak{C}_\gamma^+ =: 
	p' \left( \left(0, \infty \right)\right)$. 
	\\
	Recalling the coordinates $\vec x_-$, as given 
	in \eqref{eq:pmcoordinates} on page 
	\pageref{eq:pmcoordinates}, we find that 
	\begin{equation*}
		\mathfrak{C}_\gamma^- \to 
		\mathcal I \cross \left( \R^n \setminus 
		\lbrace 0 \rbrace \right)
		\colon 
		\left( \tau, K \right) \to  
	\left( c \tau, \vec x_- \left( K\right) \right)
	\end{equation*}
	is also a global coordinate map, which is 
	compatible with the smooth structure.
	According to 
	\thref{Prop:domexp}, the set $\dom \exp$ is open 
	and thus $\mathcal M^\gamma = \mathfrak{C}_\gamma^- 
	\cap \dom \exp \neq \emptyset$ is an open 
	submanifold of $\mathfrak{C}_\gamma^-$ with global 
	coordinates $x'$, as given by \eqref{eq:chartMgam}. \\ 
	As each slice $\lbrace \tau \rbrace \cross 
	\mathcal M _{\gamma \left( \tau \right)}$ is 
	connected, so is $\mathcal M^\gamma$. As 
	$\mathcal M^\gamma$ is 
	a submanifold of $\dom \exp$
	and $\exp$ is smooth, the 
	restriction $\xi^\gamma$ is also smooth.
\end{Proof}
Obviously, each slice $\lbrace \tau \rbrace \cross 
\mathcal M^\gamma_\tau$ at proper time $\tau \in \mathcal I$
is to be considered the observer space 
at $\gamma \left( \tau \right)$ induced by the 
observer vector $\dot \gamma_\tau$. Yet we will 
postpone the definition of an `observer spacetime', 
carrying the physically appropriate geometric structures, 
to the next subsection. Roughly speaking, the reason is 
that it is not possible to assign mutual distances to 
events seen at different times without the choice of a
particular frame of reference at those times. Nonetheless, 
the crude mathematical construction is set up at this point 
and an example is depicted in figure \ref{fig:kinematic}. 
\par
The measurement of temporal distances does, however, not 
require any choice of reference frame. As the observer $\gamma$ is 
parametrized with respect to proper time, the time passed 
between two observed events 
$\left( \tau_1, K_1\right), \left( \tau_2, K_2\right) 
\in \mathcal M^\gamma$ is simply 
$\abs {\tau_2 - \tau_1}$. In addition, 
the sign of $\tau_2 - \tau_1$ gives information on
which observed event precedes the other one in an obvious 
way, so we also have a `temporal orientation'. 
Infinitesimally, both can be encoded in the $1$-form 
$\d \tau$, where $x^0 = c \tau$ denotes the zeroth 
coordinate function in \eqref{eq:chartMgam}. 
Hence a tangent vector 
$Y \in \CapT \mathcal M ^\gamma$ may be called 
\emph{future-directed relative to $\gamma$}, if 
$\d \tau \left( Y\right) > 0$ and 
\emph{past-directed relative to $\gamma$}, if 
$\d \tau \left( Y\right) < 0$. For 
$\d \tau \left( Y \right) = 0$, it is 
\emph{spatial relative to $\gamma$}. Since 
$\mathcal M^\gamma \subset \mathcal I 
\cross \CapT \spti$, this happens precisely 
when it is tangent to the respective past 
tangent light cone. \\
In principle, if 
\begin{equation}
	\vartheta \colon \mathcal J \to \mathcal M^\gamma
	\colon s \to \vartheta \left( s \right) = 
	\left( \tau \left( s\right), K_s\right)
	\label{eq:varthet}
\end{equation}
is a curve in $\mathcal M_\gamma$ with 
$\dot \tau := \d \tau / \d s > 0$, we may measure 
the time passed between the endpoints by integration: 
\begin{equation*}
	\int_{\vartheta} \d \tau 
	= \int_{\mathcal J} \vartheta^* \d \tau
	= \int_{\mathcal J} \frac{\d \tau}{ \d s} \, \d s
	\, .
\end{equation*}
On categorical grounds, we call such a 
curve \emph{future-directed relative to $\gamma$}. 
Analogously, we have 
past-directed and spatial curves relative to $\gamma$. 
A priori, it is not guaranteed, that a future-directed 
curve $\vartheta$ relative to $\gamma$, is also 
future-directed on $\spti$ under the kinematic 
observer mapping $\xi^\gamma$. Indeed, it may happen 
that $\xi^\gamma \circ \vartheta$ is not even 
time- or lightlike. A counterexample is provided 
later in \thref{Ex:rest}. 
The next theorem gives a direct relation between 
the time directions on $\spti$ and 
$\mathcal M^\gamma$. 
\begin{Theorem}[Consistency of time directions]
	\label{Thm:consitencytime}
	Let $\left( \spti, g, \mathcal O\right)$ be a 
	spacetime and $\xi^\gamma$ be the kinematic 
	observer mapping for the observer $\gamma$. 
	Further, let $\pr_2 \colon \gamma^* \CapT 
	\spti \to \CapT \spti$ be the 
	projection on the second factor. 
	\begin{enumerate}[i)]
	\item 	If the curve $\vartheta$, as in 
			\eqref{eq:varthet}, is a future-directed 
			curve relative to $\gamma$ such that 
			$\xi^\gamma \circ \vartheta$ is either 
			time- or lightlike on $\spti$, 
			then $\xi^\gamma \circ \vartheta$ 
			is future-directed. 
			In the former case, there exists a 
			smooth (orientation-preserving) 
			reparametrization such that the reparametrized 
			curve is an observer. 
			\label{itm:timeconsist1}
	\item	Conversely, if 
			$\vartheta'\colon \mathcal J \to \spti$
			is an observer on $\spti$, such 
			that there exists a smooth curve 
			$\vartheta$ with $\vartheta' = 
			\xi^\gamma \circ \vartheta$, then 
			$\vartheta$ is future-directed relative 
			to $\gamma$. 
			\label{itm:timeconsist2}
	\item	\label{itm:timeconsist3}
			If instead $\vartheta'\colon \mathcal J 
			\to \spti$ is a future-directed, 
			lightlike curve, such that there 
			exists a smooth curve 
			$\vartheta$ with $\vartheta' = 
			\xi^\gamma \circ \vartheta$,
			then for each $s \in \mathcal J$ the 
			tangent vector $\dot \vartheta_s$ 
			is either future-directed or spatial 
			relative to $\gamma$. If it is 
			spatial, then it is tangent 
			to the geodesic 
			$r \to \exp \left( r K_s \right)$ 
			with $K_s := \pr_2 \circ \, 
			\vartheta \left( s \right)$. 
	\end{enumerate}
\end{Theorem}
\begin{Proof}
\begin{subequations}  
	`` \ref{itm:timeconsist1} '' 
	We borrowed the idea of proof from Perlick 
	\cite{Perlick1}. As 
	$K_s := 
	\pr_2 \circ \, \vartheta \left( s\right)$ is lightlike 
	for each $s \in \mathcal J$ 
	\begin{equation*}
		0 = \frac{\d}{\d s} 
		g \left(K, K \right) = 
		2 \, g \left( \frac{\nabla K}{\d s}, K \right) \, .
	\end{equation*}
	We now write $\xi^\gamma \circ \, \vartheta 
	= \exp \circ K$ and thus $\dot{\left(   
	\xi^\gamma \circ \vartheta \right)} = \exp_* \dot K$. 
	For $s \in \mathcal J$, apply \thref{Thm:diffexp} 
	from page \pageref{Thm:diffexp} to get 
	$\exp_* \dot K = J^s_1$ for the respective Jacobi 
	field $J^s$ along the geodesic 
	$\tilde \gamma \colon r \to \tilde \gamma 
	\left( r\right) = \exp
	\left( r K_s\right)$. 
	Afterwards, we use \thref{Lem:Jac} from 
	page \pageref{Lem:Jac} to compute 
	\begin{align}
		g \left( \dot {\tilde {\gamma}}^s_1 , 
		J^s_1 \right) & = 
		g \left( \dot {\tilde {\gamma}}^s_0 , \left( \frac{ 
		\nabla J^s}{\d r}\right)_0 \right)  
		+ g \left(\dot {\tilde {\gamma}}^s_0 , J^s_0 \right) 
		\notag \\
		&= g	\left( K_ s , \left( \frac{\nabla K}{\d s} \right)_s
		\right) + g \left(K_s, \dot {
		\left( \gamma \circ \tau \right)}_s \right) 
		\notag \\
		&= \dot \tau \left( s \right) 
		g \left(K_s, \dot \gamma _{\tau \left(s\right)} \right) \, .
		\label{eq:Perequ}
	\end{align}
	The expression is negative, since $\dot \tau > 0$, 
	$\dot \gamma _{\tau \left(s\right)}$ is 
	future-directed timelike and 
	$K_s$ is past-directed lightlike. On the other 
	hand $\dot {\tilde {\gamma}}^s_1$ is the parallel 
	transport $\Par_{\tilde \gamma}^{0,1} 
	\left( K_s \right)$ of $K_s$ to $\exp(K_s)$ along 
	$\tilde \gamma$, hence past-directed timelike. 
	As $J_1 = \dot{\left( \xi^\gamma \circ 
	\vartheta\right)}$ is time- or lightlike, it must be 
	future-directed. Finally, every smooth, 
	future-directed, timelike curve $\vartheta' \colon 
	\mathcal J \to \spti$ 
	can be reparametrized to a smooth observer 
	curve by calculating 
	\begin{equation*}
		s' \left( s \right) 
		= \frac{1}{c} \int_{\inf \mathcal J}^s 
		\sqrt{g \left(\dot \vartheta'_{s''}, 
		\dot \vartheta'_{s''} \right)} \, 
		\d s'' 
	\end{equation*}
	for each $s \in \mathcal J$. \\
	`` \ref{itm:timeconsist2} '': Again we have for 
	all $s \in \mathcal J$
	\begin{equation}
		g \left(	\Par_{\tilde \gamma}^{0,1} 
					\left( K_s \right), \dot{
					\vartheta}'_s
					\right)
					=
		\dot \tau \left( s \right) 
		g \left(K_s, \dot \gamma _{\tau \left(s\right)} \right) 
		\, ,
		\label{eq:Perl1}
	\end{equation}
	as in \eqref{eq:Perequ} above. As $
	\dot{\vartheta}'_s$ is future-directed timelike, 
	it must `lie above' the lightlike hyperplane 
	$\left( \R \Par_{\tilde \gamma}^{0,1} 
	\left( K_s \right) \right)^\perp$, so the left 
	hand side is negative. Since the second factor on the 
	right hand side is also negative, we conclude 
	$\dot \tau >0$. \\
	`` \ref{itm:timeconsist3} '': Again we use 
	\eqref{eq:Perl1}. $\dot \vartheta_s$ is 
	future-directed lightlike implies that it 
	lies either in or `above' the lightlike hyperplane
	$\left( \R \Par_{\tilde \gamma}^{0,1} 
	\left( K_s \right) \right)^\perp$. In the former
	case, there exists a constant $\alpha >0$ such 
	that 
	\begin{equation*}
		\dot{\vartheta}'_s= - \alpha 
	\Par_{\tilde \gamma}^{0,1} \left( K_s \right) \, ,
	\end{equation*}
	which implies that it is tangent to the geodesic 
	$r \to \exp \left( r K_s \right)$. Moreover, 
	\eqref{eq:Perl1} yields $\dot \tau = 0$. In the other
	case, the left hand side of \eqref{eq:Perl1} is 
	negative and thus again $\dot \tau >0$. 
\end{subequations}
\end{Proof}
Roughly speaking, \thref{Thm:consitencytime} states 
that the time directions on $\spti$ and 
$\mathcal M ^\gamma$ are mutually consistent. The timelike 
case in \ref{itm:timeconsist1} and \ref{itm:timeconsist2} 
effectively describes the situation that one observer 
$\gamma$ `sees' another observer $\vartheta'$. Then 
the function $\tau \colon s 
\to \tau \left( s \right)$ from \eqref{eq:varthet} 
gives the relations between the respective proper 
times measured and its derivative $\dot \tau$
may be understood 
as the redshift measured by $\gamma$ 
(cf. \citelist{ \cite{Perlick1}*{\S IV} 
\cite{Brill} \cite{Straumann}*{p. 108 sq.;}
 \cite{SchroedingerB}*{p. 48 sqq.;}}). Since $\dot \tau > 0$, the
function $\tau$ is strictly increasing and has a smooth 
inverse. Therefore, the theory predicts that 
it is physically impossible to see another observer 
whose clock stands still or even runs backwards. 
The lightlike case in 
\ref{itm:timeconsist1} and \ref{itm:timeconsist3} 
includes the physical situation that an observer `sees' 
the movement of `light'. In the spatial case of 
\ref{itm:timeconsist3}, 
the `light ray' is pointed at the observer and thus no 
time needs to pass for the observer to see it moving.
\par
We now conclude our discussion of the measurement of time 
in the splitting and continue with the question whether 
the kinematic observer mapping is invertible. 
\par 
As stated in the previous section \ref{sec:space}, the 
static observer mapping $\xi_{\gamma \left( \tau\right)}$ 
at the point $\gamma \left( \tau \right) \in \spti$ 
need neither be injective nor an immersion. So we conclude 
that the same holds true for the kinematic observer mapping 
for the observer $\gamma$. Nonetheless, we were able to 
prove two propositions, namely \thref{Lem:sinverse0} on 
page \pageref{Lem:sinverse0} and \thref{Thm:sobsinv} on page 
\pageref{Thm:sobsinv}, that gave us at least a local 
form of invertibility. The next theorem is the kinematic 
analogue of \thref{Thm:sobsinv}. 
\begin{Theorem}[Kinematic observer mapping inversion theorem]
	\label{Thm:kobsinv}
	Let $\left( \spti, g, \mathcal O\right)$ be a 
	spacetime and $\xi^\gamma$ be the kinematic 
	observer mapping for the observer 
	$\gamma \colon \mathcal I \to \spti$. \\
	Then $\xi^\gamma$ is almost everywhere 
	locally invertible. The 
	critical set is 
	\begin{equation}
		\crit \xi ^\gamma 
		= \bigsqcup_{\tau \in \mathcal I}
		\crit \xi_{\gamma \left( 
		\tau \right)} \quad .
		\label{eq:critxigamma}
	\end{equation}
\end{Theorem}
\begin{Proof}
\begin{subequations}
	For each $Z \in \CapT \mathcal 
	M^\gamma$ there exists a curve 
	$\vartheta$, as in \eqref{eq:varthet}, with 
	$\dot \vartheta _0 =Z$. Recalling equation 
	\eqref{eq:Perl1} from above for $s=0$ with 
	$\dot \vartheta'_0 = \xi^\gamma_* Z$, we see that 
	$\dot \tau \left( 0\right) \neq 0$ implies that 
	$\xi^\gamma_* Z \neq 0$. Hence we may restrict 
	ourselves to the case 
	$\dot \tau \left( 0 \right) = 0$, i.e. $Z$ is
	spatial with respect to $\gamma$ and thus 
	tangent to 
	$\lbrace \tau \left( 0 \right) \rbrace 
	\cross \mathfrak{c}^-_{(\gamma \circ \tau) 
	\left( 0\right)}$. Consequently, 
	$\xi^\gamma_* Z = 0$ if and only if 
	$\left(\xi_{(\gamma \circ \tau) \left( 0\right)}\right)_* 
	\dot K_0 = 0$. This proves \eqref{eq:critxigamma}. \\ 
	Recalling \thref{Lem:locinv} on page \pageref{Lem:locinv}
	and that $\dim \mathcal M^\gamma = \dim \spti$, we need 
	to show that the 
	set $\crit \xi ^\gamma$ has measure zero. So we 
	introduce coordinates $x'$ on $\mathcal M^\gamma$
	in accordance with \eqref{eq:chartMgam}, define the
	canonical Borel product measure $B:= B_{\mathcal I} 
	\cross B_{\R^n}$ on $\mathcal I \cross \R^n$, 
	identify $\mathcal M^\gamma$ with its image 
	in $\mathcal I \cross \R^n$ under $x'$ and 
	restrict $B$ to $\mathcal M^\gamma$. 
	\eqref{eq:critxigamma} now gives a splitting of 
	$\crit \xi^\gamma$ into sets 
	$\crit \xi_{\gamma \left( \tau \right)}$ of measure zero 
	with respect to $B_{\R^n}$. 
	In analogy to the proof in \thref{Prop:measzero}, 
	we apply `Fubini's theorem in measure theory' 
	(cf. \cite{Bogachev}*{Thm. 3.4.1}) to obtain the result.  
\end{subequations}
\end{Proof}
Therefore, for each time $\tau \in \mathcal I$ and every 
`position' $K \in \mathcal M^\gamma _\tau$, which is 
not a conjugate value, \thref{Thm:kobsinv} states that there
exists a neighborhood of the observed event 
$\left( \tau, K \right)$ in $\mathcal M^\gamma$ that can be
identified with a subset of the spacetime. Figuratively 
speaking, we can then find a direct correspondence between a small 
part of the `spacetime', as the observer sees it, and a small part of 
the `actual' spacetime. 
\par 
We believe that there also exists a kinematic analogue of 
\thref{Lem:sinverse0}, but so far we have
been unable to prove it. 
\begin{Conjecture}
	\label{Conj:ninvert}
	Let $\left( \spti, g, \mathcal O\right)$ be a 
	spacetime and $\xi^\gamma$ be the kinematic 
	observer mapping for the observer 
	$\gamma \colon \mathcal I \to \spti$. \\ 
	Then for every $\tau \in \mathcal I$ 
	there exists an open neighborhood $\mathcal V$ 
	of $\left( \tau, 0 \right)$ in 
	$\gamma^* \CapT \spti$, such that the 
	restriction of $\xi^\gamma$ to 
	$\mathcal V \cap \mathcal M^\gamma$ 
	is a diffeomorphism onto its image. 
\end{Conjecture}
\thref{Conj:ninvert} would make it possible to view 
(restricted) observer coordinates as coordinates on 
$\spti$, in analogy to what 
Klein et al. did in their article \cite{Klein}. 

\subsection{Moving Frames of Reference}
\label{ssec:movframes}

In order to give a sensible definition of the word
`observer spacetime', we first need to make mathematical 
sense of the words `frame of reference for an observer', 
which are often colloquially used in mechanics textbooks. 
We will also give a rigorous, invariant definition of what it 
means for a frame of reference to be inertial. 
\par
The underlying idea of `moving frames' in general relativity 
is that each tangent vector 
$\dot \gamma_\tau$ of the observer $\gamma$ at time 
$\tau \in \mathcal I$ may be normalized and taken as the 
zeroth vector $\left( X_0\right)_\tau$ of a frame of 
reference $(X)_\tau$ at 
$\gamma \left( \tau \right)$. Again, the frame of reference 
represents the `orientation' of the physical observer in space. 
Letting $\tau$ vary, we 
obtain a frame of reference field over $\gamma$ that is 
`adapted' to the observer in this sense and thus gives 
the physical observer's orientation in space at each time. 
The following definition formalizes the idea in terms 
of principal bundles. 
\begin{Definition}[Frame of reference bundles for observers]
	\label{Def:forbgamm}
\begin{subequations}
	Let $\left( \spti, g, \mathcal O\right)$ be a 
	spacetime of dimension $n+1$, 
	$\left(\mathcal P, \tilde \pi, \spti, 
	\Lor_{n+1} \right)$ 
	be the corresponding frame of reference bundle
	(\thref{Def:FR} on page \pageref{Def:FR}) 
	and $\gamma \colon \mathcal I \to \spti$ be an 
	observer. \\ 
	The \emph{frame of reference bundle for $\gamma$}, 
	or, equivalently, 
	\emph{reference frame bundle for 
	$\gamma$}, is the tuple $\left(\mathcal P^ \gamma, 
	\pr_1 \evat{\mathcal P ^ \gamma}, \mathcal I, 
	\LieSO_n \right)$, where 
	\begin{equation}
		\mathcal P^\gamma := 
		\set{\left( \tau, X \right) \in 
		\gamma^* \mathcal P}{X_0 = \frac{1}{c} \, 
		\dot \gamma_\tau} \, , 
	\end{equation}
	$\pr_1 \evat{\mathcal P^\gamma}$ is the restriction of 
	the projection 
	$\pr_1 \colon \mathcal I \cross \mathcal P \to \mathcal I$
	to $\mathcal P ^ \gamma$ and $\LieSO_n$ is the special 
	orthogonal group in $n$ dimensions acting canonically on 
	$\mathcal P^\gamma$ from the 
	right via the representation 
	\begin{equation}
		\rho \colon \LieSO_n \to \LieGL_{n+1} 
		\colon A \to \rho \left( A \right)
		:= 	\begin{pmatrix}
				1 & 0 \\
				0 & A
			\end{pmatrix} 
			\, .
			\label{eq:rhoSO}
	\end{equation}
	A \emph{frame 
	of reference for the observer $\gamma$ (at 
	$\tau \in \mathcal I$}) is the second component 
	$X$ of an element $\left( \tau, X \right)$
	of the fiber 
	$\mathcal P^\gamma _ \tau := 
	\left(\pr_1 \evat{\mathcal P ^ \gamma} 
	\right)^{-1} \left( \tau \right)$.   
\end{subequations}
\end{Definition}
Of course, we still need to show the frame of reference bundle 
for an observer is indeed a principal $\LieSO_n$-bundle. 
\begin{Lemma}
	\label{Lem:smoothforbgamm}
\begin{subequations}
		Let $\left( \spti, g, \mathcal O\right)$ be a 
		spacetime of dimension $n+1$ and let
		$\gamma \colon \mathcal I \to \spti$ be an 
		observer. \\
		Then there exists a unique manifold structure 
		on the frame of reference 
		bundle $\mathcal P^\gamma$ for $\gamma$ such that it is
		an embedded $\LieSO_n$-reduction of the frame 
		of reference bundle $\gamma^* \mathcal P$ 
		along $\gamma$. 
\end{subequations}
\end{Lemma}
\begin{Proof}
	Since $\mathcal P$ is a principal $\Lor_{n+1}$-bundle, 
	so is the pullback bundle $\gamma^* \mathcal P$. 
	We wish to apply \thref{Thm:Baum2.14} from page 
	\pageref{Thm:Baum2.14}. \\
	`` \ref{itm:3Baum2.14} '': Since 
	$\mathcal I$ is an open subset of $\R$, 
	$\gamma^* \mathcal P$ admits a global section 
	\begin{equation*}
		\left(. , X \right) 
		\colon \mathcal I \to \gamma^* \mathcal P
		\colon \tau \to \left( \tau, X_\tau \right)
		\, .
	\end{equation*} 
	As $\frac{1}{c} \dot \gamma$ is smooth, 
	future-directed timelike and has unit `length', 
	we may chose $X$ such that 
	$X_0 = \frac{1}{c} \dot \gamma$. Hence $X$ takes 
	values in $\mathcal P ^\gamma$. \\
	`` \ref{itm:1Baum2.14} '': This is trivial. \\ 
	`` \ref{itm:2Baum2.14} '': Let $X, Y$ be 
	frames of reference for $\gamma$ at $\tau \in \mathcal I$, 
	then $X, Y \in \mathcal P_{\gamma \left( \tau \right)}$
	and so 
	there exists a unique $\Lambda \in \Lor_{n+1}$ with 
	$Y = X \cdot \Lambda$. As $X_0 = Y_0$, the remaining 
	vectors lie in the same hyperplane 
	$\left( \R X_0\right)^\perp$ and thus 
	there exists an $A \in \LieGL_n$ with 
	\begin{equation*}
		\Lambda =
				\begin{pmatrix}
					1 & 0 \\
					0 & A
				\end{pmatrix} \, .
	\end{equation*}
	Since $\Lambda \transp \cdot \eta \cdot \Lambda = \eta$ 
	and $\det \Lambda = 1$, $A$ is indeed in $\LieSO_n$. \\
	We conclude that, by \thref{Thm:Baum2.14}, $\mathcal P^\gamma$
	is an $\LieSO_n$-reduction of $\gamma^* \mathcal P$. 
	\\
	Finally, we need to show that $\mathcal P^\gamma$ is 
	embedded in $\gamma^* \mathcal P$. Since 
	$\Lor_{n+1}$ is embedded in $\LieGL_{n+1}$, 
	it is sufficient to show that the restricted map
	$\rho \colon \LieSO_n \to \Lor_{n+1}$ 
	of \eqref{eq:rhoSO} is a topological 
	embedding. Yet as the map 
	\begin{equation*}
		\chi \colon \Lor_{n+1} \to \R^{n+1} 
		\colon \Lambda \to \Lambda \cdot \baseR_0 
		= \Lambda^i{}_0 \, \baseR_i
	\end{equation*}
	is continous, $\LieSO_n = \chi^{-1} \left( 
	\lbrace \baseR_0 \rbrace \right)$ is closed in 
	$\Lor_{n+1}$ and, by Cartan's theorem, indeed embedded. 
\end{Proof}
We call a map $X \colon \mathcal I \to \mathcal P$ a 
\emph{frame of reference (field) for $\gamma$}, if
\begin{equation*}
		\left(. , X \right) 
		\colon \mathcal I \to \mathcal P^\gamma
		\colon \tau \to \left( \tau, X_\tau \right)
		\, .
\end{equation*}
is a section of $\mathcal P^\gamma$. Of course, one may 
also define \emph{dual frame of reference (fields) for 
$\gamma$} or, equivalently, \emph{coframe of reference (fields) 
for $\gamma$} by using sections of the dual bundle 
$\mathcal P ^{\gamma *}$ instead. 
\par
If the spacetime is $4$-dimensional, the choice of 
a frame of reference 
$X$ for $\gamma$ is unique up to a smooth map 
$A \colon \mathcal I \to \LieSO_3 \colon
\tau \to A (\tau)$. Physically, 
the map $A$ corresponds to rotation of the frame of reference 
$\left( X\right)_\tau$ at each time $\tau \in \mathcal I$ to
a new frame of reference $\left( Y\right)_\tau = \left( X\right)_\tau
\cdot \, \rho (A (\tau) )$. This naturally 
raises the question whether it is possible to tell when a frame 
of reference for an observer is rotating or not. 
\par
Indeed, this may be done via the so called Fermi-Walker derivative, 
named after the physicist Thomas Fermi and mathematician 
Arthur G. Walker for their original works \cite{Fermi} 
and \cite{Walker0}. 
We refer to \citelist{\cite{Maluf} \cite{Sachs}*{\S 2.2} 
\cite{Friedmann} \cite{Hawking}*{p. 80 sqq.;}   
\cite{Straumann}*{\S 1.10} } for further reading. 
The following definition is borrowed from the book by Sachs and Wu 
\cite{Sachs}*{Prop. 2.2.1}. 
\begin{Definition}[Fermi-Walker derivative \& inertial frames]
	\label{Def:FW}
\begin{subequations}
	Let $\left(\mathcal Q, g, \mathcal {O} \right)$ 
	be a spacetime of dimension $n+1$, 
	$\gamma \colon \mathcal I \to \mathcal Q
	\colon s \to \gamma \left( s\right)$ be 
	a time- or spacelike curve and let
	$Y$ be a vector field over $\gamma$. Denote by 
	$\parallel$ and $\perp$ the projection onto 
	the parallel and orthogonal subspaces with respect to
	$\dot \gamma$, respectively. \\
	Then the \emph{Fermi-Walker derivative of $Y$ 
	(along $\gamma$)} 
	is given by
	\begin{equation}
		\frac{\Fder Y}{\d s} =  
		\left( \frac{\nabla}{\d s} \negthinspace 
		\left( Y^\parallel \right) \right)^\parallel + 
		\left( \frac{\nabla}{\d s} \negthinspace 
		\left( Y^\perp \right) \right)^\perp \, . 
		\label{eq:FWdef}
	\end{equation}
	$Y$ is called \emph{Fermi-Walker 
	transported along $\gamma$}, or 
	\emph{non-rotating (along $\gamma$)}, if 
	$\left({\Fder Y}/{\d s}\right)_s = 0$ 
	for all $s \in \mathcal I$. For timelike $\gamma$, 
	$Y$ is called \emph{rotating (along $\gamma$)}, if it is 
	non-non-rotating, i.e. there exists an $s \in \mathcal I$ 
	such that $\left({\Fder Y}/{\d s}\right)_s \neq 0$. 
	A tangent vector $Y_{s'}
	\in T_{\gamma \left( s' \right)} \spti$ over $\gamma$ 
	is called the \emph{Fermi-Walker transport of $Y_s 
	\in T_{\gamma \left( s\right)} \spti$ to $s'$}, 
	if there exists a Fermi-Walker transported vector field 
	$Y$ over $\gamma$ taking the respective values. \\
	If $X$ is a frame field over $\gamma$, then its Fermi-Walker 
	derivative is given by differentiating the individual 
	vector fields, i.e. 
	\begin{equation}
		\frac{\Fder X}{\d s} := \frac{\Fder X_i}{\d s} 
		\tp \cbaseR ^i \, .
		\label{eq:FWframe}
	\end{equation}
	The aforementioned terminology carries over to frames 
	and frame fields over $\gamma$, that is $X_i$ satisfies 
	the required conditions for each $i \in \lbrace 0, \dots, 
	n \rbrace$. \\ 
	In particular, if $X$ is a frame of reference field for an 
	observer $\gamma$, it is called non-rotating if
	$\Fder X / \d \tau = 0$. Else it is rotating. \\
	Additionally, if $\gamma$ is a non-accelerating observer 
	and $X$ is a non-rotating frame of reference field 
	for $\gamma$, then $X$ is called an 
	\emph{inertial frame of reference (field for $\gamma$)}. 
\end{subequations}
\end{Definition} 
For timelike curves the 
Fermi-Walker derivative may be physically interpreted as 
a tool to detect the presence of (infinitesimal) rotation. 
For spacelike curves it 
detects twisting of the vector field $Y$ or change of 
(hyperbolic) angle relative to $\dot \gamma$.
Both 
interpretations are derived from the definition, 
which roughly states that the Fermi-Walker 
derivative respects the parallel and orthogonal 
subspaces with respect to $\dot \gamma$.% 
\footnote{One may indeed show that Fermi-Walker transport 
preserves angles along arbitrary time- or spacelike curves 
$\gamma$. 
The idea of proof is to first show `metricity' 
\eqref{eq:Fermmetric} of the derivative 
for curves of constant length and then apply 
\eqref{eq:Fermrepara} to conclude that it holds for arbitrary 
parametrizations. Deriving the angle formula 
(cf. 
\eqref{eq:confang} on page \pageref{eq:confang}) for 
$\dot \gamma$ and non-rotating $Y$, and observing that 
$g \left( Y,Y\right)$ is constant, gives the 
result.}  
\begin{Remark}[Basic properties of Fermi-Walker derivative]
\begin{subequations}
	Recall that for a time- or spacelike curve 
	$\gamma \colon \mathcal I \to \mathcal Q$ and each 
	$s \in \mathcal I$ the projectors are given by
		\begin{equation}
			\left(\pi^\parallel \right)_s = \frac{\dot 
			\gamma _ s \tp \dot \gamma_s \cdot g}
			{\left(\gamma^*g\right)_s 
			\left( \partd{}{s}, \partd{}{s}  \right)} 
			\quad \text{and} 
			\quad \left(\pi^\perp\right)_s = \Id_s - 
			\left(\pi^\parallel \right)_s \, , 
			\label{eq:prgamma}
		\end{equation}
	where $\Id$ is the identity endomorphism field 
	over $\gamma$, i.e. $\Id \cdot Y = Y$ for all 
	$Y \in \CapT \spti$ with base point in 
	$\gamma\left(\mathcal I\right) \subset \spti$. 
	Thus the Fermi-Walker derivative of a vector field 
	$Y$ along $\gamma$ is well-defined, smooth and again 
	yields a vector field along $\gamma$. As the 
	projectors and the Levi-Civita connection are linear, 
	so is the Fermi-Walker derivative. The usage of 
	the word derivative is justified by the fact that it 
	satiesfies the \emph{Leibniz rule}:  
	\begin{equation*}
		\frac{\Fder }{\d s} \left( f Y \right) 
		= \frac{\d f}{\d s}  \, Y + f 
		\, \frac{\Fder Y}{\d s} \, ,
	\end{equation*}
	for all $f \in C^\infty \left(\mathcal I, \R \right)$
	and vector fields $Y$ along $\gamma$. A straightforward 
	calculation also shows that under a change of 
	parametrization of $\gamma$ we get 
	\begin{equation}
		\frac{\Fder \phantom{.}}{\d s} = \frac{\d s'}{\d s} \, 
		\frac{\Fder \phantom{.}}{\d s'} \quad . 
		\label{eq:Fermrepara}
	\end{equation}
	We conclude that the Fermi-Walker derivative is a 
	covariant derivative along $\gamma$ 
	(cf. \cite{Poor}*{Def. 2.51}) and thus it yields a 
	vector bundle connection on the pullback bundle 
	$\gamma^* \CapT \spti$ with corresponding parallel 
	transport (cf. \cite{Poor}*{Cor. 2.59}). However, by 
	\eqref{eq:prgamma} it is impossible to extend 
	the Fermi-Walker derivative to curves that are 
	neither space- or timelike. Therefore the derivative 
	does not give rise to a connection on the tangent 
	bundle, so the terminology `Fermi-Walker connection' 
	is only admissible in the appropriate context. 
\end{subequations} 
\end{Remark}
\par
In our case of interest, $\gamma$ is an observer and then the 
Fermi-Walker derivative takes a particularly simple form. 
\begin{Lemma}[Fermi-Walker derivative for observers]
	\label{Lem:FWO}
	Let $\left(\mathcal Q, g, \mathcal {O} \right)$ 
	be a spacetime, $\gamma \colon \mathcal I \to \mathcal Q
	\colon \tau \to \gamma \left( \tau\right)$ be 
	an observer and $Y$ be a vector field over $\gamma$. Then 
	\begin{equation}
		\frac{\Fder}{\d \tau} Y = \frac{\nabla}{\d \tau} Y 
		- \frac{1}{c^2} \, g\left(\dot \gamma, Y \right) 
		\frac{\nabla \dot \gamma}{\d \tau} 
		+ \frac{1}{c^2} \, g\left(\frac{\nabla \dot \gamma}
		{\d \tau}, Y \right) \dot \gamma \quad .
		\label{eq:FWobs}
	\end{equation}
\end{Lemma}
\begin{Proof}
	For observers $\left(\gamma^*g \right) 
	\left( \partd{}{\tau}, \partd{}{\tau} \right) = c^2$ 
	and thus $\dot \gamma \perp (\nabla \dot \gamma /\d \tau)$.  
	Using the definitions \eqref{eq:FWdef}, 
	\eqref{eq:prgamma} and the metricity of the Levi-Civita 
	connection, we obtain   
	\begin{align*}
		\frac{\Fder Y}{\d \tau} &= \pi^\parallel \cdot
		\left(\frac{\nabla \left(\pi^\parallel \cdot Y\right)}
		{\d \tau} \right) + 
		\pi^\perp \cdot
		\left(\frac{\nabla 
		\left(\pi^\perp \cdot Y\right)}{\d \tau} \right) \\ 
		&= \frac{1}{c^2} \, \pi^\parallel \cdot 
		\left( \frac{\nabla }{\d \tau} 
		\left( g \left( \dot \gamma, Y\right) 
		\dot \gamma \right) \right) +  
		\pi^\perp \cdot \left(\frac{\nabla }{\d \tau} 
		\left( Y - \frac{1}{c^2} \, 
		g \left( \dot \gamma, Y \right) 
		\dot \gamma \right) \right)   \\
		&= \frac{1}{c^2} \, \pi^\parallel \cdot \left( 
		\frac{\d }{\d \tau} 
		\left( g \left( \dot \gamma, Y\right) \right) 
		\dot \gamma  + 
		g \left( \dot \gamma, Y\right) 
		\frac{\nabla \dot \gamma }{\d \tau} 
		 \right) \\
		& \phantom{.=} 
		+ \pi^\perp \cdot \left(
		\frac{\nabla Y}{\d \tau}  - 
		\frac{1}{c^2} \, \frac{\d }{\d \tau} 
		\left( g \left( \dot \gamma, Y \right)\right) 
		\dot \gamma  
		- \frac{1}{c^2} \, g \left( \dot \gamma, Y \right)  
		\frac{\nabla \dot \gamma }{\d \tau}  
		\right)    \\ 
		& = \frac{1}{c^2} \, 
		\left( g \left( \frac{\nabla \dot \gamma }{\d \tau}, 
		Y\right) 
		+  g \left( \dot \gamma , 
		\frac{\nabla Y }{\d \tau} \right) 
		\right) \dot \gamma  \\ 
		& \phantom{.=} 
		+ \left( 
		\frac{\nabla Y}{\d \tau}  - \frac{1}{c^2} \, 
		g \left( \dot 
		\gamma, \frac{\nabla Y}{\d \tau} \right) \, 
		\dot \gamma \right) 
		- \frac{1}{c^2}\,  g \left( \dot \gamma, Y \right)  
		\frac{\nabla \dot \gamma }{\d \tau}  \quad . 
	\end{align*}
\end{Proof}
Equation \eqref{eq:FWobs} implies that for 
non-accelerated observers the Fermi-Walker derivative 
reduces to the Levi-Civita connection. Moreover, it also 
shows that the Fermi-Walker connection for observers 
is metric: 
\begin{equation}
	\frac{\d}{\d \tau}  \, 
	g \left( Y, Z \right) 
	= g \left( \frac{\Fder Y}{\d \tau}, Z \right) + 
	g \left( Y,  \frac{\Fder Z }{\d \tau} \right) 
	\, ,
	\label{eq:Fermmetric}
\end{equation}
where $Y$ and $Z$ are vector fields along $\gamma$. 
Therefore, if $X_{\tau_0}$ 
is a frame of reference for $\gamma$ 
at $\tau_0 \in \mathcal I$ and we consider the 
Fermi-Walker transport of each frame vector along 
$\gamma$, the resulting collection of vector fields 
will be an orthonormal frame field along $\gamma$. 
Since the Fermi-Walker derivative of $\dot \gamma$ 
vanishes and the Fermi-Walker transport map is 
continuous, this orthonormal frame field will be 
a frame of reference field for $\gamma$. 
\par
Mathematically, this proves that the Fermi-Walker 
derivative induces a principal bundle connection on 
the frame of reference bundle $\mathcal P^\gamma$ 
for the observer $\gamma$. We refer to the English 
books by Poor \cite{Poor} and Rudolph \cite{Rudolph1}, 
as well as the German one by Baum \cite{Baum} for an 
in-depth discussion of principal bundle connections. 
\par
We finish our discussion of frame of reference fields 
for observers with physically relevant 
examples. 
\begin{Example}[Frame of reference fields in Minkowski 
				spacetime]
	\label{Ex:RefMink}
\begin{subequations}
	We continue \thref{Ex:obsMink} from page 
	\pageref{Ex:obsMink}. 
	\begin{enumerate}[i)]
	\item 	\label{itm:obsMink1}
			For our prototypical, unaccelerated 
			observer $\gamma$ we have 
			\begin{equation*}
				\left(X_0 \right)_\tau 
				:= \frac{1}{c} \dot \gamma _\tau
				= \Evat{\partd{}{y^0}}{\gamma 
				\left( \tau \right)} 
				= \left( \partial_0 \right)_ 
				{\gamma \left( \tau \right)}
			\end{equation*}
			for all $\tau \in \R$. Since the coordinate 
			frame field $\partial$ on $\R^4$ is a frame of 
			reference field, any frame of reference field 
			$X$ for $\gamma$ may be written as 
			\begin{equation*}
				X = \left( \partial \right)_{
					\gamma} \cdot \rho 
					\left( A \right) \, ,
			\end{equation*}
			where $\rho$ is the representation 
			from \eqref{eq:rhoSO} for $n=3$ and 
			$A \colon \R \to \LieSO_3$ is a 
			smooth curve. By \eqref{eq:FWobs}, 
			the Fermi-Walker derivative reduces to the 
			Levi-Civita derivative for unaccelerated 
			observers. Thus, as $\partial$ is parallel, 
			$X$ is inertial if and only if 
			$A$ is constant. In particular, the frame of 
			reference $X$, satisfying  
			\begin{equation}
				\label{eq:stdrdX}
				\left(X\right)_\tau = \left( \partial \right)_
				{\gamma \left( \tau \right)} 
				= 
				\begin{pmatrix}
					1 & & & \\ 
					& 1 & & \\
					& & 1 & \\
					& & & 1 
				\end{pmatrix} \, ,
			\end{equation}
			is inertial. By a  
			Poincar\'e transformation as in 
			\thref{Ex:statoMink}, any inertial 
			frame of reference can be 
			brought into this standard form. 
	\item 	For our constantly accelerated observer 
			$\gamma\colon \R \to \R^4$, as given by 
			\eqref{eq:accelgam} on page 
			\pageref{eq:accelgam}, we may assume that 
			it `faces in the direction of acceleration'. 
			In mathematical terms, we have for all 
			$\tau \in \R$: 
			\begin{equation*}
				\left( X_1 \right)_\tau := \frac{1}{a} \, 
				\left( \frac{\nabla \dot \gamma}{\d \tau} \right)
				_\tau = 
				\sinh \left( \frac{a \tau}{c} \right)
				\, \left( \partial_0 \right)_{\gamma 
				\left( \tau \right)} + 
				\cosh \left( \frac{a \tau}{c} \right) \, 
				\left( \partial_1 \right)_{\gamma 
				\left( \tau \right)} \, ,
			\end{equation*}
			by a simple calculation using 
			\eqref{eq:accelgamdot}. One checks via \eqref{eq:FWobs}
			that this vector is non-rotating, 
			which means that not only the magnitude, but 
			also the \emph{direction of acceleration of 
			$\gamma$} is \emph{constant}. This is known 
			as \emph{uniform acceleration}. If we 
			choose the smooth curve  
			\begin{equation*}
				A \colon \R \to \LieSO_3 
				\colon 
				\tau \to A \left(\tau \right):= 
					\begin{pmatrix}
						1 & 0 & 0 \\
						0 & \cos \left(\omega \tau \right) 
						& - \sin \left(\omega \tau \right) \\ 
						0 & \sin \left(\omega \tau \right) 
						& \phantom{-} 
						\cos \left(\omega \tau \right)
					\end{pmatrix}  \, ,
			\end{equation*}
			representing a rotation around the first axis 
			in positive direction for $\omega \in \R_+$ 
			and in negative direction for 
			$\omega \in \R_- := \left(- \infty, 0 \right)$, 
			then $X$, given by
			\begin{equation}
				\label{eq:arX}
				\left( X \right)_\tau 
				:= 
				\begin{pmatrix}
					\cosh \left( \frac{a \tau}{c} \right) &
					\sinh \left( \frac{a \tau}{c} \right) &
					 &  \\
					\sinh \left( \frac{a \tau}{c} \right) &
					\cosh \left( \frac{a \tau}{c} \right) & 
					 &  \\
					& & 
					\cos \left(\omega \tau \right) & 
					- \sin \left(\omega \tau \right) \\ 
					& & 
					\sin \left(\omega \tau \right) 
					& \phantom{-} 
					\cos \left(\omega \tau \right)
				\end{pmatrix} \, 
			\end{equation}
			in standard coordinates for each $\tau \in \R$, 
			is a frame of reference field for $\gamma$. 
			Unless $\omega= 0$, it is rotating. Accordingly, 
			$X$ represents a physical observer 
			accelerating uniformly with $a$ and 
			rotating around its axis of acceleration with 
			constant angular velocity $\omega$. 
	\end{enumerate}
\end{subequations}
\end{Example}

\subsection{Observer Spacetime and relative Motion}
\label{ssec:ospti}

After having defined the kinematic observer mapping 
in section \ref{ssec:kingen} and frames of references 
for observers in section \ref{ssec:movframes}, we now 
give mathematical meaning to the word `observer spacetime' 
and the physical concept of relative motion. 
\par
In section \ref{ssec:kingen}, we have already discussed 
the measurement of time and the assignment of time directions
on the domain $\mathcal M^\gamma$ 
of the kinematic observer mapping $\xi^\gamma$. We 
concluded that the $1$-form $\d \tau$ is sufficient to 
define both rigorously. Now additional 
`spatial' geometric structures need to be 
carried over from the static splitting 
(cf. \thref{Def:obsspace} on page 
\pageref{Def:obsspace}) to the kinematic case, i.e. we 
need to find natural definitions of spatial distances 
and right-handedness on $\mathcal M^\gamma$.
\par
To define spatial distances between events observed at 
different times, we require a frame of 
reference field $X$ for the observer $\gamma$. 
The underlying idea is, that taking the 
components of a vector $K \in \mathcal M^\gamma_{\tau_0}$
with respect to $X_{\tau_0}$ for some time 
${\tau_0} \in \mathcal I$ makes it possible to 
identify the event at every other time 
$\tau \in \mathcal I$. So if 
$\left( \tau_1, K_1 \right), \left( \tau_2, K_2 \right) 
\in \mathcal M^\gamma$ are two observed events, their 
spatial distance is just the distance from section 
\ref{sec:space} between the vectors $K_1$, $K_2$ 
`transported' to some time $\tau_0$. Since this 
distance is invariant under the action of the rotation 
group $\LieSO_n$ on $\mathcal P^\gamma$, the 
definition is independent of the choice of $\tau_0$. 
\par
The issue of orientations is more subtle. 
Of course, if 
$\dim \spti = n+1$, $\tau \in \mathcal I$ and 
$Y_1, \dots, Y_n$ is a right-handed
basis of the tangent space $\CapT_K \mathcal M^\gamma_\tau$
with respect to the observer space orientation $O^\tau$, 
then there should be at least one `temporal vector' $Y_0$ in 
$\CapT_{\left(\tau, K \right)} \mathcal M^\gamma$ such that 
$Y = Y_i \tp \cbaseR^i$ is a `right-handed' basis in 
$\CapT_{\left(\tau, K \right)} \mathcal M^\gamma$. 
Intuitively, we take the vector field 
$\partial/ \partial \tau$ evaluated at the point 
$\left(\tau, K \right)$ to obtain $Y_0$. 
Yet this vector field is not well-defined, as geometrically, 
it depends on how one `attaches' the $\mathcal M^\gamma_\tau$ 
to each other. For this reason, we require a 
frame of reference $X$ for $\gamma$ in this case as well. 
\par
In the following, we define both 
structures in terms of coordinates $x$ with respect to $X$. 
Invariant definitions in terms of $X$ do exist, but are 
complicated and not needed here. 
\begin{Definition}[Observer spacetime]
	\label{Def:omapping}
		Let $\left( \spti , g , \mathcal O\right)$ be a 
		spacetime of dimension $n+1$, 
		$\gamma \colon \mathcal I \to \spti$ be an observer 
		and $\mathcal M^\gamma$ be the domain of 
		the respective kinematic observer mapping 
		$\xi^\gamma$. Further, let $X$ be a 
		frame of reference field for $\gamma$. \\
		Then the coordinates $x$ on $\mathcal M^\gamma$ 
		with respect to $X$, as given by equation 
		\eqref{eq:chartMgam} on page 
		\pageref{eq:chartMgam}, are called
		\emph{(kinematic) observer coordinates 
		(with respect to $X$)}. The tensor field 
		\begin{equation}
			h = h_{ij} \, \d x^i \tp \d x^j 
			:= \delta_{ab} \, \d x^a \tp \d x^b =
			\begin{pmatrix}
				0 & & & \\
				& 1 & & \\
				& & \ddots & \\
				& & & 1 
			\end{pmatrix}
		\end{equation}
		on $\mathcal M^\gamma$ is the 
		\emph{(kinematic) observer metric 
		(induced by $X$)} and the $1$-form 
		$\d \tau$ on $\mathcal M^\gamma$ is the 
		\emph{time form (with respect to $\gamma$)}. 
		The \emph{observer spacetime orientation $O$ 
		(induced by $X$)} is 
		the $\LieGL^+_n$-structure on 
		$\mathcal M ^\gamma$ induced by the 
		coordinate frame field $\partial$ and the 
		representation 
		\begin{equation*}
			\rho \colon \LieGL_n^+ \to \LieGL_{n+1}
			\colon A \to \rho \left( A\right) :=
			\begin{pmatrix}
				1 & 0 \\ 
				0 & A 
			\end{pmatrix} \, .
		\end{equation*}
		The tuple 
		$\left( \mathcal M^\gamma, \d \tau, h, O\right)$
		is called the 
		\emph{observer spacetime with respect to $X$}. 
\end{Definition}
As opposed to observer spaces, observer spacetimes 
are not Riemannian manifolds and hence are not 
by default equipped with an intrinsic connection. For this
reason, we need to explicitly define a connection, which, 
as the Levi-Civita connection for Riemannian manifolds, 
is to be viewed as intrinsic to the observer spacetime. 
Obviously, the connection should coincide with the 
Levi-Civita connection, 
if restricted to the observer spaces. Physically, 
this assures that its auto-parallels correspond to 
`straight-line motion', as seen by the observer. 
\begin{Definition}[Observer connection]
	\label{Def:obsconn}
	\NoEndMark
		Let $\left( \spti , g , \mathcal O\right)$ be a 
		spacetime and $X$ be a frame of reference field 
		for the observer $\gamma$. \\
		If $\left( \mathcal M^\gamma, \d \tau, h, O\right)$
		is the observer spacetime with respect to $X$, then 
		a covariant derivative $\Nder$ on 
		$\CapT \mathcal M^\gamma$ is called \emph{(kinematic) 
		observer 
		connection (with respect to $X$)}, if $\Nder$ is 
		torsion-free, and compatible with $\d \tau$ and 
		$h$ in the following sense: \\ 
		\phantom{
		$\overset{\alpha}{A}$ 
		\hspace{0.27 \textwidth}
		}
			$\Nder \left( \d \tau \right) = 0 
			\quad , \quad \Nder h = 0 \quad .$
		\hfill \DefinitionSymbol
		%%%it's ugly in code but looks better in print :-) 
\end{Definition}
\begin{Lemma}
	\label{Lem:obsconn}
\begin{subequations}
	The observer connection exists and is unique. 
\end{subequations}
\end{Lemma}
\begin{Proof}
In full analogy to \cite{Reddiger0}*{Lem. 2.2}, 
\thref{Lem:obsconn} is proven by constructing the standard 
Riemannian metric out of the time form and the observer metric. 
\end{Proof}
As in the static case, kinematic observer coordinates are 
adapted to all geometric structures on $\mathcal M^\gamma$ 
for the given frame of reference field $X$ for $\gamma$ and 
thus $\Nder$ is simply the standard flat connection with 
respect to the kinematic observer coordinates. 
\par
So in 
practice, we choose a particular observer 
$\gamma$, a frame
of reference field $X$ for $\gamma$, which may be rotating 
or not, and compute the kinematic observer mapping 
$\xi ^\gamma$ in the kinematic observer coordinates 
$x$ with respect to $X$. 
Explicitly, this \emph{kinematic observer mapping with 
respect to $X$} is the map 
$\varphi_X^\gamma := \xi ^\gamma \circ (x^{-1})$, i.e. 
\begin{align}
	\varphi_X^\gamma  
	&\colon &  
	x \left( \mathcal M ^\gamma \right)
	\quad &\to & \mathcal Q \notag \\
	&\colon & 
	x = \left(c \tau, \vec x  \right) 
	&\to & \varphi^\gamma_X 
	&\left( c \tau, \vec x \right) := 
	\exp \left( - \abs{\vec x} \,
	\left( X_0\right)_{\tau} + 
	x^a \, \left(X_a \right)_\tau  \right)  
	\, .
\end{align}	
Then, to relate events in $\spti$ to events in the observer spacetime 
$\mathcal M^\gamma$, this map needs to be inverted in the sense of 
\thref{Thm:kobsinv} on page \pageref{Thm:kobsinv}. The 
computation of $\varphi \equiv \varphi_X^\gamma$ and its inverse 
is usually a mathematically very challenging task, 
but once this has been achieved, the computation of temporal 
and spatial distances between the events on $\spti$, as seen 
by the observer $\gamma$, is very simple. 
We simply identify the points in $\mathcal M^\gamma$ with 
their coordinate values and then, in accordance with the defined
geometric structures, the temporal 
distances between any two observed events 
$x= \left( c \tau, \vec{x}\right), 
x'= \left( c \tau', \vec{x}'\right)$ 
are $\abs{\tau - \tau'}$ and the spatial distances 
are $\abs{ \vec{x} - \vec{x}'}$. The orientation is also 
as we would intuitively expect. Hence the geometric 
structures from \thref{Def:omapping} and 
\thref{Def:obsconn} are only implicitly used in 
practice, since distances, orientations, etc. 
accord with `Newtonian intuition'. 
\par
To get at ease with the construction and also as a check of 
physical consistency, let us continue with our two examples 
from \thref{Ex:obsMink} on page \pageref{Ex:obsMink} and 
\thref{Ex:RefMink} on page \pageref{Ex:RefMink}. 
\begin{Example}[Observer spacetimes in Minkowski spacetime]
	\label{Ex:SR}
\begin{subequations}
\begin{enumerate}[i)]
	\item 	\label{itm:SR1}
			We first compute the kinematic observer 
			mapping $\varphi$ with respect to the standard 
			intertial frame of reference field $X$, as 
			given by \eqref{eq:stndrtgam} and \eqref{eq:stdrdX}. 
			Recalling the exponential map in Minkowski 
			spacetime \eqref{eq:expMinkow} on page 
			\pageref{eq:expMinkow}, 
			the values of $\varphi$ are 
			\begin{equation}
			\varphi \left(c \tau, \vec{x}\right) = 
				\begin{pmatrix}
					c \tau - \abs{\vec x} \\ 
					\vec x
				\end{pmatrix} 
			\end{equation}
			in standard coordinates $y = 
			\left( c t, \vec y \right)$ on $\R^4$ and for all 
			$x = \left(c \tau, \vec{x}\right) \in \R \cross 
			\left( \R^3 \setminus \lbrace 0 \rbrace \right)$. 
			If we restrict $\varphi$ to its image, it  
			has the smooth inverse $\ubar \varphi$ with values
			\begin{align}
				\ubar{\varphi} \left( ct, \vec y 
				\right) = 
				\begin{pmatrix}
					c t + \abs{\vec y}  \\
					\vec y 
				\end{pmatrix} 
			\end{align} 
			and is therefore a diffeomorphism. 
			In addition, the space-time splitting is 
			\emph{global} in the sense that 
			$\spti \setminus \left( \gamma 
			\left(\mathcal I \right) \right)$ is contained 
			in the image of the kinematic observer mapping. 
			In fact, $\xi^\gamma \left( \mathcal 
			M^\gamma\right) = \spti \setminus \left( \gamma 
			\left(\mathcal I \right) \right)$ here. 
			\\
			In the physics literature the expression 
			$t = \tau - \abs{\vec x} / c$ is known as the 
			retarded time and commonly occurs in the 
			special-relativistic theory of electrodynamics. 
			The coordinate $t = y^0 / c$ is also a measure 
			of time in the sense that it is the proper time of 
			the family of 
			observers given by $\tau \to \gamma 
			\left( \tau \right) + \left(0,\vec y_0 \right)$
			with $\vec y_0 \in \R^3$. It may be called 
			\emph{Einstein-synchronized time}, since it is 
			the result of a clock synchronization among these 
			observers and is discussed already in 
			Einstein's original paper
			on special relativity \cite{Einstein0}. 
			It was famously argued by Reichenbach 
			\cite{Reichenbach} that this choice is 
			pure convention. Independent of one's position on 
			this issue, the proper time $\tau$ is not 
			conventional. \\ 
			As a diffeomorphism, we may view $\varphi$ 
			as a coordinate transformation from $x$ to $y$, so 
			we effectively `identify' the observer spacetime 
			with the spacetime itself. Thus, given any two 
			events $\left( ct, \vec y \right)$ 
			and $\left( ct', \vec y \right)$ in the image of 
			$\varphi$, their temporal distance is simply 
			$\abs{t-t'}$ and their spatial distance is
			$\abs{\vec y- \vec y '}$. Hence the general theory 
			applied to inertial frame of reference fields 
			in Minkowski spacetime indeed reproduces 
			the temporal and spatial 
			distances from special relativity. 
	\item 	\label{itm:SR2}
			For the uniformly accelerated observer 
			$\gamma$ given by \eqref{eq:accelgam}
			on page \pageref{eq:accelgam} with 
			frame of reference field $X$ from 
			\eqref{eq:arX} above, we may again compute 
			the kinematic observer mapping $\varphi$ 
			with respect to $X$: 
			\begin{align}
				\varphi \left( c \tau, \vec x \right)
				&= \gamma \left( \tau \right) + 
				\left( - \abs{\vec x } \, 
				X^i{}_0 \left( \tau \right)
				 + x^a \, 
				X^i{}_a \left( \tau \right) 
				\right) \, \baseR_i
				\notag \\
				&= 
				\begin{pmatrix}
					\frac{c^2}{a} 
					\sinh \left( \frac{a \tau}{c} \right)
					- \abs{\vec x} 
					\cosh \left( \frac{a \tau}{c} \right)
					+ x^1 \sinh \left( \frac{a \tau}{c} \right) 
					\\
					\frac{c^2}{a} 
					\cosh \left( \frac{a \tau}{c} \right)
					- \frac{c^2}{a} - \abs{\vec x} 
					\sinh \left( \frac{a \tau}{c} \right)
					+ x^1 \cosh \left( \frac{a \tau}{c} \right) \\ 
					x^2 \cos \left( \omega \tau  
					\right) 
					- x^3  \sin \left(\omega \tau \right) \\ 
					x^2 \sin \left(\omega \tau \right) 
					+  x^3 \cos \left(\omega \tau \right)
				\end{pmatrix}
				\, .
				\label{eq:acrotX}
			\end{align}
			This directly shows that the kinematic observer 
			mapping may be very complicated, even in the 
			absence of curvature. \\
			As the past light cones 
			$\mathcal C_{\gamma \left( \tau \right)}^-$
			should not intersect each other, we conjecture 
			that in Minkowski spacetime the kinematic 
			observer mapping is always a diffeomorphism 
			onto its image. The difficulty in the proof 
			of this statement is injectivity, as for each 
			$\tau \in \mathcal I = \dom \gamma$ the 
			map $\exp_{\gamma 
			\left( \tau \right)}$ has full rank and one may
			then argue as in the proof of 
			\thref{Thm:kobsinv} to conclude that 
			$\xi^\gamma$ has full rank. Therefore, 
			at least locally $\varphi$ is a diffeomorphism
			and one may use asymptotic expansions of 
			$\varphi$ in each variable and formal 
			series inversion to obtain an approximate local
			inverse. Without such an inverse, however, 
			we can only fragmentarily relate the observer 
			spacetime to the `actual' spacetime. \\
			We also wish to remark that, in the context of 
			uniform acceleration in special relativity, the 
			work of Rindler \cite{Rindler}*{\S 2.16}
			is frequently cited. The approach, however, 
			significantly differs from ours 
			(for $\omega = 0$). 
\end{enumerate}
\end{subequations}
\end{Example}
\par
As for the static splitting in section 
\ref{sec:space}, we may compute 
the differential of the kinematic observer mapping 
$\varphi$ with respect to the frame of reference 
field $X$ in terms of Jacobi fields. Since the 
`spatial part' has already been computed in 
\thref{Prop:diffsobm}, we only need to consider the 
`temporal part'. 
\begin{Proposition}[Differential of kinematic observer mapping]
	\label{Prop:diffkobm}
\begin{subequations}
	Let $\left( \spti, g, \mathcal O\right)$ be 
	a spacetime of dimension $n+1$ 
	and $\varphi$ be the kinematic observer 
	mapping with respect to the frame of reference field 
	$X$ for the observer $\gamma \colon \mathcal I \to \spti 
	\colon \tau \to \gamma \left( \tau\right)$. 
	\\
	Then for all $x = \left( c \tau, \vec x \right) 
	\in \dom \varphi \subseteq \left(c\mathcal I \right)
	\cross 
	\left(\R^n \setminus \lbrace 0 \rbrace\right)$: 
	\begin{equation}
		\left(\varphi_* \partd{}{\tau} 
		\right)_{x} = J_1 \left( x \right)
		\, , 
	\end{equation}
	where $J \left( x \right) 
	\colon s \to J_s \left( x \right)$ 
	is the unique Jacobi field along the geodesic 
	\begin{equation*}
		s \to \exp_{\gamma \left(\tau \right)} 
		\left( s \left( - \abs{\vec x} \left(X_0\right)_\tau 
		+ x^a \, \left(X_a \right)_\tau \right)
		\right) 
	\end{equation*} 
	with $J_0 \left( x \right) = \dot \gamma_\tau$ and 
	\begin{equation}
		\left( \frac{\nabla J \left( x \right)
		}{\d s} \right)_0 = 
		- \abs{\vec x} \left(\frac{\nabla X_0}{\d \tau} 
		\right)_\tau 
		+ x^a \, \left(\frac{\nabla X_a}{\d \tau} 
		\right)_\tau \, .
		\label{eq:phistau1}
	\end{equation}
	In particular, if $X$ is Fermi-Walker transported along 
	$\gamma$: 
	\begin{equation}
		\left( \frac{\nabla J \left( x \right)
		}{\d s} \right)_0 = 
		\left( \frac{x^a}{c} \, \delta_{ab} \,
		{\frac{\nabla \dot \gamma}{\d \tau}}
		^b \left( \tau \right)
		\right) \, \left( X_0 \right)_\tau - 
		\frac{\abs{\vec x}}{c} \,
		{\frac{\nabla \dot \gamma}{\d \tau}}^a 
		\left( \tau \right) \, \left( X_a \right)_\tau 
		\, ,
		\label{eq:phistau2}
	\end{equation}
	where we defined
	\begin{equation*}
		{\frac{\nabla \dot \gamma}{\d \tau}}^b := 
		\ubar{X}^b \cdot \frac{\nabla \dot \gamma}{\d \tau} 
	\end{equation*}
	for all $b \in \lbrace 1, \dots, n \rbrace$. 
\end{subequations}
\end{Proposition}
\begin{Proof}
	In full analogy to the proof of \thref{Prop:diffsobm}, 
	we need to consider the curve 
	\begin{equation*}
		Y \colon \mathcal I \to \CapT \spti \colon
		\tau \to Y_\tau := - \abs{\vec x} \left( X_0 \right)_\tau
		 + x^a \, \left( X_a \right)_\tau
	\end{equation*}
	and apply 
	\thref{Thm:diffexp} from page \pageref{Thm:diffexp} 
	to its tangent vector field. Now observe that its 
	base curve is $\gamma$ and compute the covariant 
	derivative to obtain \eqref{eq:phistau1}. From the 
	expression of the Fermi-Walker derivative for observers
	\eqref{eq:FWobs} follows \eqref{eq:phistau2}. 
\end{Proof}
\par
The vector field $\partial/ \partial \tau$ has a particular 
physical significance for the theory, as already 
suggested by the definition of observer spacetimes. This 
significance is revealed within the subject of relative motion. 
\begin{Definition}[Relative motion]
	\label{Def:relmot}
	Let $\left( \spti, g, \mathcal O\right)$ be 
	a spacetime of dimension $n+1$, 
	$\xi^\gamma \colon \mathcal M^\gamma \to \spti$ 
	be the kinematic observer mapping for the observer 
	$\gamma \colon \mathcal I \to \spti 
	\colon \tau \to \gamma \left( \tau\right)$
	and $X$ be a frame of reference field for $\gamma$. 
	Further, let $x= \left( c \tau, \vec x 
	\right)$ be observer coordinates with respect 
	to $X$ and $\Nder$ be the observer connection. If 
	\begin{equation*}
		\vartheta \colon \mathcal J \to 
		\mathcal M^\gamma 
		\colon
		s \to \vartheta \left( s \right) = 
		\left( \tau \left( s \right), K_s \right)
	\end{equation*}
	is a (smooth) curve such that $\xi^\gamma \circ 
	\vartheta$ is an observer, denote by 
	$\breve{\vartheta}$ the reparametrized curve 
	with respect to the coordinate $\tau$. Its tangent 
	vector field in observer coordinates is 
	\begin{equation*}
		\dot{\breve{\vartheta}} = c 
		\Evat{\partd{}{x^0}}{\breve{\vartheta}} 
		+ v^a \, 
		\Evat{\partd{}{x^a}}{\breve{\vartheta}} 
	\end{equation*}
	The spatial part of $\dot{\breve{\vartheta}}$
	is called the \emph{velocity (field) of $\vartheta$
	relative to $X$} and, if evaluated at $\tau 
	\in \dom \breve{\vartheta}$, it 
	is called the \emph{velocity of $\vartheta$ 
	relative to $X$ at $\tau$}. Moreover, 
	\begin{equation*}
		\frac{\Nder \breve{\vartheta}}{\d \tau}
	\end{equation*}
	is called the \emph{acceleration field of 
	$\vartheta$ relative to $X$} and, if evaluated 
	at $\tau$, it is 
	called the \emph{acceleration of $\vartheta$ 
	relative to $X$ at $\tau$}. We say 
	\emph{$\vartheta$ is at rest with 
	respect to $X$}, if its velocity field vanishes. 
\end{Definition}
The definition is sensible as, by 
\thref{Thm:consitencytime}\ref{itm:timeconsist2}, 
the reparametrization of $\vartheta$
exists. \thref{Def:relmot} may appear very technical, but in 
fact it simply reproduces the definitions of velocity and 
acceleration given by `Newtonian intuition'. That is, if
we write $\left( x^a \circ \vartheta \right) \left( \tau\right)
\equiv x^a \left( \tau \right)$ in 
observer coordinates, then the components of the relative 
velocity at $\tau$ are 
\begin{equation*}
	v^a \left( \tau \right) = \frac{\d x^a} {\d \tau} 
	\left( \tau \right)	
\end{equation*}
and the components of the 
acceleration at $\tau$ are 
\begin{equation*}
	\dot v^a \left( \tau \right) 
	= \frac{\d^2 x^a}{ \d \tau^2} \left( \tau \right) 
	\, .	
\end{equation*}
\begin{Remark}[Relative motion]
	\label{Rem:relmot}
\begin{enumerate}[i)]
	\item Sometimes it is convenient to parametrize the 
	functions $v^a$ with respect to the proper time $s$ 
	of the `observed observer'. By a slight abuse of 
	terminology, we also speak of relative velocity 
	and acceleration in this context. 
	\item In the context of relative motion, we stress the 
	fact that the kinematic observer mapping is usually not 
	injective. Physically, one observer at an instant of time 
	may see another one at different (observed) events and thus 
	the relative state of motion can differ vastly. 
	\item Instead of considering just a curve in 
	$\mathcal M^\gamma$, that yields a future directed timelike 
	curve under 
	the kinematic observer mapping, one may attach to 
	it a `frame of reference' in the sense of Newtonian 
	mechanics to model the orientation of the 
	observed physical observer in space. 
	Conversely, one may use local inverses of 
	the kinematic observer mapping to relate a frame of 
	reference field for a second observer to such a 
	`Newtonian frame of reference'. A priori, it 
	should also be possible to determine an 
	`infinitesimal length contraction' 
	from this. However, we decided not to treat this problem 
	here.
\end{enumerate}
\end{Remark}
We conclude that any integral curve $\breve{\vartheta}$ 
of the vector field $\partial / \partial \tau$ corresponds 
to a physical observer at rest relative to the reference 
frame field $X$, provided 
$\xi^\gamma \circ \breve{\vartheta}$ is timelike (see
\thref{Thm:consitencytime}\ref{itm:timeconsist1}). 
As the next example shows, the latter condition is not 
guaranteed in general. 
\begin{Example}[Non-existence of physical observers at rest]
	\label{Ex:rest}
	We consider \thref{Ex:SR}\ref{itm:SR2} for $a \to 0$, 
	i.e. where we have an unaccelerated observer 
	with rotating frame of reference field. The map $\varphi$
	is then given by 
	\begin{equation*}
		\varphi \left( c \tau, \vec x \right)
		= 
			\begin{pmatrix}
				c \tau 
				- \abs{\vec x} \\
				x^1 \\ 
				x^2 \cos \left( \omega \tau  
				\right) 
				- x^3  \sin \left(\omega \tau \right) \\ 
				x^2 \sin \left(\omega \tau \right) 
				+  x^3 \cos \left(\omega \tau \right)
			\end{pmatrix} 
		= 
			\begin{pmatrix}
			y^0 \\ 
			\vec y
			\end{pmatrix} 
			\, .
	\end{equation*}
	Since $\abs{\vec y} = \abs {\vec x}$, we may easily 
	compute a global inverse, but this is not needed here. 
	The integral curves of $\partial /\partial \tau$ under 
	$\varphi$ are timelike if and only if
	$\varphi_* \left( \partial /\partial \tau \right)$ is timelike.  
	It is thus sufficient to compute 
	\begin{align*}
		g_{\varphi \left( x \right)} 
		\left( \varphi_* \partd{}{\tau}, 
		\varphi_* \partd{}{\tau} \right) 
		&= \partd{\varphi^i}{\tau} \left( x\right)
		\, \eta_{ij} \, 
		\partd{\varphi^j}{\tau} \left( x \right) \notag \\
		&= c^2 - \left( \partd{\vec y}{\tau} \left( x\right) 
		\right)^2 \notag \\
		&=  c^2 - \omega^2 \left( \left(x^2\right)^2 
		+\left(x^3\right)^2 \right)
	\end{align*}
	for $x \in \dom \varphi$. This is negative for large enough 
	values of $x^2$ and $x^3$, and indeed to be 
	expected: To keep up with the 
	rotation of the `observing' physical observer 
	for increasing spatial distances (perpendicular to the 
	axis of rotation), a far away `observed' physical observer
	would eventually need to move faster than light.
	An impossibility. 
	\par 
	The subject of non-accelerating, constantly 
	rotating `observers' in Minkowski spacetime has been 
	widely discussed in the relativity literature, 
	see e.g. \citelist{ \cite{DieksA0} \cite{HillA0} 
	\cite{Hendriksen} \cite{Corum}}. 
	To our knowledge, however, the approach followed here 
	has not been pursued elsewhere. 
\end{Example}
We will continue our discussion of the subject of 
relative motion in the following chapter.  

\chapter{The Newtonian Limit}
\label{chap:nlimit}

Any physical theory, which is not able to reproduce 
empirically supported results in their domain of validity, must 
necessarily be false. This statement, which might appear as a 
platitude at first sight, reveals itself as a powerful tool of 
falsifying physical theories already on the theoretical level. 
In our case, this implies that a mathematical theory 
of separating space 
and time in general relativity has to be able to reproduce 
Newtonian mechanics within its domain of validity 
in some mathematically admissible approximation. Such an
approximation procedure for obtaining Newtonian mechanics out of 
general relativity via a theory of space-time splitting is what 
we philosophically define as the \emph{Newtonian limit}. 
\par
Einstein himself was well aware of the fact that the existence 
of the Newtonian limit would be a crucial requirement for the 
physical feasibility of his general theory of relativity. In the 
special theory of relativity, the proof of the existence of the 
Newtonian limit, though naive, was straightforward  
(see e.g. \cite{Reddiger0}*{\S 2} for a detailed discussion), but 
for the general one more sophisticated reasoning had to be 
applied. Einstein decided that he had to generalize the 
Gau\ss' law for gravity \cite{Einstein0}*{p. 87} to arrive at 
a law for the spacetime curvature and then find a procedure to 
rederive the Gau\ss' law as an 
approximation. His success in this endeavor in 1915 
\cite{EinsteinA0} marks a historic event: 
The sought-after equation is nowadays known as the Einstein 
(field) equation and the procedure is called the 
weak-field approximation. Accounts of his reasoning 
can be found in many introductory books on general 
relativity, see e.g.
the books by Carroll \cite{Carroll}*{\S 4.1 \& 4.2} and Wald 
\cite{Wald}*{\S 4.4(a)}. 
\par
Unfortunately, the standard approaches to the Newtonian limit 
both in the special and the general theory of relativity 
are worthy of criticism, the main defect being a 
reliance on coordinates rather than on geometric structures. 
In particular, from a mathematician's perspective, the weak field 
approximation is a heuristic rather than a rigorous method. 
Though these coordinate methods might be shown to be
justifiable under a more careful mathematical analysis, 
by themselves they constitute inadmissible evidence for the 
existence of the Newtonian limit. Hence 
for the general theory other approaches 
have been found trying to address these issues, notably 
due to \'Elie Cartan and J\"urgen Ehlers. A discussion thereof, 
including references to the original works, can be 
found in Maren Reimold's work \cite{Reimold}. Our discussion here, 
however, takes a different route and relates the Newtonian limit 
to the construction in the previous chapter. On one hand 
this provides a rigorous approach to the Newtonian limit without
relying on somewhat arbitrary coordinates or the ad-hoc 
introduction of geometric structures, 
on the other hand the existence of the Newtonian limit is 
required to give physical credibility to the splitting 
formalism itself. Scientific care
dictates that the Newtonian limit is to be derived
by mathematically and philosophically sound methods. Its existence
may not be taken for granted a priori, as doing so would deprive 
the theorist of one of the primary means of falsifying 
the theory. 
\par
Having this discussion in mind, in section \ref{sec:Nlimitgen} 
we start with a general a priori approach 
to the Newtonian limit employing the splitting construction. 
Important problems required for a proof of the existence 
of the Newtonian limit (in our sense) for the general 
theory of relativity are brought forward. In section 
\ref{sec:Nlimitsr} we voice criticism towards the 
standard approach to the Newtonian limit in the special theory 
of relativity and show that for inertial 
frame of reference fields in Minkowski spacetime the Newtonian 
limit indeed exists. In addition, we 
give lower order correction terms. 
\par
We wish to remark that our discussion only applies to the 
Newtonian limit of point masses for a given spacetime, observer 
and frame of reference field. The issue of the Newtonian limit 
of general-relativistic field theories 
(e.g. magneto-hydrodynamics, quantum theories) depends 
highly on the theory under consideration 
and a discussion thereof would go beyond the scope of this thesis. 
It should, however, be said that for field theories the 
relevant geometric structures have to be identified and a coherent, 
critical understanding of what precisely constitutes a Newtonian 
limit in the respective case needs to be attained.

\section{General Newtonian Limit}
\label{sec:Nlimitgen}

The two main ingredients of Newtonian mechanics are a 
model of space and time with associated geometric 
structures, called Newtonian spacetime, as well as a 
law determining the dynamics of mass points therein, that is 
Newton's second law of motion. A careful mathematical 
axiomatization of the physical concept of 
Newtonian spacetime has been given 
in \cite{Reddiger0}*{\S 2} and shall not be repeated here. 
For our purposes, it is sufficient to recall that the observer
spacetime models `the world as the physical observer sees it' and
thus the Newtonian spacetime needs to be somehow related to 
$4$-dimensional observer spacetimes. Indeed, it directly follows 
from the respective definitions that, mathematically speaking, 
every observer 
spacetime $\left( \mathcal M^\gamma, \d \tau, h, O \right)$ 
in a spacetime $\left(\spti, g, \mathcal O \right)$ of 
dimension $4$ is a Newtonian spacetime under the 
identification of $\mathcal M^\gamma$ with the 
domain of the respective kinematic observer coordinates 
$x$, i.e. an open subset of $\R^4$. In particular, this 
statement is independent of the spacetime, observer
$\gamma$ or the frame of reference field $X$. 
\par
If we transcribe the implicit law of Newtonian mechanics 
that all clocks run at the same rate to this 
setting, then for a curve 
\begin{equation*}
	\vartheta \colon \mathcal J \to \mathcal M^\gamma 
	\colon s \to \vartheta \left( s \right) 
	= \left( \tau \left( s\right), K_s \right) \, ,
\end{equation*}
representing physical motion, we must demand 
$\dot \tau \equiv \d \tau ( \dot 
\vartheta ) = 1$. If this holds, the curve 
parameters $s$ and $\tau$ are equal (up to a shift) 
and then the observer 
connection $\Nder$ may be used to reformulate
Newton's second law 
\begin{equation}
	\vec F = m \, \frac{\Nder \dot \vartheta}{\d \tau} \, ,
	\label{eq:N2limit}
\end{equation}
where $m \in \R_+$ is the mass of the `observed object' 
traveling along $\vartheta$ and 
$\vec F$ is a (spatial) vector field along 
$\vartheta$. That is, given $\vartheta$ such that 
$\gamma' := \xi^\gamma \circ \vartheta$ is an observer, 
equation \eqref{eq:N2limit} yields the 
\emph{relative force $\vec F$}. Expressions for $\vec F$ 
in observer coordinates ought to be sensible 
within the Newtonian theory.
\par
We have thus obtained two necessary conditions for the 
existence of the Newtonian limit. Since both conditions 
in conjunction with respective initial conditions specify 
the motion entirely, the two conditions are in fact 
sufficient. However, as the spacetime $\spti$ 
represents `the world as it is', neither the condition 
$\dot \tau = 1$ nor equation \eqref{eq:N2limit} ought 
to be viewed as equations of motion in relativity 
theory, but rather serve 
as indicators of how well the relativistic model 
may be cast into the framework of Newtonian mechanics.
Their approximate validity proves the 
existence of the Newtonian limit mathematically - under the 
assumptions necessary to make these approximations. 
\par
Let us further specify the model 
case. In practical situations, we first 
determine the spacetime $\left( \spti, g, 
\mathcal O \right)$, second we require an observer 
$\gamma$ and an appropriate frame of reference field $X$ for 
$\gamma$. In the third step, we compute the kinematic 
observer mapping $\varphi$ with respect to $X$. In the fourth 
step, a second observer $\gamma'\colon \mathcal J \to \spti$ 
needs to be found, which satisfies the dynamical law 
\begin{equation}
	F' = m \, \frac{\nabla \dot \gamma'}{\d s} \, ,
	\label{eq:realNewt}
\end{equation}
where $m \in \R_+$ is the mass of $\gamma'$ and $F'$ denotes
the `actual' force acting on $\gamma'$. 
If $\gamma'$ does not lie in the image of $\varphi$, then 
$\gamma$ does not `see' $\gamma'$ and thus the question
of the existence of the Newtonian limit is meaningless. 
Therefore, we are only interested in the case, where 
$\gamma'$ lies fully in the image of $\varphi$. 
Indeed, there is 
a simple sufficient condition for the local
existence of a smooth curve $s \to x \left( s\right) = 
\left(c \tau \left( s\right), \vec x \left(s \right) 
\right)$ in $\dom \varphi \subset \R^4$
with $\left(\varphi \circ x \right)\left( 
s\right) = \gamma' \left( s\right)$ for all admissible 
$s \in \R$.
\begin{Lemma}
	\label{Lem:Nlimcurve}
	Let $\left( \spti, g, \mathcal O\right)$ be a spacetime, 
	and let $\xi^\gamma \colon \mathcal M^\gamma \to \spti$ 
	be the kinematic observer mapping 
	for the observer $\gamma$. Further, let 
	$\gamma' \colon \mathcal J \to \spti$ be another
	observer and let there exist an $s_0 \in 
	\mathcal J$ such that $\gamma' \left( s_0\right)$ lies
	in the image of $\xi^\gamma$ and is a regular value of 
	$\xi^\gamma$. \\
	Then there exists an open neighborhood 
	$\mathcal J'$ of $s_0$ in $\mathcal J \subseteq \R$
	and a smooth curve 
	\begin{equation*}
	\vartheta \colon \mathcal J' \to \mathcal M^\gamma 
	\colon s \to \vartheta \left( s \right) 
	= \left( \tau \left( s\right), K_s \right) \, ,
	\end{equation*}
	such that $\gamma' \evat{\mathcal J'} 
	= \xi^\gamma \circ \vartheta$ and $\d \tau / \d s > 0$. 
\end{Lemma}
\begin{Proof}
	This is a direct corollary of the kinematic observer 
	mapping inversion theorem (\thref{Thm:kobsinv}): Since 
	$\gamma' \left( s_0\right)$ is a regular value lying 
	in the image, there exists an 
	$\left( \tau', K' \right) \in \mathcal M^\gamma$ 
	with open neighborhood $\mathcal V$ such that 
	$\xi^\gamma \evat{\mathcal V}$ is a diffeomorphism 
	onto its image. Since $\xi^\gamma\left( \mathcal V 
	\right)$ is open and $\gamma$ is continuous, 
	$\mathcal J'$ exists and 
	$\vartheta := 
	\left( \xi^\gamma \evat{\mathcal V}\right)^{-1} 
	\circ \left( \gamma' \evat{\mathcal J'} \right)$ does the 
	job. $\d \tau / \d s > 0$ follows from 
	the consistency of time directions 
	(\thref{Thm:consitencytime}\ref{itm:timeconsist2}). 
\end{Proof}
As the proof suggests, finding the curve $x$ in practice 
requires inverting the kinematic observer mapping $\varphi$ 
and thus the curve need neither be unique nor can we 
guarantee that $\mathcal J = \mathcal J'$. Moreover, 
$\gamma' \left( s_0\right)$ need not be a regular value of
$\varphi$. In fact, it is 
even possible that 
$\gamma' \left( s\right)$ is a critical value 
for every $s \in \mathcal J$. An example of such a curve 
can be constructed in the plane wave 
spacetimes, again we refer to the articles by 
Perlick \cite{Perlick2}*{\S 5.11} and the original one by 
Penrose \cite{PenroseA0}. 
\par
Without the existence of a smooth curve $x$ satisfying 
$\varphi \circ x = \gamma' \evat {\mathcal J'}$, the 
question of the existence of the Newtonian limit is 
again superfluous. Indeed, Newton's second law and the 
condition $\dot \tau \approx 1$ are only sensible in this 
setting, if such a curve exists. 
\par 
We may therefore continue with the assumption 
that a curve $x \colon \mathcal J \to \dom \varphi$ is given 
and that $\gamma' := \varphi \circ x$ is an observer. 
Then, defining 
$\alpha_s := \left( \varphi^* g\right)_{x \left( s \right)}$ 
for any $s \in \mathcal J'$ and taking the components of 
the relative velocity field to be 
$v^i := \d x^i / \d \tau$, we may find a general expression 
for $\dot \tau >0$ by computing 
\begin{equation*}
	c^2 = g \left( \dot{\left( \varphi \circ x \right)},
	\dot{\left( \varphi \circ x \right)} \right) 
	= \alpha \left( \dot x, \dot x \right) 
	= \alpha_{ij} \, \dot x^i  \dot x^j 
	= \alpha_{ij} \, {\dot \tau}^2 \, v^i v^j \, ,
\end{equation*}
which yields
\begin{equation}
	\dot \tau = \frac{1}{\sqrt{\alpha_{00} 
		+ 2 \alpha_{0a} \, \frac{v^a}{c}  
		+ \alpha_{ab}\, \frac{v^a}{c} \, \frac{v^b}{c}
	}} 
	\label{eq:dottaugen}
\end{equation}
with $a, b \in \lbrace 1, 2, 3 \rbrace$. It is obvious that
this expression is not identically $1$ in most cases, yet 
we only require an approximate validity. \par
To determine which approximation to use, we recall that, 
empirically, Newtonian 
mechanics is known to provide an adequate description of 
phenomena at length and time scales familiar to everyday 
human experience. With respect to these scales the 
speed of light $c$ is in general very large and we may thus
consider $\varepsilon := 1/c$ as a perturbation parameter. 
Yet mindlessly expanding all equations in $\varepsilon$ will 
yield wrong results as the occurrence or non-occurrence of 
$c$ in a physical equation depends on the particular 
conventions used. Philosophically, the problem of what 
to expand boils down to the question which convention 
is `most natural' in the sense that it does not 
`artificially' introduce factors of $c$. We claim that a 
natural convention is the one where all expressions are written 
in terms of the coordinates $\left( \tau, \vec x \right)$. 
In order to avoid 
philosophically deep discussions on naturalness, we justify 
this choice by observing that it yields 
reasonable results in the example treated in the 
concluding section. Nonetheless, the approximation 
can only be made if the dependence of the $\alpha_{ij}$ on 
$c$ is known, so additional assumptions are needed to 
make statements on the existence of the Newtonian limit 
in the general case. 
\begin{Example}
	\label{Ex:limdottau}
	In the particular case where $\alpha_{00} > 0$ and all 
	$\alpha_{ij}$ are independent of $\varepsilon$, a 
	second-order Taylor expansion of the above expression 
	\eqref{eq:dottaugen} in $\varepsilon=1/c$ 
	around $0$ yields: 
	\begin{equation}
	\dot \tau = \frac{1}{\left(\alpha_{00}\right)^{1/2}}
	- \frac{\alpha_{0a}}{\left(\alpha_{00}\right)^{3/2}} 
	\, \frac{v^a}{c} 
	+ \frac{\left( 3 \alpha_{0a} \alpha_{0b} - \alpha_{00} 
	\alpha_{ab} \right)} 
	{2 \left(\alpha_{00}\right)^{5/2}} \, 
	\frac{v^a}{c} \, \frac{v^b}{c} + 
	\bigo \left(1/c^3 \right)
	\quad .
	\end{equation} 
	Here we have used the common `big O notation' to indicate 
	the order of the approximation 
	(cf. \cite{Roessel}*{\S 1.B}). 
	Therefore in this case the Newtonian limit can only exist, 
	if $\alpha_{00} = 1$. Note that additional 
	approximations may be needed for this to hold. 
	Unless all $\alpha_{0a}$ vanish, the Newtonian 
	limit needs to be attained in the zeroth order approximation 
	in $1/c$. 
\end{Example}
Expansion terms that give corrections to the expression 
$\dot \tau$ as well as $F^c$ in the Newtonian limit are
correspondingly called \emph{relativistic correction 
terms}. 
\par
For the second condition \eqref{eq:N2limit}, we need to relate 
the `actual dynamics' \eqref{eq:realNewt} of $\gamma' = 
\varphi \circ x$ to the `observed dynamics' of 
$x$. In general this can only be done in a $1$-$1$ manner, 
if $\varphi$ restricted to an open neighborhood 
$\mathcal V$ of the curve $x$ in 
$\dom \varphi \subset \R^4$ is a diffeomorphism onto its image. 
For convenience, we assume that $\varphi 
\left( \mathcal V\right) \subseteq \spti$ is contained 
in the domain of the coordinate
map $\kappa$. Then we may understand $\varphi$ as 
a coordinate transformation from observer coordinates 
$x$ with respect to $X$ to the coordinates $\kappa$. 
We denote the inverse of $\varphi \evat{\mathcal V}$ 
by $\ubar \varphi$ and assume that 
$\partial/ \partial \kappa^0$ is timelike. \par
In the most common situations, we are given the functions
$F'^a := \d \kappa^a \cdot F'$ with 
$a \in \lbrace 1, 2, 3 \rbrace$ and we still 
need to calculate $F'^0 := \d \kappa^0 \cdot 
F'$. This is obtained from 
\begin{align*}
	0 &= g \left(\dot \gamma', F' \right) 
	= \alpha \left( \dot x, \ubar \varphi_* F'\right) 
	= \alpha_{ij} \, \dot x^i \, {\ubar \varphi}^j{}_{,k} 
	F'^k  \\
	&= \dot \tau \left( F'^0 
	\left(\alpha_{0i} \,c \, {\ubar \varphi}^i{}_{,0} 
	+ \alpha_{ai} \, v^a \, {\ubar \varphi}^i{}_{,0} 
	\right)
	+ \left( 
	\alpha_{0i} \,c \, {\ubar \varphi}^i{}_{,a} F'^a  
	+ \alpha_{ai} \, v^a \, {\ubar \varphi}^i{}_{,a} F'^a  
	\right)
	\right) \, ,
\end{align*}
where we used the notation $\ubar \varphi ^i {}_{,j} := 
\partial \kappa^i /\partial x^j$. Since 
$\partial /\partial \kappa^0$ is timelike: 
\begin{equation*}
	g \left(\gamma',  \partial /\partial \kappa^0 \right) 
	= \alpha_{ij} \, \dot x^i \,  {\ubar \varphi}^j{}_{,0} 
	=
	\dot \tau \left(\alpha_{0i} \,c \, {\ubar \varphi}^i{}_{,0} 
	+ \alpha_{ai} \, v^a \, {\ubar \varphi}^i{}_{,0} 
	\right) \neq 0 \, ,
\end{equation*}
and therefore 
\begin{equation}
	F'^0 = - \frac{
	\left( \alpha_{0i} 
	+ \alpha_{bi} \, \frac{v^b}{c}  
	\right) {\ubar \varphi}^i{}_{,b} F'^b
	}{
	\left(\alpha_{0j} 
	+ \alpha_{cj} \, \frac{v^c}{c}  
	\right) {\ubar \varphi}^j{}_{,0}
	} \, .
	\label{eq:F'0}
\end{equation}
In accordance with \eqref{eq:N2limit}, the `observed dynamics' 
in observer coordinates is given by 
\begin{equation*}
	 m \frac{\d^2 x^c}{\d \tau^2} = m 
	 \frac{\d v^c}{\d \tau} = F^c
\end{equation*}
and we need to determine the components $F^c$ of the 
relative force $\vec F$ from the `actual dynamics' 
\eqref{eq:realNewt} in observer coordinates: 
\begin{equation}
	\frac{\d^2 x^c}{\d s^2} + 
	\Upsilon^c{}_{ij} \,  \frac{\d x^i}{\d s} 
	\frac{\d x^j}{\d s} = 
	\frac{1}{m} \, {\ubar \varphi}^c{}_{,i} F'^i 
	\, ,
	\label{eq:actualforce}
\end{equation}
where the $\Upsilon^c{}_{ij}$ are the relevant Christoffel 
symbols. These are obtained 
from the Christoffel symbols $\Gamma^k{}_{ij}$ in 
coordinates $\kappa$ by the usual transformation formula: 
\begin{equation}
	\Upsilon^c{}_{ij} = \partd{x^c}{\kappa^{l_3}} 
	\, \Gamma^{l_3}{}_{{l_1}{l_2}} \, 
	\partd{\kappa^{l_1}}{x^i} \, \partd{\kappa^{l_2}}{x^j}
	+ \partd{x^c}{\kappa^{l}} \, \frac{\partial^2 
	\kappa^{l}}{\partial x^i \partial x^j}
	\, .
	\label{eq:UpsGam}
\end{equation}
Of course, they can also be calculated from the pullback 
$\varphi^* g$ restricted $\mathcal V$. If we define 
$\ubar \alpha := \left( \alpha\right)^{-1}$
and denote partial derivatives by a comma, this is done via
\begin{equation}
	\Upsilon^c{}_{ij} = \frac{{\ubar {\alpha}}^{cl}}{2} 
	\left( \alpha_{li,j} + \alpha_{lj,i} 
	-\alpha_{ij,l} \right)
	\, , 
	\label{eq:Upsalpha}
\end{equation}
but here one requires the algebraic 
inverse of $\varphi^* g$. 
Once all $\Upsilon^c{}_{ij}$ have been found, we compute
\begin{equation*}
	\frac{\d^2 x^c}{\d s^2} = \frac{\d}{\d s} 
	\left( \dot \tau \, \frac{\d x^c}{\d \tau}\right)
	= \ddot \tau \, \frac{\d x^c}{\d \tau} 
	+ \dot \tau^2 \, \frac{\d^2 x^c}{\d \tau^2} \, ,
\end{equation*}
and after splitting the left hand side of 
\eqref{eq:actualforce} into `temporal' and `spatial' 
parts along with some rearrangement of terms, we
ultimately get an expression for $F^c$:
\begin{equation}
	m \frac{\d v^c}{\d \tau} = 
	\underbrace{ 
	\frac{1}{ \dot \tau^2} \, {\ubar \varphi}^c{}_{,i} F'^i }
	_{\text{`actual' forces}} 
	\underbrace{
	- m c^2 \, \Upsilon^c{}_{00} 
	- m \frac{ \ddot \tau}{\dot \tau^2} \, v^c 
	- m \, 2 c \, \Upsilon^c{}_{0a}\, v^a 
	- m \Upsilon^c{}_{ab}\, v^a \, v^b} 
	_{\text{pseudo-forces}}
	\, .
	\label{eq:NewtF}
\end{equation}
Thus the relative force is always the sum of the `actual' 
forces acting on the observer $\gamma'$ and the pseudo-forces, 
which, by definition, are of purely geometric origin and so
always contain the mass of $\gamma'$ as a simple factor. Physically, 
(actual) forces have the property that they cause an 
absolute acceleration, while pseudo-forces 
only lead to relative accelerations. In physics textbooks 
one sometimes reads the claim that `pseudo-forces only occur 
in non-inertial frames of reference', but, at least in the 
presence of curvature, equation \eqref{eq:NewtF} 
shows that this is incorrect. In particular, 
\eqref{eq:NewtF} states that gravity is a pseudo-force. 
\par
In analogy to the expansion of $\dot \tau$ in 
$\epsilon = 1/c$, we may expand the right hand 
side of \eqref{eq:NewtF} to check for the 
existence of the Newtonian limit. As before, we need 
additional assumptions on the $c$-dependence of 
the $\alpha_{ij}$ to make statements on the general case. 
\begin{Example}
	\label{Ex:Ex:limdotF}
	We continue \thref{Ex:limdottau} from above, i.e. 
	we assume $\alpha_{00} = 1$ and all $\alpha_{ij}$, 
	if written in terms of $\tau$ and $\vec x$,  
	are independent of $c$. According to equation
	\eqref{eq:Upsalpha} and since $\alpha_{00,l} \equiv 0$, 
	the relevant Christoffel symbols turn out to be
	\begin{align*}
		\Upsilon^c{}_{00} & = 
		\frac{1}{c} \, {\ubar{\alpha}}^{ca} \,
		\partd{\alpha_{a0}}{\tau}
		\\
		\Upsilon^c{}_{0a} &	= \frac{1}{2} \,{\ubar {\alpha}}^{cb}
		\left(   
		\alpha_{b0,a} - \alpha_{0a,b} 
		\right) + 
		\frac{1}{2 c} \, 
		{\ubar {\alpha}}^{cb} \, \partd{\alpha_{ab}}{\tau}
		\\
		\Upsilon^c{}_{ab} &= \frac{1}{2}\, {\ubar {\alpha}}^{cl} 
		\left( \alpha_{la,b} + \alpha_{lb,a} 
		- \delta_l^d \alpha_{ab,d} \right)
		- 
		\frac{1}{2 c} \, 
		{\ubar {\alpha}}^{c0} \, \partd{\alpha_{ab}}{\tau}
		\, .
	\end{align*}
	For the existence of the Newtonian limit in the force-free 
	case $F'=0$, we require that the pseudo-forces in
	\eqref{eq:NewtF} do not diverge for $1/c \to 0$. Therefore,
	if we 
	plug in the Christoffel symbols and group the terms by 
	their $c$-dependence, we find that
	\begin{align*}
		0 &\approx {\ubar{\alpha}}^{ca} \,
		\partd{\alpha_{a0}}{\tau} \\
		0 &\approx {\ubar {\alpha}}^{cb}
		\left(   
		\alpha_{b0,a} - \alpha_{0a,b} 
		\right)
	\end{align*}
	needs to hold for the Newtonian limit to exist for 
	various values of the $v^a < c$. Again, unless all 
	$\alpha_{0a}$ vanish, the Newtonian 
	limit needs to be attained in the zeroth order 
	approximation in $1/c$, so taking the limit 
	$1/c \to 0$, our force equation reads:
	\begin{equation*}
		m \frac{\d v^c}{\d \tau} \approx - m \,
		{\ubar {\alpha}}^{cb} \, \partd{\alpha_{ab}}{\tau} v^a
		- \frac{m}{2}\, {\ubar {\alpha}}^{cl} 
		\left( \alpha_{la,b} + \alpha_{lb,a} 
		- \delta_l^d \alpha_{ab,d} \right) \, v^a\, v^b
	\, .
	\end{equation*}
\end{Example}
\par
Summing up, the existence of the Newtonian limit depends on 
the following choices: 
\begin{enumerate}[1)]
	\item 	the spacetime $\left( \spti, g, 
			\mathcal O \right)$, 
	\item 	the observer $\gamma$,
	\item 	the frame of reference field $X$ for $\gamma$,
	\item 	the chosen (maximal) set $\mathcal V$ in the 
			domain of the kinematic observer mapping 
			$\varphi$ with respect to $X$, where 
			$\varphi \evat{\mathcal V}$ is a diffeomorphism 
			onto its image, and
	\item 	the allowed forces $F'$, which may also depend on 
			$c$. 
\end{enumerate}
\par
Fortunately, the consistency of the general theory of 
relativity with Newtonian gravitational theory only 
requires a proof of the existence of the Newtonian limit 
for a few particular cases, namely those where Newtonian mechanics 
makes statements on gravity. Here care must be taken 
regarding the fundamentally different conceptions of 
gravity in the two theories, which are to a certain extent  
incommensurable.%
\footnote{We refer to the work by Kuhn \cite{Kuhn}
and the synopsis by Pajares \cite{Pajares} for a discussion of 
the concept of mutual incommensurability of scientific theories.} 
 To our assessment, the cases relevant for Newtonian gravitational 
theory are:
\begin{enumerate}[i)]
	\item 	\label{itm:nlim1}
			inertial frame of reference field
			for a (non-accelerated) observer in 
			Minkowski spacetime, 
	\item 	\label{itm:nlim2}
			arbitrarily rotating frame of reference 
			field for a constantly accelerated observer
			in Minkowski spacetime,
	\item 	\label{itm:nlim3}
			non-rotating frame of reference field
			for certain observers in Schwarzschild 
			spacetime sufficiently `far away' from the 
			gravitating mass. 
\end{enumerate}
\par
Case \ref{itm:nlim1} is required to show that 
special relativity in the Newtonian limit 
agrees with Newtonian mechanics in the 
absence of gravity. In particular, we require 
$F^c \approx 0$ if and only if $F'=0$. The case is 
considered in section \ref{sec:Nlimitsr}. 
\par
Case \ref{itm:nlim2} should reproduce the constant 
gravitational force, as well as the coriolis, centrifugal 
and Euler pseudo-forces in the Newtonian theory 
(see e.g. \cite{Bradbury}*{\S 10.4} for formulas). 
We emphasize that the constant `downward' gravitational 
force acting on the `observed' observer $\gamma'$ in 
the Newtonian ontology needs to be described by a 
constant `upward' acceleration on the observer 
$\gamma$ in the relativistic ontology. 
\par 
Number \ref{itm:nlim3} is the mathematically most 
challenging case, but also the most interesting one from a 
physical perspective. The Newtonian limit should 
partially reproduce the pseudo-force (field)
\begin{equation}
	\vec F = - m \, \frac{G \,M }{r^2} \, \partd{}{r} \, ,
	\label{eq:Ngrav}
\end{equation}
where $G$ is the gravitational constant, $M$ the `active' 
mass of the `gravitational source' and $r$ is an adapted 
coordinate representing the 
distance of the observer $\gamma'$ from the 
`source' - as viewed by the observer $\gamma$. 
Note that $r$ need not be the respective Schwarzschild 
coordinate. We do not know whether $\gamma$ should be 
taken to be \emph{static}, i.e. its tangent vector is 
parallel to 
$(\partial/\partial t)_{\gamma}$, or unaccelerated and `moving 
around the source' for a derivation of the Newtonian limit. 
We say the Newtonian limit should ``partially 
reproduce'' the above Newtonian force, because formally the 
(exterior) Schwarzschild spacetime from 
\thref{Ex:exschwspti}, as a solution of 
the vacuum equation $\Ric =0$, depends on the parameter $R > 
0$, not on the factor $G \,M$. If one assumes that 
$R$ ought to depend on $M$, then an argument 
of physical dimensions implies that 
\begin{equation*}
	R \propto \frac{G\, M}{c^2} \, ,
\end{equation*}
i.e. they need to be proportional, if only the physical 
constants $G$ and $c$ are allowed. Indeed, the weak field 
approximation claims that $R= 2 GM /c^2$ 
(see e.g. \cite{Wald}*{p. 124}). 
\begin{Remark}[On the Einstein equation and Newton's law of gravity]
	\label{Rem:Einst}
	As stated before, the Einstein equation is a generalization 
	of the Gau\ss' law for gravity, where the relation is given 
	by the weak field approximation. However, the Gau\ss' law 
	itself is an abstraction from Newton's law 
	\eqref{eq:Ngrav} of gravity to the continuous case. 
	A derivation can be found, for instance, in the 
	book by Bradbury \cite{Bradbury}*{\S 5.4}. So if 
	the Newtonian limit in case \ref{itm:nlim2} exists and 
	it is also possible in case \ref{itm:nlim3} 
	to derive Newton's law (in terms of $R$)
	in an approximation where $\dot \tau \approx 1$, 
	the vacuum Einstein equation $\Ric= 0$ alone would reproduce 
	the bulk of Newtonian gravitational theory. 
	\par 
	This raises the important question whether the equation 
	$\Ric =0$ is enough to explain the empirical data. In fact, 
	Einstein himself, together with Infeld and Hoffmann,
	raised this question in a 1937 article 
	\cite{EinsteinA1}: 
	\begin{quotation}
	\noindent
	$[\dots]$ energy-momentum tensors, however, must be 
	regarded as purely temporary and more or less 
	phenomenological devices for representing 
	the structure of matter, and their entry into the 
	equations makes it impossible to determine how far the 
	results obtained are independent of the particular
	assumption made concerning the constitution of matter. 
	\par
	Actually, the only equations of gravitation which 
	follow without ambiguity from the fundamental 
	assumptions of the general theory of relativity are the
	equations for empty space, and it is important 
	to know whether they \emph{alone} are
	capable of determining the motion of bodies.
	\end{quotation}
	We leave it to the reader to judge whether their 
	argument in favor of the hypothesis is convincing and 
	proceed with our own discussion. 
	\par
%	Within the Newtonian ontology, the `active' mass $M$, 
%	which we identify as the inertial mass of the gravitational 
%	source, has to be proportional to $R'$ in the force law 
%	\begin{equation*}
%		\vec F = - m \, \frac{R'}{r^2} \, \partd{}{r} \, , 
%	\end{equation*}
%	due to Newton's 
%	third law: To every force $\vec F$ (at a point) 
%	there is associated a reactive force $- \vec F$. 
%	So in Newtonian theory, the situation that the mass 
%	$M$ attracts $m$ via the gravitational force $\vec F$ 
%	is equivalent to $m$ attracting $M$ via $- \vec F$ 
%	parallel translated to the spatial point where $M$ is 
%	located (at the same time). This implies that 
%	in Newtonian gravitational theory `active mass' and 
%	`passive mass' constitute the same concept.%
%	\footnote{The parallel translation of the force vector 
%	here is the mathematical implementation of so called 
%	`action at a distance', which is generally viewed as an 
%	undesirable property of Newtonian graviational theory.} 
%	Within general relativity, however, no analogue of 
%	Newton's third law exists and 
	%	\begin{equation*}
%		\vec F = - m \, \frac{R'}{r^2} \, \partd{}{r} \, , 
%	\end{equation*}
	We first observe that modeling purely gravitational 
	interactions between two objects 
	indeed only requires solutions to 
	the vacuum equation, not 
	solutions to the full Einstein equation:  
	If the influence of one of the (inertial) 
	masses on the overall spacetime geometry is negligible, 
	the model employing non-accelerating observers in 
	the Schwarzschild spacetime is sufficient. 
	In order to describe the simplest situation, 
	where two objects `interact gravitationally', we require 
	a `non-rotating two black hole solution' of the 
	equation $\Ric = 0$. To our knowledge, no such explicit 
	solution has been found so far, but, formally, this is 
	where the full law \eqref{eq:Ngrav} needs to be 
	derivable in some `Newtonian limit'. 
	So the appropriate identification of the integration 
	constants 
	with physical parameters 
	(e.g. expressing $R$ in terms of $M$ in the 
	Schwarzschild model) 
	by `gluing' the $(\Ric \neq 0)$-solutions to the 
	respective $(\Ric = 0)$-solutions
	is the only point, 
	where an argument employing the full Einstein 
	equation would be needed to fully reproduce Newtonian 
	gravitational theory. For an example of such a `gluing', 
	we refer to the book by Wald \cite{Wald}*{\S 6.2}. 
	\par
	Of course, this discussion underlies the implicit 
	assumption that \eqref{eq:Ngrav} actually describes 
	the empirical data, where the masses $M$, $m$ and 
	distance $r$ are obtained by an independent procedure 
	and not simply matched to fit the law. Due to a lack of 
	knowledge on the subject, we cannot 
	make definitive statements on 
	this. According to a review article by Gillies 
	\cite{Gillies} on the measurement of the gravitational 
	constant $G$, Newton's law does seem to be a good 
	approximation in a variety of instances. Yet, 
	due to mutually contradicting values of $G$ appearing in 
	the literature, the experimental issue is not entirely 
	settled. 
\end{Remark}
\par
In the next section we show that the Newtonian limit for 
\ref{itm:nlim1} indeed exists. For the cases \ref{itm:nlim2} 
and \ref{itm:nlim3} we have not obtained a proof so far 
and expect additional approximations to be necessary. These
additional approximations would give further qualitative 
and quantitative constraints on the validity of the 
Newtonian theory. 
\par
We conclude this section with the remark that, if one would 
like to go beyond Newtonian gravitational 
theory (e.g. introduce electromagnetic fields), one first 
needs to postulate the corresponding 
(invariant) force $F'$ acting on the second observer 
$\gamma'$ and then show that it reduces to the correct 
Newtonian force $\vec F$ under the Newtonian limit. We 
recommend to do this first for inertial frames of 
reference in Minkowski spacetime and then consider 
more complicated situations, if necessary. 

\section{Newtonian Limit in Special Relativity} 
\label{sec:Nlimitsr}

Before we proceed with our proof of the existence 
of the Newtonian limit in special relativity, 
we state two reasons why the common derivation of the 
Newtonian limit in special relativity is naive: 
\begin{enumerate}[i)]
	\item 	It philosophically assumes the standard observer 
			$\gamma$, as defined by equation 
			\eqref{eq:stndrtgam} on 
			page \pageref{eq:stndrtgam}, together with the 
			standard inertial frame of reference field for 
			$\gamma$, as given by \eqref{eq:stdrdX} 
			on page \pageref{eq:stdrdX}, but these mathematical 
			objects do not appear explicitly in the definition 
			of the Newtonian limit. 
	\item	For the definition of the velocity of the `observed 
			observer' $\gamma'$, it employs the 
			Einstein-synchronized time $t$ 
			(see \thref{Ex:SR}\ref{itm:SR1}) with respect
			to $\gamma$, but this is not the time referred to in 
			Newtonian physics. The Einstein-synchronized time 
			may be understood to take account of the `finiteness 
			of the speed of light', whereas the time in 
			Newtonian physics coincides with the time 
			$\tau$ as measured by $\gamma$ at each point in 
			space. Thus the Newtonian limit needs to 
			employ $\tau$ rather than $t$. 
\end{enumerate}
Both points ought to be remedied by the approach to the 
Newtonian limit discussed in the preceding section. Since 
we are interested in the Newtonian limit for the special 
theory of relativity, the spacetime we need to consider is 
Minkowski spacetime from \thref{Ex:Mspti}. As the ontology 
of special relativity requires `inertial observers', 
we conclude that we have to 
mathematically consider an arbitrary inertial frame of 
reference field $X$ for an arbitrary (non-accelerating) 
observer $\gamma$. Historically, 
the importance of inertial frames of reference for the 
laws of Newtonian mechanics has first been deduced 
by Ludwig Lange in 1885. By successfully ridding the theory 
of the notion of `absolute space' 
(cf. \cite{Jammer}*{p. 140 sq.}), Lange paved the way for 
the development of the special theory of relativity. 
\par
As argued in \thref{Ex:obsMink}\ref{itm:obsMink1}, 
it is sufficient to consider the standard frame of reference 
field $X:= (\partial)_\gamma$ for the standard observer 
$\gamma$. This is the mathematical justification of the 
physical statement that `all inertial observers are 
mechanically equivalent'. From \thref{Ex:SR}\ref{itm:SR1} on 
page \pageref{Ex:SR}, we recall that the kinematic observer 
mapping $\varphi$ with respect to $X$ is given by 
	\begin{equation*}
		\varphi \left(c \tau, \vec{x}\right) = 
				\begin{pmatrix}
					c \tau - \abs{\vec x} \\ 
					\vec x
				\end{pmatrix} 
	\end{equation*}
for $x = \left( c \tau, \vec x\right) \in 
\R^4 \setminus \left(\R \cross 
\lbrace 0 \rbrace \right)$ 
and is a diffeomorphism onto its image. Thus for the 
Newtonian limit we do not need to restrict ourselves to 
a particular subset of the observer spacetime, contrary to what 
one might need to do in different instances. We may 
therefore consider $\varphi$ (restricted to its image) as a coordinate
transformation from observer coordinates $x$ to standard 
coordinates $y$ on $\R^4$. Its Jacobian is given by the 
expressions 
	\begin{align*}
		\partd{y^0}{x^0} \left( x\right) 
		&= 1 \quad , & 
		\partd{y^0}{x^a} \left( x\right) 
		&= - \frac{\delta_{ab} x^b}
		{\abs{\vec x}}  \, , \\
		\partd{y^a}{x^0} \left( x\right) &= 0 \quad , &
		\partd{y^a}{x^b} \left( x\right)
		&= \delta^a_b \, ,
	\end{align*}
where $a, b \in \lbrace 1, 2, 3 \rbrace$. Writing now 
$s \to x \left( s \right)$ for the second observer 
$\gamma'$ viewed as a curve in the domain 
$\R^4 \setminus \left(\lbrace 0 \rbrace \cross \R^3 \right)$ 
of $\varphi$, we may compute the components of $\alpha := 
\left(\varphi^*g\right)_{x}$: 
\begin{align*}
	\alpha_{00} &= 	\partd{y^i}{x^0} \, \eta_{ij} \,   
					\partd{y^j}{x^0} = 1	\\
	\alpha_{0a} &= 	\partd{y^i}{x^0} \, \eta_{ij} \, 
					\partd{y^j}{x^a} = 
		- \delta_{ab} \hat{x}^b	\\
	\alpha_{ab} &= 	\partd{y^i}{x^a} \, \eta_{ij} \, 
					\partd{y^j}{x^b} 
				=  \delta_{ac} {\hat{x}}^c \, \delta_{bd} 
				{\hat{x}}^d - \delta_{ab} \, . 
\end{align*}
Here we defined $\hat x := \vec x / \abs{\vec x}$ with
respective components $\hat x^a$ for convenience. 
As in \thref{Ex:limdottau}, all $\alpha_{ij}$ are 
independent of $c$, $\alpha_{00}=1$ and the $\alpha_{0a}$ 
do not vanish. Hence the Newtonian limit needs to be 
attained in the limit $c \to \infty$. \par
From our general expression \eqref{eq:dottaugen} of
$\dot \tau$ we find 
\begin{equation*}
	\dot \tau = \frac{1}{\sqrt{1
	-  \frac{2 \delta_{ab} \, \hat x^a v^b}{c} 
	+ \frac{\left( \delta_{ab} \, \hat x^a v^b\right)^2}{c^2}
	- \frac{\delta_{ab}\, v^a v^b}{c^2} 
	}} \, .
\end{equation*}
To simplify this we may write 
$\vec u \cdot \vec w := \delta_{ab} \, u^a w^b$ for 
any $\vec u, \vec w \in \R^3$, $\vec v := v^a \, \baseR_a$, 
$v :=\abs{\vec v}$ and $\hat v := \vec v / v$: 
\begin{equation}
	\dot \tau = \frac{1}{\sqrt{1-2 \left( \hat x \cdot \hat v 
	\right) \, 
	\frac{v}{c} + \left( \left( \hat x \cdot \hat v\right)^2 
	-1 \right) 
	\left( \frac{v}{c} \right)^2
	}} \, .  
	\label{eq:SRdottau}
\end{equation}
\par
It is worthwhile to look 
at the admissible values of $v$. If the velocity $\vec v$ 
of the second observer is orthogonal to its location 
vector $\vec x$, we must have $v < c$. 
Yet if $\vec v$ is pointing away from the observer, i.e.  
$\hat x \cdot \hat v = 1$, $v$ is less than $c / 2$ and, 
if $\vec v$ points towards the observer, $v$ may assume any 
positive value. Though this seems problematic at first sight, 
it is in fact a reasonable prediction of special relativity, 
if one measures the velocity $\vec v$ with respect to 
$\gamma$'s time $\tau$ instead of the synchronized time $t$. 
\par
To obtain the Newtonian limit together with the 
next two relativistic correction terms, we expand 
expression \eqref{eq:SRdottau} in terms of $1/c$ up to second 
order: 
\begin{equation}
	\dot \tau = 1 +  \left( \hat x \cdot \hat v\right) 
	\, \frac{v}{c} + 
	\left(  \left( \hat x \cdot \hat v \right)^2 
	+ \frac{1}{2} \right)\, \left( \frac{v}{c}\right)^2 
	+ \bigo \left(1/c^3 \right)
	\, .
	\label{eq:appdtau}
\end{equation}
\par
As a `real-world comparison', assume a jet fighter reaches a 
speed $v$ of $7000$ kilometers per hour relative to the 
observer $\gamma$ and recall that the speed 
of light $c$ is approximately $300.000$ kilometers per second. 
Then, if we neglect the influence of the upward 
acceleration on the clocks (`constant gravity'), 
equation \eqref{eq:appdtau} states that the first 
order relativistic correction of their respective clock rates 
$\dot \tau$ is at most $6.5 \cdot 10^{-6}$ in absolute value. 
\par
Let us continue with the assumption that the second observer 
$\gamma'$ moves under the influence of a force $F'$, 
which is independent of the speed of light $c$ and small in the
sense that the Newtonian limit, if it exists, remains 
appoximately valid. In 
coordinates $y$ all Christoffel symbols 
$\Gamma^k{}_{ij}$ vanish and thus, 
according to \eqref{eq:UpsGam} on page \pageref{eq:UpsGam}, 
\begin{equation*}
	\Upsilon^c{}_{ij} = \partd{x^c}{y^k}
	\frac{\partial^2 y^k}{\partial x^i \partial x^j} 
	= \frac{\partial^2 y^c}{\partial x^i \partial x^j} \equiv 0
	\, .
\end{equation*}
By our relative force equation \eqref{eq:NewtF}, we 
therefore have 
\begin{equation}
	m \frac{\d \vec v}{\d \tau} = 
	\frac{1}{ \dot \tau^2} \,  \vec{F}' 
	- m \frac{ \ddot \tau}{\dot \tau^2} \, \vec v  \, , 
	\label{eq:SRForce}
\end{equation}
where $\vec{F}'$ is the spatial part of $F'$. 
Hence there does appear a pseudo-force here, despite the fact 
that the frame of reference is inertial. 
If $\dot \tau \approx 1$, then 
$\ddot \tau \approx 0$ and thus we obtain Newton's second law:  
\begin{equation*}
	m \frac{\d \vec v}{\d \tau} \approx {\vec F'} \quad \, .
\end{equation*}
We conclude that the Newtonian limit of the 
special theory of relativity indeed exists and that 
$\vec F = \vec F'$ in this limit. 
\par
In the remainder of this section, we calculate the first two
relativistic correction terms of the force equation 
\eqref{eq:SRForce} by using the approximation 
\eqref{eq:appdtau} from above. Consequently, we 
require expansions of $1/ {\dot \tau}^2$ and of
\begin{equation}
	\frac{\ddot \tau}{{\dot \tau}^2}
	= \frac{\dot \tau \, \frac{\d }{\d \tau} \dot \tau}
	{{\dot \tau}^2}
	= \frac{\frac{\d \dot \tau}{\d \tau}}{\dot \tau}
	\label{eq:ddottau}
\end{equation}
up to second order in $1/c$. For the 
derivative $\d \dot \tau /\d \tau$ we find 
\begin{equation*} 
	\left( \frac{\d \hat x }{\d \tau} \cdot \vec v 
	+ \hat x \cdot \frac{\d \vec v}{\d \tau} \right)
	\, \frac{1}{c} 
	+ \left( \vec v \cdot \frac{\d \vec v}{\d \tau} 
	+ 2 \left( \hat x \cdot \vec v \right) 
	\left( \frac{\d \hat x }{\d \tau} \cdot \vec v 
	+ \hat x \cdot \frac{\d \vec v}{\d \tau} \right)
	\right)
	\, \frac{1}{c^2} 
	+ \bigo \left( 1/c^3 \right)
	\, , 
\end{equation*}
and so we compute 
\begin{equation}
	\frac{\d \hat x}{\d \tau} = \frac{1}{\abs{\vec{x}}} 
	\left( \vec v - \left( \hat x \cdot \vec v\right) 
	\hat x\right)
	\, . 
\end{equation}
Combining them yields 
\begin{multline}
	\frac{\d \dot \tau}{ \d \tau} = \left( 
	\hat x \cdot \frac{\d \vec v}{\d \tau} 
	+ \frac{v^2}{\abs{\vec{x}}} \left( 1 - 
	\left( \hat x \cdot \hat v \right)^2 \right)
	\right) 
	\, \frac{1}{c} \\
	+ \left( \vec v \cdot \frac{\d \vec v}{\d \tau} 
	+ 2 \left( \hat x \cdot \hat v \right) 
	\left( \frac{v^2}{\abs{\vec{x}}} \left( 1 - 
	\left( \hat x \cdot \hat v \right)^2 \right) 
	+\hat x \cdot \frac{\d \vec v}{\d \tau} \right)
	\right)
	\, \frac{1}{c^2} 
	+ \bigo \left( 1/c^3 \right)
	\, . 
	\label{eq:expddottau}
\end{multline}
Now, if $f$ and $g$ are (real) polynomial expansions in a 
real perturbation parameter $\varepsilon$ around $0$, we 
may write 
\begin{equation*}
	f \left( \varepsilon \right) 
	= f_0 + f_1 \, \varepsilon + f_2 \, \varepsilon^2
	 + \bigo \left( \varepsilon^3 \right)
	\, , \quad 
	g \left( \varepsilon \right) 
	= g_0 + g_1 \, \varepsilon + g_2 \, \varepsilon^2
	+ \bigo \left( \varepsilon^3 \right) \, . 
\end{equation*}
Their product $fg$ is given by
\begin{equation}
	f \left( \varepsilon \right) g \left( \varepsilon \right) 
	= f_0 g_0 + \left( f_0 g_1+ f_1 g_0 \right) \varepsilon
	+ \left(f_0 g_2 + f_1 g_1 + f_2 g_0 \right) \varepsilon^2
	+ \bigo \left( \varepsilon^3 \right) \, ,
	\label{eq:epsprod}
\end{equation}
and so the algebraic inverse $g=1/f$ is obtained 
by demanding $g f = 1$, i.e. 
\begin{equation}
	\frac{1}{f \left( \varepsilon\right)} 
	= \frac{1}{f_0}
	- \frac{f_1}{f_0}\, \varepsilon 
	+ \left( - 
	\frac{f_2}{(f_0)^2} + \frac{(f_1)^2}{(f_0)^3}
	\right) \, \varepsilon^2 
	+ \bigo \left( \varepsilon^3 \right)
	\, ,
	\label{eq:epsinv}
\end{equation}
provided it exists. See e.g. \cite{Roessel}*{Thm. 1.6} 
for general formulas. From 
\eqref{eq:ddottau} we find, that we first need to invert 
$\dot \tau$ via \eqref{eq:epsinv} (to first order) and then 
multiply by the 
expansion \eqref{eq:expddottau} of $\d \dot \tau / \d \tau$ 
via the rule \eqref{eq:epsprod}. After some labor we obtain 
\begin{multline}
	\frac{\ddot \tau}{{\dot \tau}^2} 
	= 
	\left( 
	\hat x \cdot \frac{\d \vec v}{\d \tau} 
	+ \frac{v^2}{\abs{\vec{x}}} \left( 1 - 
	\left( \hat x \cdot \hat v \right)^2 \right)
	\right) 
	\, \frac{1}{c}
	\\
	+ \left( 
	\vec v \cdot \frac{\d \vec v}{\d \tau} 
	+ \left( \hat x \cdot \hat v \right)
	\left( 
	\hat x \cdot \frac{\d \vec v}{\d \tau} 
	+ \frac{v^2}{\abs{\vec{x}}} \left( 1 - 
	\left( \hat x \cdot \hat v \right)^2 \right)
	\right) 
	\right) 
	\, \frac{1}{c^2}
	+ \bigo \left(\frac{1}{c^3} \right)
	\, . 
	\label{eq:f2}
\end{multline}
Similarly, to obtain an expansion of $1/ \dot \tau^2$, 
we first compute $\dot \tau^2$
by plugging the expansion \eqref{eq:appdtau} for $\dot \tau$
into the multiplication rule \eqref{eq:epsprod}, and then 
invert it via \eqref{eq:epsinv}:
\begin{equation*}
	\frac{1}{\dot \tau^2} 
	= 1 - 2 \left( \hat x \cdot \hat v \right) \, 
	\frac{v}{c} + 
	\left( \left( \hat x \cdot \hat v \right)^2 -1 
	\right) \, \left( \frac{v}{c}\right)^2 
	+ \bigo \left( \frac{1}{c^3} \right) \, . 
\end{equation*}
Finally both expressions need to be put into the relative force law 
\eqref{eq:SRForce} from above. It should be noted, that this 
force law is more adequate for comparing the predictions 
of Newtonian mechanics with those of special relativity than for 
calculating trajectories. In particular, \eqref{eq:f2} shows, that 
for very small $\abs{\vec x}$ relative to the chosen length scale,
the approximation in $1/c$ may break down.

\addchap{References}

\addchap{Acknowledgements}

Without my parent's continuous support and faith in me, 
I would not stand where I stand today. 
As many other people in their age and many other people before 
them, they have spent their lives working in the hope to 
provide a better future for their children - despite a countless 
number of obstacles along the way. Thank you so much, Ina and Heiko! 
I shall pass on 
the love you gave me to my children and the people around me, 
in the hope that they also pass it on and mankind has a future. 
\par
Of course, they are not the only ones to credit: I would also like 
to thank my grandparents, family Pankratz, my brother and his family, 
my friends, and all the people who have given me insight 
into themselves and the world throughout the course of my life. 
Please forgive me for not mentioning each one of you, but I am 
thankful for every one of you and that you are so numerous. 
\par 
I also thank Heike Makk and Elisabeth T\"urk for their 
support during rough times. Moreover, thanks goes to Sina Bergmann for 
her help in getting a foot into this city. It was certainly not easy. 
\par
In addition, I would like to mention and thank some of 
the people in academia, who also have their fair share in making this 
work possible. In this spirit, I would like to express my gratitude 
to Dr. Wolfgang Hasse for taking his time to talk and discuss with me, as 
well as for providing me with critical feedback and helpful references, 
despite his lack of time. Without him this thesis would 
most likely have a different topic, be of less quality, 
and I would still believe that
the affine parameter distance and the parallax distance are the same. 
My gratitude also goes to Prof. Yuri B. Suris and Prof. Horst-Heino von Borzeszkowski for giving me the opportunity 
to write this thesis. Wading through and evaluating it 
has certainly been a lot of work. Of course, I also thank them for 
the good grade. ;-) 
Dr. Gil Cavalcanti, Prof. Marius 
Crainic, Prof. Eric van den Ban, Prof. Dorothee Sch\"uth and also 
Prof. Suris deserve credit for their ability and patience in teaching me 
differential geometry - it was definitely more physical interest than 
mathematical ability that got me into this subject. On the physics  
side of academia, I thank Dr. Thoralf Chrobok, Gerold O. Schellstede, 
and again 
Prof. Borzeszkowski for teaching me general relativity, especially 
their openness towards other (not always correct) points of view 
voiced by myself. Ultimately, our own mistakes are our greatest teachers. 
I also thank Prof. Dennis Dieks, Prof. Gleb Arutyunov and Prof. Crainic
for listening to me at Utrecht University when noone else would. 
Furthermore, 
I am indebted to my teachers at the University of Leipzig, who showed 
me the path into academia. Among them are Prof. Thomas Kuhn, Dr. 
Arwed Schiller, Prof. Gerd Rudolph, Prof. Michael Ziese and 
Dr. Roland Kirschner. 
\par
Moreover, we often forget to mention those that shaped our 
paths from very early on and whose influence has been the 
greatest: When I was a child in middle school, my grandfather 
taught me how to balance equations -- I just could not wrap my head 
around it. That I still remember this so well is certainly due to the 
fact, that it was one of my first steps towards a career in 
mathematics. This should serve as a reminder that this path 
can be blocked from very early on by unfortunate circumstances 
and that this path should not just be open to the brightest among 
us. Later in high school I benefited from such great teachers as Mrs. 
Makk, Mrs. Barner, Mr. Schn\"urpel, Mr. Dalsten, Mr. 
R\"osner and Mr. Klotz, who, among many others, also 
contributed to this path. 
\par
Last but not least, I am grateful to Dr. Hasse, Dr. Chrobok, 
Christof Tinnes and Benedict Wenzel for reading (parts of) the 
manuscript. 
\par
Finally, I wish to note that this thesis was written 
entirely with open source software running on Ubuntu. 
Figures were drawn with `gimp 2' and the text was written in 
\LaTeX \, using `gummi git', `TeX Live' and `pdflatex'. The Taylor 
expansions in chapter \ref{chap:nlimit} were checked with Wolfram 
Alpha.         

\end{document}